\documentclass[twocolumn,aps,prl,superscriptaddress,tightenlines]{revtex4-2}
\usepackage{graphicx}
\usepackage{dcolumn}
\usepackage{bm}
\usepackage{epsfig}
\usepackage{amsmath}
\usepackage{amsfonts}
\usepackage{amssymb}
\usepackage{empheq}
\usepackage{color}
\usepackage{bbm}
\usepackage[caption=false]{subfig}
\usepackage{placeins}
\usepackage{soul}
\usepackage{physics}

\usepackage{setspace}
\usepackage{mathtools}
\usepackage{cases}
\usepackage{esint}
\usepackage{extarrows}
\usepackage{slashed}
\usepackage{pifont}
\renewcommand{\arraystretch}{1}

\definecolor{mybo}{cmyk}{0,0,0.10,0}

\usepackage{mathrsfs}
\usepackage{changes} 
\definechangesauthor[name={Per cusse}, color=red]{per} 
\usepackage[colorlinks,citecolor=blue,linkcolor=blue]{hyperref}
\usepackage{orcidlink}
\newcommand{\customsqrt}[1]{\sqrt{\smash[b]{#1}}}
\makeatletter
\newcommand*\bigcdot{\mathpalette\bigcdot@{.8}}
\newcommand*\bigcdot@[2]{\mathbin{\vcenter{\hbox{\scalebox{#2}{$\m@th#1\bullet$}}}}}
\makeatother

\begin{document}

\title{Topological Quantum Batteries}
\author{Zhi-Guang Lu~\orcidlink{0009-0007-4729-691X}}
\affiliation{School of Physics and Institute for Quantum Science and Engineering, Huazhong University of Science and Technology and Wuhan Institute of Quantum Technology, Wuhan 430074, China}
\author{Guoqing Tian}
\affiliation{School of Physics and Institute for Quantum Science and Engineering, Huazhong University of Science and Technology and Wuhan Institute of Quantum Technology, Wuhan 430074, China}
\author{Xin-You Lü}
 \email{xinyoulu@hust.edu.cn}
	\affiliation{School of Physics and Institute for Quantum Science and Engineering, Huazhong University of Science and Technology and Wuhan Institute of Quantum Technology, Wuhan 430074, China}
\author{Cheng Shang~\orcidlink{0000-0001-8393-2329}}
 \email{c-shang@iis.u-tokyo.ac.jp}
\affiliation{\mbox{Department of Physics, The University of Tokyo, 5-1-5 Kashiwanoha, Kashiwa, Chiba 277-8574, Japan}}
 \affiliation{Analytical quantum complexity RIKEN Hakubi Research Team, RIKEN Center for Quantum Computing (RQC), 2-1 Hirosawa, Wako, Saitama 351-0198, Japan}

\date{\today}

\begin{abstract}
We propose an innovative design for quantum batteries (QBs) that involves coupling two-level systems to a topological photonic waveguide.~Employing the resolvent method, we analytically explore the thermodynamic performance of QBs.~First, we demonstrate that in the long-time limit, only bound states significantly contribute to the stored energy of QBs.~We observe that near-perfect energy transfer can occur in the topologically nontrivial phase.~Moreover, the maximum stored energy exhibits singular behavior at the phase boundaries, where the number of bound states undergoes a transition. Second, when a quantum charger and a quantum battery are coupled at the same sublattice within a unit cell, the ergotropy becomes immune to dissipation at that location, facilitated by a dark state and a topologically robust dressed bound state. Third, we show that as dissipation intensifies along with the emergence of the quantum Zeno effect, the charging power of QBs experiences a temporary boost. Our findings offer valuable guidance for improving quantum battery performance in realistic conditions through structured reservoir engineering.
\end{abstract}

\maketitle

\emph{Introduction}.---With the decline of fossil fuels and the worsening of the global energy crisis, conventional chemical batteries that charge and discharge through chemical reactions will gradually be phased out.~Instead, driven by the potential power of quantum effects and the demands for nanotechnological miniaturization, the size of energy storage and conversion devices has shrunk to atomic scales. With this background, Alicki and Fannes first proposed the concept of quantum battery (QB) in 2013~\cite{PhysRevE.87.042123}. Fundamentally distinct from conventional batteries, quantum batteries (QBs) exploit unique quantum features for energy storage and release, potentially outperforming classical counterparts with enhanced charging power~\cite{PhysRevLett.118.150601,PhysRevLett.127.100601,PhysRevLett.131.240401,PhysRevA.108.052213,PhysRevLett.125.236402,gyhm2023beneficial,PhysRevLett.128.140501,rosa2020ultra,PhysRevA.108.042618,PhysRevA.107.032218,PhysRevA.109.042207,PhysRevE.99.052106}, increased capacity~\cite{PhysRevLett.131.030402,PhysRevResearch.2.023113,PhysRevResearch.4.043150,PhysRevA.109.042424,tirone2024quantum}, and superior work extraction~\cite{PhysRevLett.129.130602,PhysRevLett.122.047702,PhysRevE.109.044119,PhysRevLett.124.130601,PhysRevLett.131.060402,tirone2023quantum}. Since then, a variety of possible QBs have been constructed, including Dicke type, spin chain type, central-spin type, etc~\cite{PhysRevLett.120.117702,PhysRevB.100.115142,PhysRevA.103.052220,PhysRevE.94.052122,PhysRevA.109.022210,PhysRevB.102.245407,PhysRevLett.132.210402,PhysRevA.105.L010201,Shaghaghi_2022,batteries9040197,andolina2024dicke,PhysRevA.106.022618,PhysRevB.104.245418,PRXQuantum.5.030319}. In particular, a minimal yet favorite QB model based on a two-level system has been extensively studied both in theory~\cite{binder2015quantacell,PhysRevLett.125.040601,PhysRevE.102.052109,PhysRevA.108.062402,PhysRevA.109.022226,PhysRevE.103.042118} and experimental implementation~\cite{bruzewicz2019trapped,forn2017ultrastrong,baumann2010dicke,devoret2013superconducting}.

From an engineering perspective, QBs offer a practical way to incorporate quantum effects into thermodynamics~\cite{RevModPhys.96.031001,binderthermodynamics2018, southwell2008quantum,PhysRevResearch.4.L022017,saha2023harnessing}. Extensive studies have focused on the performance of QBs in terms of their charging power and stored energy~\cite{PhysRevA.109.012204}. Notably, the concept of ergotropy—another crucial performance indicator for QBs that describes the maximum extractable energy—was introduced by Allahverdyan, Balian, and Nieuwenhuizen~\cite{allahverdyan2004maximal}.~Recent research indicates that coupling a quantum charger and a QB to a reservoir, such as a rectangular hollow metal waveguide, facilitates efficient remote charging of the QB but inevitably results in low stored energy and diminished ergotropy~\cite{PhysRevLett.132.090401,PhysRevLett.127.083602,song2017dissipation}. A related challenge is whether a structured reservoir exists that can effectively enhance the stored energy and the ergotropy of QBs. Inspired by the advantages of topological baths~\cite{PhysRevLett.61.2015, RevModPhys.90.015001, PhysRevApplied.15.044041,bello2019unconventional,PhysRevLett.131.073602,tian2024powerlawexponentialinteractioninducedquantum} and building on recent experimental advancements~\cite{lu2014topological,RevModPhys.91.015006,PhysRevX.6.041043,10.1093/acprof:oso/9780198509141.001.0001,RevModPhys.87.347,RevModPhys.90.031002,doi:10.1126/science.aaq0327,PhysRevLett.64.2418,liu2017quantum,PhysRevX.11.011015,PhysRevX.9.011021,lodahl2017chiral,mirhosseini2018superconducting,PhysRevX.11.041043,gu2017microwave}, we have proposed a scheme in this work that enhances the stored energy and the ergotropy by coupling two-level systems (TLSs) with topological photonic waveguide to fully overcome this challenge. Towards implementing QB in practical applications, another natural obstacle is environment-induced decoherence caused by inevitable dissipation, which in general, decreases the performance of the QB, such as the energy loss and aging of QB~\cite{PhysRevApplied.20.044073,PhysRevA.100.043833,PhysRevE.104.054117,PhysRevE.104.064143,PhysRevApplied.21.044048,carrega2020dissipative}.\!\!\! Recently, the study of QB dynamics in the presence of an environment has attracted a deal of attention, and several schemes have been proposed to mitigate the effects of decoherence, including feedback control~\cite{PhysRevE.106.014138}, exploiting non-Markovian effects~\cite{PhysRevLett.122.210601,kamin2020non}, Floquet engineering~\cite{PhysRevA.102.060201}, frequency modulation~\cite{hadipour2025amplifiedquantumbatterydynamical}, etc. However, does a configuration exist that can completely isolate the QB from the effects of dissipation? Our work offers a substantial answer. We have discovered that directly coupling the quantum charger and QB enables the performance of the QB to resist decoherence, stemming from the presence of a dark state and vacancy-like dressed bound state.

In this letter, by leveraging topological properties, we develop a novel design named topological quantum batteries, which consist of TLSs coupled to a topological photonic waveguide. In this setting, we simultaneously address two major challenges related to QBs. One involves achieving long-range perfect charging for QBs, and the other focuses on dissipation-immune engineering. Additionally, we demonstrate that the charging power of QBs can be temporarily enhanced as dissipation increases with the emergence of the quantum Zeno effect.

\begin{figure}
  \centering
  \includegraphics[width=\linewidth]{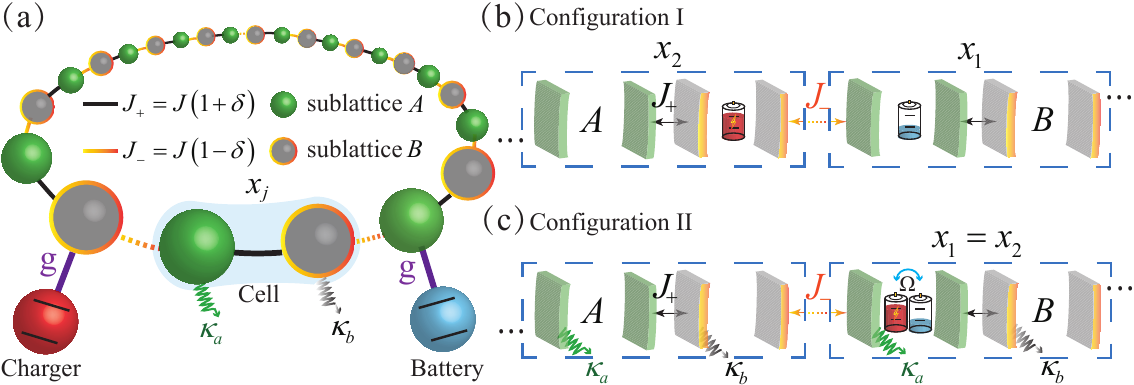}
  \caption{Schematic of the topological quantum battery setup. (a) General setup: The system consists of a quantum charger and a QB, each modeled as a two-level system, connected to a 1D periodic SSH photonic lattice. This setup is categorized into two main configurations: (b) Configuration I, where the charger and QB without direct interaction are coupled to sublattice $B$ of unit cell $x_2$ and sublattice $A$ of unit cell $x_1$, respectively; and (c) Configuration II, where the charger and QB with direct interaction are coupled to the same sublattice $A$ of unit cell $x_1=x_2$. Here, configuration I features long-range perfect charging in the non-dissipative topological waveguide, while configuration II supports short-range perfect charging in the dissipative topological waveguide.}
  \label{Fig-setup}
\end{figure}

\emph{Model and Hamiltonian}.---As illustrated in Fig.\,\ref{Fig-setup}(a), the system consists of a quantum charger and a QB, each modeled as a TLS, with the raising (lowering) operator defined as $\hat{\sigma}_{+}^{\rm C}$ ($\hat{\sigma}_{-}^{\rm C}$) for the charger and $\hat{\sigma}_{+}^{\rm B}$ ($\hat{\sigma}_{-}^{\rm B}$) for the QB. These TLSs are connected to a one-dimensional (1D) periodic Su–Schrieffer–Heeger (SSH)~\cite{SSH,PhysRevB.22.2099} photonic lattice designed with engineered photon loss~\cite{nhbath,PhysRevA.106.053517,PhysRevA.96.043811,PhysRevLett.118.200401}. Under the Markovian~\cite{PhysRevA.40.4077} and rotating-wave approximations~\cite{supp}, the equation of motion for the total system is given by the Lindblad master equation ($\hbar=1$)
\begin{align}\label{II-1}
	\dot{\hat{\rho}}_t=-i[\hat{H}_{\rm sys}+\hat{H}_{\rm ssh}+\hat{H}_{\rm int},\hat{\rho}_t]+\mathcal{L}_a\hat{\rho}_t+\mathcal{L}_b\hat{\rho}_t.
\end{align}
Here the coherent part consists of three terms. The system Hamiltonian is given by
\begin{align}
	\hat{H}_{\rm sys}=\omega_e\hat{\sigma}_+^{\rm B}\hat{\sigma}_{-}^{\rm B}+\omega_e\hat{\sigma}_+^{\rm C}\hat{\sigma}_{-}^{\rm C}+\Omega_{12}^{\alpha\beta}(\hat{\sigma}_+^{\rm B}\hat{\sigma}_-^{\rm C}+{\rm H.c.} ),\label{II-2}
\end{align}
where $\omega_e$ is the atomic transition frequency and $\Omega_{12}^{\alpha\beta}\equiv\Omega\delta_{x_1,x_2}\delta_{\alpha,\beta}$ denotes the direct interaction strength between the charger and QB. The bath Hamiltonian reads
\begin{align}
	\hat{H}_{\rm ssh}\!=\!\sum_{j=1}^N{\omega_c(\hat{n}^{a}_{j}\!+\!\hat{n}^{b}_{j})+(J_+\hat{a}_j^\dagger \hat{b}_j^{}\!+\!J_-\hat{b}_j^\dagger \hat{a}_{j+1}^{}\!+\!{\rm H.c.})},\label{II-3}
\end{align}
where $\hat{a}_j\,(\hat{b}_j)$ is the photon annihilation operator for sublattice $A\,(B)$ at the $j$th cell, and $\hat{n}_j^{a}=\hat{a}_j^\dagger\hat{a}_j^{}$, $\hat{n}_j^{b}=\hat{b}_j^\dagger\hat{b}_j^{}$. Here, $\omega_c$ is the cavity eigenfrequency, and $J_\pm=J(1\pm\delta)$ denotes the intracell and intercell hopping, with $J$ being the average hopping amplitude and $\delta\in[-1,1]$ setting the strength and sign of the dimerization. The system-bath interaction is assumed to be local (on site),
\begin{align}
	\hat{H}_{\rm int}&=\text{g}(\hat{\sigma}_-^{\rm B}\hat{o}_{x_{1,\alpha}}^\dagger+\hat{\sigma}_-^{\rm C}\hat{o}_{x_{2,\beta}}^\dagger+{\rm H.c.}),\label{II-4}
\end{align}
where $\text{g}$ denotes the interaction strength, and $x_{1,\alpha}\,(x_{2,\beta})$ labels the unit cell $x_1\,(x_2)$ and sublattice $\alpha\,(\beta)$, in which the QB (charger) is located, with $\hat{o}_{x_{1(2),A}}=\hat{a}_{x_{1(2)}}$ and $\hat{o}_{x_{1(2),B}}= \hat{b}_{x_{1(2)}}$. The dissipative part consists of photon loss in each cavity, modeled by single-lattice dissipators: $\mathcal{L}_{a(b)} = \kappa_{a(b)}\sum_{j}\mathcal{D}[\hat{a}_j\,(\hat{b}_j)]$, where $\kappa_a$ and $\kappa_b$ control the photon loss rates for sublattices $A$ and $B$, respectively, and $\mathcal{D}[\hat{L}]\hat{\rho} \equiv \hat{L}\hat{\rho}\hat{L}^\dagger - \{ \hat{L}^\dagger\hat{L},\hat{\rho}\}/2$ is the Lindblad superoperator. For configuration I [see Fig.\,\ref{Fig-setup}(b)], the dissipative part is absent, with $\kappa_a=\kappa_b=0$. For configuration II [see Fig.\,\ref{Fig-setup}(c)], dissipation is present and categorized into two types: (i) single-sublattice dissipation, where either $\kappa_a\neq0, \kappa_b=0$ ($\mathbb{Q}=1$) or $\kappa_a=0, \kappa_b\neq0$ ($\mathbb{Q}=-1$); (ii) two-sublattice dissipation, where both $\kappa_a\neq0$ and $\kappa_b\neq0$ $(\mathbb{Q}=0)$, where $\mathbb{Q}\equiv(\delta_{\kappa_a,0}-\delta_{\kappa_b,0})/(\delta_{\kappa_a,0}+\delta_{\kappa_b,0}-2)$. \nocite{qt, PT, pa, LRedge, edge, lu2024, PhysRevLett.115.153901, PhysRevLett.110.243603, PhysRevLett.107.070601, PhysRevB.94.165116}

The effective non-Hermitian Hamiltonian derived from Eq.\,(\ref{II-1}) is expressed as $\hat{H}_{\rm eff}^{} = \hat{H}_{\rm sys}^{} + \hat{H}_{\rm ssh}^{\rm eff} + \hat{H}_{\rm int}^{}$, where $\hat{H}_{\rm ssh}^{\rm eff} = \hat{H}_{\rm ssh} - (i/2)\sum_{j=1}^{N}{(\kappa_a\hat{n}_j^{a} + \kappa_b\hat{n}_j^{b})}$. We assume that the initial state is in the single-excitation sector. Specifically, the charger is fully charged to the excited state, whereas the QB is depleted to the ground state and the bath is in the vacuum state. Then, the solution to Eq.\,(\ref{II-1}) reads $\hat{\rho}_t = e^{-i\hat{H}_{\rm eff}t}\hat{\rho}_0e^{i\hat{H}_{\rm eff}^\dagger t} + p_t| g,g;\text{vac} \rangle \langle g,g;\text{vac}|$, where $p_t = 1 -\text{Tr}[e^{-i\hat{H}_{\rm eff}t}\hat{\rho}_0e^{i\hat{H}_{\rm eff}^\dagger t}]$~\cite{supp}, in which the initial density matrix is written as $\hat{\rho}_0 = |\psi(0)\rangle\langle \psi(0)|$ with $| \psi(0)\rangle  = |e,g;\text{vac}\rangle$. Therefore, by limiting our analysis to the single-excitation sector, we can concentrate on studying the effective non-Hermitian Hamiltonian. Further, by defining $\hat{\textbf{o}}_k = [\hat{a}_k,\hat{b}_k]^T$ with $\hat{a}_k^\dagger = \sum_{j=1}^N e^{ikj}\hat{a}_j^\dagger/\sqrt{N}$ and $\hat{b}_k^\dagger = \sum_{j = 1}^N e^{ikj}\hat{b}_j^\dagger/\sqrt{N}$, where $k = 2\pi n/N$ for $n \in (-N/2, N/2]$ within the lattice of cell size $N$, the effective Hamiltonian of the bath, when moved to the momentum space, is expressed as $\sum_k \hat{\textbf{o}}_k^\dagger \widetilde{\textbf{h}}_k^{}\hat{\textbf{o}}_k^{}$, with~\cite{supp}
 \begin{align} 
\widetilde{\textbf{h}}_k = \Re[f_k]\sigma_x - \Im[f_k]\sigma_y - i\kappa_-\sigma_z + (\omega_c-i\kappa_+)\sigma_0,\label{II-5}
\end{align}
where $f_k = J_+ + J_-e^{-ik}$ is the coupling in momentum space between the $A\,(B)$ modes, $\hat{a}_k\,(\hat{b}_k)$, and $\kappa_\pm = (\kappa_a\pm\kappa_b)/4$. Then, by incorporating Eq.\,(\ref{II-5}) and applying the definitions of $\hat{a}_k$ and $\hat{b}_k$, we can derive the effective non-Hermitian Hamiltonian expressed in momentum space.

\begin{figure*}
	\centering
	\includegraphics[width=\textwidth]{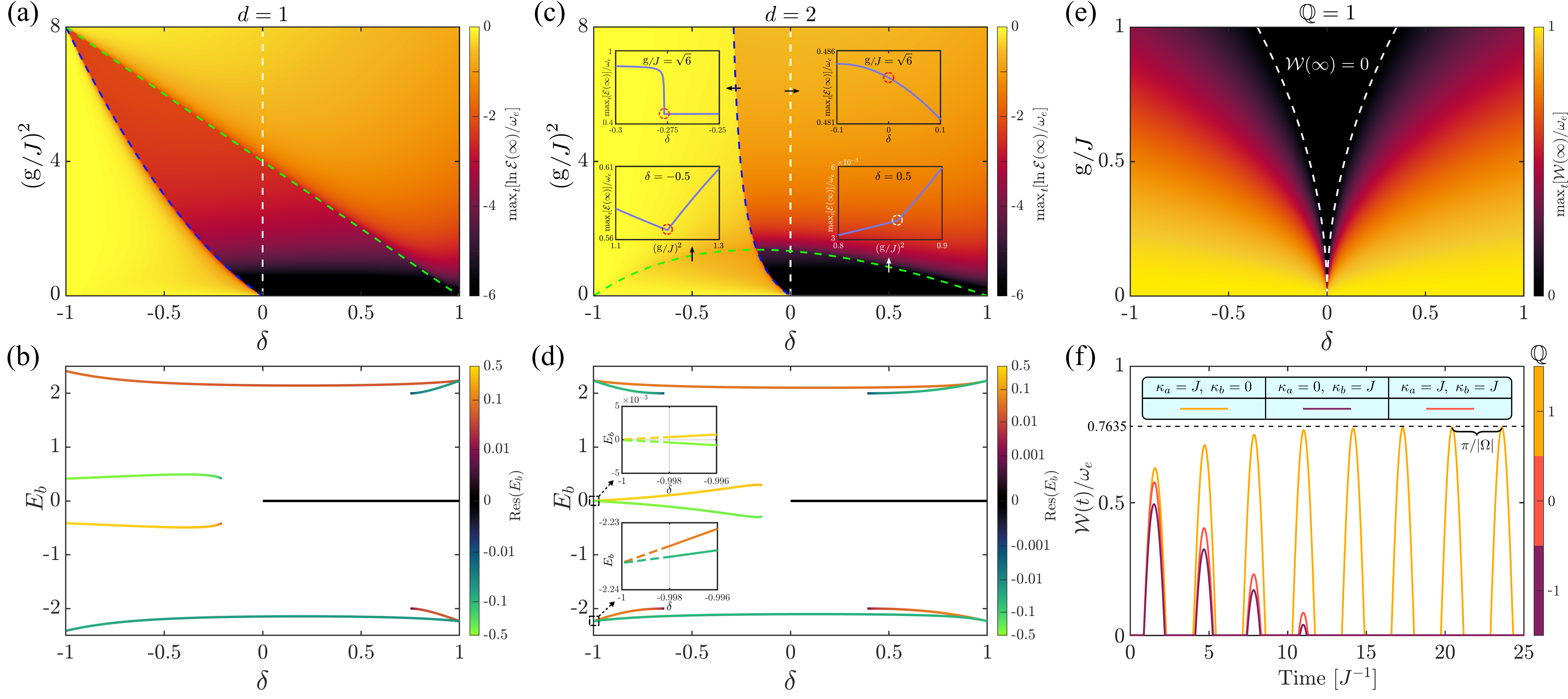}
	\caption{Configuration I: (a, c) Maximum stored energy $\max_t[\mathcal{E}(\infty)]$ in different $\text{g}$, and (b, d) bound-state energies $E_b$ in $\text{g}=J$ as a function of $\delta$. In (a, c), the white $(\ell_0)$ dashed line is the topological phase boundary $(\delta = 0)$, while the blue $(\ell_1)$ and green $(\ell_2)$ dashed curves mark the MSE phase boundaries. Insets in (c) show the MSE behavior near these phase boundaries. In (b, d), residues at the BSEs are color-coded to highlight bound-state contributions. The unit-cell distances are $d=1$ in (a, b) and $d=2$ in (c, d). Configuration II:  (e) Maximum ergotropy $\max_t[\mathcal{W}(\infty)]$ under single-sublattice dissipation $(\mathbb{Q}=1)$ as a function of $\delta$ and $\text{g}$, and (f) ergotropy evolution $\mathcal{W}(t)$ for different types of dissipation at $\text{g}=J/2$ and $\delta=0.9$. The dashed curves in (e) mark the boundary where $\mathcal{W}(\infty)= 0$. Other parameters are set as $\Delta=0$ for (a-d) and $\Delta=-\Omega=J$ for (e-f).}
	\label{Performance} 
\end{figure*}

\emph{Dynamics}.---Let us move on to study the dynamics of the QB. Specifically, we focus on the nonunitary evolution $|\psi(t)\rangle = \text{exp}(-i\hat{H}_{\rm eff}t)|\psi(0)\rangle$ starting from the initial state $|\psi(0)\rangle = |e,g;\text{vac}\rangle$. To analytically solve the dynamics of the QB in this scenario, we assume that the bath is in the thermodynamic limit $(N \to \infty)$. By using the resolvent method~\cite{PhysRevLett.119.143602,cohen1998atom}, the probability amplitude for the QB to be excited at any time can be calculated~\cite{supp}
\begin{equation}
\setlength\abovedisplayskip{0.5pt}
\setlength\belowdisplayskip{8pt}
c_{\rm B}(t) = \int_{\mathcal{C}}\frac{\dd{z}}{2\pi i} \frac{\Sigma_{12}^{\alpha \beta}(z) + \Omega_{12}^{\alpha \beta }}{\mathscr{D}(z)}e^{-izt},\label{II-6}
\end{equation}
where $\Sigma_{mn}^{\alpha \beta}(z) = \text{g}^2G(x_{m,\alpha},x_{n,\beta};z)$ refers to the self-energy of TLSs. The single-particle Green's function of the bath is denoted as $G(x_{m,\alpha},x_{n,\beta};z) = \langle \text{vac}|\hat{o}_{x_{m,\alpha}}(z-\sum_k\hat{\textbf{o}}_k^\dagger \widetilde{\textbf{h}}_k^{}\hat{\textbf{o}}_k^{})^{-1}\hat{o}_{x_{n,\beta}}^\dagger |\text{vac}\rangle$, and $\mathscr{D}(z) = {[z - \omega_e - \Sigma _{11}^{\alpha \alpha}(z)]}[z - \omega_e - \Sigma_{22}^{\beta \beta}(z)] - [\Omega_{12}^{\alpha \beta} + \Sigma_{12}^{\alpha \beta}(z)]^2$.

\emph{Bound state}.---Now, we introduce the bound states, a key quantity characterizing the long-time performance of the QB. By analytically solving the probability amplitude in Eq.\,(\ref{II-6}), we find that the time evolution of the TLSs is fully contributed by three parts: bound states, branch-cut detours, and unstable poles. Since the contributions from the branch-cut detours and the unstable poles decay quickly over time, only the bound-state contributions survive in the long-time limit, and we obtain~\cite{supp} 
\begin{align}
c_{\rm B}(\infty) = \sum_{z_k \in E_b}{\rm{Res}}\bigg[\frac{\Sigma_{12}^{\alpha \beta}(z)+\Omega_{12}^{\alpha \beta}}{\mathscr{D}(z)},z_k\bigg]e^{-iz_kt},\label{II-7}
\end{align} 
where ${E_b}$ denotes the bound-state energies (BSEs)~\cite{menard2015coherent}, i.e., the real eigenenergies of the bound states, which can be obtained by solving the real roots of the pole equation $\mathscr{D}(E_b) = 0$, or equivalently, by solving the equation $\hat{H}_{\rm eff}|\psi_{b}\rangle = E_b|\psi_{b}\rangle$. Henceforth, we shall use the abbreviation $\Res(z_k)$ to denote $\text{Res}\{[\Sigma_{12}^{\alpha \beta}(z) + \Omega_{12}^{\alpha \beta}]/\mathscr{D}(z),z_k\}$.

\emph{QB performance indicators}.---Next, to quantify the performance of QBs, we introduce three crucial thermodynamic indicators, starting with the stored energy. The stored energy of the QB at time $t$ is given by
\begin{align}
\mathcal{E}(t) = {\rm{Tr}}\big[\hat{\rho}_{\rm{B}}(t)\hat{H}_{\rm{B}}\big]-{\rm{Tr}}\big[\hat{\rho}_{\rm{B}}(0)\hat{H}_{\rm{B}}\big] = \omega_e\left| c_{\rm{B}}(t)\right|^2, \label{III-1}  
\end{align}
where $\hat{H}_{\rm B} = \omega_e\hat{\sigma}_+^{\rm B}\hat{\sigma}_-^{\rm B}$ and $\hat{\rho}_{\rm B}(t)={\rm{Tr}}_{\bigcdot}[\ketbra{\psi(t)}]$ are the free Hamiltonian and the reduced density matrix of the QB. Based on the stored energy, we can define the second thermodynamic indicator, the charging power of the QB, as $P(t) = \mathcal{E}(t)/t$, its performance will be discussed later. Moreover, the third key indicator is called ergotropy, which is used to describe the maximum energy that can be extracted at time $t$, defined by
\begin{equation}
    \setlength\abovedisplayskip{8pt}
    \setlength\belowdisplayskip{6pt}
	\mathcal{W}(t)={\rm{Tr}}\big[\hat{\rho}_{\rm B}(t)\hat{H}_{\rm B}\big]-{\rm{Tr}}\big[\hat{\tilde{\rho}}_{\rm B}(t)\hat{H}_{\rm B}\big], \label{III-2}
\end{equation}
where $\hat{\tilde{\rho}}_{\rm B}(t) = \sum_s r_s(t)|\varepsilon_s\rangle\langle\varepsilon_s|$ is the passive state, $r_s(t)$ are the eigenvalues of $\hat{\rho}_{\rm B}(t)$ arranged in descending order, while $|\varepsilon_s\rangle $ are the eigenstates of $\hat{H}_{\rm B}$ with the corresponding egienvalues $\varepsilon_s$ sorted in ascending order.

\emph{QB phase diagram}.---By substituting Eq.~(\ref{II-7}) into Eq.~(\ref{III-1}), we demonstrate that the maximum stored energy (MSE) of the QB in the long-time limit is only determined by the bound-state contributions. Thus, both the BSEs and the residues are essential for QB performance. As shown in Figs.\,\ref{Performance}(a) and \ref{Performance}(c), we observe that under the resonance condition $\Delta=\omega_e-\omega_c=0$, the MSE of the QB varies significantly with the unit-cell distance $(d = x_1 - x_2\in\mathbb{Z}^+)$ between the charger and QB, exhibiting a singular behavior (derivative discontinuity) precisely at the MSE phase boundaries~\cite{supp}
\begin{equation}
    \setlength\abovedisplayskip{6pt}
    \setlength\belowdisplayskip{6pt}
    \ell_1:\frac{\text{g}^2}{J^2} = \frac{4(1-\delta^2)\delta}{(1-2d)\delta-1},\ \ \ell_2:\frac{\text{g}^2}{J^2} = \frac{4(1-\delta^2)}{\delta-(1-2d)}.\label{III-3} 
\end{equation} 
Specifically, this behavior arises from a jump in the number of bound states at the phase boundaries and the corresponding discontinuity in their residues. However, despite such the jump at the topological phase boundary $\ell_0$, the MSE remains continuous at the boundary [see inserts in Fig.\,\ref{Performance}(c)] due to the vanishing residue of the double degenerate zero-energy bound states~\cite{supp}, as shown in Figs.\,\ref{Performance}(b) and \ref{Performance}(d). Notably, on the left side of the phase boundary $\ell_1$, most of the energy from the charger can transfer to the QB, whereas on the right side, energy transfer is nearly suppressed. In particular, under the Markovian limit $(\text{g}\ll 2J\abs{\delta})$, energy transfer between the charger and QB is perfect in the topologically nontrivial phase $(\delta<0)$ but completely suppressed in the topologically trivial phase $(\delta>0)$. Physically, when these TLSs are connected to the topological waveguide [see Fig.\,\ref{Fig-setup}(b)], the parity-positive and parity-negative bound states $\ket{E_\pm}$ emerge, which satisfy $\hat{H}_{\rm eff}\ket{E_\pm}=E_\pm\ket{E_\pm}$ and $\braket{E_\mu}{E_\nu}=\delta_{\mu,\nu}$ with $\mu,\nu\in\{+,-\}$. These bound states are composed of two orthonormal chiral edge-like states $\ket{L}$ and $\ket{R}$~\cite{supp1}, expressed as $\ket{E_\pm}\propto \ket{L}\pm\ket{R}$, where the charger and QB serve as the respective edges, satisfying $\ket{L}\simeq\ket{e,g;{\rm{vac}}}$ and $\ket{R}\simeq\ket{g,e;{\rm{vac}}}$. In the topologically trivial phase, the spatial envelopes of the two edge-like states do not overlap, leading to a degeneracy of the bound states ($E_+=E_-$), which indicates $\hat{H}_{\rm eff}\ket{L}\propto\ket{L}$. Consequently, the excitation in the charger is confined to the left edge, completely suppressing energy transfer. However, in the topologically nontrivial phase, the spatial envelopes of the two edge-like states overlap, breaking the degeneracy of the bound states ($E_+\neq E_-$), which implies $\hat{H}_{\rm eff}\ket{L}\not\propto\ket{L}$. As a result, the excitation oscillates between the two edges with a period of $2\pi/|E_+-E_-|$, enabling perfect long-range energy transfer even for $d\gg1$~\cite{supp}. Note that for an ordinary non-topological waveguide (i.e., $\delta=0$), the two bound states $\ket{E_\pm}$ vanish, yet the system can still support a dressed bound state in the continuum (BIC) under certain parametric conditions~\cite{vbs}. In this case, although the presence of the BIC prevents the complete leakage of the excitation into the waveguide, the charger can nevertheless transfer at most one fourth of its energy to the QB~\cite{supp}. Additionally, we observe that for $d=1$ and $d=2$, the intersection of the phase boundaries ($\ell_1$ and $\ell_2$) shifts, significantly expanding the parameter space for efficient energy transfer, with the intersection determined by Eq.~(\ref{III-3}).

\emph{QB dissipation immunity}.---When photon loss is considered in the topological waveguide, energy transfer in configuration I is fully suppressed in the long-time limit. In contrast, we find that energy transfer in configuration II can be completely immune to single-sublattice dissipation ($\mathbb{Q}=1$), as depicted by the orange curve in Fig.\,\ref{Performance}(f). This is evident from the analytical expression~\cite{supp} of the maximum ergotropy in the long-time limit,
\begin{align}
\frac{\max_t[\mathcal{W}(\infty)]}{\omega_e}=\frac{4J^4\delta^2-\text{g}^4/2}{( 2J^2\abs{\delta} + \text{g}^2)^2}\Theta\left(2^{\frac{3}{4}}J\sqrt{\abs{\delta}}-\abs{\text{g}}\right)\label{III-4} 
\end{align}
with $\Delta=-\Omega\neq0$. Physically, introducing the direct interaction $\Omega$ gives rise to a dark state~\cite{PhysRevApplied.14.024092}, while further satisfying $\Delta+\Omega=0$ leads to the formation of a vacancy-like dressed state (VDS)~\cite{vbs}, also referred to as a topologically robust dressed state in the SSH model. The coexistence of these two bound states ensures that the QB's performance is immune to single-sublattice dissipation, allowing perfect energy transfer with appropriately chosen parameters $\text{g}$ and $\delta$, as shown in Fig.\,\ref{Performance}(e). Moreover, we emphasize that the strong robustness of the VDS against disorder contributes to the QB's performance in resisting the effects of disorder~\cite{supp}. For single-sublattice $(\mathbb{Q}=-1)$ and two-sublattice $(\mathbb{Q}=0)$ dissipation, photon loss in sublattice $B$ completely prevents the formation of the VDS, leading to the extractable energy (ergotropy) reducing to zero in the long-time limit, as depicted by the purple and red curves in Fig.\,\ref{Performance}(f). 

\begin{figure}
  \centering
  \includegraphics[width=\linewidth]{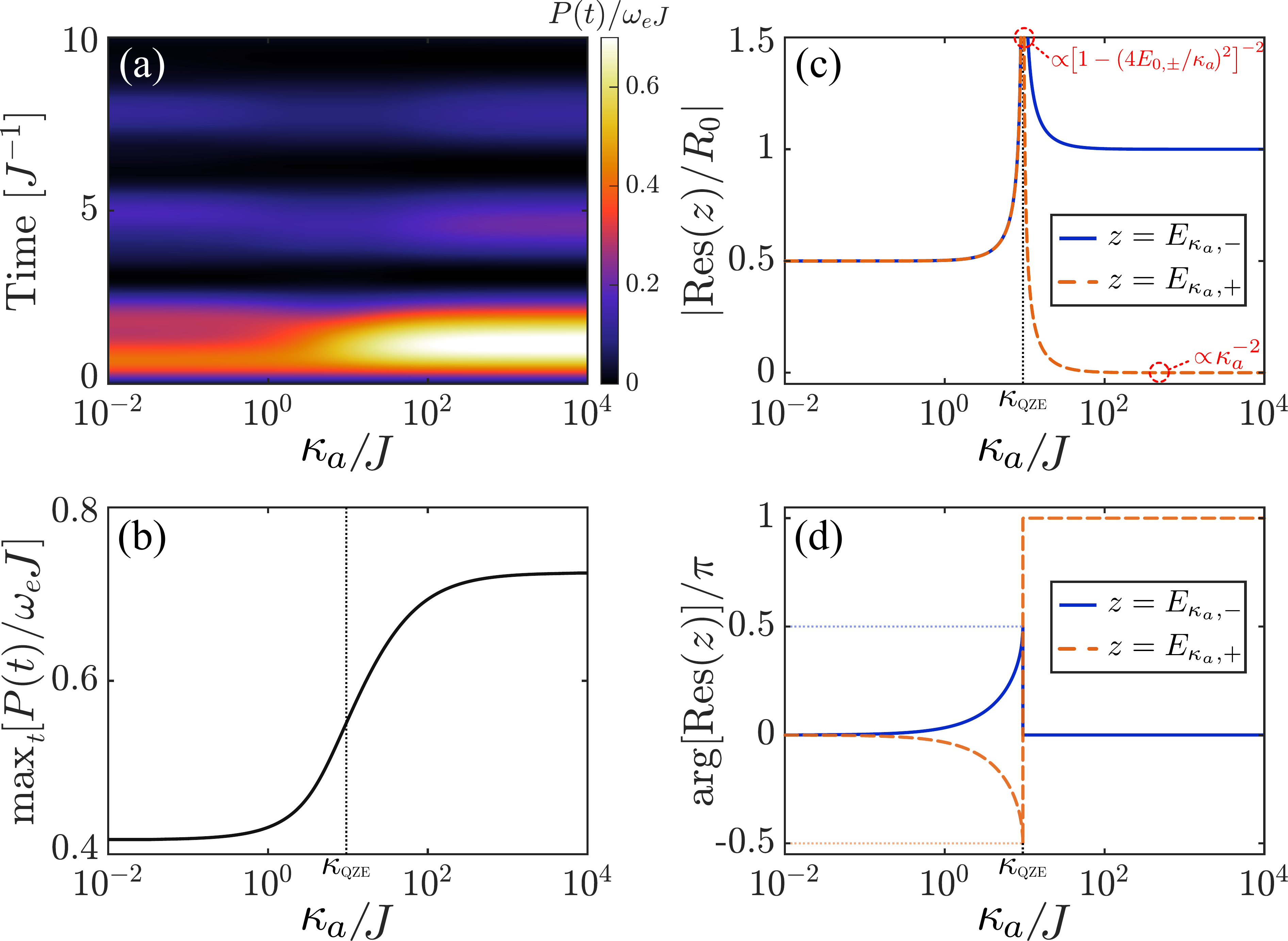}
  \caption{(a) Charging power $P(t)$ in different time $t$, and (b) maximal $P(t)$ via optimizing $t$ as a function of the dissipation strength $\kappa_a$. (c) Modulus $\abs{\Res(z)}$ and (b) phase $\arg[\Res(z)]$ of the residues at the dissipative BSEs $E_{\kappa_a,\pm}$ as a function of $\kappa_a$. The critical point for the occurrence of the quantum Zeno effect is marked as $\kappa_{\rm QZE}$. The system parameters are chosen as $\Delta=-\Omega=J$, $\text{g}=J$, $\delta=0.9$, and $\mathbb{Q}=1$.}
  \label{Zeno effect}
\end{figure}

\emph{QB performance boost in short time}.---To assess the performance of QBs, it is crucial to consider not only the stored energy and the ergotropy as key physical quantities but also the charging power as an indispensable indicator. In addition to these indicators, charging time and charging protocol are discussed in the Supplemental Material~\cite{supp}. Since the stored energy of a minimal QB is bounded, i.e., $0 \le \mathcal{E}(t) \le \omega_e$, the charging power inevitably approaches zero in the long-time limit. Thus, discussions on the charging power are primarily focused on a short timescale. Comparing the curves of $\mathbb{Q}=-1$ and $\mathbb{Q}=0$ in Fig.\,\ref{Performance}(f), we observe that an increasing in $\kappa_a$ effectively enhances the ergotropy and stored energy on the short timescale. This indicates that dissipation can be leveraged to enhance the charging power. For concreteness, we focus on single-sublattice dissipation ($\mathbb{Q}=1$) in configuration II as a starting point. In this case, in addition to the two dissipation-independent bound states (i.e., the dark state and the VDS), the system also possesses two dissipative bound states with imaginary energies~\cite{supp}
\begin{align}
E_{\kappa_a, \pm} = -i(\kappa_a/4) \pm \sqrt{E_{0, +}^2 - (\kappa_a/4)^2}, \label{III-6} 
\end{align}
where $E_{0,\pm}$ denotes the real energies of the other two bound states in the non-dissipative topological waveguide. When $\kappa_a \gg \kappa_{\rm QZE} \equiv  4|E_{0, +}|$, the lifetimes of the two dissipative bound states exhibit starkly opposite scaling behaviors, i.e., $-\Im[E_{\kappa_a, \pm}] \propto\kappa_a^{\pm1}$. Here, we note that the lifetime of the dissipative bound state with energy $E_{\kappa_a,-}$ is inversely proportional to $\kappa_a$, and thus we may refer to this phenomenon as the quantum Zeno effect~\cite{PhysRevLett.126.190402}. Simultaneously, the contributions from the dissipative bound states also satisfy $\Res(E_{\kappa_a, +})\propto\kappa_a^{-2} $ and $\Res(E_{\kappa_a,-})\approx R_0=\Res(E_{0,+})+\Res(E_{0,-})$, where $R_0=2\text{g}^4/[E_{0,+}^4-2J^2(1+\delta^2)E_{0,+}^2]$, and we see that $\Res(E_{\kappa_a, +})$ significantly decreases as $\kappa_a$ increases, while $\Res(E_{\kappa_a,-})$ remains unchanged. Notably, in Figs.\,\ref{Zeno effect}(c) and \ref{Zeno effect}(d), we observe that the modulus and phase of $\Res(E_{\kappa_a,\pm})$ exhibit sharp transitions near $\kappa_{\rm QZE}$. As a result, we can identify ${\kappa_{\rm QZE}}$ as the critical point at which the quantum Zeno effect starts to manifest. In Fig.\,\ref{Zeno effect}(a), we show that when $\kappa_a$ is present, $P(t)$ exhibits a rapid periodic oscillatory decay over time. On the short timescale, $t \sim \pi/\abs{\Omega}$, increasing $\kappa_a$ is accompanied by an enhancement of $P(t)$. In Fig.\,\ref{Zeno effect}(b), as $\kappa_a$ continues to increase and goes beyond $\kappa_{\rm QZE}$, the emergence of the quantum Zeno effect results in a substantial increase in $\max_t[P(t)]$. In addition, it is worth noting that the stroboscopic dynamics of the non-dissipative system at $t = 2\pi\mathbb{Z}^+/\abs{E_{0,+}}$ closely resembles the dynamics of the dissipative system~\cite{supp}.

\emph{Summary and outlook}.---To summarize, we have developed a general framework for analyzing the atomic dynamics of two-level systems coupled to a topological photonic waveguide. First, we have demonstrated that only the contributions from bound state energies are retained in the long-time limit. We have pointed out that topological properties determine the charging process from the quantum charger to the QB, and near-perfect transfer may occur in a topologically nontrivial phase. Moreover, we have discovered that the maximum stored energy exhibits singular behavior at the phase boundary. Second, we have highlighted that even with extremely weak coupling between the quantum charger and the QB, the performance of the QB, such as ergotropy, can resist environment-induced decoherence due to the presence of a dark state and a topologically robust dressed bound state. Third, we have shown that an increase in dissipation significantly enhances the charging power of the QB over a short time due to the emergence of the quantum Zeno effect. Looking forward, one intriguing possibility is to delve deeper into QB performance within generalized open quantum systems, such as by exploring it from the perspective of the Hatano-Nelson model~\cite{PhysRevLett.77.570}. Another promising practical avenue is the investigation of multi-excitation dynamics~\cite{PhysRevX.6.021027,shi2018effective,PhysRevA.84.063803}, collective effects, and quantum coherence~\cite{PhysRevA.105.062203} in QB charging by using numerical techniques~\cite{schollwock2011density, tanimura2020numerically}.

\emph{Acknowledgments}.---We are grateful to Naomichi Hatano for carefully reading the manuscript.~C. Shang thanks Ken-Ichiro Imura, Tomotaka Kuwahara, and Takano Taira for their valuable discussions.~C. Shang would also like to thank the committee members in the Ph.D. defense, Yuto Ashida, Masahito Ueda, Mio Murao, Kiyotaka Aikawa, and Kuniaki Konishi, for their careful review and insightful comments. This work is supported by the National Key Research and Development Program of China grant 2021YFA1400700, the National Science Fund for Distinguished Young Scholars of China (Grant No.~12425502), and the Fundamental Research Funds for the Central Universities (Grant No.~2024BRA001).~C. Shang acknowledges the financial support by the China Scholarship Council, the Japanese Government (Monbukagakusho-MEXT) Scholarship under Grant No. 211501, the RIKEN Junior Research Associate Program, and the Hakubi projects of RIKEN.


%

\clearpage
\setcounter{secnumdepth}{2}
\onecolumngrid
\begin{center}
	{\Large \textbf{ Supplemental Material for\\
			``Topological Quantum Batteries"}}
\end{center}

\begin{center}
	Zhi-Guang Lu$^{1}$, Guoqing Tian$^{1}$, Xin-You L\"{u}$^{1, *}$, and Cheng Shang$^{2, 3, \dagger}$
\end{center}

\begin{center}
\begin{minipage}[]{16cm}
    \small{\it
    \centering $^{1}$School of Physics and Institute for Quantum Science and Engineering, Huazhong University of Science and Technology and Wuhan institute of quantum technology, Wuhan, 430074, China\\}
    \small{\it\centering $^{2}$Department of Physics, The University of Tokyo, 5-1-5 Kashiwanoha, Kashiwa, Chiba 277-8574, Japan\\}
        \small{\it\centering $^{3}$Analytical quantum complexity RIKEN Hakubi Research Team, RIKEN Center for Quantum Computing (RQC), 2-1 Hirosawa, Wako, Saitama 351-0198, Japan\\}
\end{minipage}
\end{center}

\setcounter{equation}{0}
\setcounter{figure}{0}
\setcounter{table}{0}
\setcounter{section}{0}
\makeatletter
\renewcommand{\theequation}{S\arabic{equation}}
\renewcommand{\thefigure}{S\arabic{figure}}
\renewcommand{\bibnumfmt}[1]{[S#1]}


\vspace{8mm}

This supplement material contains eight parts: I. Exact dynamics of quantum charger-battery system coupled to structured bosonic bath; II. Dynamics of quantum battery in topological bath; III. Quantum battery performance in different configurations. IV. Physical interpretation of near-perfect energy transfer between charger and battery. V. The charging time in Markovian and non-Markovian regimes. VI. Full cycle of quantum charging protocol in the scheme. VII. Effects of symmetry-preserving and symmetry-breaking disorder to the performance of a quantum battery. VIII. The validity of the rotating wave approximation in our system.

\tableofcontents
\newpage
\section{Exact Dynamics of Quantum Charger-Battery System Coupled to Structured Bosonic Bath}\label{I}
In this section, we present the exact dynamical expression for the charging of a quantum battery (QB) by a quantum charger through an external bath. Firstly, we consider that a quantum battery $\text{B}$ and a quantum charger $\text{C}$ are concurrently coupled with a structured bosonic bath. The Hamiltonian of the total system under the rotating-wave approximation reads $H_{\text{tot}}=H_{\text{sys}}+H_{\text{bath}}+H_{\text{int}}$, where
\begin{align}
	H_{\text{sys}}=\omega_e\sigma_+^{\text{B}}\sigma_-^{\text{B}}+\omega_e\sigma_+^{\text{C}}\sigma_-^{\text{C}}+\Omega_{12}^{\alpha\beta}(\sigma_+^{\text{B}}\sigma_-^{\text{C}}+\sigma_+^\text{C}\sigma_-^\text{B}),\quad H_{\text{int}}=\text{g}(\sigma_-^{\text{B}} o^\dagger_{x_{1,\alpha}}+\sigma_+^{\text{B}}o^{}_{x_{1,\alpha}}+ \sigma_-^{\text{C}}o^\dagger_{x_{2,\beta}}+\sigma_+^{\text{C}}o^{}_{x_{2,\beta}}),\label{S1}
\end{align}                                                                                                                                         
and $\Omega_{12}^{\alpha\beta}=\Omega\delta_{x_1,x_2}\delta_{\alpha,\beta}$ with $\Omega\in\mathbb{R}$. Here, $H_{\text{sys}}$ represents the Hamiltonian of the QB and the quantum charger, and $H_{\text{bath}}$ is the Hamiltonian of the structured bosonic bath. The interaction Hamiltonian between the system and the bath is represented by $H_{\text{int}}$, where $\sigma_+^{\mathrm{B}}$ ($\sigma_+^{\mathrm{C}}$) and $\sigma_-^{\mathrm{B}}$ ($\sigma_-^{\mathrm{C}}$) represent the raising and lowering Pauli operators of QB (quantum charger), respectively, while $o^{}_{x_{j,\alpha}}$ and $o^{\dagger}_{x_{j,\alpha}}$ denote the annihilation and creation operators at the position $x_{j,\alpha}$ of bath, respectively. Without loss of generality, we assume that the spectrum of $H_{\text{bath}}$ possesses energy band structure. Note that, for consistency with the main text, the second subscript $\alpha\in\{A, B\}$ in $o_{x_{j,\alpha}}$ is used to emphasize different sublattices, i.e., $o_{x_{j,A}}\equiv a_{x_j}$ and $o_{x_{j,B}}\equiv b_{x_j}$. 

To analytically solve the dynamics of the two-level system, we assume that the bath is in the thermodynamic limit in the following derivations. Consequently, the time-evolution operator of the system can be obtained by the Inverse-Fourier transform of the Green function\,\cite{cohen1998atom}
\begin{align}\label{S2}
	U(t)=e^{-iH_{\text{tot}}t}=\frac{1}{2\pi i}\int_{\mathcal{C}}G_{\text{tot}}(z)e^{-izt}\dd{z}=\frac{1}{2\pi i}\int_{\mathcal{C}}\frac{1}{z-H_{\text{tot}}}e^{-izt}\dd{z},
\end{align}
where the integration path $\mathcal{C}$ lies immediately above the real axis in the complex plane, extending infinitely from right to left. In the single-excitation subspace, to explore the dynamics of QB, we need to project the evolution operator $U(t)$ onto the subspace of the system involving QB and quantum charger. Thus, we define 
\begin{align}\label{S3}
	P\equiv (\ketbra{e,g}+\ketbra{g,e})\otimes\ketbra{\text{vac}},\quad Q\equiv\ketbra{g,g}\otimes\sum_{j,\alpha}o^\dagger_{x_{j,\alpha}}\ketbra{\text{vac}}o_{x_{j,\alpha}}^{},
\end{align}
which satisfy $P+Q=\mathbb{I}_1$, where $\mathbb{I}_1$ is the identity operator in the single-excitation subspace. For the sake of simplicity, we define $\ket{e_1}\equiv\ket{e,g}=\sigma_{+}^{\rm C}\ket{g,g}$, $\ket{e_2}\equiv\ket{g,e}=\sigma_{+}^{\rm B}\ket{g,g}$, and $\ket{x_{j,\gamma}}\equiv o_{x_{j,\gamma}}^\dagger\ket{\rm vac}$ in the following steps. As a result, the evolution operator projected onto the subspace of the system can be written as
\begin{align}\label{S4}
	PU(t)P=\frac{1}{2\pi i}\int_{\mathcal{C}}PG_{\text{tot}}(z)Pe^{-izt}\dd{z},
\end{align}
where 
\begin{align}
	PG_{\text{tot}}(z)P=\frac{P}{z-PH_{\text{sys}}P-P\Sigma(z)P},\quad
	P\Sigma(z)P=PH_{\text{int}}P+PH_{\text{int}}\frac{Q}{z-QH_{\text{tot}}Q}H_{\text{int}}P.\label{S5}
\end{align}
Since $PH_{\text{int}}P=QH_{\text{int}}Q=QH_{\text{sys}}Q=0$, the last term in~(\ref{S5}) can be further simplified as
\begin{align}\label{S6}
	P\Sigma(z)P=PH_{\text{int}}\frac{Q}{z-H_{\text{bath}}}H_{\text{int}}P=PH_{\text{int}}G_{\text{bath}}(z)H_{\text{int}}P,\quad \text{with } G_{\text{bath}}(z)=Q(z-H_{\text{bath}})^{-1}Q.
\end{align}
In the basis $\{\ket{e_1;\text{vac}},\ \ket{e_2;\text{vac}}\}$, the first term in~(\ref{S5}) can be written in a matrix form as
\begin{align}\label{S7}
	\mqty[G_{11}(z)&G_{12}(z)\\G_{21}(z)&G_{22}(z)]=\mqty[z-\omega_e-\Sigma_{11}(z)&-\Omega_{12}^{\alpha\beta}-\Sigma_{12}(z)\\-\Omega_{12}^{\alpha\beta}-\Sigma_{21}(z)&z-\omega_e-\Sigma_{22}(z)]^{-1},
\end{align}
where 
\begin{align}
	G_{mn}(z)=\mel{e_m;\text{vac}}{PG_{\text{tot}}(z)P}{e_n;\text{vac}},\quad \Sigma_{mn}(z)=\mel{e_m;\text{vac}}{P\Sigma(z)P}{e_n;\text{vac}}.
\end{align}
According to Eq.~(\ref{S6}), by inserting $H_{\text{int}}$ into $\Sigma_{mn}(z)$, we have
\begin{align}
	\Sigma_{11}^{\alpha\alpha}(z)=\Sigma_{11}(z)&=\text{g}^2\mel{\text{vac}}{o^{}_{x_{1,\alpha}}(z-H_{\text{bath}})^{-1}o_{x_{1,\alpha}}^{\dagger}}{\text{vac}}=\text{g}^2\mel{x_{1,\alpha}}{(z-H_{\text{bath}})^{-1}}{x_{1,\alpha}}\equiv \text{g}^2G(x_{1,\alpha},x_{1,\alpha};z),\label{S9}\\
	\Sigma_{12}^{\alpha\beta}(z)=\Sigma_{12}(z)&=\text{g}^2\mel{\text{vac}}{o^{}_{x_{1,\alpha}}(z-H_{\text{bath}})^{-1}o_{x_{2,\beta}}^{\dagger}}{\text{vac}}=\text{g}^2\mel{x_{1,\alpha}}{(z-H_{\text{bath}})^{-1}}{x_{2,\beta}}\equiv \text{g}^2G(x_{1,\alpha},x_{2,\beta};z),\label{S10}\\
	\Sigma_{21}^{\beta\alpha}(z)=\Sigma_{21}(z)&=\text{g}^2\mel{\text{vac}}{o^{}_{x_{2,\beta}}(z-H_{\text{bath}})^{-1}o_{x_{1,\alpha}}^{\dagger}}{\text{vac}}=\text{g}^2\mel{x_{2,\beta}}{(z-H_{\text{bath}})^{-1}}{x_{1,\alpha}}\equiv \text{g}^2G(x_{2,\beta},x_{1,\alpha};z),\\
	\Sigma_{22}^{\beta\beta}(z)=\Sigma_{22}(z)&=\text{g}^2\mel{\text{vac}}{o^{}_{x_{2,\beta}}(z-H_{\text{bath}})^{-1}o_{x_{2,\beta}}^{\dagger}}{\text{vac}}=\text{g}^2\mel{x_{2,\beta}}{(z-H_{\text{bath}})^{-1}}{x_{2,\beta}}\equiv \text{g}^2G(x_{2,\beta},x_{2,\beta};z),\label{S12}
\end{align}
where $\Sigma_{12}^{\alpha\beta}(z)$ refers to the self-energy of the two-level systems and $G(x_{1,\alpha},x_{2,\beta};z)$ represents the single-particle Green function of the bath. As a result, according to Eq.~(\ref{S7}), the projected evolution operator in Eq.~(\ref{S4}) is given by
\begin{align}\label{S13}
	PU(t)P=\frac{1}{2\pi i}\int_{\mathcal{C}}\dd{z}\frac{e^{-izt}}{\mathscr{D}(z)}\mqty[\ket{e_1;\text{vac}}\\\ket{e_2;\text{vac}}]^T\mqty[z-\omega_e-\Sigma_{22}^{\beta\beta}(z)& \Omega_{12}^{\alpha\beta}+\Sigma_{21}^{\beta\alpha}(z)\\\Omega_{12}^{\alpha\beta}+ \Sigma_{12}^{\alpha\beta}(z)&z-\omega_e^{}-\Sigma_{11}^{\alpha\alpha}(z)]\mqty[\bra{e_1;\text{vac}}\\\bra{e_2;\text{vac}}],
\end{align}
where
\begin{align}\label{S14}
	\mathscr{D}(z)=[z-\omega_e-\Sigma_{11}^{\alpha\alpha}(z)][z-\omega_e-\Sigma_{22}^{\beta\beta}(z)]-[\Omega_{12}^{\alpha\beta}+\Sigma_{12}^{\alpha\beta}(z)][\Omega_{12}^{\alpha\beta}+\Sigma_{21}^{\beta\alpha}(z)].
\end{align}
Finally, let us assume that the total system is prepared in the initial state $\ket{\psi(0)}=\ket{e_1;\text{vac}}$, i.e., the quantum charger is in the excited state, QB is in the ground state, and the bath is in the vacuum state. According to Eq.~(\ref{S13}), the probability amplitude for QB to be in the excited state $\ket{e_2;\text{vac}}$ at $t$ time is given by
\begin{align}\label{S15}
	c_\text{B}(t)=\mel{e_2;\text{vac}}{PU(t)P}{e_1;\text{vac}}=\frac{1}{2\pi i}\int_{\mathcal{C}}\frac{\Sigma_{12}^{\alpha\beta}+\Omega_{12}^{\alpha\beta}}{\mathscr{D}(z)}e^{-izt}\dd{z}\equiv\frac{1}{2\pi i}\int_{\mathcal{C}}\mathscr{C}(z)e^{-izt}\dd{z},
\end{align}
and the reduced density matrix of QB is computed as
\begin{align}\label{S16}
	\rho_{\text{B}}(t)=\Tr_{\bigcdot}[\ketbra{\psi(t)}]=\Tr_{\text{charger}\otimes \text{bath}}[\ketbra{\psi(t)}]=\abs{c_{\text{B}}(t)}^2\ketbra{e}+[1-\abs{c_\text{B}(t)}^2]\ketbra{g}.
\end{align}
\section{Dynamics of Quantum Battery in Topological Bath}\label{II}

In this section, we show the full derivation of the setup in the main text. We exploring the QB dynamics in a topological bath, both with and without dissipation. We begin with a detailed discussion of the topological bath in subsection A and B. Next, we derive an analytical expressions for self-energy in the dissipative topological bath and demonstrate the connection between bound-state energies and QB dynamics in the long-time limit.

\subsection{Su-Schrieffer-Heeger model without dissipation}\label{IIA}
Here, we choose the simplest topological model, the Su-Schrieffer-Heeger (SSH) model\,\cite{SSH}, as a topological bath. For simplicity, we employ two abbreviations $o_{x_{j,A}}\equiv a_j$ and $o_{x_{j,B}}\equiv b_j$. By setting $\hbar=1$, the Hamiltonian of the topological bath is given by
\begin{align}\label{S17}
	H_{\text{bath}}=\sum_{j=1}^{N}\omega_c(a_j^\dagger a_j^{}+b_j^\dagger b_j^{} )+J_+\sum_{j=1}^N (a_j^\dagger b_j^{} +  b_j^{\dagger} a_j^{} )+J_-\sum_{j=1}^{N}(b_{j}^\dagger a_{j+1}^{}+a_{j+1}^\dagger b_j^{}),
\end{align}
where $a_j(a_j^\dagger)$ and $b_j(b_j^\dagger)$ are the annihilation (creation) operators of boson on the sites $a$ and $b$ at position $j$, respectively. The resonant frequency of these modes is $\omega_c$. The topological waveguide consists of two interspersed photonic lattices with alternating nearest-neighbor hopping $J_\pm=J(1\pm\delta)$ between bosonic modes. Here, $J$ defines the hopping strength, and $\delta$, known as the dimerization parameter, controls the asymmetry between the lattices. Under the periodic boundary conditions (i.e., $a_{N+j}=a_j$ and $b_{N+j}=b_j$) and in the momentum space with
\begin{align}\label{S18}
	a_k^\dagger=\frac{1}{\sqrt{N}}\sum_{j=1}^N e^{ikj}a_j^\dagger,\quad 	b_k^\dagger=\frac{1}{\sqrt{N}}\sum_{j=1}^N e^{ikj}b_j^\dagger,\quad k=\frac{2\pi}{N}n,\quad n\in(-N/2,N/2],
\end{align}
the bath Hamiltonian in the momentum space can be written as $H_{\text{bath}}=\sum_k\mathbf{o}_k^\dagger \mathbf{h}_k\mathbf{o}_k$, with $\mathbf{o}_k=[a_k,b_k]^T$, and the corresponding Bloch Hamiltonian reads
\begin{equation}
    \setlength\abovedisplayskip{6pt}
    \setlength\belowdisplayskip{6pt}
\mathbf{h}_k=\mqty[\omega_c&f_k\\f^*_k&\omega_c]=\Re[f_k]\sigma_x-\Im[f_k]\sigma_y+\omega_c\sigma_0,
\end{equation}
where $f_k=J(1+\delta)+J(1-\delta)e^{-ik}\equiv\omega_k e^{i\phi_k}$ (with $\omega_k>0$) is the coupling coefficient in the momentum space between the bosonic modes of $a_k$ and $b_k$. Hereafter, we set $\omega_c$ as the energy reference, causing $\omega_e$ in Eqs.~(\ref{S13}, \ref{S14}) to become $\omega_e-\omega_c$. By simply diagonalizing $\mathbf{h}_k$, the Hamiltonian $H_{\text{bath}}$ in Eq.~(\ref{S17}) can be further written as
\begin{align}\label{S19}
	H_{\text{bath}}=\sum_k\mqty[u_k^\dagger & l_k^\dagger]\mqty[\omega_k&0\\0& -\omega_k]\mqty[u_k\\l_k]=\sum_k[\omega_k u_k^\dagger u_k^{}-\omega_kl_k^\dagger l_k^{}],
\end{align}
where $u_k/l_k=(\pm a_k + b_k e^{i\phi_k} )/\sqrt{2}$, $\omega_k = J\sqrt{2(1+\delta^2)+2(1-\delta^2)\cos(k)}$, and $\phi_k=\arctan[\Im(f_k)/\Re(f_k)]$. The corresponding dispersion relations are given by $\epsilon_{k,\pm}=\pm\omega_k$, where the subscript $+$ ($-$) denotes the upper (lower) energy band of the SSH bath.

\begin{figure}
	\includegraphics[width=17cm]{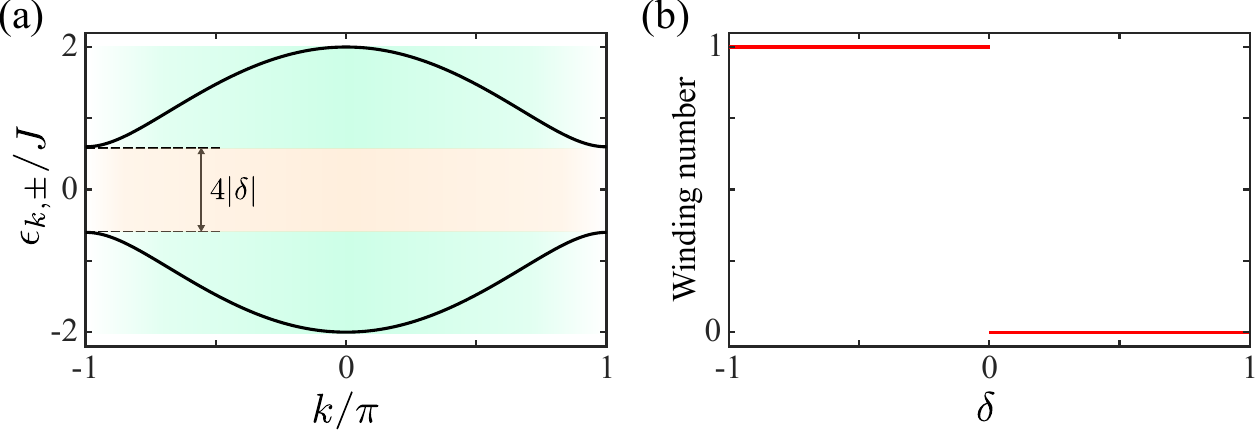}\\
	\caption{(a) Dispersion relations $\epsilon_{k,\pm}$ of the Su-Schrieffer-Heeger bath with periodical boundary conditions. The parameter is set as $\abs{\delta}=0.3$. (b) The winding number as function of the dimerization strength $\delta$. When $\delta<0$, the model is in a topologically nontrivial phase with a winding number of one. Conversely, when $\delta>0$, the model is in a topologically trivial phase with a winding number of zero.
	}\label{fig1}
\end{figure}

In Fig.\,\ref{fig1}(a), we present the dispersion relation for a dimerization parameter $\abs{\delta}=0.3$. The energy bands of the bath are observed to be symmetric with respect to th  cavity resonant frequency $\omega_c$. The energy bands span the range $[-2J, -2\abs{\delta}J]\cup [2\abs{\delta}J, 2J]$, featuring a central bandgap of $4\abs{\delta}J$. These energy bands can be adjusted by varying the dimerization strength $\delta$. In the SSH model, the topological properties of the system are characterized by the winding number, which takes values of either one or zero, depending on the parameters of the system. In Fig.\,\ref{fig1}(b), we depict the winding number of the SSH bath. In the case in which the intracell hopping strength outweighs the intercell hopping strength (i.e., $\delta > 0$), the winding number equals to zero, corresponding to the so-called topologically trivial phase. Conversely, when the intercell hopping strength dominates over the intracell hopping strength (i.e., $\delta<0$), the winding number is one, indicating a topologically nontrivial phase.

\subsection{Su-Schrieffer-Heeger model with dissipation}\label{IIB}
\begin{figure}
	\centering
	\includegraphics[width=17cm]{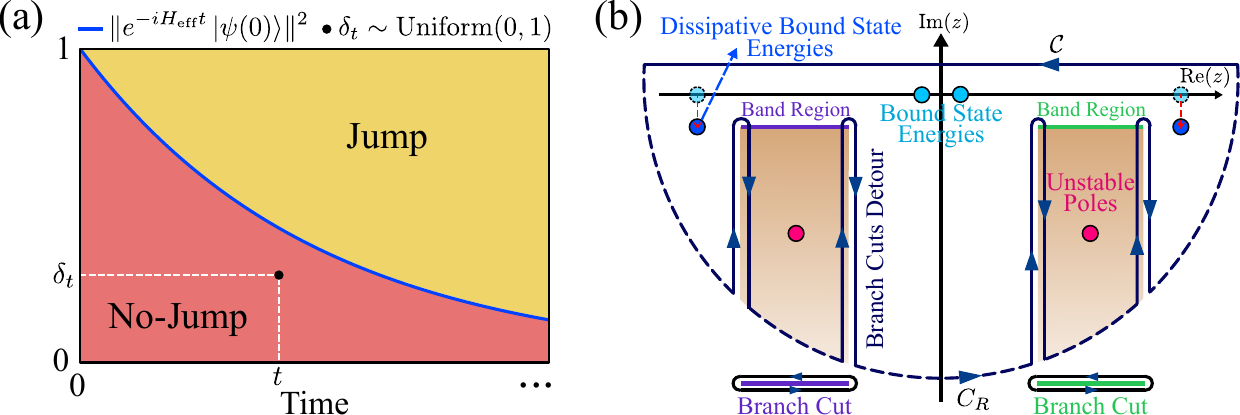}\\
	\caption{(a) The norm squared of the evolved state governed by the effective NH Hamiltonian $H_{\text{eff}}$ as a function of time. For a single trajectory, the occurrence of a quantum jump at time $t$ can be determined by comparing a random number $\delta_t$ with the norm squared at that time, as indicated by the black dot falling within the red region. (b) An integration contour (horizontal dark blue line) to calculate Eq.~(\ref{S15}). One needs to close the contour of integration in the lower half of the complex plane (dashed and vertical dark blue line) to evaluate the integration. Here, the information including (dissipative) bound-state energies, branch cut detours, unstable poles, and band regions, within the lower half of the complex plane is given by the effective NH Hamiltonian described by Sec.~\ref{IIB}. At the band edges the path changes from the first to the second Riemann sheet of the integrand $\mathscr{C}(z)$ (brown areas).
	}\label{fig2}
\end{figure}

Let us consider a realistic scenario: a one-dimensional SSH photonic lattice with engineered photon loss\,\cite{nhbath}. Under the Born-Markov and rotating-wave approximations, the equation of motion reads
\begin{align}\label{S20}
	\dot{\rho}_t=-i[H_{\text{sys}}+H_{\text{bath}}+H_{\text{int}},\rho_t]+\mathcal{L}_a\rho_t+\mathcal{L}_b\rho_t,
\end{align}
where $H_{\text{sys}}$, $H_{\text{int}}$, and $H_{\text{bath}}$ are defined by Eqs.~(\ref{S1}) and (\ref{S17}). The photon dissipators for different sublattices are given by $\mathcal{L}_a=\kappa_a\sum_j\mathcal{D}[a_j]$ and $\mathcal{L}_b=\kappa_b\sum_j\mathcal{D}[b_j]$, where $\kappa_a$ ($\kappa_b$) controls the photon loss rates of sublattice $A$ ($B$), and $\mathcal{D}[L]\rho\equiv L\rho L^\dagger-\{L^\dagger L,\rho\}/2$ is the Lindblad superoperator. To find the solution to Eq.~(\ref{S20}) in the single-excitation sector, we rewrite the Lindblad master equation (\ref{S20}) as
\begin{equation}\label{S21}
    \setlength\abovedisplayskip{6pt}
    \setlength\belowdisplayskip{6pt}
	\dot{\rho}_t=-i(H_{\text{eff}}^{}\rho_t-\rho_t H_{\text{eff}}^\dagger)+\kappa_a\sum_{j=1}^{N}a_j^{}\rho_ta_j^\dagger+\kappa_b\sum_{j=1}^Nb_j^{}\rho_tb_j^\dagger,
\end{equation}
where $a_j$ and $b_j$ within the last two terms on the right-hand side (RHS) of Eq.~(\ref{S21}) are the ``jump" operators associated with the sublattices dissipation resulting from emission into free space, and $H_{\text{eff}}$ is the effective non-Hermitian (NH) Hamiltonian for the dissipative system, i.e., $H_{\text{eff}}=H_{\text{sys}}+H_{\text{bath}}^{\text{eff}}+H_{\text{int}}$ with $H_{\text{bath}}^{\text{eff}}=H_{\text{bath}}-(i/2)\sum_{j}(\kappa_aa_j^\dagger a_j^{}+\kappa_bb_j^\dagger b_j^{})$. In this form, the terms $\kappa_a\sum_ja_j^{}\rho_ta_j^\dagger$ and $\kappa_b\sum_jb_j^{}\rho_tb_j^\dagger$ are often called the recycling terms, as it recycles the population that is lost from certain states due to the effective NH Hamiltonian, placing it in other states. For the initial state $\ket{\psi(0)}$ in the single-excitation subspace, on the one hand, the time evolution under the effective NH Hamiltonian is given by  $\ket{\psi(t)}=\exp(-iH_{\text{eff}}t)\ket{\psi(0)}$, resulting in a non-normalized final state with a norm squared that monotonically decreases over time, as shown by the blue line in Fig.\,\ref{fig2}(a). On the other hand, once the recycling terms work, i.e., when a jump process occurs, the final state deterministically transitions to the zero-excitation state $\ket{g,g;\text{vac}}$. According to quantum trajectory method\,\cite{qt}, whether a jump process occurs at time $t$ is determined by comparing a random number $\delta_t$ between 0 and 1 with the norm squared $\braket{\psi(t)}=\|\exp(-iH_{\text{eff}}t)\ket{\psi(0)}\|^2$. Specifically, if $\delta_t>\braket{\psi(t)}$, the jump occurs; otherwise, it does not, as illustrated by the red and yellow regions in Fig.\,\ref{fig2}(a). 

Then, we focus on the state at time $t$ and generate $N$ random numbers uniformly distributed between 0 and 1. For the sake of simplicity, we define $N_1$ and $N_2$ (satisfying $N_1+N_2=N$) as the counts of no-jump and jump occurrences, respectively. As a result, based on the condition for the occurrence of jumps, the solution to Eq.~(\ref{S20}) reads
\begin{align}\label{S22}
	\rho_t=\lim\limits_{N\to\infty}[|\tilde{\psi}(t)\rangle\langle\tilde{\psi}(t)|\times N_1+\ketbra{g,g;\text{vac}}\times N_2]/N,
\end{align}
where $|\tilde{\psi}(t)\rangle=\ket{\psi(t)}/\sqrt{\braket{\psi(t)}}$ representing the normalized state. Provided our random number generators are well behaved, these two ratios satisfy 
\begin{align}\label{S23}
	\lim\limits_{N\to\infty}\frac{N_1}{N}=\braket{\psi(t)},\quad \lim\limits_{N\to\infty}\frac{N_2}{N}=1-\lim\limits_{N\to\infty}\frac{N_1}{N}=1-\braket{\psi(t)}=p_t.
\end{align}
Finally, by plugging Eq.~(\ref{S23}) into Eq.~(\ref{S22}), we have
\begin{align}\label{densityM}
	\rho_t=\ketbra{\psi(t)}+p_t\ketbra{g,g;\text{vac}}=e^{-iH_{\text{eff}}t}\rho_0 e^{iH^\dagger_{\text{eff}}t}+p_t\ketbra{g,g;\text{vac}}.
\end{align}
If the initial state is a mixed state $\rho_0$, the norm squared $\braket{\psi(t)}$ mentioned above should be rewritten as $\mathrm{Tr}[e^{-iH_{\text{eff}}t}\rho_0 e^{iH^\dagger_{\text{eff}}t}]$. Notice that $\mathrm{Tr}[e^{-iH_{\text{eff}}t}\rho_0 e^{iH^\dagger_{\text{eff}}t}]=\braket{\psi(t)}$ when $\rho_0=\ketbra{\psi(0)}$. In fact, when we replace the total Hamiltonian $H_{\text{tot}}$ in Eq.~(\ref{S2}) with the effective NH Hamiltonian $H_{\text{eff}}$, i.e., $H_{\text{bath}}\to H_{\text{bath}}^{\text{eff}}$, the derivation procedures from Eq.~(\ref{S2}) to Eq.~(\ref{S16}) remain valid.

Therefore, we only focus on studying the effective NH Hamiltonian when we restrict ourselves to the single-excitation subspace. Following Eqs.~(\ref{S17}-\ref{S19}), we express the corresponding NH Bloch Hamiltonian as
\begin{align}
	\widetilde{\textbf{h}}_k=\mqty[-i\kappa_a/2 & f_k\\ f^*_k & -i\kappa_b/2]=\Re[f_k]\sigma_x-\Im[f_k]\sigma_y-i\kappa_-\sigma_z-i\kappa_+\sigma_0,
\end{align}
whose energy dispersion reads $\tilde{\epsilon}_{k,\pm}=-i\kappa_+\pm\tilde{\omega}_k$, where $\tilde{\omega}_k=\customsqrt{\omega_k^2-\kappa_-^2}$ and  $\kappa_\pm=(\kappa_a\pm\kappa_b)/4$. Remarkably, we find that the system exhibits a passive parity-time symmetry\,\cite{PT}
\begin{align}
	\sigma_x \left(\widetilde{\textbf{h}}_k+i\kappa_+\sigma_0\right)^*\sigma_x=\widetilde{\textbf{h}}_k+i\kappa_+\sigma_0
\end{align}
and has two exceptional points (EPs) at $k_{\text{EP}}=\pm\arccos[(\kappa_-^2-2J^2(1+\delta^2))/(2J^2(1-\delta^2))]$ in the Brillouin zone for $\abs{\delta}\le\abs{\kappa_-/(2J)}<1$. Then, by diagonalizing $\widetilde{\textbf{h}}_k$, the effective NH Hamiltonian $H_{\text{bath}}^{\text{eff}}$ in Eq.~(\ref{S21}) can be further written as
\begin{align}\label{S25}
	H_{\text{bath}}^{\text{eff}}=\sum_{k}\mqty[u_{k,L}^\dagger & l_{k,L}^\dagger]\mqty[\tilde{\epsilon}_{k,+} & 0\\ 0 &\tilde{\epsilon}_{k,-}]\mqty[u_{k,R}\\l_{k,R}]=\sum_k[\tilde{\epsilon}_{k,+}u_{k,L}^\dagger u_{k,R}^{}+\tilde{\epsilon}_{k,-}l_{k,L}^\dagger l_{k,R}^{}],
\end{align}
where
\begin{align}
	u_{k,L}^\dagger&=\frac{1}{\sqrt{2}}\left[\frac{\tilde{\omega}_k+i\kappa_-}{\omega_ke^{i\phi_k}}b_k+a_k^\dagger\right],\quad  u_{k,R}=\frac{1}{\sqrt{2}}\left[\frac{\omega_ke^{i\phi_k}}{\tilde{\omega}_k}b_k+\frac{\tilde{\omega}_k- i\kappa_-}{\tilde{\omega}_k}a_k\right],\label{S26-1}\\
	l_{k,L}^\dagger&=\frac{1}{\sqrt{2}}\left[\frac{\tilde{\omega}_k- i\kappa_-}{\omega_ke^{i\phi_k}}b_k-a_k^\dagger\right],\quad\  l_{k,R}=\frac{1}{\sqrt{2}}\left[\frac{\omega_ke^{i\phi_k}}{\tilde{\omega}_k}b_k-\frac{\tilde{\omega}_k+ i\kappa_-}{\tilde{\omega}_k}a_k\right].\label{S26-2}
\end{align}
It is evidently that when $\kappa_a=\kappa_b=0$, we have $u_{k,R}=u_{k,L}=u_k$, $l_{k,R}=l_{k,L}=l_k$, and $\tilde{\epsilon}_{k,\pm}=\epsilon_{k,\pm}$.

\subsection{The calculation of self-energy in the dissipative topological bath}
To evaluate the integral~(\ref{S15}), these expressions~(\ref{S9})-(\ref{S12}), referred to as the self-energies of two quantum emitters, require further computation. When $\kappa_a=\kappa_b=0$, the dissipative bath degenerates into a non-dissipative bath. For the sake of generality, we compute the self-energy directly within the dissipative topological bath. Firstly, we transform $Q$ in Eq.~(\ref{S3}) into momentum space, i.e., 
\begin{align}\label{S27}
	Q=\ketbra{g,g}\otimes\sum_{j}(a_j^\dagger\ketbra{\text{vac}}a_j^{}+b_j^\dagger\ketbra{\text{vac}}b_j^{})=\ketbra{g,g}\otimes\sum_k\left(\ketbra{u_{k,L}}{u_{k,R}}+\ketbra{l_{k,L}}{l_{k,R}} \right),
\end{align}
where 
\begin{align}
	\ket{u_{k,L}}=u_{k,L}^\dagger\ket{\text{vac}},\quad \ket{l_{k,L}}=l_{k,L}^\dagger\ket{\text{vac}},\quad\bra{u_{k,R}}=\bra{\text{vac}}u_{k,R},\quad\bra{l_{k,R}}=\bra{\text{vac}}l_{k,R},
\end{align}
which satisfy $\braket{u_{k,R}}{u_{k',L}}=\braket{l_{k,R}}{l_{k',L}}=\delta_{k,k'}$ and $\braket{u_{k,R}}{l_{k',L}}=\braket{l_{k,R}}{u_{k',L}}=0$. 

For the sake of simplicity, we define a new notation as $\tilde{Q}=\sum_k\left(\ketbra{u_{k,L}}{u_{k,R}}+\ketbra{l_{k,L}}{l_{k,R}} \right)$. According to Eqs.~(\ref{S10}, \ref{S18}, \ref{S25}, \ref{S27}), we have
\begin{align}\label{S28}
	G(x_{1,\alpha}, x_{2,\beta};z)&=\mel{\text{vac}}{o_{x_{1,\alpha}}^{}(z-H_{\text{bath}}^{\text{eff}})^{-1}o_{x_{2,\beta}}^\dagger}{\text{vac}}=\mel{\text{vac}}{o_{x_{1,\alpha}}^{}\tilde{Q}(z-H_{\text{bath}}^{\text{eff}})^{-1}\tilde{Q}o_{x_{2,\beta}}^\dagger}{\text{vac}}\nonumber\\
	&=\mel{\text{vac}}{o_{x_{1,\alpha}}^{}\sum_k\left[\frac{\ketbra{u_{k,L}}{u_{k,R}}}{z+i\kappa_+-\tilde{\omega}_k}+\frac{\ketbra{l_{k,L}}{l_{k,R}}}{z+i\kappa_++\tilde{\omega}_k} \right]o_{x_{2,\beta}}^\dagger}{\text{vac}}.
\end{align}
For $\alpha=\beta=A$, according to Eqs.~(\ref{S26-1}) and (\ref{S26-2}), Eq.~(\ref{S28}) can be further written as
\begin{align}
	G({x_{1,A},x_{2,A};z})&=\frac{1}{N}\sum_{k,k',k''}e^{ik'x_1-ik''x_2}\mel{\text{vac}}{ a_{k'}^{}\left[\frac{\ketbra{u_{k,L}}{u_{k,R}}}{z+i\kappa_+-\tilde{\omega}_k}+\frac{\ketbra{l_{k,L}}{l_{k,R}}}{z+i\kappa_++\tilde{\omega}_k} \right] a_{k''}^\dagger}{\text{vac}}\nonumber\\
	&=\frac{1}{2N}\sum_{k,k',k''}e^{ik'x_1-ik''x_2}\left[\frac{\tilde{\omega}_{k''}-i\kappa_-}{z+i\kappa_+-\tilde{\omega}_k}+\frac{\tilde{\omega}_{k''}+i\kappa_-}{z+i\kappa_++\tilde{\omega}_k} \right]\frac{\delta_{k,k'}\delta_{k,k''}}{\tilde{\omega}_{k''}}\nonumber\\
	&=\frac{1}{N}\sum_k\frac{(z+i\kappa_b/2)e^{ik(x_1-x_2)}}{z_{\text{nh}}^2-\omega_k^2}=\int_{-\pi}^{\pi}\frac{\dd{k}}{2\pi}\frac{(z+i\kappa_b/2)e^{ikd}}{z_{\text{nh}}^2-\omega_k^2},\label{S29}
\end{align}
where $z_{\text{nh}}^2=(z+i\kappa_+)^2+\kappa_-^2$ and $d=x_1-x_2$. Similarly, for $\alpha=\beta=B$, we have
\begin{align}\label{S30}
	G({x_{1,B},x_{2,B};z})=\frac{1}{N}\sum_k\frac{(z+i\kappa_a/2)e^{ikd}}{z_{\text{nh}}^2-\omega_k^2}=\int_{-\pi}^{\pi}\frac{\dd{k}}{2\pi}\frac{(z+i\kappa_a/2)e^{ikd}}{z_{\text{nh}}^2-\omega_k^2},
\end{align}
whereas for other cases of $\alpha$ and $\beta$, we have
\begin{align}\label{S31}
	G(x_{1,A},x_{2,B};z)=\int_{-\pi}^{\pi}\frac{\dd{k}}{2\pi}\frac{\omega_k e^{ikd+i\phi_k}}{z_{\text{nh}}^2-\omega_k^2},\quad G(x_{1,B},x_{2,A};z)=\int_{-\pi}^{\pi}\frac{\dd{k}}{2\pi}\frac{\omega_k e^{ikd-i\phi_k}}{z_{\text{nh}}^2-\omega_k^2}.
\end{align}
By substituting the given dispersion relation, these two integrals in Eqs.~(\ref{S29}) and (\ref{S30}) can be evaluated as
\begin{align}
	G(x_{1,A},x_{2,A};z)&=\int_{-\pi}^{\pi}\frac{\dd{k}}{2\pi}\frac{(z+i\kappa_b/2)e^{ikd}}{z_{\text{nh}}^2-J^2[2(1+\delta^2)+2(1-\delta^2)\cos(k)]}
	=\ointctrclockwise_{\abs{y}=1}\frac{\dd{y}}{2\pi i}\frac{(z+i\kappa_b/2)y^{d}}{J^2(\delta^2-1)(y^2+1)+[z_{\text{nh}}^2-2J^2(1+\delta^2)]y}\nonumber\\
	&=-\frac{(z+i\kappa_b/2)\sum_{p=\pm}p\tilde{y}_p^{\abs{d}}\Theta[p(1-\abs{\tilde{y}_+})]}{\sqrt{z_{\text{nh}}^4-4J^2z_{\text{nh}}^2(1+\delta^2)+16J^4\delta^2}}=G(x_{2,A},x_{1,A};z),\label{S32}\\
	G(x_{1,B},x_{2,B};z)&=\int_{-\pi}^{\pi}\frac{\dd{k}}{2\pi}\frac{(z+i\kappa_a/2)e^{ikd}}{z_{\text{nh}}^2-J^2[2(1+\delta^2)+2(1-\delta^2)\cos(k)]}
	=\ointctrclockwise_{\abs{y}=1}\frac{\dd{y}}{2\pi i}\frac{(z+i\kappa_a/2)y^{d}}{J^2(\delta^2-1)(y^2+1)+[z_{\text{nh}}^2-2J^2(1+\delta^2)]y}\nonumber\\
	&=-\frac{(z+i\kappa_a/2)\sum_{p=\pm}p\tilde{y}_p^{\abs{d}}\Theta[p(1-\abs{\tilde{y}_+})]}{\sqrt{z_{\text{nh}}^4-4J^2z_{\text{nh}}^2(1+\delta^2)+16J^4\delta^2}}=G(x_{2,B},x_{1,B};z),\label{S33}
\end{align}
where 
\begin{align}
	\tilde{y}_\pm=\frac{z_{\text{nh}}^2-2J^2(1+\delta^2)\pm\sqrt{z_{\text{nh}}^4-4J^2z_{\text{nh}}^2(1+\delta^2)+16J^4\delta^2}}{2J^2(1-\delta^2)},\quad \Theta[x]=\begin{cases}
		1 & x\ge0\\
		0 & x<0
	\end{cases}.
\end{align}
Given $f_k=\omega_ke^{i\phi_k}$ and $f^*_k=\omega_ke^{-i\phi_k}$,  these two integrals in Eq.~(\ref{S31}) are evaluated as
\begin{align}
	G(x_{1,A},x_{2,B};z)&=\int_{-\pi}^{\pi}\frac{\dd{k}}{2\pi}\frac{J[1+\delta+(1-\delta)e^{-ik}]e^{ikd}}{z_{\text{nh}}^2-J^2[2(1+\delta^2)+2(1-\delta^2)\cos(k)]}=\ointctrclockwise_{\abs{y}=1}\frac{\dd{y}}{2\pi i}\frac{J[(1+\delta)y^{d}+(1-\delta)y^{d-1}]}{z_{\text{nh}}^2-J^2[2(1+\delta^2)+2(1-\delta^2)\cos(k)]}\nonumber\\
	&=-\frac{J\sum_{p=\pm}pF_{d}(\tilde{y}_p,\delta)\Theta[p(1-\abs{\tilde{y}_+})]}{\sqrt{z_{\text{nh}}^4-4J^2z_{\text{nh}}^2(1+\delta^2)+16J^4\delta^2}}=G(x_{2,B},x_{1,A};z),\label{S35}\\
	G(x_{1,B},x_{2,A};z)&=\int_{-\pi}^{\pi}\frac{\dd{k}}{2\pi}\frac{J[1+\delta+(1-\delta)e^{ik}]e^{ikd}}{z_{\text{nh}}^2-J^2[2(1+\delta^2)+2(1-\delta^2)\cos(k)]}=\ointctrclockwise_{\abs{y}=1}\frac{\dd{y}}{2\pi i}\frac{J[(1+\delta)y^{d}+(1-\delta)y^{d+1}]}{z_{\text{nh}}^2-J^2[2(1+\delta^2)+2(1-\delta^2)\cos(k)]}\nonumber\\
	&=-\frac{J\sum_{p=\pm}pF_{d+1}(\tilde{y}_p,-\delta)\Theta[p(1-\abs{\tilde{y}_+})]}{\sqrt{z_{\text{nh}}^4-4J^2z_{\text{nh}}^2(1+\delta^2)+16J^4\delta^2}}=G(x_{2,A},x_{1,B};z),\label{S36}
\end{align}
where $F_{d}(\tilde{y}_p,\delta)=(1+\delta)\tilde{y}_p^{\abs{d}}+(1-\delta)\tilde{y}_p^{\abs{d-1}}$. Note that in the last step of Eqs.~(\ref{S35}) and (\ref{S36}), we employ the following relation, i.e., $F_{d}(\tilde{y}_p,\delta)=F_{-d+1}(\tilde{y}_p,-\delta)$.
\subsection{The calculation of probability amplitude for quantum battery}
To further determine the dynamics of the QB, i.e., evaluating the integral in Eq.~(\ref{S15}), we can utilize the residue theorem by closing the contour in the lower half of the complex plane, as illustrated in Fig.\,\ref{fig2}(b). Since the presence of sublattices dissipation makes the distribution of band regions, namely branch cuts, in the complex plane exceedingly intricate in certain cases, the subsequent calculations will focus solely on the non-dissipative SSH bath (see Sec.~\ref{IIA}). Accordingly, Fig.\,\ref{fig2}(b) should be slightly modified for the non-dissipative bath as follows: (1) The dissipative bound-state energies should be replaced by non-dissipative bound-state energies; (2) The branch cuts should be shifted to the real axis, $\Re[z]$, since in the absence of dissipation, the spectrum of SSH bath are real-valued.

Let us now focus on the integral in Eq.~(\ref{S15}). In addition to the poles corresponding to the bound-state energies, the integrand also has branch cuts along the real axis, specifically in the regions $z\in[-2J, -2J|\delta|]\cup [2J|\delta|, 2J]$, which correspond to the continuous spectrum of $H_{\rm bath}$. Therefore, according to the residue theorem, the integral can be simplified to handling the poles and branch cuts. The treatment of the poles is straightforward: one simply needs to compute the residue of the integrand at these poles. However, the handling of the branch cuts is more complex, and here we present two main approaches. The first method is to choose a small contour that tightly encloses the branch cuts (as shown by the small contours around the branch cuts in Fig.\,\ref{fig2}(b)) and directly integrate along this contour. The second method is detour at the band edges to other Riemann sheets of the integrand, as depicted by the branch cut detours in the modified Fig.\,\ref{fig2}(b). Thus, the integrand may have unstable poles in the second Riemann sheet. As a result, the contribution from the branch cuts will be divided into the contributions from the unstable poles and the branch cut detours. In the following derivations, we prioritize the second method, while the first method mainly applies to strongly dissipative SSH bath. For convenience, we introduce a new notation, $\mathscr{C}(z)$, to represent the integrand excluding the term $\exp(-izt)$. The analytical expressions for the self-energies presented in Eqs.(\ref{S32}), (\ref{S33}), (\ref{S35}), and (\ref{S36}) correspond to the integrand in the first Riemann sheet, denoted $\mathscr{C}^{\mathrm{i}}(z)$. We can analytically continue this expression to the second Riemann sheet, $\mathscr{C}^{\mathrm{ii}}(z)$ (depicted by the brown areas), by replacing $\Theta[p(\cdots)]$ with $\Theta[-p(\cdots)]$ in the self-energy expressions. Consequently, the integrand in the first and second Riemann sheets is
\begin{equation}\label{S37}
    \setlength\abovedisplayskip{6pt}
    \setlength\belowdisplayskip{4pt}
	\mathscr{C}^{\mathrm{i}/\mathrm{ii}}(z)=\frac{\mathrm{g}^2G^{\mathrm{i}/\mathrm{ii}}(x_{1,\alpha},x_{2,\beta};z)+\Omega_{12}^{\alpha\beta}}{\mathscr{D}^{\mathrm{i}/\mathrm{ii}}(z)}
\end{equation}
with 
\begin{align}
	G^{\color{blue}{\mathrm{i}/\mathrm{ii}}}(x_{1,A},x_{2,A};z)&=-\frac{z\sum_{p=\pm}p{y}_{ p}^{\abs{x_1-x_2}}\Theta[{\color{blue}{\pm}}p(1-|{y}_+|)]}{\sqrt{z^4-4J^2z^2(1+\delta^2)+16J^4\delta^2}}=G^{\color{blue}{\mathrm{i}/\mathrm{ii}}}(x_{1,B},x_{2,B};z),\label{G11}\\
	G^{\color{blue}{\mathrm{i}/\mathrm{ii}}}(x_{1,A},x_{2,B};z)&=-\frac{J\sum_{p=\pm}pF_{x_1-x_2}({y}_{p}, \delta)\Theta[{\color{blue}{\pm}}p(1-|{y}_+|)]}{\sqrt{z^4-4J^2z^2(1+\delta^2)+16J^4\delta^2}},\label{G12}
\end{align}
where ${y}_\pm=\tilde{y}_\pm|_{z_{\mathrm{nh}}\to z}$ representing the non-dissipative SSH bath. Here, since $\Sigma_{11}^{\alpha\alpha}(z)=\Sigma_{11}^{\beta\beta}(z)$ and $\Sigma_{12}^{\alpha\beta}(z)=\Sigma_{21}^{\beta\alpha}(z)$, the denominator in Eq.~(\ref{S37}) can be further simplified as
\begin{align}\label{key}
	\mathscr{D}^{\mathrm{i}/\mathrm{ii}}(z)=[z-\Delta-\mathrm{g}^2G^{\mathrm{i}/\mathrm{ii}}(x_{1,\alpha}, x_{1,\alpha};z)]^2-[\Omega_{12}^{\alpha\beta}+\mathrm{g}^2G^{\mathrm{i}/\mathrm{ii}}(x_{1.\alpha},x_{2,\beta};z)]^2,
\end{align}
where $\Delta=\omega_e-\omega_c$ is the detuning between the atomic transition frequency and the cavity resonance frequency. On the one hand, since the imaginary part of $\mathscr{C}^\mathrm{i}(z+i0^+)$ and $\mathscr{C}^\mathrm{ii}(z-i0^+)$ is nonzero in the band regions, we should only take into account the real poles (i.e., the roots of $\mathscr{D}^{\mathrm{i}}(z)=0$) of $\mathscr{C}^\mathrm{i}(z)$ outside the band regions, corresponding to the bound-state energies (BSEs), and the complex poles (i.e., the roots of $\mathscr{D}^{\mathrm{ii}}(z)=0$) of $\mathscr{C}^\mathrm{ii}(z)$ with real part inside band regions, corresponding to the unstable poles (UPs). On the other hand, aside from the integral path $\mathcal{C}$ (which corresponds to Eq.~(\ref{S15})) and the semicircular path $C_R$ (which vanished as the radius of the semicircle approaches infinity, according to Jordan's lemma), we need to add eight additional integral paths parallel to the imaginary axis, corresponding to the branch cut detours (BCDs), so that these paths form a closed loop on the complex plane, as shown in the modified Fig.\,\ref{fig2}(b). According to the residue theorem, the sum of the integrals along these paths should equal the sum of the residues at the aforementioned poles. Thus, we have\,\cite{pa}
\begin{align}\label{S41}
	c_{\mathrm{B}}(t)&=\sum_{z_k\in \mathrm{BSEs}}\Res[\mathscr{C}^\mathrm{i}(z), z_k]e^{-iz_kt}+\sum_{z_k\in \mathrm{UPs}}\Res[\mathscr{C}^\mathrm{ii}(z), z_k]e^{-iz_kt}-\sum_{k=1}^{4}C_{\mathrm{BCD}}^{(k)}(t).
\end{align}
For simplicity, we introduce the notations $c_{\rm B}^{\rm BS}(t)$, $c_{\rm B}^{\rm UP}(t)$, and $c_{\rm B}^{\rm BCD}(t)$ to represent the contributions from the bound states (BSs), UPs, and BCDs, respectively. Therefore, the probability amplitude of the QB can be expressed as $c_{\rm B}(t)=c_{\rm B}^{\rm BS}(t)+c_{\rm B}^{\rm UP}(t)+c_{\rm B}^{\rm BCD}(t)$, with each term given explicitly by Eq.~(\ref{S41}). Note that the contributions in the last two terms arise from the branch cuts (BCs), which are denoted as $c_{\rm B}^{\rm BC}(t)=c_{\rm B}^{\rm UP}(t)+c_{\rm B}^{\rm BCD}(t)$, and the last term specifically represents the contributions from the BCDs, which can be computed as
\begin{align}\label{S42}
	C_{\text{BC},k}(t)&=(-1)^{k}\int_{0}^{\infty}\frac{\dd y}{2\pi}\left[\mathscr{D}^{\text{i}}(c_k-iy)-\mathscr{D}^{\text{ii}}(c_k-iy) \right]e^{-ic_kt-yt},
\end{align}
where $c_1=2J$, $c_2=2J\abs{\delta}$, $c_3=-2J\abs{\delta}$, and $c_4=-2J$, corresponding to the four band edges. Based on the form of Eq.~(\ref{S41}), we note that the QB dynamics is fully described by the contributions from BSEs, UPs, and BCDs. Given that the integrand in Eq.~(\ref{S42}) contains $\exp(-yt)$ with $y\ge0$, the contribution from BCDs evidently diminishes as time increases. Besides, since the imaginary parts of the UPs are negative, their contribution also decay over time. Therefore, only the bound state contributions survial in the long-time limit $t\gg 1/\text{g}$, i.e.,
\begin{equation}\label{LTL}
    \setlength\abovedisplayskip{6pt}
    \setlength\belowdisplayskip{4pt}
	c_{\mathrm{B}}(\infty)\equiv\lim\limits_{t\gg 1/\text{g}} c_{\mathrm{B}}(t)=\sum_{z_k\in \mathrm{BSEs}}\Res[\mathscr{C}^\mathrm{i}(z), z_k]e^{-iz_kt}=\lim\limits_{t\gg 1/\text{g}} c_{\mathrm{B}}^{\rm BS}(t).
\end{equation}

\begin{figure}
	\centering
	\includegraphics[width=18cm]{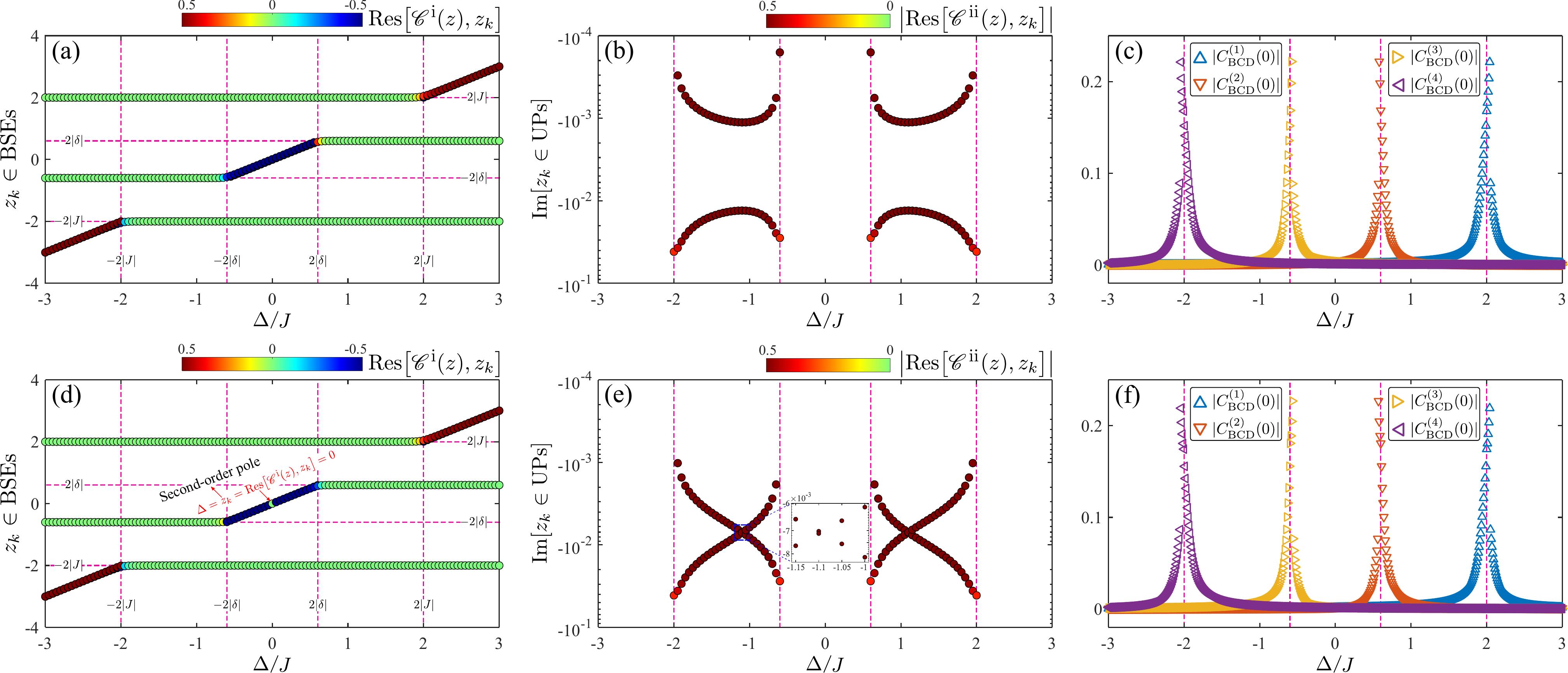}\\
	\caption{First row (a)-(c): topologically nontrivial phase ($\delta=-0.3$). Second row (d)-(f): topologically trivial phase ($\delta=0.3$). Panels (a) and (d) describe the bound-state energies $z_k$ and its corresponding residue $\Res[\mathscr{C}^{\mathrm{i}}(z),z_k]$ as a function of the emitter detuning $\Delta$. Panels (b) and (e) describe the imaginary part of the unstable poles $\Im[z_k]$ and the absolute value of its corresponding residue $\big|\Res[\mathscr{C}^{\mathrm{ii}}(z),z_k]\big|$ as a function of the emitter detuning $\Delta$. Panels (c) and (f) describe the contributions from the branch cut detours $|C_{\text{BCD}}^{(k)}(0)|$ at time $t=0$ as a function of the detuning $\Delta$. The system parameters are $\mathrm{g}/J=0.1$, $\alpha=A$, $\beta=B$, and $d=x_1-x_2=1$.
	}\label{fig3}
\end{figure}

Here, we show how the contributions from the bound states (corresponding to residues at the real poles), the unstable poles (corresponding to absolute values of residues at the complex poles), and the branch cut detours (corresponding to absolute values given by Eq.~(\ref{S42})) vary with the detuning $\Delta$ in both topologically trivial and non-trivial phases. First, as shown in Figs.\,\ref{fig3}(a) and~\ref{fig3}(d), the number of bound states and their corresponding residues vary similarly with detuning in both topologically trivial and non-trivial phases, except at zero detuning (i.e., $\Delta=0$). However, when $\Delta=0$, the two non-zero BSEs in the topologically nontrivial phase, with residues approaching $\pm0.5$ and opposite in sign, will merge into a degenerate zero-energy bound state in the topologically trivial phase, where its residue equals zero. Moreover, the contributions from the bound states become most significant only when the detuning falls within the band gap regions, and the number of bound states also increases by one compared to those within the band regions, except at $\Delta=0$ in the topologically trivial phase. Second, as shown in Figs.\,\ref{fig3}(b) and~\ref{fig3}(e), the UPs appear in pairs only when the detuning is within the band regions, and the effective dissipation (i.e., $-\Im[z_k]$) is significantly enhanced in the topologically trivial phase compared to the topologically non-trivial phase. Finally, as shown in Figs.\,\ref{fig3}(c) and~\ref{fig3}(f), the contributions from the BCDs at time $t=0$ is almost identical in both topologically trivial and non-trivial phases and is most significant only when the detuning is near the band edges. 
\section{Quantum Battery Performance in Different Configurations}
To support the phase boundaries outlined in Eq.(10) and the maximum stored energy described in Eq.(11) of the main text, we provide detailed derivations and discussions in the following subsections: subsection A outlines the relationships between various indicators used to characterize the performance of QB, subsection B covers the phase diagram of QB, subsection C addresses the dissipation immunity of QB, and subsection D demonstrates how environmental dissipation can be utilized to enhance QB performance over a short time.

\subsection{The relationships among key performance indicators of quantum batteries}\label{IIIAn}
In the following, we will provide a detailed derivation of the general analytical expressions for various QB performance indicators within our system, specifically including stored energy, ergotropy, the ratio between ergotropy and stored energy, and charging power, and reveal their interrelationships. 

\begin{figure}
	\centering
	\includegraphics[width=17cm]{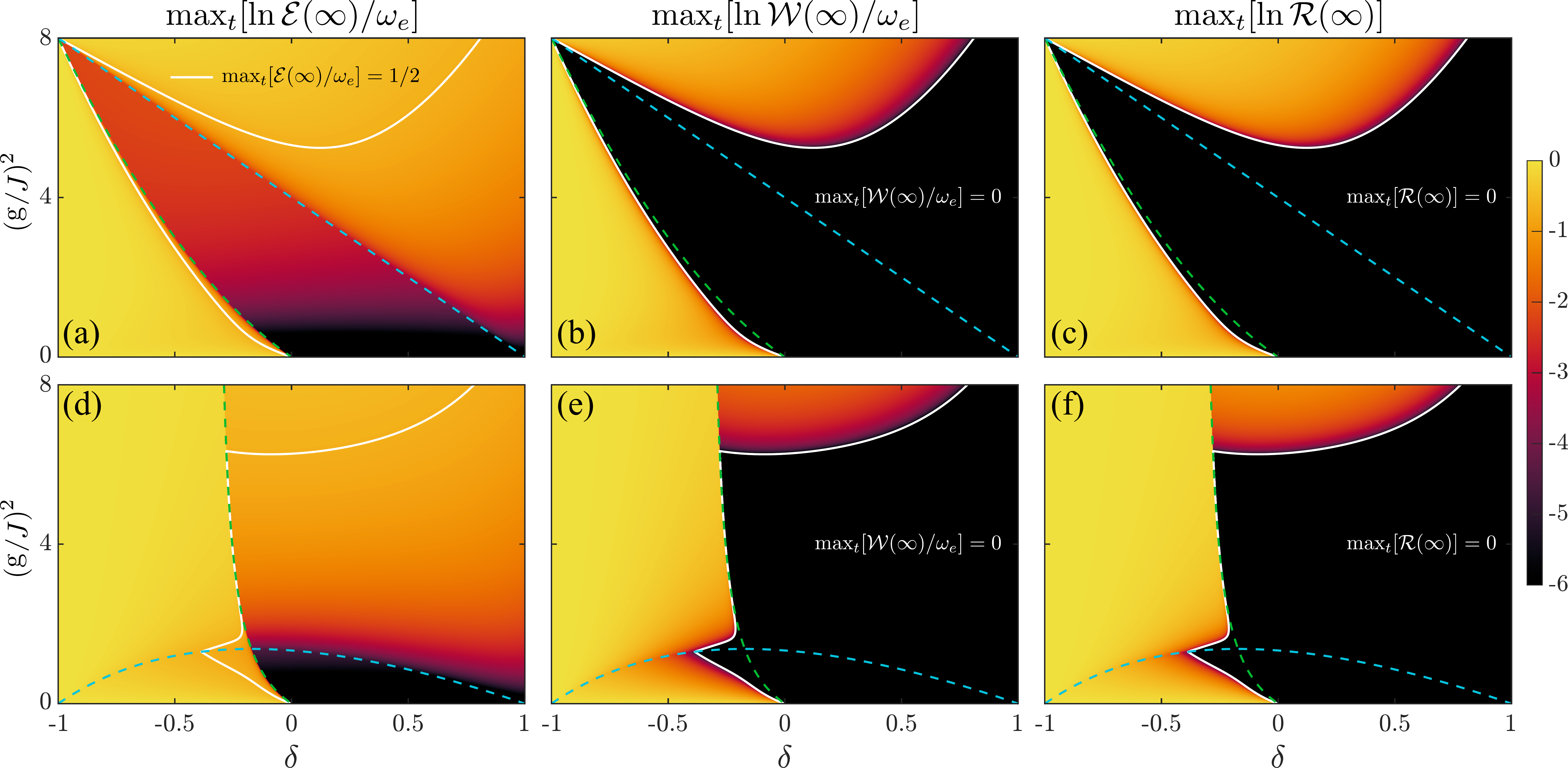}\\
	\caption{Panels (a, d), (b, e), and (c, f) depict the maximum stored energy, maximum ergotropy, and maximum ergotropy-to-stored-energy ratio as a function of the dimerization strength $\delta$ and the coupling strength $\text{g}$, respectively. The plots in the first and second rows correspond to $d=1$ and $d=2$, respectively. The solid and dashed lines represent $\max[\mathcal{E}(\infty)]/\omega_e=1/2$ and the phase boundaries for the maximum stored energy, respectively. All parameters are the same as in Fig.\,2 of the main text.
	}\label{fign4}
\end{figure}

Here, the definition of the stored energy for QB is given by $\mathcal{E}(t)=\Tr[\rho_\mathrm{B}(t)H_{\mathrm{B}}]$, where $H_\mathrm{B}=\omega_e\sigma_+^\mathrm{B}\sigma_-^\mathrm{B}$ and $\rho_{\text{B}}(t)$ are the free Hamiltonian and the reduced density matrix of QB, respectively. For the system described in Section \ref{I}, the reduced density matrix of QB can be obtained from the steps of Eqs.~(\ref{S2}-\ref{S16}), and the matrix representation for the free Hamiltonian and reduced density matrix of QB can be written as
\begin{align}\label{IIIAn-1}
	H_{\rm B}\coloneqq\omega_e\begin{bmatrix}
		0&0\\0&1
	\end{bmatrix},\quad\rho_{\rm B}(t)\coloneqq\begin{bmatrix}
	1-\abs{c_{\text{B}}(t)}^2 & 0\\ 0 & \abs{c_{\text{B}}(t)}^2
	\end{bmatrix}.
\end{align}
By plugging Eq.~(\ref{IIIAn-1}) into the definition of the stored energy, we have
\begin{equation}\label{IIIAn-2}
    \setlength\abovedisplayskip{6pt}
    \setlength\belowdisplayskip{6pt}
	\mathcal{E}(t)=\Tr[\rho_{\rm B}(t)H_{\rm B}]=\omega_e\abs{c_{\text{B}}(t)}^2,
\end{equation}
and the charging power is also computed as
\begin{align}\label{IIIAn-3}
	P(t)=\mathcal{E}(t)/t=\omega_e\abs{c_{\text{B}}(t)}^2/t.
\end{align}
Besides, the definition of ergotropy is given by $\mathcal{W}(t)=\Tr[\rho_{\rm B}(t)H_{\rm B}]-\Tr[\tilde{\rho}_{\rm B}(t)H_{\rm B}]$, where $\tilde{\rho}_{\rm B}(t)=\sum_sr_s(t)\ketbra{\varepsilon_s}$ is the passive state, $r_s(t)$ are the eigenvalues of $\rho_{\rm B}(t)$ arranged in descending order, while $\ket{\varepsilon_s}$ are the eigenstates of $H_{\rm B}$ with the corresponding eigenvalues $\varepsilon_s$ sorted in ascending order. Since $\rho_{\rm B}(t)$ and $H_{\rm B}$ both are diagonal matrices, the matrix representation of the passive state can be readily solved, i.e.,
\begin{align}\label{IIIAn-4}
	\tilde{\rho}_{\rm B}(t)\coloneqq\begin{bmatrix}
		\abs{c_{\text{B}}(t)}^2 & 0\\ 0 & 1-\abs{c_{\text{B}}(t)}^2
	\end{bmatrix}\ \text{for}\ \abs{c_{\text{B}}(t)}^2>\frac{1}{2},\quad \tilde{\rho}_{\rm B}(t)\coloneqq\begin{bmatrix}
	1-\abs{c_{\text{B}}(t)}^2 & 0\\ 0 & \abs{c_{\text{B}}(t)}^2
	\end{bmatrix}\ \text{for}\ \abs{c_{\text{B}}(t)}^2<\frac{1}{2}.
\end{align}
By plugging Eq.~(\ref{IIIAn-4}) into the definition of ergotropy, we have
\begin{align}\label{IIIAn-5}
	\mathcal{W}(t)=\Tr[\rho_{\rm B}(t)H_{\rm B}]-\Tr[\tilde{\rho}_{\rm B}(t)H_{\rm B}]=2\omega_e[\abs{c_{\rm B}(t)}^2-1/2]\Theta[\abs{c_{\rm B}(t)}^2-1/2].
\end{align}
Based on Eqs.~(\ref{IIIAn-2}) and (\ref{IIIAn-5}), the ratio between ergotropy and stored energy can be written as
\begin{equation}\label{IIIAn-6}
    \setlength\abovedisplayskip{6pt}
    \setlength\belowdisplayskip{6pt}
	\mathcal{R}(t)=\mathcal{W}(t)/\mathcal{E}(t)=[2-1/\abs{c_{\rm B}(t)}^{2}]\Theta[\abs{c_{\rm B}(t)}^2-1/2].
\end{equation}
Finally, according to these analytical expressions (\ref{IIIAn-2}, \ref{IIIAn-3}, \ref{IIIAn-5}, \ref{IIIAn-6}), we find that charging power, ergotropy, and the ratio can each be expressed as functions of the stored energy. This suggests that the stored energy indicator serves as a comprehensive proxy, effectively capturing the behavior of other QB performance indicators. As illustrated in Fig.\,\ref{fign4}, we can observe that in regions where $\max_t[\mathcal{E}(\infty)/\omega_e]\sim 1$, the trends in maximum ergotropy and the ratio closely follow that of the maximum stored energy, further validating this point. Notably, in our model, the maximum stored energy shows phase transition-like behavior, as illustrate in Figs.\,\ref{fign4}(a, d). However, for indicators such as ergotropy and the ratio, this behavior is largely eliminated by the presence of a Heaviside's step function, i.e., $\Theta[\abs{c_{\rm B}(t)}^2-1/2]$, as the maximum population $\max_t\abs{c_{\rm B}(\infty)}^2$ does not necessarily exceed $1/2$ at the phase boundaries. Specifically, in Figs.\,\ref{fign4}(b, c) for $d=1$, the region inside the white curve ($\max_t[\mathcal{E}(\infty)/\omega_e]<1/2$) fully encompasses both phase boundaries, thereby eliminating the phase-transition-like behavior in the maximum ergotropy and the ratio. By contrast, in Figs.\,\ref{fign4}(e, f) for $d=2$, the region inside the white curve does not fully cover the phase boundaries, allowing the maximum ergotropy and the ratio to exhibit phase transition-like behavior at the phase boundaries.

\subsection{Phase diagram of quantum battery}\label{IIIA}
\begin{figure}[b]
	\centering
	\includegraphics[width=17cm]{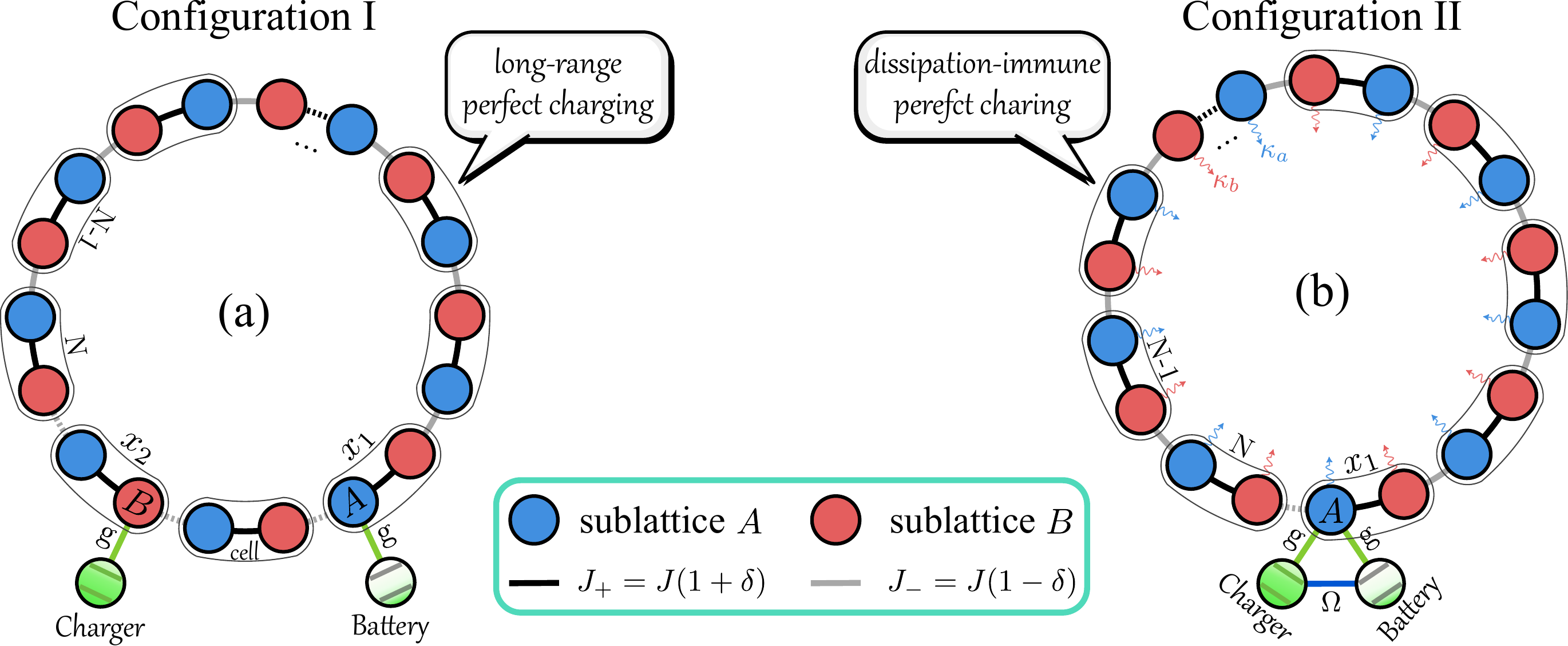}\\
	\caption{(a) Configuration I: The quantum charger and the QB without direct interaction ($\Omega_{12}^{\alpha\beta}=0$), are coupled to sublattice $B$ of unit cell $x_2$ and sublattice $A$ of unit cell $x_1$, respectively. (b) Configuration II: the charger and the QB with direct interaction ($\Omega_{12}^{\alpha\beta}=\Omega$), are coupled to the same sublattice $A$ of unit cell $x_1=x_2$. Here, configuration I features long-range perfect charging in the non-dissipative topological waveguide, i.e., $\kappa_a=\kappa_b=0$, while configuration II supports perfect charging in the dissipative topological waveguide, i.e., $\kappa_a\neq0$ and $\kappa_b\neq0$. In the following calculation, we focus on the single-lattice dissipation (i.e., $\kappa_a\neq0$ and $\kappa_b=0$) in configuration II.
	}\label{fig5}
\end{figure}
{\color{black}According to Eq.~(\ref{LTL}), in the long-time limit, only the bound states contribute to the probability amplitude, which, combined with Eq.~(\ref{IIIAn-2}), yields $\mathcal{E}(\infty)=\omega_e\abs{c_{\rm B}(\infty)}^2$.} Thus, to determine the maximum stored energy $\max_t[\mathcal{E}(\infty)]$, we need to identify the poles within the band-gap regions of the integrand in Eq.~(\ref{S15}) by solving the poles equation
\begin{align}\label{S47}
	\mathscr{D}(E_i)=[E_i-\Delta-\Sigma_{11}^{\alpha\alpha}(E_i)][E_i-\Delta-\Sigma_{22}^{\beta\beta}(E_i)]-[\Omega_{12}^{\alpha\beta}+\Sigma_{12}^{\alpha\beta}(E_i)][\Omega_{12}^{\alpha\beta}+\Sigma_{21}^{\beta\alpha}(E_i)]=0.
\end{align}
Subsequently, we focus on the system presented in Fig.2(a-d) of the main text, as depicted by the configuration I in Fig.\,\ref{fig5}(a). The corresponding total Hamiltonian is denoted as
\begin{align}
	H_{\text{tot}}=\omega_e\sigma_+^{\text{B}}\sigma_-^{\text{B}}+\omega_e\sigma_+^{\text{C}}\sigma_-^{\text{C}}+H_{\text{bath}}+\mathrm{g}(\sigma_+^\mathrm{B} o_{x_{1,A}}+\sigma_+^\mathrm{C} o_{x_{2,B}}+\text{H.c.}),
\end{align}
where $H_{\text{bath}}$ is given by Eq.~(\ref{S17}). For simplicity, we assume that the quantum charger is always positioned on the left side of the QB, which implies $d=x_1-x_2\in\mathbb{Z}^{+}$. According to these expressions of the self-energies in Eqs.~(\ref{S32}, \ref{S33}, \ref{S35}, \ref{S36}) without environmental dissipation (i.e., $\kappa_a=\kappa_b=0$), the poles equation~(\ref{S47}) can be further simplified as
\begin{align}
	\mathscr{D}(E_i)=[E_i-\Delta-\mathrm{g}^2G(x_{1,A},x_{1,A};E_i)]^2-[\mathrm{g}^2G(x_{1,A},x_{2,B};E_i)]^2=\mathrm{g}^4\prod_{p=\pm}[(E_i-\Delta)/\mathrm{g}^2-G_{p}(E_i)]=0,
\end{align}
where 
\begin{align}
	G_{{\color{blue}{\pm}}}(E_i)=G(x_{1,A},x_{1,A};E_i){\color{blue}{\pm}} G(x_{1,A},x_{2,B};E_i)=-\frac{\sum_{p=\pm}[z{\color{blue}{\pm}} JF_{d}({y}_{p}, \delta)]p\Theta[p(1-|{y}_+|)]}{\sqrt{z^4-4J^2z^2(1+\delta^2)+16J^4\delta^2}}\bigg|_{z=E_i},
\end{align}
which satisfies $G_\pm(E_i)=-G_\mp(-E_i)$ since both $y_\pm$ are even function with respect to $z$. To solve this poles equation, we first need to analyze the characteristics of $G_\pm(E_i)$ within the band-gap region and at the band edges. This analysis will help us determine the number of intersection points between $G_\pm(E_i)$ and $(E_i-\Delta)/\mathrm{g}^2$, i.e., the number of bound states. For simplicity, we divide the bandgap into three intervals, i.e., $\mathcal{R}_l=(-\infty, -2J)$, $\mathcal{R}_m=(-2J\abs{\delta},2J\abs{\delta})$, and $\mathcal{R}_r=(2J,+\infty)$. It is not difficult to observe that when $z\in \mathcal{R}_m$, $-1<y_+<0$, while when $z\in\mathcal{R}_l\cup\mathcal{R}_r$, $y_+>1$.

\begin{figure}
	\centering
	\includegraphics[width=17cm]{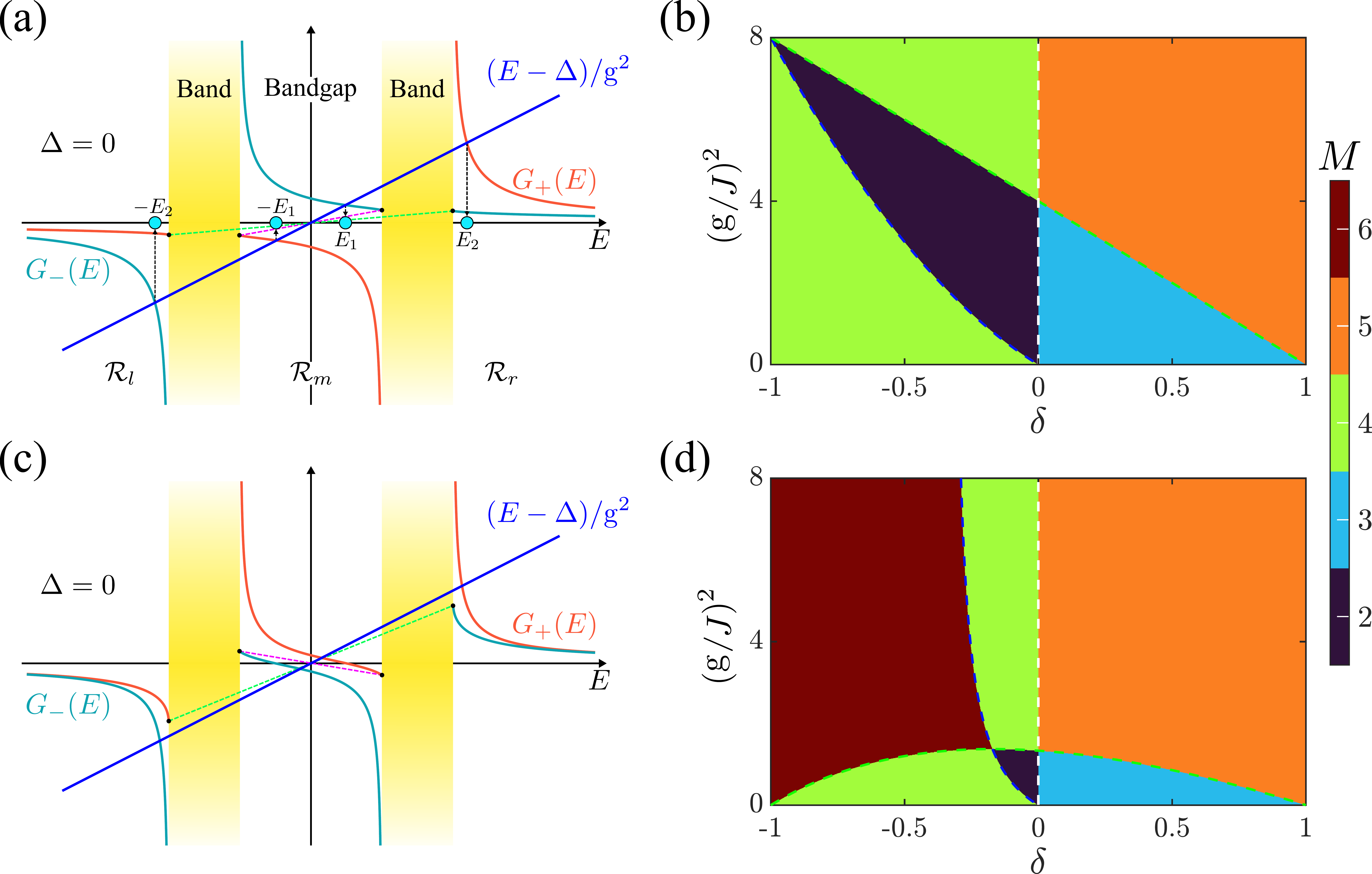}\\
	\caption{Panels (a) and (c) describe the difference $G_\pm(E)$ as a function of $E$ in the SSH bath within the topologically nontrivial phase, for $d=1$ and $d=2$ respectively. Roots of the poles equation (i.e., the BSEs) are obtained from the intersection points between $(E-\Delta)/\mathrm{g}^2$ (blue solid line) and $G_\pm(E)$ (teal and orange solid lines). Panels (b) and (d) show the number of bound states $M$ as a function of the dimerization parameter $\delta$ and the atom-bath coupling strength $\mathrm{g}$, for $d=1$ and $d=2$ respectively. The blue and green dashed lines are given by Eq.~(\ref{PB}). All parameters are the same as in Fig.2 of the main text.
	}\label{fig6}
\end{figure}

First, we will start our analysis with the middle region of the bandgap $\mathcal{R}_m$, and we have
\begin{align}\label{S51}
	G_\pm(z)=\frac{-[z\pm JF_{d}({y}_{+},\delta)]}{\sqrt{z^4-4J^2z^2(1+\delta^2)+16J^4\delta^2}}=-\frac{z\pm J[(1+\delta)y_+^{\abs{d}}+(1-\delta)y_+^{\abs{d-1}}]}{\sqrt{z^4-4J^2z^2(1+\delta^2)+16J^4\delta^2}}=-G_\mp(-z),\quad {\rm for}\ z\in \mathcal{R}_m.
\end{align}
According to Eq.~(\ref{S51}), we only need to analyze the behavior of $G_+(z)$ because $G_+$ and $G_-$ are mutually symmetric with respect to the ordinate origin, as shown in Figs.\,\ref{fig6}(a) and ~\ref{fig6}(c). Additionally, we find that $G_+(z)$ is a monotonic function with respect to $z$ in the interval $\mathcal{R}_m$, i.e., $[\mathrm{d}G_{+}/\mathrm{d}z]|_{z\in\mathcal{R}_m}<0$. Finally, determining whether the number of bound states can be change abruptly mainly depends on the behavior (divergence or convergence) of $G_\pm(z)$ at the band edges. As $z\to \pm 2J\abs{\delta}$, the denominator of $G_\pm(z)$ evidently approaches zero with a behavior proportional to $\sqrt{2J\abs{\delta}\pm z}$. Thus, we only need to analyze the Taylor expansion of $F_{d}(y_+,\delta)$ at $z=\pm2J\abs{\delta}$, i.e., 
\begin{align}\label{S52}
	F_{d}(y_+,\delta)|_{z=\pm2J\abs{\delta}}=2(-1)^d\{\delta-\sqrt{\frac{\abs{\delta}/J}{1-\delta^2}}[\delta(\abs{d}+\abs{d-1})+\abs{d}-\abs{d-1}](2J\abs{\delta}\mp z)^{\frac{1}{2}}\}+\order{2J\abs{\delta}\mp z}.
\end{align}
By plugging Eq.~(\ref{S52}) into $G_+(z)$ in Eq.~(\ref{S51}), we have
\begin{align}\label{S53}
	\lim\limits_{z\to\pm2J\abs{\delta}}G_+(z)=\frac{-1}{2J\sqrt{1-\delta^2}}\lim\limits_{z\to\pm2J\abs{\delta}}\frac{z+2J(-1)^d\delta}{\sqrt{4J^2\delta^2-z^2}}+\mathcal{G}_1(d,\delta),
\end{align}
where, for later convenience, we define 
\begin{align}
	\mathcal{G}_1(d,\delta)=\frac{(-1)^{d+\Theta[-d]}}{2J(1-\delta^2)}[1-(1-2d)\delta].
\end{align}
For the first term in Eq.~(\ref{S53}), it is evident that different values of $d$ and $\delta$ will yield distinctly different results, i.e., 
\begin{align}
	\lim\limits_{z\to-2J\abs{\delta}}\frac{z+2J(-1)^d\delta}{\sqrt{4J^2\delta^2-z^2}}=\begin{cases}
		-\infty & \Theta[(-1)^d\delta]=0\\
		0 & \Theta[(-1)^d\delta]=1
	\end{cases},\qquad \lim\limits_{z\to2J\abs{\delta}}\frac{z+2J(-1)^d\delta}{\sqrt{4J^2\delta^2-z^2}}=\begin{cases}
		0 & \Theta[(-1)^d\delta]=0\\
		+\infty & \Theta[(-1)^d\delta]=1
	\end{cases},
\end{align}
which result in
\begin{align}\label{GpRm}
	\lim\limits_{z\to-2J\abs{\delta}}G_+(z)=\begin{cases}
		+\infty & \Theta[(-1)^d\delta]=0\\
		\mathcal{G}_1(d,\delta) & \Theta[(-1)^d\delta]=1
	\end{cases},\qquad \lim\limits_{z\to2J\abs{\delta}}G_+(z)=\begin{cases}
		\mathcal{G}_1(d,\delta) & \Theta[(-1)^d\delta]=0\\
		-\infty & \Theta[(-1)^d\delta]=1
	\end{cases}.
\end{align}
In addition, the difference $G_+(z)-G_-(z)$ also plays an important role in the analysis of the aforementioned intersection points. According to $\max(y_+)=y_+|_{z=0}=[(\delta-1)/(\delta+1)]^{\mathrm{sign}(\delta)}$ and Eq.~(\ref{S51}), the difference is computed as
\begin{align}
	\text{sign}[ G_+(z)-G_-(z)]=\text{sign}[G_+(z)+G_+(-z)]=\text{sign}\left\{\frac{-2J[(1+\delta)y_+^{\abs{d}}+(1-\delta)y_+^\abs{d-1}]}{\sqrt{z^4-4J^2z^2(1+\delta^2)+16J^4\delta^2}}\right\}=(-1)^{(d-1)}\text{sign}(\delta).
\end{align}
Therefore, within the band-gap region $\mathcal{R}_m$, $G_+(z)$ is consistently positioned either entirely above or below $G_-(z)$, and only when $(d\ge1, \delta>0)$ or $(d\le0, \delta<0)$, these two functions have a unique intersection point, i,e., $G_\pm(0)=0$.

Second, for the other two band-gap regions, $\mathcal{R}_l$ and $\mathcal{R}_r$, we also have
\begin{align}\label{S55}
	G_\pm(z)=\frac{z\pm JF_{d}({y}_{-},\delta)}{\sqrt{z^4-4J^2z^2(1+\delta^2)+16J^4\delta^2}}=\frac{z\pm J[(1+\delta)y_-^{\abs{d}}+(1-\delta)y_-^{\abs{d-1}}]}{\sqrt{z^4-4J^2z^2(1+\delta^2)+16J^4\delta^2}},\quad {\rm for}\ z\in \mathcal{R}_l\cup\mathcal{R}_r.
\end{align}
Similarly, we only need to analyze the behavior of $G_\pm(z)$ in one of these regions due to $G_\pm(z)=-G_\mp(-z)$. As a consequence, we focus on the case of $z\in\mathcal{R}_r$. We find that $G_{\pm}(z)$ both are a monotonic function in the interval $\mathcal{R}_l$, i.e., $[\mathrm{d}G_{\pm}/\mathrm{d}z]|_{z\in\mathcal{R}_r}<0$, and the difference $G_{+}(z)-G_-(z)$ is always more than zero, i.e., $G_+(z)>G_-(z)$ for $z\in\mathcal{R}_r$. It is evident that when $z\to+\infty$, $G_{\pm}(z)$ both approach zero. As $z\to 2J$, the denominator of $G_{\pm}(z)$ evidently approaches zero with a behavior proportional to $\sqrt{z-2J}$. Thus, the Taylor expansion of $F_{d}(y_-,\delta)$ at $z=2J$ is 
\begin{align}\label{S56}
	F_{d}(y_-,\delta)|_{z=2J}=2-\frac{2[\delta(\abs{d}-\abs{d-1})+\abs{d}+\abs{d-1}]}{\sqrt{J(1-\delta^2)}}(z-2J)^{\frac{1}{2}}+\order{z-2J}.
\end{align}
By plugging Eq.~(\ref{S56}) into $G_\pm(z)$ of Eq.~(\ref{S55}), we have
\begin{align}\label{GpRlr}
	\lim\limits_{z\to 2J}G_+(z)=+\infty,\quad \lim\limits_{z\to 2J}G_-(z)=\mathcal{G}_2(d,\delta)\equiv\frac{(-1)^{\Theta[-d]}[\delta-(1-2d)]}{2J(1-\delta^2)}>0.
\end{align}

Now, as shown in Figs.\,\ref{fig6}(a) and~\ref{fig6}(c), for $\Delta=0$, we find that the critical point, where the number of intersection points (BSEs) between $(E-\Delta)/\mathrm{g}^2$ and $G_\pm(E)$ undergoes an abrupt change, occurs then the slope of the blue solid line equals the slope the pink dashed line or the green dashed line. For the pink dashed line within the bandgap $\mathcal{R}_m$, according to Eq.~(\ref{GpRm}), its slope is given by
\begin{align}
	k_1=(-1)^{\Theta[(-1)^d\delta]}\frac{\mathcal{G}_1(d,\delta)-[-\mathcal{G}_1(d,\delta)]}{2J\abs{\delta}-[-2J\abs{\delta}]}=\frac{(-1)^{\Theta[-d]}[(1-2d)\delta-1]}{4J^2\delta(1-\delta^2)}.
\end{align}
Similarly, for the green dashed line, according to Eq.~(\ref{GpRlr}), its slope is given by
\begin{align}
	k_2=\frac{\mathcal{G}_2(d,\delta)-[-\mathcal{G}_2(d,\delta)]}{2J-[-2J]}=\frac{(-1)^{\Theta[-d]}[\delta-(1-2d)]}{4J^2(1-\delta^2)}.
\end{align}
Thus, the two phase boundaries are given by 
\begin{align}\label{PB}
	\ell_1:\ k_1=\frac{1}{\mathrm{g}^2}\Longrightarrow \abs{\mathrm{g}}=2J\sqrt{\frac{(-1)^{\Theta[-d]}\delta(1-\delta^2)}{(1-2d)\delta-1}},\quad \ell_2:\ k_2=\frac{1}{\mathrm{g}^2}\Longrightarrow \abs{\mathrm{g}}=2J\sqrt{\frac{(-1)^{\Theta[-d]}(1-\delta^2)}{\delta-(1-2d)}},
\end{align}
which recover Eq.(10) in the main text for $d>0$ and the number of bound states changes in pairs on either side of these two phase boundaries. Additionally, when $(d\ge1, \delta>0)$ or $(d\le0, \delta<0)$ are satisfied, the number of bound states also changes on either side of the topological phase boundary ($\delta=0$) due to $G_\pm(0)=0$, but these changes do not occur in pairs, as shown in Figs.\,\ref{fig6}(b) and~\ref{fig6}(d). 

Finally, let us return to the long-time behavior of the maximum stored energy, i.e., $\max_t[\mathcal{E}(\infty)]=\omega_e\max_t[\abs{c_\mathrm{B}(\infty)}^2]$. After analyzing the situation of changes in the number of bound states, according to Eq.~(\ref{LTL}), we also need to calculate the residues corresponding to these bound states, i.e., $\Res[\mathcal{\mathscr{C}}(z), E_i]$. When $(d\ge1, \delta>0)$ or $(d\le0, \delta<0)$, based on the above discussion, we know that the number of bound states is odd. Among these, there must be a zero-energy bound state (corresponding to a second-order pole), and the remaining BSEs are non-zero and occur in pairs with opposite signs (corresponding to first-order pole). Meanwhile, the zero-energy bound-state residue equals zero due to $G(x_{1,B},x_{2,A};E_i=0)=0$, whereas the residues of the other bound states are non-zero due to $G(x_{1,B},x_{2,A};E_i\neq 0)\neq0$, specifically given by:
\begin{equation}\label{S64}
    \setlength\abovedisplayskip{4pt}
    \setlength\belowdisplayskip{8pt}
	\Res[\mathscr{C}(z), E_i=0]=0, \quad \Res[\mathscr{C}(z),  E_i\neq0]=-\Res[\mathscr{C}(z), -E_i\neq0]\neq0.
\end{equation}
The last term in Eq.~(\ref{S64}) also holds when the above condition $(d\ge1, \delta>0)$ or $(d\le0, \delta<0)$ is not satisfied. As a consequence, the maximum stored energy in the long-time limit is computed as
\begin{align}
	{\text{max}_t[\mathcal{E}(\infty)]}/\omega_e=\text{max}_t\abs{\sum_{E_k\in \text{BSEs}}\Res[\mathscr{C}(z), E_k]e^{-iE_kt}}^2=\text{max}_t\abs{\sum_{E_i>0}2\Res[\mathscr{C}(z), E_i]\sin(E_it)}^2,
\end{align}
which can reach values arbitrarily close to $4\{\sum_{i}\abs{\Res[\mathscr{C}(z), E_i]}\}^2$, where $E_i>0$. Because these self-energies and their derivatives are continuous across different bandgaps (i.e., $\mathcal{R}_l$, $\mathcal{R}_m$, and $\mathcal{R}_r$), the residues corresponding to the bound-state energies distributed in these bandgaps should also be continuous. Furthermore, since $\mathrm{d}G_{\pm}/\mathrm{d}z$ diverges at the band-gap edges, $\Res[\mathscr{C}(z),E_i]$ approaches zero when the bound-state energy $E_i$ is close to the band-gap edges. Therefore, integrating the above discussions and Eq.~(\ref{S64}), we conclude that the derivative of the maximum stored energy (itself is always continuous), $\max_t[\mathcal{E}(\infty)]$, remains continuous across the topological phase boundary, while it becomes discontinuous across the phase boundaries $\ell_1$ and $\ell_2$. This conclusion explains the phenomenon shown in the insets of Fig.2(c) in the main text.

\subsection{Dissipation immunity of quantum battery}\label{IIIB}
For the configuration I considered in Sec.~\ref{IIIA}, as shown in Fig.\,\ref{fig5}(a), regardless of the distance between the quantum charger and QB, at the appropriate parameters, the quantum charger can always transfer almost all of its energy to QB through the topological bath, i.e., $\max_t[\mathcal{E}(\infty)]/\omega_e\approx1$. However, once there is photon loss in the SSH photonic lattice, all the coherent bound states (i.e., $\Im[E_i]=0$) will transform into dissipative bound states (i.e., $\Im[E_i]<0$), and consequently, all the energy in the quantum charger and QB will be lost in the long-time limit, which implies $\mathcal{E}(\infty)=0$. In fact, there exists another configuration with a dark state, different from the configuration I, whose energy transfer from the quantum charger to QB remains unaffected by the single-sublattice dissipation ($\kappa_a=\kappa\neq0$ and $\kappa_b=0$), as shown in Fig.\,\ref{fig5}(b). For the configuration II, the corresponding pole equation is
\begin{equation}\label{S66}
	\mathscr{D}(E_i)=[E_i-\Delta-\Sigma_{11}^{ AA}(E_i)]^2-[\Omega+\Sigma_{11}^{AA}(E_i)]^2
	=[E_i-\Delta-\Omega-2\times\Sigma_{11}^{ AA}(E_i)][E_i-\Delta+\Omega]=0
\end{equation}
with the single-emitter self-energy [see Eq.~(\ref{S32})]
\begin{align}
	\Sigma_{11}^{AA}(z)=\mathrm{g}^2G(x_{1,A},x_{1,A};z)=-\frac{z\mathrm{g}^2\sum_{p=\pm}p\Theta[p(1-|\tilde{y}_+|)}{\sqrt{z_\text{nh}^4-4J^2z_{\text{nh}}^2(1+\delta^2)+16J^4\delta^2}}=\frac{z\mathrm{g}^2\times\text{sign}(|\tilde{y}_+|-1)}{\sqrt{z_\text{nh}^4-4J^2z_{\text{nh}}^2(1+\delta^2)+16J^4\delta^2}},
\end{align}
where $z_{\text{nh}}=\customsqrt{z(z+i\kappa/2)}$. According to the pole equation, it is evident that this dissipative system can have at most two coherent bound states. One is an environment-independent bound state, also known as the dark state, with energy $E_{\text{dark}}=\Delta-\Omega$. The other, known as the vacancy-like dressed state\,\cite{vbs}, appears only when $\Delta+\Omega=0$, and has the energy $E_{\text{vds}}=0$. Subsequently, when $\Delta=-\Omega\neq0$, the residues at these bound-state energies are computed as
\begin{align}\label{S68}
	\Res[\mathscr{C}(z), E_{\text{dark}}]=\frac{\Omega+\Sigma_{11}^{AA}(z)}{\mathrm{d}\mathscr{D}(z)/\mathrm{d}z}\bigg|_{z=E_\text{dark}}=-\frac{1}{2},\qquad
	\Res[\mathscr{C}(z), E_{\text{vds}}]=\frac{\Omega+\Sigma_{11}^{AA}(z)}{\mathrm{d}\mathscr{D}(z)/\mathrm{d}z}\bigg|_{z=E_\text{vds}}=\frac{J^2\abs{\delta}}{\mathrm{g}^2+2J^2\abs{\delta}}.
\end{align}
Besides, the dark state $\ket{\Psi_{\text{dark}}}$ and the vacancy-like dressed state $\ket{\Psi_{\text{vds}}}$ can be obtained by solving the secular equation $H_{\text{eff}}\ket{\Psi}=E\ket{\Psi}$, where $E=E_{\rm dark}$ and $E=E_{\rm vds}$, respectively. Consequently, these bound states can be written as
\begin{equation}\label{states}
    \setlength\abovedisplayskip{2pt}
    \setlength\belowdisplayskip{2pt}
	\ket{\Psi_{\text{dark}}}=\frac{1}{\sqrt{2}}(\sigma_+^\mathrm{C}-\sigma_+^\mathrm{B})\ket{g,g;\text{vac}},\quad
	\ket{\Psi_{\text{vds}}}=\mathcal{N}\big[(\sigma_+^\mathrm{C}+\sigma_+^\mathrm{B})+\sum_{j} (c_{j,A}^{}a_j^\dagger+c_{j,B}^{}b_j^\dagger)\big]\ket{g,g;\text{vac}},
\end{equation}
where
\begin{equation}\label{IIIC76}
    \setlength\abovedisplayskip{2pt}
    \setlength\belowdisplayskip{2pt}
	\mathcal{N}=\sqrt{\frac{J^2\abs{\delta}}{\mathrm{g}^2+2J^2\abs{\delta}}},\quad c_{j,A}=0, \quad c_{j,B}=-\frac{2\mathrm{g}}{J(1+\delta)}\left(\frac{\delta-1}{\delta+1}\right)^{\mathrm{x}_j-x_1}\text{sign}(\delta)\times\Theta[(\mathrm{x}_j-x_1+0^+)\delta].
\end{equation}
Note that $\mathrm{x}_j$ in Eq.~(\ref{IIIC76}) is a new index used to denote the cell position of the environmental bosonic modes, distinct from $x_j$, and is typically assumed to be $\mathrm{x}_j=j$. Finally, combining Eqs.~(\ref{LTL}) with (\ref{S68}), the probability amplitude for the QB to be in the excited state in the long-time limit is given by
\begin{align}
	c_\mathrm{B}(\infty)=\frac{J^2\abs{\delta}}{\mathrm{g}^2+2J^2\abs{\delta}}-\frac{1}{2}e^{2i\Omega t},
\end{align}
which shows beyond doubt that even in the dissipative bath, the energy in the quantum charger can always be transferred to the QB. Particularly in the Markov regime, i.e., $\mathrm{g}\ll 2J\abs{\delta}$, we have $c_{\rm B}(t)\approx[1-\exp(2i\Omega t)]/2$, implying that all of the energy from the quantum charger can be transferred to the QB. The two bound states play a crucial role in the energy transfer process. Additionally, according to Eq.~(\ref{IIIAn-5}), the maximum ergotropy in the long-time limit is computed as
\begin{align}\label{analy_ergo}
	\text{max}_t[\mathcal{W}(\infty)]/\omega_e=\text{max}_t[2\abs{c_\mathrm{B}(\infty)}^2-1]\Theta[\abs{c_\mathrm{B}(\infty)}^2-1/2]=\frac{8J^4\delta^2-\mathrm{g}^4}{2(2J^2\abs{\delta}+\mathrm{g}^2)^2}\Theta\left(2^{\frac{3}{4}}J\sqrt{\abs{\delta}}-\abs{\mathrm{g}}\right),
\end{align}
which supports Eq.(11) in the main text. We emphasize that both bound states are indispensable; the absence of either one would result in the maximum extractable energy vanishing in the long-time limit, i.e., $\max_t[\mathcal{W}(\infty)]=0$. This is because the absence of either bound state reduces the atomic population in the long-time limit to below 1/4. 
\subsection{The effect of strong dissipation on quantum battery performance}\label{IIIC}
\begin{figure}[b]
	\centering
	\includegraphics[width=17cm]{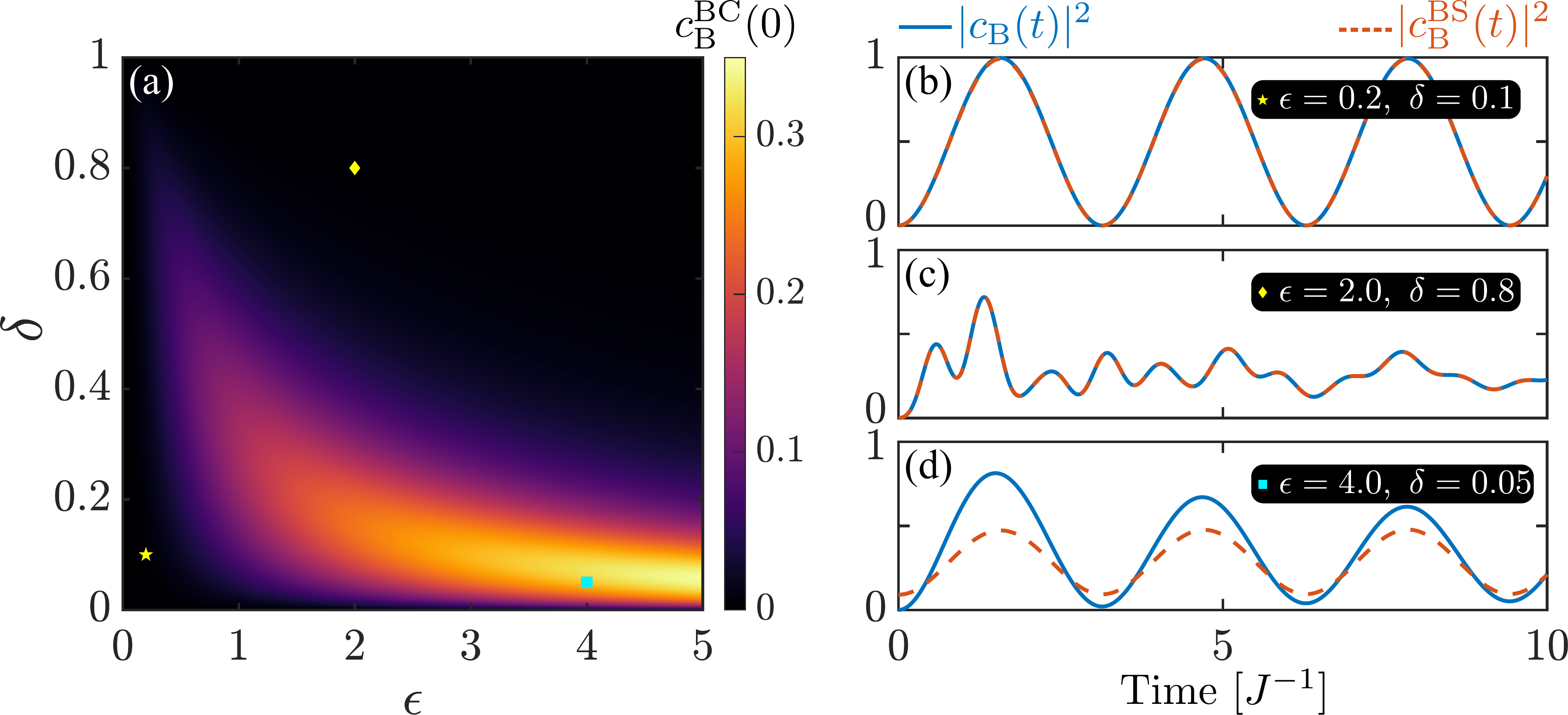}\\
	\caption{(a) The initial contribution $c_{\rm B}^{\rm BC}(0)$ from the branch cuts as a function of the dimerization strength $\delta$ and the dimensionless quantity $\epsilon=\text{g}/(2J\abs{\delta})$. (b-d) Atomic dynamics with full contributions (solid lines) and bound state contributions (dashed lines) in different parameters $(\epsilon, \delta)$. The system parameters are chosen as $\Delta=-\Omega=\kappa=J$.
	}\label{fign9}
\end{figure}

Before proceeding, we turn to the pole equation~(\ref{S66}) and take $\Delta=-\Omega$, and the pole equation can be further simplified as
\begin{align}
	\mathscr{D}(E_i)=E_i(E_i+2\Omega)\left[1+\frac{\mathrm{g}^2\sum_{p=\pm}p\Theta[p(1-\abs{\tilde{y}_+})]}{\sqrt{z_\text{nh}^4 +4J^2z_\text{nh}^2(1+\delta^2)+16J^4\delta^2}}\right]_{z=E_i}=0.
\end{align}
Apart from the two bound state mentioned in Sec.~\ref{IIIB}, according to the pole equation, we can also find two dissipative bound states, and the corresponding energies can be obtained by solving 
\begin{align}\label{S74}
	z_\text{nh}|_{z=E_{i}}=\pm\sqrt{2J^2(1+\delta^2)+2\sqrt{\mathrm{g}^4+J^4(1-\delta^2)^2}}\equiv E_{0,\pm},
\end{align}
where $E_{0,\pm}$ are the bound-state energies in the non-dissipative bath, which satisfy $\mathscr{D}(E_{\kappa,\pm})|_{\kappa=0}=0$. The solution in Eq.~(\ref{S74}) is given by
\begin{align}\label{DBSEs}
	E_{\kappa,\pm}=-\frac{i}{4}\kappa\pm\sqrt{E_{0,\pm}^2-(\kappa/4)^2},
\end{align}
which represent the dissipative bound-state energies in the dissipative bath. Subsequently, let us analyze the dissipative bound state contributions in a short time when $\kappa/E_{0,+}\gg1$. First, following Eq.~(\ref{S68}), the corresponding residues are computed as
\begin{align}\label{S76}
	\Res[\mathscr{C}(z), E_{\kappa,\pm}]=\frac{\mathrm{g}^4}{E_{\kappa,\pm}(E_{\kappa,\pm}+i\kappa/4)[E_{0,\pm}^2-2J^2(1+\delta^2)]}=\frac{1\pm [1-(4E_{0,\pm}/\kappa)^2]^{-1/2}}{E_{0,\pm}^2[E_{0,\pm}^2-2J^2(1+\delta^2)]}\times\text{g}^4,
\end{align}
which satisfies $\sum_{p=\pm}\Res[\mathscr{C}(z), E_{\kappa,p}]=\sum_{p=\pm}\Res[\mathscr{C}(z), E_{\kappa,p}]|_{\kappa=0}$, implying that, regardless of whether the bath is dissipative or non-dissipative, the contributions from these two types of bound states are always equal at time $t=0$. Besides, as the contribution from the branch cuts generally decreases over time, it becomes predominantly determined by its initial contribution. For simplicity, we rewrite Eq.~(\ref{S41}) as $c_{\rm B}(t)=c_{\rm B}^{\rm BS}(t)+c_{\rm B}^{\rm BC}(t)$, where
\begin{align}\label{BS-DBS}
	c_{\rm B}^{\rm BS}(t)=\Res[\mathscr{C}(z), E_{\text{vds}}]e^{-iE_{\rm vds}t}+\Res[\mathscr{C}(z), E_{\text{dark}}]e^{-iE_{\rm dark}t}+{\textstyle \sum_{p=\pm}}\Res[\mathscr{C}(z), E_{\kappa,p}]e^{-iE_{\kappa,p}t}
\end{align}
represents the contributions from the aforementioned two bound states and the dissipative bound states. Here, $c_{\rm B}^{\rm BC}(t)$ denotes the contribution from the branch cuts, expressed as $c_{\rm B}^{\rm BC}(t)=c_{\rm B}^{\rm UP}(t)+c_{\rm B}^{\rm BCD}(t)$. Since the QB is in the ground state at $t=0$, the initial contribution from the branch cuts can be derived as
\begin{align}
	c_{\rm B}^{\rm BC}(0)=0-\frac{J^2\abs{\delta}}{\text{g}^2+2J^2\abs{\delta}}+\frac{1}{2}-\frac{2\text{g}^2}{E_{0,\pm}^2[E_{0,\pm}^2-2J^2(1+\delta^2)]}=\frac{\text{g}^2J^2[\text{g}^2(1+\delta^2)-2\abs{\delta}\sqrt{\text{g}^4+J^4(1-\delta^2)^2}]}{2(\text{g}^4-4J^4\delta^2)\sqrt{\text{g}^4+J^4(1-\delta^2)^2}}.
\end{align}
This expression shows that the initial contribution from the branch cuts is unaffected by the single-sublattice dissipation. Notably, when the parameters $(\text{g}, \delta)$ or $(\epsilon, \delta)$ are carefully chosen such that the initial contribution from the branch cuts becomes negligible, the bound-states contribution [see Eq.~(\ref{BS-DBS})] fully governs the dynamics of the QB, as shown in Figs.\,\ref{fign9}(b) and \ref{fign9}(c). Conversely, when the initial contribution from the branch cuts cannot be not neglected, as illustrated in in Fig.\,\ref{fign9}(d), we observe that the atomic dynamics with full contributions and bound-states contribution exhibit significant discrepancies on short time scales. For configuration II, it is generally advisable to select parameters that render the initial branch-cut contribution negligible. Specific guidance on parameters selection can be found in Fig.\,\ref{fign9}(a). This condition is crucial for obtaining an analytical expression for the atomic dynamics. Note that the parameters used in Fig.~4 of the main text also satisfy this condition. 

Returning to the main topic, we consider a strong single-sublattice dissipation (i.e., $\kappa/E_{0,+}\gg1$) and select suitable parameters such that the contribution from the branch cuts can be negligible. Accordingly, the dissipative bound-state energies (\ref{DBSEs}) can be expressed as a Taylor series expansion, i.e.,
\begin{equation}\label{DBSEsT}
    \setlength\abovedisplayskip{4pt}
    \setlength\belowdisplayskip{4pt}
	E_{\kappa,\pm}=-\frac{i\kappa}{4}\left[1\mp\sqrt{1-(4E_{0,\pm}/\kappa)^2} \right]=-\frac{i\kappa}{4}\left\{1\mp \left[1-8(E_{0,\pm}/\kappa)^2+\order{E_{0,\pm}^4/\kappa^4}\right]  \right\}\approx(\pm 1-1)\frac{i\kappa}{4}\mp\frac{2iE_{0,\pm}^2}{\kappa},
\end{equation}
and the corresponding residues are given by
\begin{align}
	\Res[\mathscr{C}(z), E_{\kappa,+}]&=\frac{2\mathrm{g}^4}{[E_{0,+}^2-2J^2(1+\delta^2)]E_{0,+}^2}+\frac{8\mathrm{g}^4}{[E_{0,+}^2-2J^2(1+\delta^2)]\kappa^2}+\order{E_{0,+}^4/\kappa^4},\label{DBS1}\\
	\Res[\mathscr{C}(z), E_{\kappa,-}]&=\frac{-8\text{g}^4}{[E_{0,-}^2-2J^2(1+\delta^2)]\kappa^2}-\order{E_{0,-}^4/\kappa^4}.\label{DBS2}
\end{align}
Since the contribution from the branch cuts can be negligible, by plugging Eqs.~(\ref{S68}, \ref{DBSEsT}-\ref{DBS2}) into Eq.~(\ref{BS-DBS}), the dynamics of the QB can be expressed as
\begin{align}
	c_\mathrm{B}(t; \kappa)&\approx c^{\mathrm{BS}}_\mathrm{B}(t; \kappa)=\Res[\mathscr{C}(z), E_{\text{vds}}]e^{-iE_{\rm vds}t}+\Res[\mathscr{C}(z), E_{\text{dark}}]e^{-iE_{\rm dark}t}+\sum_{p=\pm}\Res[\mathscr{C}(z),E_{\kappa,p}]e^{-iE_{\kappa,p}t}\nonumber\\
	&=\frac{J^2\abs{\delta}}{\mathrm{g}^2+2J^2\abs{\delta}}-\frac{\exp(2i\Omega t)}{2}+\frac{2\mathrm{g}^4[(\kappa^2/E_{0,+}^{2}+4)\text{exp}(-2E_{0,+}^2t/\kappa)-4\exp(-\kappa t/2)]}{[E_{0,+}^2-2J^2(1+\delta^2)]\kappa^2}+\order{E_{0,+}^4/\kappa^4},\label{S80}
\end{align}
and for a short time, i.e., $\kappa t\ll (\kappa/E_{0,+})^2$, we have $\exp[-2E_{0,+}^2t/\kappa]\approx 1$. Thus, we discard second-order and higher-order small terms and obtain
\begin{align}
	c_\mathrm{B}(t;\kappa\gg E_{0,+})\approx-\frac{\exp(2i\Omega t)}{2}+\frac{J^2\abs{\delta}}{\mathrm{g}^2+2J^2\abs{\delta}}+\frac{2\mathrm{g}^4}{[E_{0,+}^2-2J^2(1+\delta^2)]E_{0,+}^2}=c^{\mathrm{BS}}_{\mathrm{B}}(t_n;\kappa=0)\approx c_{\mathrm{B}}(t_n;\kappa=0),
\end{align}
where $t_n=2n\pi/E_{0,+}\ll \kappa/E_{0,+}^2$ and $c_\mathrm{B}^{\text{BS}}(t_n; \kappa=0)$ represents the contributions from all the bound states in the configuration II with $\kappa=0$. As a result, for configuration II, with increasing dissipation $\kappa$, we find that the dynamics of the QB in the dissipative system---dominated by the dissipative bound states, dark state, and vacancy-like dressed state---converge to the stroboscopic dynamics of the QB at $t=2\pi\mathbb{Z}/|E_{0,\pm}|$ in the non-dissipative system, which are governed by the bound states. This established a possible short-time atomic dynamics connection between the strongly dissipative non-Hermitian system and the Hermitian system.

\begin{figure}[t]
	\centering
	\includegraphics[width=17cm]{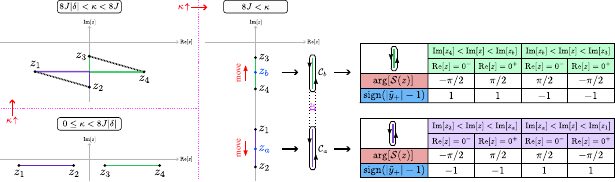}\\
	\caption{The distribution of the two branch cuts in the complex plane as a function of the dissipation strength $\kappa$. The two tables exhibit the behavior of $\arg[\mathcal{S}(z)]$ and $\text{sign}(|\tilde{y}_+|-1)$ on both sides near the two branch cuts in the strongly dissipative regime $\kappa\gg E_{0,+}$.
	}\label{Fig_Suppn8}
\end{figure}

However, when the selected parameters make the initial contribution from the branch cuts significant, as depicted in Fig.\,\ref{fign9}(d), it becomes crucial to account for all contributions---namely, the dissipative bound states, dark state, vacancy-like dressed state, and branch cuts---to accurately capture the full dynamics of the QB over short timescales. Therefore, we will analytically derive the contribution from the branch cuts to the dynamics of the QB in the strong dissipation case. Specifically, for the non-dissipative SSH bath (i.e., $\kappa\neq0$), the branch points, i.e., the endpoints of the branch cut, correspond to the band edges on the real axis, while in the single-sublattice dissipative SSH bath, the corresponding branch points are located in the complex plane. In fact, the branch points are determined by the roots of the following equation:
\begin{align}\label{ROOT}
	\mathcal{S}^2(z)\equiv z_{\rm nh}^4-4J^2z_{\rm nh}^2(1+\delta^2)+16J^4\delta^2=(z_{\rm nh}^2-4J^2)(z_{\rm nh}^2-4J^2\delta^2)=0,
\end{align}
where $z_{\rm nh}^2=z(z+i\kappa/2)$. By solving Eq.~(\ref{ROOT}), we can obtain the corresponding roots, i.e.,
\begin{align}
	z_1&=-\frac{i\kappa}{4}-\sqrt{4J^2-(\kappa/4)^2},\quad z_2=-\frac{i\kappa}{4}-\sqrt{4J^2\delta^2-(\kappa/4)^2},\\ z_3&=-\frac{i\kappa}{4}+\sqrt{4J^2\delta^2-(\kappa/4)^2},\quad z_4=-\frac{i\kappa}{4}+\sqrt{4J^2-(\kappa/4)^2}.
\end{align}
The analytical expressions of the branch points derived above reveal that the distribution of branch cuts in the complex plane varies significantly with dissipation strength, as shown in Fig.\,\ref{Fig_Suppn8}, making their contributions more complex to calculate compared to those from the dissipative bound states. Next, we maintain our focus on the strongly dissipative regime, i.e., $\kappa/E_{0,+}\gg1$. Notably, for $\kappa>8J$, which lies within the strongly dissipative regime, we can find that the two branch cuts move along the imaginary axis in opposite directions as $\kappa$ increases. Therefore, these four branch points in the strongly dissipative regime can be also expressed as a Taylor series expansion, i.e.,
\begin{align}
	z_1&=-\frac{i\kappa}{4}\left[1+\sqrt{1-(8J/\kappa)^2} \right]=-i\kappa\left[1/2+8J^2/\kappa^2+\order{8^4J^4/\kappa^4} \right],\label{BP1}\\ z_2&=-\frac{i\kappa}{4}\left[1+\sqrt{1-(8J\delta/\kappa)^2} \right]=-i\kappa\left[1/2+8J^2\delta^2/\kappa^2+\order{8^4J^4\delta^4/\kappa^4} \right],\\ z_3&=-\frac{i\kappa}{4}\left[1-\sqrt{1-(8J\delta/\kappa)^2} \right]=-i\kappa\left[8J^2\delta^2/\kappa^2+\order{8^4J^4\delta^4/\kappa^4} \right],\\
	z_4&=-\frac{i\kappa}{4}\left[1-\sqrt{1-(8J/\kappa)^2} \right]=-i\kappa\left[8J^2/\kappa^2+\order{8^4J^4/\kappa^4} \right].\label{BP4}
\end{align}
For simplicity, we will adopt the first method to calculate the contributions from the branch cuts by choosing two small contours that tightly enclose the branch cuts and directly integrating along them, and the corresponding contours are denoted as $\mathcal{C}_a$ and $\mathcal{C}_b$, as illustrated in Fig.\,\ref{Fig_Suppn8}. Besides, the corresponding integrand of configuration II can be written as (assuming $\Delta+\Omega=0$) 
\begin{align}\label{INT2}
	\mathscr{C}(z)=\frac{\Sigma_{11}^{AA}(z)+\Omega}{\mathscr{D}(z)}=\frac{\Sigma_{11}^{AA}(z)+\Omega}{[z-2\times\Sigma_{11}^{AA}(z)][z+2\Omega]}=\frac{1}{2}\left[\frac{1}{z-2\times\Sigma_{11}^{AA}(z)}-\frac{1}{z+2\Omega}\right],
\end{align}
where 
\begin{align}
	\Sigma_{11}^{AA}(z)=\text{g}^2G(x_{1,A},x_{1,A};z)=\frac{z\text{g}^2\text{sign}(|\tilde{y}_+|-1)}{\mathcal{S}(z)},\quad \tilde{y}_+=\frac{z_{\rm nh}^2-2J^2(1+\delta^2)+\mathcal{S}(z)}{2J^2(1-\delta^2)}.
\end{align}
Thus, according to the two tables in Fig.\,\ref{Fig_Suppn8}, the contributions from the two branch cuts are computed as
\noindent
\begin{align}
	c_{{\rm B}, a}^{\rm BC}(t)&\equiv\int_{\mathcal{C}_a}\frac{\dd{z}}{2\pi i}\mathscr{C}(z)e^{-izt}=\lim\limits_{\varepsilon\to 0^+}\int_{\Im[z_1]}^{\Im[z_2]}\left[\mathscr{C}(ix-\varepsilon)e^{xt+i\varepsilon t}-\mathscr{C}(ix+\varepsilon)e^{xt-i\varepsilon t} \right]\frac{\dd{x}}{2\pi}\nonumber\\
	&=\lim\limits_{\varepsilon\to 0^+}\left[\int_{\Im[z_1]}^{\Im[z_a]}+\int_{\Im[z_a]}^{\Im[z_2]}\right]\left[\frac{\exp(xt+i\varepsilon t)/(ix-\varepsilon)}{1-2\text{g}^2\text{sign}(|\tilde{y}_+|-1)/\mathcal{S}(ix-\varepsilon)}-\frac{\exp(xt-i\varepsilon t)/(ix+\varepsilon)}{1-2\text{g}^2\text{sign}(|\tilde{y}_+|-1)/\mathcal{S}(ix+\varepsilon)}\right]\frac{\dd{x}}{4\pi}\nonumber\\
	&=\int_{\Im[z_1]}^{\Im[z_2]}\frac{\exp(xt)}{x}\left[\frac{1}{1+2i\text{g}^2/|\mathcal{S}(ix)|}-\frac{1}{1-2i\text{g}^2/|\mathcal{S}(ix)|}\right]\frac{\dd{x}}{4\pi i}=\int_{\Im[z_2]}^{\Im[z_1]}\frac{\exp(xt)}{x}\frac{\text{g}^2|\mathcal{S}(ix)|}{|\mathcal{S}(ix)|^2+4\text{g}^2}\frac{\dd{x}}{\pi }\nonumber\\
	&=\int_{\Im[z_2]}^{\Im[z_1]}\frac{\exp(xt)}{x}\frac{\text{g}^2\sqrt{[\Im(z_1)-x][x-\Im(z_2)][x-\Im(z_3)][x-\Im(z_4)]}}{[\Im(z_1)-x][x-\Im(z_2)][x-\Im(z_3)][x-\Im(z_4)]+4\text{g}^4}\frac{\dd{x}}{\pi},\label{BC1}\\
	c_{{\rm B}, b}^{\rm BC}(t)&\equiv\int_{\mathcal{C}_b}\frac{\dd{z}}{2\pi i}\mathscr{C}(z)e^{-izt}=\lim\limits_{\varepsilon\to 0^+}\int_{\Im[z_3]}^{\Im[z_4]}\left[\mathscr{C}(ix-\varepsilon)e^{xt+i\varepsilon t}-\mathscr{C}(ix+\varepsilon)e^{xt-i\varepsilon t} \right]\frac{\dd{x}}{2\pi}\nonumber\\
	&=\lim\limits_{\varepsilon\to 0^+}\left[\int_{\Im[z_3]}^{\Im[z_b]}+\int_{\Im[z_b]}^{\Im[z_4]}\right]\left[\frac{\exp(xt+i\varepsilon t)/(ix-\varepsilon)}{1-2\text{g}^2\text{sign}(|\tilde{y}_+|-1)/\mathcal{S}(ix-\varepsilon)}-\frac{\exp(xt-i\varepsilon t)/(ix+\varepsilon)}{1-2\text{g}^2\text{sign}(|\tilde{y}_+|-1)/\mathcal{S}(ix+\varepsilon)}\right]\frac{\dd{x}}{4\pi}\nonumber\\
	&=\int_{\Im[z_3]}^{\Im[z_4]}\frac{\exp(xt)}{x}\left[\frac{1}{1-2i\text{g}^2/|\mathcal{S}(ix)|}-\frac{1}{1+2i\text{g}^2/|\mathcal{S}(ix)|}\right]\frac{\dd{x}}{4\pi i}=\int_{\Im[z_4]}^{\Im[z_3]}\frac{\exp(xt)}{x}\frac{-\text{g}^2|\mathcal{S}(ix)|}{|\mathcal{S}(ix)|^2+4\text{g}^2}\frac{\dd{x}}{\pi }\nonumber\\
	&=\int_{\Im[z_4]}^{\Im[z_3]}\frac{\exp(xt)}{x}\frac{-\text{g}^2\sqrt{[x-\Im(z_1)][x-\Im(z_2)][x-\Im(z_3)][\Im(z_4)-x]}}{[x-\Im(z_1)][x-\Im(z_2)][x-\Im(z_3)][\Im(z_4)-x]+4\text{g}^4}\frac{\dd{x}}{\pi},\label{BC2}
\end{align}
where $z_a\approx(z_1+z_2)/2$ and $z_b\approx(z_3+z_4)/2$ in the strongly dissipative regime. To solve Eqs.~(\ref{BC1}) and (\ref{BC2}) more effectively, we apply approximations derived from Eqs.~(\ref{BP1}-\ref{BP4}) under the conditions: $\kappa\gg 8J$ and $\kappa t\ll[\kappa/(8J)]^2$, as outlined below:
\begin{align}\label{approxi}
	\begin{cases}
		[x-\Im(z_3)][x-\Im(z_4)]\approx\kappa^2/4,\ \exp(xt)\approx\exp(-\kappa t/2) & \text{for}\ \Im[z_2]<x<\Im[z_1]\\
		[x-\Im(z_1)][x-\Im(z_2)]\approx\kappa^2/4,\ \exp(xt)\approx1 & \text{for}\ \Im[z_4]<x<\Im[z_3]
	\end{cases}.
\end{align}
Additionally, we need to employ a definite integral, which is given by
\begin{align}\label{integral}
	\int_{a}^{b}\frac{x^{-1}\sqrt{(x-a)(b-x)}}{(x-a)(b-x)+c}\dd{x}=\frac{(a+b)\sqrt{c}+\sqrt{ab[(a-b)^2+4c]}}{(c-ab)\sqrt{(a-b)^2+4c}}\times\pi\quad\text{for}\quad a<b<0\le c.
\end{align}
Consequently, by applying the approximation in Eq.~(\ref{approxi}) and the integral in Eq.~(\ref{integral}), we obtain
\begin{align}
	c_{{\rm B}, a}^{\rm BC}(t)&\approx2\text{g}^2e^{-\kappa t/2}\int_{\Im[z_2]}^{\Im[z_1]}\frac{x^{-1}\sqrt{[\Im(z_1)-x][x-\Im(z_2)]}}{[\Im(z_1)-x][x-\Im(z_2)]+16\text{g}^4/\kappa^2}\frac{\dd{x}}{\kappa\pi }\approx\frac{8\text{g}^4-4\text{g}^2[E_{0,+}^2-2J^2(1+\delta^2)]}{[E_{0,+}^2-2J^2(1+\delta^2)]\kappa^2}e^{-\kappa t/2},\\
	c_{{\rm B}, b}^{\rm BC}(t)&\approx-2\text{g}^2\int_{\Im[z_4]}^{\Im[z_3]}\frac{x^{-1}\sqrt{[\Im(z_3)-x][x-\Im(z_4)]}}{[\Im(z_3)-x][x-\Im(z_4)]+16\text{g}^4/\kappa^2}\frac{\dd{x}}{\kappa\pi}\approx\frac{\text{g}^2}{2(\text{g}^2+2J^2|\delta|)}-\frac{2\text{g}^4}{[E_{0,+}^2-2J^2(1+\delta^2)]E_{0,+}^2},
\end{align}
and the total contributions from the branch cuts are given by $c_{\rm B}^{\rm BC}(t)=c_{{\rm B},a}^{\rm BC}(t)+c_{{\rm B},b}^{\rm BC}(t)$. As a result, according to Eqs.~(\ref{S80}, \ref{BC1}, \ref{BC2}), the full dynamics of the QB is expressed as
\begin{align}\label{CBT}
	c_{\rm B}(t)=c_{\rm B}^{\rm BS}(t)+c_{\rm B}^{\rm BC}(t)\approx\frac{1-\exp(2i\Omega t)}{2}+\frac{2\mathrm{g}^4[(1+4E_{0,+}^2/\kappa^2)\text{exp}(-2E_{0,+}^2t/\kappa)-1]}{[E_{0,+}^2-2J^2(1+\delta^2)]E_{0,+}^2}-\frac{4\text{g}^2\exp(-\kappa t/2)}{\kappa^2},
\end{align}
which satisfies $\kappa\gg \max\{E_{0,+},8J\}$ and $\kappa t\ll [\kappa/(8J)]^2$. For short times, $\kappa t\ll (\kappa/E_{0,+})^2$, we have $\exp[-2E_{0,+}^2t/\kappa]\approx 1$, which further refines the above condition to $\kappa t\ll (\kappa/\max\{E_{0,+},8J\})^2$. Thus, Eq.~(\ref{CBT}) can be simplified as
\begin{align}\label{CBTF}
	c_{\rm B}(t)\approx[1-\exp(2i\Omega t)]/2+\{\mathrm{g}^2/\sqrt{\text{g}^4+J^4(1-\delta^2)^2}-\exp(-\kappa t/2)\}\times (2\text{g}/\kappa)^2\approx[1-\exp(2i\Omega t)]/2,
\end{align}
where in the last step, we use $\kappa\gg\text{g}$, as $\kappa\gg E_{0,+}>\text{g}$. This expression (\ref{CBTF}) demonstrates that strong dissipation induces a rapid transition in atomic dynamics from the non-Markovian to the Markovian regime over short timescales. This transition facilitates more energy transfer, thereby improving the performance of the QB, such as the stored energy, the charging time, and the charging power.

\section{Physical Interpretation of Near-Perfect Energy Transfer Between Charger and Battery}\label{IVA}
In this section, for concreteness, we focus on a parameter regime where the Born-Markov approximation is justified, and we elucidate the physical mechanism behind the near-perfect energy transfer from the quantum charger to the QB in the configuration I. For the configuration II, the physics underlying its energy transfer has been thoroughly discussed and demonstrated in Section \ref{IIIB}.

\begin{figure}[b]
	\centering
	\includegraphics[width=17cm]{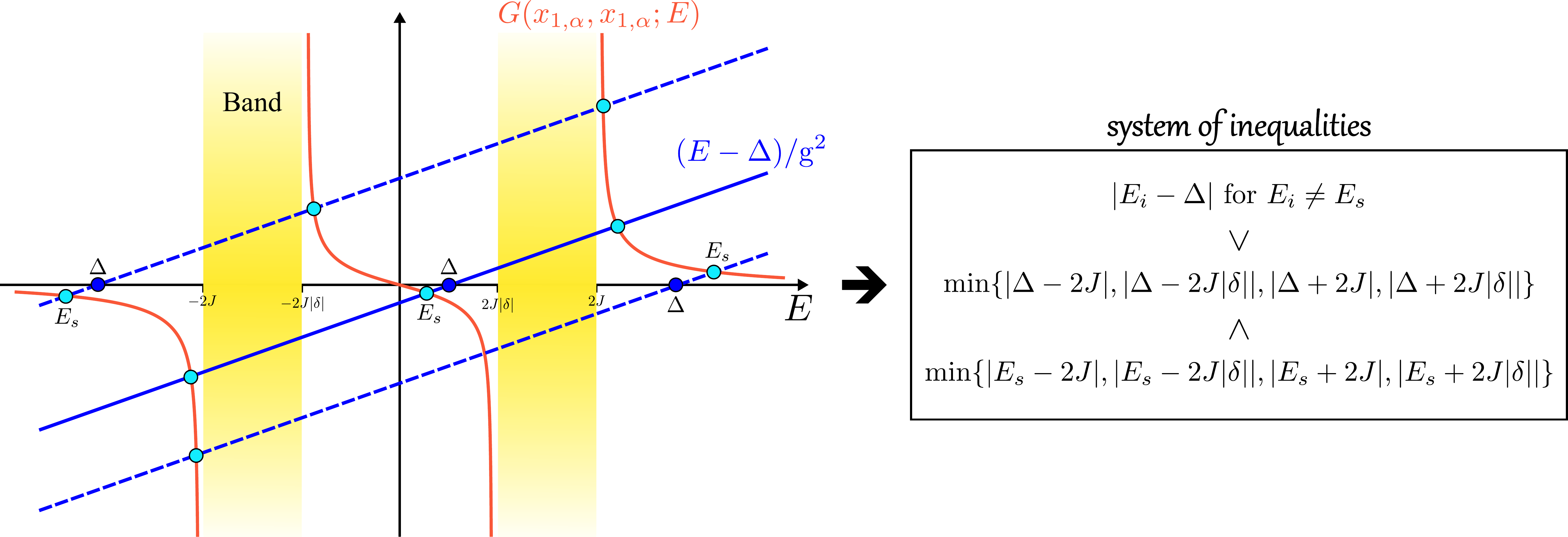}\\
	\caption{The single-particle Green function versus $E$ (orange lines) for a one-dimensional photonic SSH bath with two bands. Roots of the poles equation (i.e., the BSEs) are obtained from the intersection points (cyan circles) between $(E-\Delta)/\text{g}^2$ (blue solid or dashed line) and $G(x_{1,\alpha},x_{1,\alpha};E)$ (orange solid lines). Here, $E_s$ is defined as the bound state energy closest to $\Delta$. For $\Delta$ located within different bandgaps (corresponding three blue circles), the set of inequalities within the box consistently holds.
	}\label{Fig_Suppn2}
\end{figure}

\subsection{Single atom coupled to photonic Su-Schrieffer-Heeger model}\label{SIVA}
Here, to facilitate a more intuitive and clear understanding of energy transfer between two atoms (i.e., quantum charger and QB) in a topological waveguide, we begin by analyzing the specific case of a single atom coupled to the topological waveguide. Following these steps (\ref{S2}-\ref{S15}), we first rewrite the total Hamiltonian, i.e.,
\begin{align}
	H_{\rm tot}^{(1)}=H_{\rm sys}+H_{\rm bath}+H_{\rm int}=\omega_e\sigma_+\sigma_-+H_{\rm bath}+\text{g}(\sigma_- o^\dagger_{x_{1,\alpha}}+o_{x_{1,\alpha}}\sigma_+),
\end{align}
where $H_{\rm bath}$ is given by Eq.~(\ref{S17}). Hereafter we set $\omega_c$ as the energy reference. Therefore, the projection operators and the bath's Hamiltonian in the momentum space [see Eq.~(\ref{S19})] can be written as
\begin{equation}
    \setlength\abovedisplayskip{4pt}
    \setlength\belowdisplayskip{4pt}
	P=\ketbra{e}\otimes\ketbra{\text{vac}},\quad Q=\ketbra{g}\otimes\tilde{Q},\quad H_{\rm bath}=\sum_k[\omega_k u_k^\dagger u_k^{}-\omega_kl_k^\dagger l_k^{}],
\end{equation}
where $\tilde{Q}=\sum_{j,\beta}\ketbra{x_{j,\beta}}=\sum_k(u_k^\dagger\ketbra{\rm vac}u_k^{}+l_k^\dagger\ketbra{\rm vac}l_k^{})$. Based on the above discussion, we have
\begin{align}\label{IVA2}
	PG_{\rm tot}^{(1)}P=\frac{\ketbra{e;\text{vac}}}{z-\Delta-\Sigma_{11}^{\alpha\alpha}(z)}=\frac{\ketbra{e;\text{vac}}}{z-\Delta-\text{g}^2G(x_{1,\alpha},x_{1,\alpha};z)}, \quad \mathscr{C}^{(1)}(z)=\frac{1}{z-\Delta-\text{g}^2G(x_{1,\alpha},x_{1,\alpha};z)}.
\end{align}
Solving the poles equation $E_i-\Delta-\text{g}^2G(x_{1,\alpha},x_{1,\alpha};E_i)=0$, one can get the BSEs of $H_{\rm tot}^{(1)}$ denoted by $E_i$. Here, we focus on a special bound state with energy $E_s$, which lies closest to $\Delta=\omega_e-\omega_c$. Given these poles, the corresponding residues are
\begin{align}\label{IVA3}
	\text{Res}[\mathscr{C}^{(1)}(z), E_i]=\frac{1}{\dv*{[z-\Delta-\text{g}^2G(x_{1,\alpha},x_{1,\alpha};z)]}{z}}\Big|_{z=E_i}=\frac{1}{1+\text{g}^2\mel{x_{1,\alpha}}{(E_i-H_{\text{bath}})^{-2}}{x_{1,\alpha}}},
\end{align}
where $\mel{x_{1,\alpha}}{(E_i-H_{\text{bath}})^{-2}}{x_{1,\alpha}}=-\dv*{G(x_{1,\alpha},x_{1,\alpha};z)}{z}|_{z=E_i}\ge0$. Note that only few residues in Eq.~(\ref{IVA3}) have non-negligible contributions for the atomic dynamics. Specifically, there is one solution of the poles equation within each band-gap regime. For $E_i\neq E_s$, the derivative of the bath's single-particle Green function satisfies
\begin{align}
	-\dv*{G(x_{1,\alpha},x_{1,\alpha};z)}{z}|_{z=E_i}&=\mel{x_{1,\alpha}}{(E_i-H_{\text{bath}})^{-2}}{x_{1,\alpha}}=\mel{x_{1,\alpha}}{(E_i-H_{\text{bath}})^{-1}\tilde{Q}(E_i-H_{\text{bath}})^{-1}}{x_{1,\alpha}}\nonumber\\
	&=\sum_{j,\beta}\abs{G(x_{1,\alpha}, x_{j,\beta};E_i)}^2\ge \abs{G(x_{1,\alpha}, x_{1,\alpha};E_i)}^2=(E_i-\Delta)^2/\text{g}^4>\Delta_{\rm min}^2/\text{g}^4,\label{IVA4}
\end{align}
where $\Delta_{\rm min}=\min_{k,p=\pm}\abs{\Delta-\epsilon_{k,p}}$ represents the minimum distance from $\Delta$ to the four band edges, with the dispersion relations $\epsilon_{k,\pm}=\pm\omega_k$ provided in Section \ref{IIA}. Thus, we have 
\begin{align}\label{IVA5}
	0\le\text{Res}[\mathscr{C}^{(1)}(z), E_i\neq E_s]=\frac{1}{1-\text{g}^2\dv*{G(x_{1,\alpha},x_{1,\alpha};z)}{z}|_{z=E_i}}<\frac{1}{1+\Delta_{\rm min}^2/\text{g}^2}.
\end{align}
Similarly, for $E_i=E_s$, the derivative of the bath's single-particle Green function satisfies
\begin{align}
	-\dv*{G(x_{1,\alpha},x_{1,\alpha};z)}{z}|_{z=E_s}
	&=\sum_k\mel{x_{1,\alpha}}{(E_s-H_{\text{bath}})^{-2}(u_k^\dagger\ketbra{\rm vac}u_k^{}+l_k^\dagger\ketbra{\rm vac}l_k^{})}{x_{1,\alpha}}\nonumber\\
	&=\sum_k\frac{|\mel{x_{1,\alpha}}{u_k^\dagger}{\text{vac}}|^2}{(E_s-\omega_k)^{2}}+\sum_k\frac{|\mel{x_{1,\alpha}}{l_k^\dagger}{\text{vac}}|^2}{(E_s+\omega_k)^{2}}\nonumber\\
	&<\frac{\sum_k(|\mel{x_{1,\alpha}}{u_k^\dagger}{\text{vac}}|^2+|\mel{x_{1,\alpha}}{l_k^\dagger}{\text{vac}}|^2)}{(\min_{k,p=\pm}|E_s-\epsilon_{k,p}|)^2}\nonumber\\
	&=\frac{\mel{x_{1,\alpha}}{\tilde{Q}}{x_{1,\alpha}}}{(\min_{k,p=\pm}|E_s-\epsilon_{k,p}|)^2}=\frac{1}{(\min_{k,p=\pm}|E_s-\epsilon_{k,p}|)^2}<\frac{1}{\Delta_{\rm min}^2}.\label{IVA7}
\end{align}
Thus, we have
\begin{align}\label{IVA8}
	\frac{1}{1+\text{g}^2/\Delta_{\rm min}^2}<\text{Res}[\mathscr{C}^{(1)}(z), E_i= E_s]=\frac{1}{1-\text{g}^2\dv*{G(x_{1,\alpha},x_{1,\alpha};z)}{z}|_{z=E_i}}\le1.
\end{align}
Note that in the last step of Eqs.~(\ref{IVA4}, \ref{IVA7}), we use the system of inequalities, as shown by the box in Fig.\,\ref{Fig_Suppn2}. Under the Born-Markov approximation, i.e., $\text{g}\ll\Delta_{\rm min}$, Eqs.~(\ref{IVA5}) and (\ref{IVA8}) satisfy
\begin{align}\label{1TLSC}
	\text{Res}[\mathscr{C}^{(1)}(z), E_i\neq E_s]\to0,\quad \text{Res}[\mathscr{C}^{(1)}(z), E_i= E_s]\to1,
\end{align}
which renders the contribution of bound states with energies $E_i\neq E_s$ to the atomic dynamics negligible, while the contribution from bound state with energy $E_i=E_s$ remains significant. Note that the precise form of $E_s$ is typically challenging to determine, except in the special case where $\Delta=0$, yielding $E_s=0$. 

Next, let us direct our attention to the bound state $\ket{\Psi_s}$ with the special energy $E_s$. Following \cite{cohen1998atom}, the projected Green function of the single-atom coupled to a topological bath can be written as
\begin{align}\label{QGP-QGQ}
	QG_{\rm tot}^{(1)}P=G_{\rm bath}(z)H_{\rm int}PG_{\rm tot}^{(1)}P,\quad QG_{\rm tot}^{(1)}Q=G_{\rm bath}(z)+G_{\rm bath}(z)H_{\rm int}PG_{\rm tot}^{(1)}PH_{\rm int}G_{\rm bath}(z).
\end{align}
Then, according to Eqs.~(\ref{S6}, \ref{S9}-\ref{S12}, \ref{IVA2}), we have
\begin{align}
	\mel{g;\mathrm{x}_{i,\gamma}}{QG_{\rm tot}^{(1)}P}{e;{\rm vac}}&=\frac{\text{g}G(\mathrm{x}_{i,\gamma},x_{1,\alpha};z)}{z-\Delta-\text{g}^2G(x_{1,\alpha},x_{1,\alpha};z)},\\
	\mel{g;\mathrm{x}_{i,\gamma}}{QG_{\rm tot}^{(1)}Q}{g;\mathrm{x}_{j,\gamma'}}&=G(x_{i,\gamma},\mathrm{x}_{j,\gamma'};z)+\frac{\text{g}^2G(\mathrm{x}_{i,\gamma},x_{1,\alpha};z)G(x_{1,\alpha},\mathrm{x}_{j,\gamma'};z)}{z-\Delta-\text{g}^2G(x_{1,\alpha},x_{1,\alpha};z)},\label{QGQ}
\end{align}
Since this expression has a first-order pole at $z=E_s$, a tiny cycle integral around $E_s$, denoted as $\mathcal{C}_s$, is performed to carry out the residue, which gives nothing but the bound state component $\ket{\Psi_s}$ as
\begin{align}\label{IVbe}
	\ointctrclockwise_{\mathcal{C}_s}\mel{g;\mathrm{x}_{i,\gamma}}{QG_{\rm tot}^{(1)}P}{e;{\rm vac}}\dd{z}&=2\pi i\braket{g;\mathrm{x}_{i,\gamma}}{\Psi_s}\braket{\Psi_s}{e;{\rm vac}}=\ointctrclockwise_{\mathcal{C}_s}\frac{\text{g}G(\mathrm{x}_{i,\gamma},x_{1,\alpha};z)}{z-\Delta-\text{g}^2G(x_{1,\alpha},x_{1,\alpha};z)}\dd{z}\nonumber\\
	&=2\pi i\text{g}G(\mathrm{x}_{i,\gamma},x_{1,\alpha};E_s)\times\text{Res}[\mathscr{C}^{(1)}(z), E_s],
\end{align}
\begin{align}\label{IVbb}
	\ointctrclockwise_{\mathcal{C}_s}\mel{g; \mathrm{x}_{i,\gamma}}{QG_{\rm tot}^{(1)}Q}{g;\mathrm{x}_{j,\gamma'}}\dd{z}&=2\pi i\braket{g;\mathrm{x}_{i,\gamma}}{\Psi_s}\braket{\Psi_s}{g;\mathrm{x}_{j,\gamma'}}=\ointctrclockwise_{\mathcal{C}_s}\frac{\text{g}^2G(\mathrm{x}_{i,\gamma},x_{1,\alpha};z)G(x_{1,\alpha},\mathrm{x}_{j,\gamma'};z)}{z-\Delta-\text{g}^2G(x_{1,\alpha},x_{1,\alpha};z)}\dd{z}\nonumber\\
	&=2\pi i\text{g}^2G(\mathrm{x}_{i,\gamma},x_{1,\alpha};E_s)G^*(\mathrm{x}_{j,\gamma'},x_{1,\alpha};E_s)\times\text{Res}[\mathscr{C}^{(1)}(z), E_s].
\end{align}
In the first step of Eqs.~(\ref{IVbe}) and (\ref{IVbb}), we perform a spectral decomposition on $H_{\rm tot}^{(1)}$. For simplicity, we denote $c_e$ and $c_{i,\gamma}$ as the components of the bound state on the atom and bosonic modes of the bath, respectively, i.e., $c_e=\braket{e;{\rm vac}}{\Psi_s}$ and $c_{i,\gamma}=\braket{g; \mathrm{x}_{i,\gamma}}{\Psi_s}$. Therefore, according to Eqs.~(\ref{IVA3}, \ref{IVbe}, \ref{IVbb}), we have
\begin{align}
	c_{i,\gamma}^{}c_e^*=\text{g}G(x_{i,\gamma},x_{1,\alpha};E_s)\text{Res}[\mathscr{C}^{(1)}(z), E_s]&,\quad
	c_{i,\gamma}^{}c_{j,\gamma'}^*=\text{g}^2G(\mathrm{x}_{i,\gamma},x_{1,\alpha};E_s)G^*(\mathrm{x}_{j,\gamma'},x_{1,\alpha};E_s)\text{Res}[\mathscr{C}^{(1)}(z), E_s],\nonumber\\
	\Longrightarrow c_{i,\gamma}=\text{g}c_eG(\mathrm{x}_{i,\gamma},x_{1,\alpha};E_s)&,\quad c_e=\frac{1}{\sqrt{1-\text{g}^2\dv*{G(x_{1,\alpha},x_{1,\alpha};z)}{z}|_{z=E_s}}}.\label{IVc}
\end{align}
Thus, the concrete expression of the dressed bound state can be written as 
\begin{align}\label{IVAdbs}
	\ket{\Psi_s}=c_e\ket{e}\otimes\ket{\rm vac}+\sum_{i,\gamma}c_{i,\gamma}o_{\mathrm{x}_{i,\gamma}}^\dagger\ket{g}\otimes\ket{\rm vac}=c_e\ket{e;\rm vac}+\sum_{i}(c_{i,A}a_{\mathrm{x}_i}^\dagger+c_{i,B}b_{\mathrm{x}_i}^\dagger)\ket{g;\rm vac},
\end{align}
which satisfies $H_{\rm tot}^{(1)}\ket{\Psi_s}=(E_s+\omega_c)\ket{\Psi_s}$.

\begin{figure}
	\centering
	\includegraphics[width=17cm]{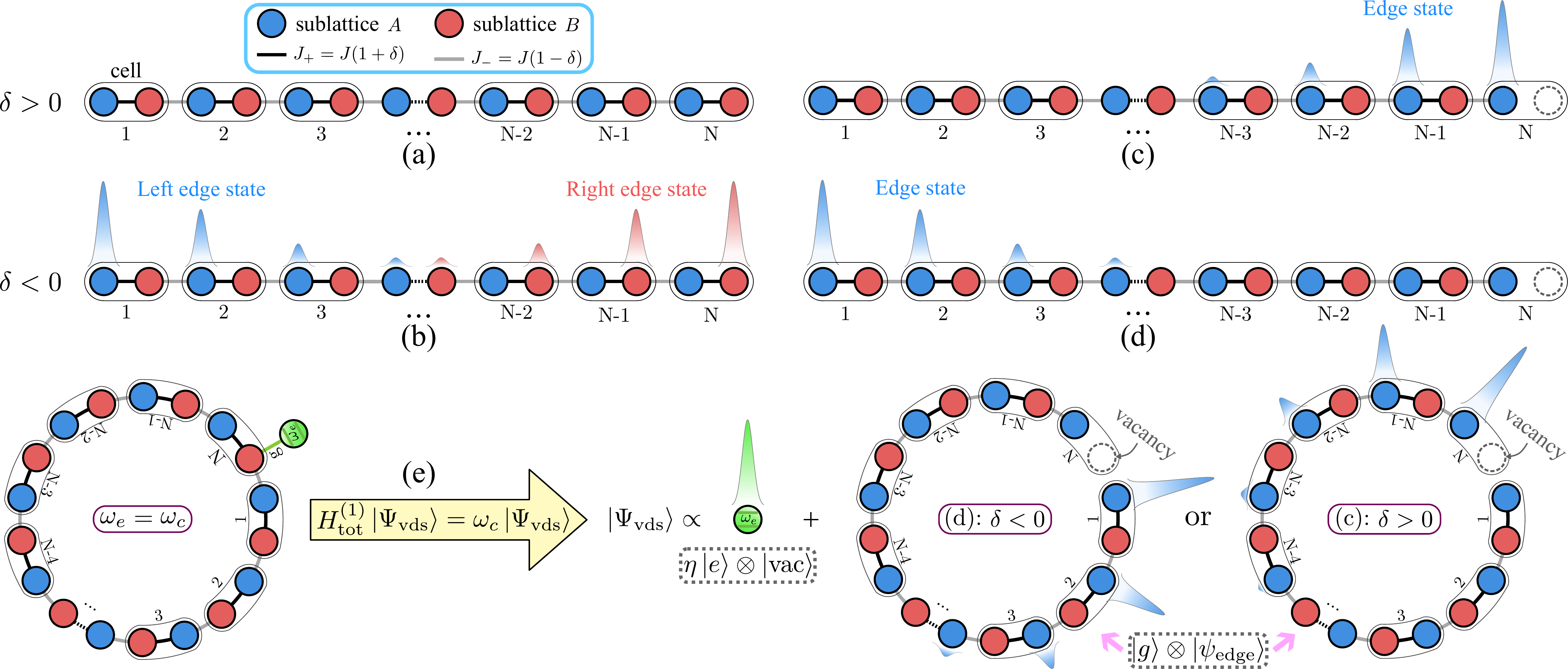}\\
	\caption{Edge state spatial distributions in an open SSH chain with an even number of sites $2N$ in (a, b) and an odd number of sites $2N-1$ in (c, d), for both topologically trivial ($\delta>0$, first row) and nontrivial ($\delta<0$, second row) phases. Note that for an even number of sites, the thermodynamic limit ($N\to\infty$) must be met. Panel (e) presents a schematic of the vacancy-like dressed state formed by a single atom coupled to a periodic SSH chain with $N$ unit cells. 
	}\label{fign3}
\end{figure}

Finally, we consider a parameter used in Fig.2(a-d) of the main text, i.e., resonance condition $\omega_e=\omega_c$. In this parameter, according to Eqs.~(\ref{S32}-\ref{S36}), we can analytically obtain the two coefficients in Eq.~(\ref{IVc}) and the dressed bound state energy $E_s=0$. However, in addition to the Green function approach discussed above, there exists a more intuitive method for solving this dress bound state. Before introducing this intuitive method, we need to understand an open SSH chain with an even and odd number of sites. For an even number of sites $2N$, as shown in Fig.\,\ref{fign3}(a-b), we observe in the thermodynamic limit of $N\to \infty$ and open boundary condition that in the topologically trivial phase ($\delta>0$), edge states are absent, whereas in the topologically nontrivial phase ($\delta<0$), two edge states appear. The corresponding left and right edge states are given by\,\cite{LRedge}
\begin{align}\label{IVedge1}
	|\psi_{\rm edge}^{L}\rangle\xlongequal{N\to\infty}\frac{2\sqrt{\abs{\delta}}}{1+\abs{\delta}}\sum_{n=1}^{N}\left(\frac{\delta+1}{\delta-1}\right)^{n-1}a_n^\dagger\ket{\rm vac},\quad
	|\psi_{\rm edge}^{R}\rangle\xlongequal{N\to\infty}\frac{2\sqrt{\abs{\delta}}}{1+\abs{\delta}}\sum_{n=1}^{N}\left(\frac{\delta+1}{\delta-1}\right)^{N-n}b_n^\dagger\ket{\rm vac}.
\end{align}
However, for an odd number of sites $2N-1$, as shown in Fig.\,\ref{fign3}(c-d), we observe in open boundary condition that, irrespective of whether the system is in a topologically trivial or non-trivial phase, a single in-gap edge state is consistently present, with its position---on the left or right---determined by the sign of $\delta$. The corresponding edge state is given by\,\cite{edge}
\begin{align}\label{IVedge2}
	\ket{\psi_{\rm edge}}=\left(\frac{\delta-1}{\delta+1}\right)^{(N-1)\Theta(\delta)}\times\frac{2\sqrt{\abs{\delta}}}{1+\abs{\delta}}\sum_{n=1}^{N}\left(\frac{\delta+1}{\delta-1}\right)^{n-1}a_n^\dagger\ket{\rm vac}.
\end{align}
Here, Eq.~(\ref{IVedge2}) holds exactly even for finite $N$. Besides, we can confirm that the coefficient of the left edge state in Fig.\,\ref{fign3}(b) corresponds to the edge state in Fig.\,\ref{fign3}(d), and the coefficient of the right edge state corresponds to that in Fig.\,\ref{fign3}(c). This correspondence can also be verified by comparing Eq.~(\ref{IVedge2}) with Eq.~(\ref{IVedge1}). Next, for the sake of simplicity, we will assume $x_1=N$ and $\alpha=B$, as shown in Fig.\,\ref{fign3}(e). Therefore, the total Hamiltonian $H_{\rm tot}^{(1)}$ can be rewritten as
\begin{align}
	H_{\rm tot}^{(1)}=\omega_e\sigma_+\sigma_-+H_{\rm bath}+\text{g}(\sigma_-^{} b_N^\dagger + b_N \sigma_+),
\end{align}
where 
\begin{align}
	H_{\rm bath}=\omega_c b_N^\dagger b_N^{} + J_+(a_{N}^\dagger b_N^{}+b_N^\dagger a_{N}^{})+J_-(a_1^\dagger b_N^{} + b_N^\dagger a_1^{})+H_{\rm bath}^{\rm vac}.
\end{align}
Here, $H_{\rm bath}^{\rm vac}$ denotes the Hamiltonian formed by removing all terms associated with the $b_N$ mode from $H_{\rm bath}$ [i.e., Eq.~(\ref{S17})], which is equivalent to replacing the $b_N$ mode cavity with a vacancy, yielding an open SSH chain with an odd number of sites $2N-1$, as shown in Fig.\,\ref{fign3}(c,d). 

A vacancy-like dressed state (VDS) is defined by $\ket{\Psi_{\rm vds}}\propto \eta\ket{e;\rm vac}+\ket{g; \psi}$, fulfilling $H_{\rm tot}^{(1)}\ket{\Psi_{\rm vds}}=\omega_c\ket{\Psi_{\rm vds}}$ with $\ket{\psi}=\sum_{n=1}^N(c_{n,A}^{}a_n^\dagger+c_{n,B}^{}b_n^\dagger)\ket{\rm vac}$. When $\omega_e=\omega_c$, we have
\begin{align}
	H_{\rm tot}^{(1)}\ket{\Psi_{\rm vds}}&=\omega_c\ket{\Psi_{\rm vds}}\nonumber\\
	\Longrightarrow[\omega_c b_N^\dagger b_N^{} + J_+(a_{N}^\dagger b_N^{}+b_N^\dagger a_{N}^{})+J_-(a_1^\dagger b_N^{} + b_N^\dagger a_1^{})+&H_{\rm bath}^{\rm vac}]\ket{g;\psi}+\text{g}[\eta b_N^\dagger\ket{g;\rm vac}+b_N^{}\ket{e;\psi}]=\omega_c\ket{g; \psi}.\label{IVAll}
\end{align}
This can be seen by projecting Eq.~(\ref{IVAll}) onto $\ket{e;\rm vac}$, yielding $\text{g}\mel{e;\rm vac}{b_N}{e;\psi}=0$; hence, $c_{N,B}=0$ and Eq.~(\ref{IVAll}) can be further simplified as
\begin{align}\label{IVAll1}
	[H_{\rm bath}^{\rm vac}-\omega_c]\ket{g;\psi}+[\eta\text{g}+c_{N,A}J_++c_{1,A}J_-]b_N^\dagger\ket{g;\rm vac}=0.
\end{align}
Since the first and second terms in Eq.~(\ref{IVAll1}) are orthogonal, we have
\begin{align}\label{IVAll2}
	H_{\rm bath}^{\rm vac}\ket{g;\psi}=\omega_c\ket{g;\psi},\quad \eta=-(c_{N,A}J_++c_{1,A}J_-)/\text{g}.
\end{align}
Here, $H_{\rm bath}^{\rm vac}$ is the Hamiltonian of an open SSH chain with an odd number of sites $2N-1$. As shown in Fig.\,\ref{fign3}(c) and \ref{fign3}(d), this chain is well known to exhibit a single in-gap topologically robust edge state $\ket{\psi_{\rm edge}}$ [see Eq.~(\ref{IVedge2})] of energy $\omega_c$ with nonzero amplitude on sites of given parity. Therefore, we have $\ket{\psi}=\ket{\psi_{\rm edge}}$, and the corresponding coefficients $c_{n,A}$ and $c_{n,A}$ can be written as
\begin{align}\label{IVAll3}
	c_{n,A}=\frac{2\sqrt{\abs{\delta}}}{1+\abs{\delta}}\left(\frac{\delta+1}{\delta-1}\right)^{n-1-(N-1)\Theta(\delta)},\quad c_{n,B}=0.
\end{align}
By plugging Eq.~(\ref{IVAll3}) into Eq.~(\ref{IVAll2}), in the thermodynamic limit of $N\to \infty$, we have
\begin{align}\label{coeffeta}
	\lim\limits_{N\to \infty}\eta=-\lim\limits_{N\to \infty}(c_{N,A}J_++c_{1,A}J_-)/\text{g}=-2\sqrt{\abs{\delta}}J/\text{g},
\end{align}
and the VDS can be written as
\begin{align}\label{IVAll4}
	\ket{\Psi_{\rm vds}}=[-2\sqrt{|\delta|}J/\text{g}\ket{e}\otimes\ket{\rm vac}+\ket{g}\otimes\ket{\psi_{\rm edge}} ]\times(-\text{g}/\sqrt{\text{g}^2+4J^2\abs{\delta}}),
\end{align}
which is equivalent to the dressed state of $\Delta=0=E_s$ in Eq.~(\ref{IVAdbs}). Now, we assume the initial state of the total system to be $\ket{\psi(0)}=\ket{e}\otimes\ket{\rm vac}$ and define the evolution operator of the system $U(t)=\text{exp}[-iH_{\rm tot}^{(1)}t]$. In order to obtain the probability amplitude of the atom being in the excited state at $t$ time under the Born-Markov approximation, we aim to calculate the difference between the initial state and the VDS in terms of the atomic and photonic components. Based on Eq.~(\ref{IVAll4}), the corresponding differences satisfy
\begin{align}
	\bra{e;\rm vac}[\ket{\psi(0)}-\ket{\Psi_{\rm vds}}]=\order{\text{g}^2/\Delta_{\rm min}^2},\quad \bra{g;\mathrm{x}_{j,\gamma}}[\ket{\psi(0)}-\ket{\Psi_{\rm vds}}]=\order{\text{g}^3/\Delta_{\rm min}^3}.
\end{align}
For simplicity, the initial state can therefore be expressed as $\ket{\psi(0)}=\ket{\Psi_{\rm vds}}+|\order{\text{g}^2/\Delta_{\rm min}^2}\rangle$, where the perturbation term satisfies $\bra{e;\rm vac}\order{\text{g}^2/\Delta_{\rm min}^2}\rangle=\order{\text{g}^2/\Delta_{\rm min}^2}$ and $\bra{g;\mathrm{x}_{j,\gamma}}\order{\text{g}^2/\Delta_{\rm min}^2}\rangle=\order{\text{g}^3/\Delta_{\rm min}^3}$. Combining the above definitions and discussions, the probability amplitude of the atom being in the excited state at $t$ time is computed as 
\begin{align}
	\mel{e;\rm vac}{U(t)}{\psi(0)}&=\mel{e;\rm vac}{U(t)}{\Psi_{\rm vds}}+\mel{e;\rm vac}{U(t)}{\order{\text{g}^2/\Delta_{\rm min}^2}}=\exp(-i\omega_c t)\braket{e;\rm vac}{\Psi_{\rm vds}}+\order{\text{g}^2/\Delta_{\rm min}^2}\nonumber\\
	&=\exp(-i\omega_c t)[1+\order{\text{g}^2/\Delta_{\rm min}^2}]+\order{\text{g}^2/\Delta_{\rm min}^2}=\exp(-i\omega_c t)+\order{\text{g}^2/\Delta_{\rm min}^2},
\end{align}
where $\Delta_{\rm min}=2J\abs{\delta}$. For $\text{g}\ll\Delta_{\rm min}$, the probability of the atom being in the excited state at any time approaches to one, i.e., $|\mel{e;\rm vac}{U(t)}{\psi(0)}|^2\to1$. As a result, the existence of the VDS keeps the atom almost permanently in the excited state under the Born-Markov approximation.

\subsection{Two atoms coupled to photonic Su-Schrieffer-Heeger model}\label{IVB}
Now, let us return to the two-atom system coupled to the topological waveguide. To ensure consistency with the single-atom discussion, we also rewritten the total Hamiltonian as follows: 
\begin{align}\label{totHamiltonian}
	H_{\rm tot}^{(2)}=H_{\rm sys}+H_{\rm bath}+H_{\rm int}=\omega_e(\sigma_+^{\rm B}\sigma_-^{\rm B}+\sigma_+^{\rm C}\sigma_-^{\rm C})+H_{\rm bath}+\text{g}(\sigma_-^{\rm B} o^\dagger_{x_{1,\alpha}}+o_{x_{1,\alpha}}\sigma_+^{\rm B}+\sigma_-^{\rm C} o^\dagger_{x_{2,\beta}}+o_{x_{2,\beta}}\sigma_+^{\rm C}),
\end{align}
and the system's projection operator, $P=(\ketbra{e,g}+\ketbra{g,e})\otimes\ketbra{\rm vac}$. Following these steps (\ref{S2}-\ref{S15}), when the total system is prepared in the initial state $\ket{\psi(0)}=\ket{e_1;\rm vac}$, i.e., the quantum charger is in the excited state, QB is in the ground state, and the bath is in the vacuum state, the probability amplitude for QB to be in the excited state $\ket{e_2;\rm vac}$ at $t$ time is given by $c_{\rm B}(t)=\int_{\mathcal{C}}\mathscr{C}^{(2)}(z)e^{-izt}\dd{z}/(2\pi i)$ with $\mathscr{C}^{(2)}(z)=\text{g}^2G(x_{1,\alpha},x_{2,\beta};z)/\mathscr{D}(z)$, where
\begin{align}\label{poleequation}
	\mathscr{D}(z)&=[z-\Delta-\mathrm{g}^2G(x_{1,\alpha},x_{1,\alpha};z)][z-\Delta-\mathrm{g}^2G(x_{2,\beta},x_{2,\beta};z)]-\mathrm{g}^4G(x_{1,\alpha},x_{2,\beta};z)G(x_{2,\beta},x_{1,\alpha};z).
\end{align}
According to Eqs.~(\ref{S32}, \ref{S33}, \ref{S35}, \ref{S36}) in the non-dissipation case, the bath Green functions obey
\begin{align}
	G(x_{1,\alpha},x_{1,\alpha};z)=G(x_{2,\beta},x_{2,\beta};z),\quad G(x_{1,\alpha},x_{2,\beta};z)=G(x_{2,\beta},x_{1,\alpha};z).
\end{align} 
Therefore, we have
\begin{align}\label{IVB3}
	\mathscr{C}^{(2)}(z)=\frac{\text{g}^2G(x_{1,\alpha},x_{2,\beta};z)}{[z-\Delta-\mathrm{g}^2G(x_{1,\alpha},x_{1,\alpha};z)]^2-[\text{g}^2G(x_{1,\alpha},x_{2,\beta};z)]^2}
	=\frac{1/2}{z-\Delta-\text{g}^2G_+(z)}-\frac{1/2}{z-\Delta-\text{g}^2G_-(z)},
\end{align}
where
\begin{align}\label{Gpm}
	G_{\pm}(z)=G(x_{1,\alpha},x_{1,\alpha};z)\pm G(x_{1,\alpha},x_{2,\beta};z)=\mel{\rm vac}{(o_{x_{1,\alpha}}\pm o_{x_{2,\beta}})(z-H_{\rm bath})^{-1}(o_{x_{1,\alpha}}\pm o_{x_{2,\beta}})^\dagger}{\rm vac}/2.
\end{align}
For the sake of simplicity, Eq.~(\ref{IVB3}) can be simplified as 
\begin{align}
	\mathscr{C}^{(2)}(z)=\mathscr{C}^{(2)}_+(z)-\mathscr{C}^{(2)}_-(z),\ \text{where}\  \mathscr{C}^{(2)}_\pm(z)=[z-\Delta-\text{g}^2G_\pm(z)]^{-1}/2.
\end{align}

Next, similar to Section \ref{SIVA}, we first solve the poles equations $E_i^{\pm}-\Delta-\text{g}^2G_\pm(E_i^{\pm})=0$ to obtain the BSEs. We then focus on two special bound states with energies $E_s^{\pm}$, which lie closest to $\Delta$. Given these poles, their residues are
\begin{align}\label{Rpm}
	\text{Res}[\mathscr{C}^{(2)}_{\pm}(z), E_i^{\pm}]=\frac{1/2}{\dv*{[z-\Delta-\text{g}^2G_\pm(z)]}{z}}\Big|_{z=E_i^\pm}.
\end{align}
According to Eqs.~(\ref{IVA4}-\ref{IVA8}), we also have
\begin{align}\label{IVB6}
	0\le\text{Res}[\mathscr{C}^{(2)}_{\pm}(z), E_i^{\pm}\neq E_s^{\pm}]<\frac{1/2}{1+\Delta_{\rm min}^2/\text{g}^2},\quad 
	\frac{1/2}{1+\text{g}^2/\Delta_{\rm min}^2}<\text{Res}[\mathscr{C}^{(2)}_{\pm}(z), E_i^\pm= E_s^\pm]\le\frac{1}{2}.
\end{align}
Under the Born-Markov approximation ($\text{g}\ll\Delta_{\rm min}$), Eq.~(\ref{IVB6}) satisfies
\begin{align}\label{atomsR}
	\text{Res}[\mathscr{C}^{(2)}_{\pm}(z), E_i^{\pm}\neq E_s^{\pm}]\to0,\quad \text{Res}[\mathscr{C}^{(2)}_\pm(z), E_i^{\pm}= E_s^{\pm}]\to1/2.
\end{align}
As indicated by the conclusion of Eq.~(\ref{1TLSC}), we find a similar result for the two atoms system: the contribution of bound states with energies $E_i^\pm\neq E_s^\pm$ to the atomic dynamics is negligible, whereas bound states with energies $E_i^\pm=E_s^\pm$ (with $E_s^+\neq E_s^-$) contribute significantly. In the following discussion, we will primarily focus on the case where $E_i^\pm\neq E_s^\pm$, as the condition $E_s^+=E_s^-$ does not change the conclusions we draw. 

Now, let us direct our attention to the bound states $\ket{\Psi_s^{\pm}}$ with the special energies $E_{s}^\pm$. For the sake of simplicity, we define a new notation based on Eq.~(\ref{Gpm}), i.e.,
\begin{align}
	G_\pm(\mathrm{x}_{i,\gamma};z)=G(\mathrm{x}_{i,\gamma}, x_{1,\alpha};z)\pm G(\mathrm{x}_{i,\gamma},x_{2,\beta};z),\ {\rm where} \ G_\pm(x_{1,\alpha};z)=G_\pm(z).
\end{align}
Similarly, following these steps (\ref{QGP-QGQ}-\ref{QGQ}), we also have
\begin{align}
	\mel{g; \mathrm{x}_{i,\gamma}}{QG_{\rm tot}^{(2)}P}{e_1;\rm vac}&=\text{g}\sum_{p=\pm}\frac{pG_p(\mathrm{x}_{i,\gamma};z)}{z-\Delta-\text{g}^2G_p(z)},\quad
	\mel{g; \mathrm{x}_{i,\gamma}}{QG_{\rm tot}^{(2)}P}{e_2;\rm vac}=\text{g}\sum_{p=\pm}\frac{G_p(\mathrm{x}_{i,\gamma};z)}{z-\Delta-\text{g}^2G_p(z)},\\
	\mel{g; \mathrm{x}_{i,\gamma}}{QG_{\rm tot}^{(2)}Q}{g; \mathrm{x}_{j,\gamma'}}
	&=G(\mathrm{x}_{i,\gamma},\mathrm{x}_{j,\gamma'};z)+\text{g}^2\sum_{p=\pm}\frac{G_p(\mathrm{x}_{i,\gamma};z)G^*_p(\mathrm{x}_{j,\gamma'};z)}{z-\Delta-\text{g}^2G_p(z)}.
\end{align}
Since this expression has a first-order pole (assuming $E_s^+\neq E_s^-$, the conclusions derived in the following steps still hold even if $E_s^+=E_s^-$) at $z=E_s^p$ with $p\in\{+,-\}$, a tiny cycle integral around $E_s^p$, denoted as $\mathcal{C}_s^p$, is performed to carry out the residue, which gives the bound state $\ket{\Psi_s^p}$ component as
\begin{align}
	\ointctrclockwise_{\mathcal{C}_s^p}\mel{g; \mathrm{x}_{i,\gamma}}{QG_{\rm tot}^{(2)}P}{e_1;{\rm vac}}\dd{z}&=2\pi i\braket{g; \mathrm{x}_{i,\gamma}}{\Psi_s^p}\braket{\Psi_s^p}{e_1;{\rm vac}}=\ointctrclockwise_{\mathcal{C}_s^p}\frac{p\text{g}G_p(\mathrm{x}_{i,\gamma};z)}{z-\Delta-\text{g}^2G_p(z)}\dd{z}\nonumber\\
	&=2\pi i p\text{g}G_p(\mathrm{x}_{i,\gamma};E_s^p)\times\text{Res}[\mathscr{C}^{(2)}_p(z), E_s^p],\label{QG2P1}\\
	\ointctrclockwise_{\mathcal{C}_s^p}\mel{g; \mathrm{x}_{i,\gamma}}{QG_{\rm tot}^{(2)}P}{e_2;{\rm vac}}\dd{z}&=2\pi i\braket{g;\mathrm{x}_{i,\gamma}}{\Psi_s^p}\braket{\Psi_s^p}{e_2;{\rm vac}}=\ointctrclockwise_{\mathcal{C}_s^p}\frac{\text{g}G_p(\mathrm{x}_{i,\gamma};z)}{z-\Delta-\text{g}^2G_p(z)}\dd{z}\nonumber\\
	&=2\pi i\text{g}G_p(\mathrm{x}_{i,\gamma};E_s^p)\times\text{Res}[\mathscr{C}^{(2)}_p(z), E_s^p],\label{QG2P2}\\
	\ointctrclockwise_{\mathcal{C}_s^p}\mel{g; \mathrm{x}_{i,\gamma}}{QG_{\rm tot}^{(2)}Q}{g; \mathrm{x}_{j,\gamma'}}\dd{z}&=2\pi i\braket{g;\mathrm{x}_{i,\gamma}}{\Psi_s^p}\braket{\Psi_s^p}{g;\mathrm{x}_{j,\gamma'}}=\ointctrclockwise_{\mathcal{C}_s^p}\frac{\text{g}^2G_p(\mathrm{x}_{i,\gamma};z)G^*_p(\mathrm{x}_{j,\gamma'};z)}{z-\Delta-\text{g}^2G_p(z)}\dd{z}\nonumber\\
	&=2\pi i\text{g}^2G_p(\mathrm{x}_{i,\gamma};E_s^p)G^*_p(\mathrm{x}_{j,\gamma'};E_s^p)\times\text{Res}[\mathscr{C}^{(2)}_p(z), E_s^p].\label{QG2Q}
\end{align}
Here, we denote $c_{e, 1(2)}^{p}$ and $c_{i,\gamma}^{p}$ as the components of the bound state $\ket{\Psi^p_s}$ on the quantum charger (battery) and bosonic modes of the bath, respectively, i.e., $c_{e,1}^p=\braket{e_1;\rm vac}{\Psi_s^p}$, $c_{e,2}^p=\braket{e_2;\rm vac}{\Psi_s^p}$, and $c_{i,\gamma}^p=\braket{g; \mathrm{x}_{i,\gamma}}{\Psi_s^p}$. Therefore, according to Eqs.~(\ref{Rpm}, \ref{QG2P1}-\ref{QG2Q}), we have
\begin{align}\label{IVBc}
	c_{i,\gamma}^p=\text{g}c_{e, 1}^p[pG(\mathrm{x}_{i,\gamma},x_{1,\alpha};E_{s}^p)+G(\mathrm{x}_{i,\gamma},x_{2,\beta};E_{s}^p)], \quad c_{e, 1}^p=\frac{1}{\sqrt{2}}\frac{1}{\sqrt{1-\text{g}^2\dv*{G_p(z)}{z}|_{z=E_s^p}}}=pc_{e, 2}^p.
\end{align}
Subsequently, we will establish a crucial link between the special bound states for a single atom ($\ket{\Psi_s}$) and two atoms ($\ket{\Psi_s^\pm}$) coupled to the SSH chain, specifically connecting Eq.~(\ref{IVBc}) with Eq.~(\ref{IVc}). Before we proceed with the discussion, we need to redefine the special bound state $|\Psi_s^{(1,2)}\rangle$ for a single atom coupled to the SSH chain, where the superscripts $(1)$ and $(2)$ represent the two coupled sites $x_{2,\beta}$ and $x_{1,\alpha}$, respectively. Therefore, according to Eqs.~(\ref{IVc}-\ref{IVAdbs}), the special bound state can be written as
\begin{align}
	|\Psi_{s}^{(1)}\rangle=c_e^{(1)}\ket{e_{1}}\otimes\ket{\rm vac}+\sum_{i,\gamma}c_{i,\gamma}^{(1)}\ket{g}\otimes\ket{\mathrm{x}_{i,\gamma}},\quad |\Psi_{s}^{(2)}\rangle=c_e^{(2)}\ket{e_{2}}\otimes\ket{\rm vac}+\sum_{i,\gamma}c_{i,\gamma}^{(2)}\ket{g}\otimes\ket{\mathrm{x}_{i,\gamma}},
\end{align}
where
\begin{figure}
	\centering
	\includegraphics[width=17cm]{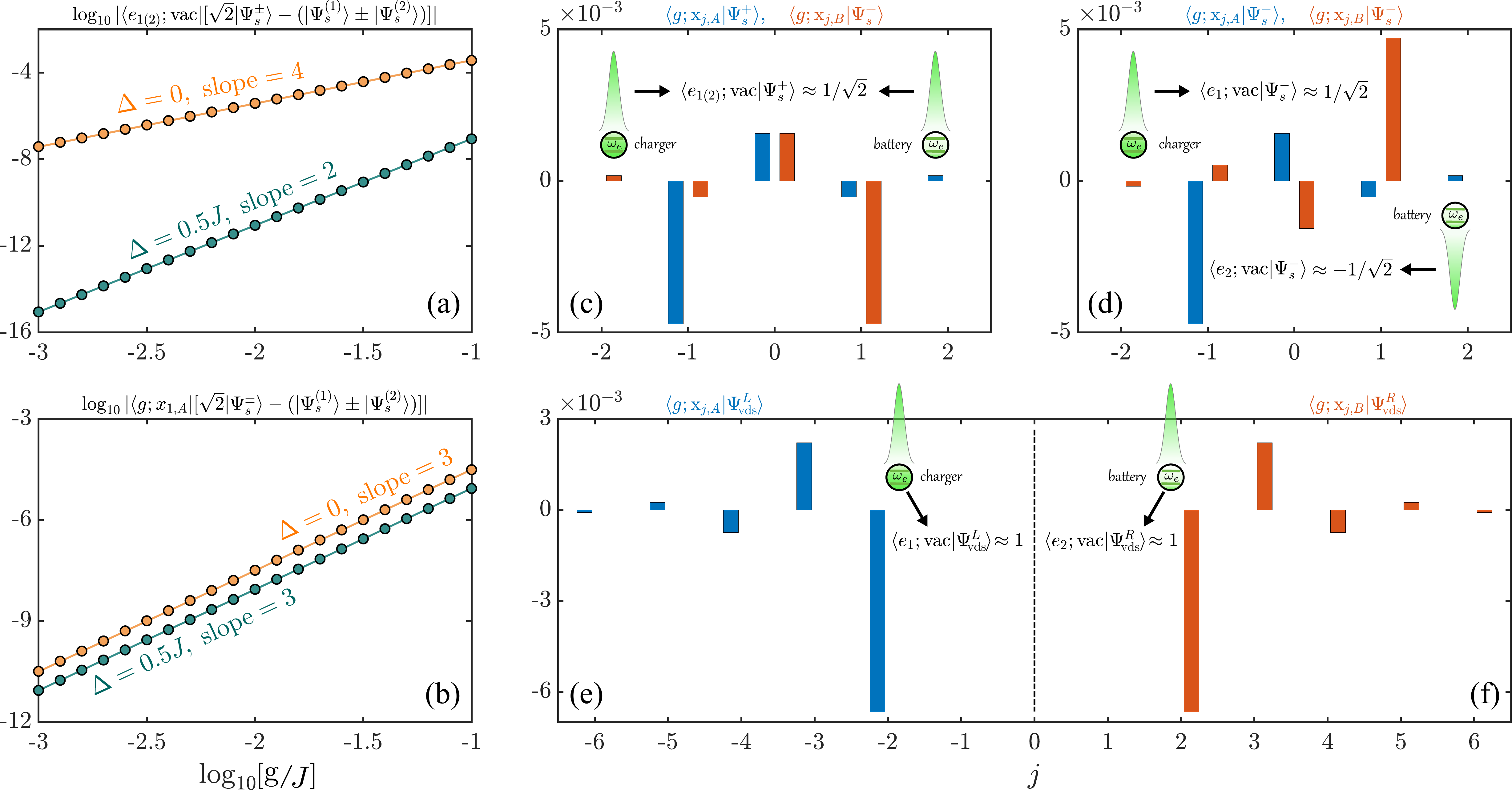}\\
	\caption{Panel (a) confirms a second-order scaling (slope = 2) for $\Delta=0.5J$ and a fourth-order scaling (slope = 4) for $\Delta=0$ in the derivation of the atomic components between the parity-positive (parity-negative) bound state $\ket{\Psi_s^\pm}$ and the positive (negative) superposition state $|\Psi_s^{(1)}\rangle\pm|\Psi_s^{(2)}\rangle$, while panel (b) shows a consistent third-order scaling (slope = 3) for the photonic components across both detuning conditions. Panels (c) and (d) illustrate the atomic and photonic distributions of the parity-positive and parity-negative bound states for $\Delta=0$ in the topologically nontrivial phase, while panels (e) and (f) depict the left and right vacancy-like dressed states for $\Delta=0$ in the topologically trivial phase. In panels (c-f), the photonic components are shown in blue for sublattice $A$ and in orange for sublattice $B$. In all plots, we set $\alpha=A$, $\beta=B$, and $d=x_1-x_2=4$. The system parameters are chosen as $\delta=-0.5$ in panels (a-d), $\delta=0.5$ in panels (e-f), and $\text{g}=0.01J$ in panels (c-f). 
	}\label{fign6}
\end{figure}\noindent
\begin{align}
	c_{i,\gamma}^{(1)}=\text{g}c_{e}^{(1)}G(\mathrm{x}_{i,\gamma},x_{2,\beta};E_s),\quad c_{i,\gamma}^{(2)}=\text{g}c_{e}^{(2)}G(\mathrm{x}_{i,\gamma},x_{1,\alpha};E_s),\quad c_e=\frac{1}{\sqrt{1-\text{g}^2G'(x_{1,\alpha},x_{1,\alpha};E_s)}}=c_{e}^{(1)}=c_e^{(2)}.
\end{align}
Besides, we also need to introduce the following inequality:
\begin{align}
	\abs{\text{d}^nG(x_{1,\alpha},x_{2,\beta};z)/\dd{z}^n}_{z=E_s}&=\abs{(-1)^nn!\mel{x_{1,\alpha}}{(E_s-H_{\rm bath})^{-n}(E_s-H_{\rm bath})^{-1}}{x_{2,\beta}}}\nonumber\\
	&\le n![\mel{x_{1,\alpha}}{(E_s-H_{\rm bath})^{-2n}}{x_{1,\alpha}}\mel{x_{2,\beta}}{(E_s-H_{\rm bath})^{-2}}{x_{2,\beta}}]^{1/2}\nonumber\\
	&<n![\Delta_{\rm min}^{-2n}\Delta_{\rm min}^{-2}]^{1/2}=n!/\Delta_{\rm min}^{n+1},\label{IVBineq}
\end{align}
where $E_s$ denotes the energy of the special bound state $\ket{\Psi_s}$ discussed in Section \ref{IVA}. In the second step above, we use the Cauchy-Schwarz inequality, i.e., $\abs{\braket{\psi}{\varphi}}^2\le\braket{\psi}\braket{\varphi}$. In the last step, we use the derivation of Eq.~(\ref{IVA7}). Now, under the Born-Markov approximation, we perform a perturbation expansion to poles equation $z-\Delta-\text{g}^2G_p(z)=0$ at $z=E_s$ up to the first order of $\delta E_s^p+\mathcal{O}[(\delta E_s^p)^2]=E_s^p-E_s$, and obtain
\begin{align}
	&E_s+\delta E_{s}^p -\Delta-\text{g}^2G_p(E_s)-\text{g}^2\delta E_{s}^pG_{p}'(E_s)+\mathcal{O}[(\delta E_s^p)^2]=0\Longrightarrow\delta E_s^p=p\frac{\text{g}^2G(x_{1,\alpha},x_{2,\beta};E_s)}{1-\text{g}^2G_p'(E_s)}+\mathcal{O}[(\delta E_s^p)^2],
\end{align}
which satisfies
\begin{align}
	\abs{\delta E_s^p/\Delta_{\rm min}}&=\abs{\frac{\text{g}^2G(x_{1,\alpha},x_{2,\beta};E_s)}{1-\text{g}^2G_p'(E_s)}\frac{1}{\Delta_{\rm min}}+\frac{\mathcal{O}[(\delta E_s^p)^2]}{\Delta_{\rm min}}}\le \abs{\frac{\text{g}^2G(x_{1,\alpha},x_{2,\beta};E_s)}{\Delta_{\rm min}}}+\abs{\mathcal{O}[(\delta E_s^p/\Delta_{\rm min})^2]}\nonumber\\
	&<\text{g}^2/\Delta_{\rm min}^2+\abs{\mathcal{O}[(\delta E_s^p/\Delta_{\rm min})^2]}<\text{g}^2/\Delta_{\rm min}^2+\abs{\mathcal{O}(\text{g}^4/\Delta^4_{\rm min})},
\end{align}
implying that the perturbation expansion at $z=E_s$ is reasonable. For concreteness, we retain only the terms up to the second-order approximation in the following derivations, i.e., $\delta E_s^\pm\approx \pm\text{g}^2G(x_{1,\alpha},x_{2,\beta};E_s)$. Therefore, we could perform a perturbation expansion to the components of the bound state $\ket{\Psi_s^p}$ [i.e., Eq.~(\ref{IVBc})] at $z=E_s$, and obtain
\begin{align}
	\sqrt{2}c_{e,1}^p=c_e+p\text{g}^2G'(x_{1,\alpha},x_{2,\beta};&E_s)/2+\order{\text{g}^4/\Delta_{\rm min}^4}=c_e+\order{\text{g}^2/\Delta_{\rm min}^2}=\bra{e_1;\rm vac}(|\Psi_s^{(1)}\rangle+p|\Psi_s^{(2)}\rangle)+\order{\text{g}^2/\Delta_{\rm min}^2}\nonumber\\
	&\Longrightarrow \langle e_{1(2)};{\rm vac}|[\sqrt{2}\ket{\Psi_s^p}-(|\Psi_s^{(1)}\rangle+p|\Psi_s^{(2)}\rangle)]=\order{\text{g}^2/\Delta_{\rm min}^2},\label{IVBee}
\end{align}
and
\begin{align}
	\sqrt{2}c_{i,\gamma}^p&=p\text{g}c_eG_p(\mathrm{x}_{i,\gamma};E_s)+\text{g}^3[G'(x_{1,\alpha},x_{2,\beta};E_s)G_p(\mathrm{x}_{i,\gamma};E_s)+2G(x_{1,\alpha},x_{2,\beta};E_s)G_p'(\mathrm{x}_{i,\gamma};E_s)]/2+\order{\text{g}^5/\Delta_{\rm min}^5}\nonumber\\
	&=\text{g}[c_e^{(2)}pG(\mathrm{x}_{i,\gamma}, x_{1,\alpha};E_s)+c_e^{(1)}G(\mathrm{x}_{i,\gamma}, x_{2,\beta};E_s)]+\order{\text{g}^3/\Delta_{\rm min}^3}=\bra{g;\mathrm{x}_{i,\gamma}}(|\Psi_s^{(1)}\rangle+p|\Psi_s^{(2)}\rangle)+\order{\text{g}^3/\Delta_{\rm min}^3}\nonumber\\
	&\qquad\qquad\qquad\qquad\qquad\ \Longrightarrow \bra{g; \mathrm{x}_{i,\gamma}}[\sqrt{2}\ket{\Psi_{s}^p}-(|\Psi_{s}^{(1)}\rangle+p|\Psi_{s}^{(2)}\rangle)]=\order{\text{g}^3/\Delta_{\rm min}^3},\label{IVBxi}
\end{align}
which the corresponding numerical validation are shown in Fig.\,\ref{fign6}(a) and \ref{fign6}(b), respectively. Note that in Fig.\,\ref{fign6}(a), the slope of the line for $\Delta=0$ is 4, which stems from $G'(x_{1,A}, x_{2,B};E_s)=0$ in Eq.~(\ref{IVBee}). Thus, this further confirms the validity of the Eqs.~(\ref{IVBee}, \ref{IVBxi}). During the derivation of Eqs.~(\ref{IVBee}, \ref{IVBxi}), we utilized inequality (\ref{IVBineq}), i.e., 
\begin{align}
	\text{g}^2 |G'(x_{1,\alpha}, x_{2,\beta};E_s)|<\text{g}^2/\Delta_{\rm min}^2\Longrightarrow &\ \text{g}^2|G'(x_{1,\alpha}, x_{2,\beta};E_s)|=\order{\text{g}^2/\Delta_{\rm min}^2},\\
	\text{g}^3|G'(x_{1,\alpha},x_{2,\beta};E_s)G_p(\mathrm{x}_{i,\gamma};E_s)+2G(x_{1,\alpha},x_{2,\beta};&E_s)G_p'(\mathrm{x}_{i,\gamma};E_s)|<\text{g}^3(2/\Delta_{\rm min}^3+4/\Delta_{\rm min}^3)=6\text{g}^3/\Delta_{\rm min}^3\nonumber\\
	\Longrightarrow \text{g}^3|G'(x_{1,\alpha},x_{2,\beta};E_s)G_p(\mathrm{x}_{i,\gamma};E_s)+&2G(x_{1,\alpha},x_{2,\beta};E_s)G_p'(\mathrm{x}_{i,\gamma};E_s)|=\order{\text{g}^3/\Delta_{\rm min}^3}.
\end{align}
Finally, by combing Eq.~(\ref{IVBee}) and Eq.~(\ref{IVBxi}), under the Markovian approximation, i.e., $\text{g}\ll \Delta_{\rm min}$, the two special bound states $\ket{\Psi_s^\pm}$ with energies $E_s^\pm$ satisfy
\begin{align}
	\ket{\Psi_s^\pm}=[|\Psi_s^{(1)}\rangle\pm|\Psi_s^{(2)}\rangle]/\sqrt{2}+\ket{\order{\text{g}^2/\Delta_{\rm min}^2}}\simeq[|\Psi_s^{(1)}\rangle\pm|\Psi_s^{(2)}\rangle]/\sqrt{2},\label{pps}
\end{align}
which we refer to as the parity-positive $\ket{\Psi_s^+}$ and parity-negative $\ket{\Psi_s^-}$ bound states, as shown in Figs.\,\ref{fign6}(c) and \ref{fign6}(d), respectively. Now, if the total system is initially prepared in the state $\ket{\psi(0)}=\ket{e_1}\otimes\ket{\rm vac}$, i.e., the quantum charger is in the excited state, QB is in the ground state, and the environment is in the vacuum state, and considering that $\ket{e_1;\rm vac}=|\Psi_{s}^{(1)}\rangle+|\order{\text{g}^2/\Delta_{\rm min}^2}\rangle$ and $\ket{e_2;\rm vac}=|\Psi_{s}^{(2)}\rangle+|\order{\text{g}^2/\Delta_{\rm min}^2}\rangle$, the wave function at $t$ time is computed as
\begin{align}
	\ket{\psi(t)}&=U(t)\ket{\psi(0)}= U(t)|\Psi_s^{(1)}\rangle+U(t)|\order{\text{g}^2/\Delta_{\rm min}^2}\rangle=\text{exp}[-iH_{\rm tot}^{(2)}t]\left[\ket{\Psi_s^+}+\ket{\Psi_s^-}\right]/\sqrt{2}+U(t)|\order{\text{g}^2/\Delta_{\rm min}^2}\rangle\nonumber\\
	&=\{\exp[-i(E_s^++\omega_c)t]\ket{\Psi_{s}^+}+\exp[-i(E_s^-+\omega_c)t]\ket{\Psi_{s}^-}\}/\sqrt{2}+U(t)|\order{\text{g}^2/\Delta_{\rm min}^2}\rangle\nonumber\\
	&= \exp[-i(E_s+\omega_c)t]\{\cos[\text{g}^2G(x_{1,\alpha},x_{2,\beta};E_s)t]|\Psi_{s}^{(1)}\rangle-i\sin[\text{g}^2G(x_{1,\alpha},x_{2,\beta};E_s)t]|\Psi_{s}^{(2)}\rangle\}+U(t)|\order{\text{g}^2/\Delta_{\rm min}^2}\rangle\nonumber\\
	&=\exp[-i(E_s+\omega_c)t]\left[\cos(J_{12}^{\alpha\beta}t)\ket{e_1}-i\sin(J_{12}^{\alpha\beta}t)\ket{e_2}\right]\otimes\ket{\rm vac}+U(t)|\order{\text{g}^2/\Delta_{\rm min}^2}\rangle,\label{psit}
\end{align}
where $J_{12}^{\alpha\beta}=\text{g}^2G(x_{1,\alpha},x_{2,\beta}; E_s)$. In the next-to-last step, it is observed that the excitation oscillates between the two special bound states $|\Psi_{s}^{(1)}\rangle$ and $|\Psi_{s}^{(2)}\rangle$, which transform into VDSs when the detuning vanishes (i.e., $\Delta=0$). Besides, under the Born-Markov approximation, from Eq.~(\ref{psit}), it is easy to see that in the bandgap regime, the interaction with the bath leads to an effective unitary dynamics governed by the following Hamiltonian
\begin{align}\label{ddint}
	H_{\rm sys}^{\rm eff}=(E_s+\omega_c)(\sigma_{+}^{\rm B}\sigma_{-}^{\rm B}+\sigma_{+}^{\rm C}\sigma_{-}^{\rm C})+J_{12}^{\alpha\beta}(\sigma_{+}^{\rm B}\sigma_{-}^{\rm C}+\sigma_{+}^{\rm C}\sigma_{-}^{\rm B}).
\end{align}
That is, the topological waveguide mediates a dipole-dipole interactions between the QB and the charger, and the interaction $J_{12}^{\alpha\beta}$ inherit many properties of the special bound states, such as exponentially localized in space. Note that, as discussed in Section \ref{IIIA} and illustrated in Figs.\,\ref{fig6}(a) and \ref{fig6}(c), the case of $E_s^+=E_s^-$ can only occur under resonance conditions. Therefore, according to Eq.~(\ref{S51}) or Eq.~(\ref{Gpm}), for the system parameters given in Fig.2(a-d) of the main text, i.e., $\Delta=0, \alpha=A, \beta=B$, and $d=x_1-x_2\in\mathbb{Z}^+$, we have 
\begin{align}
	\lim\limits_{z\to0}G_\pm(z)=\mp\lim\limits_{z\to0}[(1+\delta)y_+^d+(1-\delta)y_+^{d-1}]/(4J\abs{\delta})=\pm\xi^{d}\Theta(-\delta)/[(\delta+1)J],
\end{align}
where $\xi=(\delta+1)/(\delta-1)$. This leads to $E_s^{+}=E_s^{-}=0$ in the topologically trivial phase ($\delta>0$) and $E_s^+=-E_s^-\neq0$ in the topologically nontrivial phase ($\delta<0$). On the one hand, under the Markovian limit, the parity-positive and parity-negative bound states in the topologically nontrivial phase can be further expressed as
\begin{align}\label{Esvds}
	\ket{\Psi_s^\pm}=[|\Psi_{\rm vds}^{(1)}\rangle\pm|\Psi_{\rm vds}^{(2)}\rangle]/\sqrt{2}+|\order{\text{g}^2/\Delta_{\rm min}^2}\rangle\simeq [|\Psi_{\rm vds}^{(1)}\rangle\pm|\Psi_{\rm vds}^{(2)}\rangle]/\sqrt{2},
\end{align}
where $|\Psi_{\rm vds}^{(1)}\rangle$ and $|\Psi_{\rm vds}^{(2)}\rangle$ denote the VDS formed by a single atom coupled to the SSH chain at sites $x_{2,B}$ and $x_{1,A}$, respectively. On the other hand, considering that $E_s=0$, Eq.~(\ref{ddint}) can be simplified as
\begin{align}\label{ddintF}
	H_{\rm sys}^{\rm eff}=\omega_e(\sigma_{+}^{\rm B}\sigma_{-}^{\rm B}+\sigma_{+}^{\rm C}\sigma_{-}^{\rm C})+\text{g}^2\xi^d\Theta(-\delta)/[J(1+\delta)]\times(\sigma_{+}^{\rm B}\sigma_{-}^{\rm C}+\sigma_{+}^{\rm C}\sigma_{-}^{\rm B})\ \ \text{with}\ \ \delta\neq0.
\end{align}
In the topologically nontrivial phase, the existence of a finite interaction not only demonstrate perfect energy transfer between the charger and the QB but also opens the potential for achieving a perfect long-range charging, i.e., $\max_t[\mathcal{E}(\infty)/\omega_e]=1$ with $d\gg1$. However, the interaction vanishes in the topologically trivial phase, and this results in the excitation confined in the charger, thereby completely suppressing the energy transfer. Note that the existence of overlap in Fig.\,\ref{fign10} precisely determines whether this interaction is nonzero. In fact, in addition to the effective Hamiltonian given in Eq.~(\ref{ddintF}), which reveals the reasons for the perfect energy transfer between the charger and the QB, we will subsequently provide a more intuitive and clear physical explanation by relating the overlap of the spatial envelopes to the perfect energy transfer.

\begin{figure}
	\centering
	\includegraphics[width=17cm]{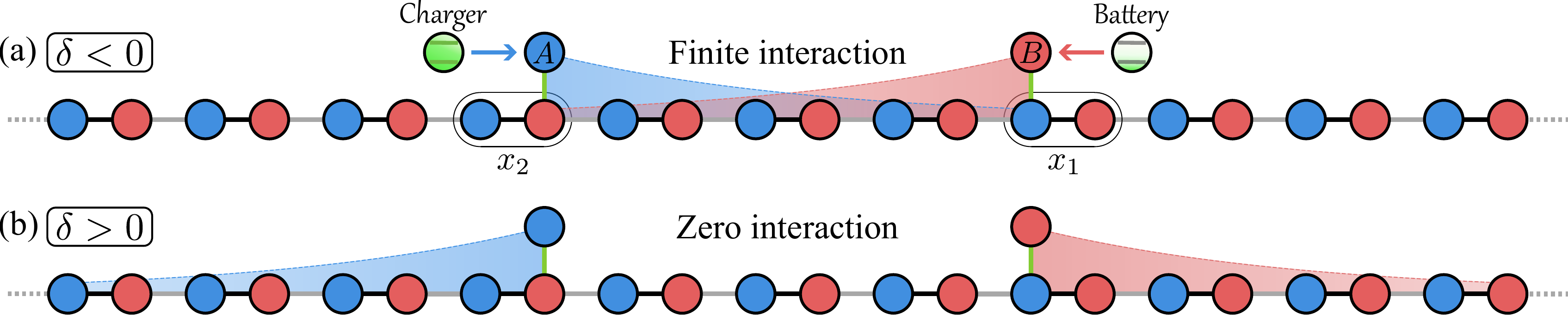}\\
	\caption{The spatial envelopes (blue and red dashed lines) of the two chiral edge-like states overlap in the topologically nontrivial phase (a), giving rise to finite interaction between the charger and the QB, whereas in the topologically trivial phase (b),  the spatial envelopes remain non-overlapping, leading to zero interaction. Here, the left TLS (quantum charger) is treated as the $A$ mode, and the right TLS (quantum battery) is treated as the $B$ mode. The blue and red shaded regions represent the areas formed by the spatial envelopes of the edge-like states confined in sublattices $A$ and $B$, respectively.
	}\label{fign10}
\end{figure}
\subsubsection*{Physical mechanism under resonance conditions in the Markovian regime}
Specifically, when we treat the atomic component as a new photonic component, the VDS can be interpreted as an edge-like state, as shown in Fig.\,\ref{fign10}. Thus, the new photon components are defined as ($\ket{g}\to\ket{\rm vac}$)
\begin{equation}
    \setlength\abovedisplayskip{5pt}
    \setlength\belowdisplayskip{5pt}
	\sigma_+^{\rm C}\ket{g}\equiv a_{x_2+\varepsilon}^\dagger\ket{\rm vac},\quad \sigma_+^{\rm B}\ket{g}\equiv b_{x_1-\varepsilon}^\dagger\ket{\rm vac},
\end{equation}
where $\varepsilon=0^+$ is introduced to support the commutation relation, e.g., $[a_{x_2+\varepsilon}^\dagger,a_{x_2}]=0$ and $[b_{x_1-\varepsilon}^\dagger,b_{x_1}]=0$. According to the explicit expression (\ref{IVAll4}), we find that the VDS exhibits its largest component on the new photonic components, due to $|\eta|\gg1$ under the Markovian limit, thereby justifying the treatment of the new photonic component as an edge. Thus, the VDSs at different states are can be written as
\begin{align}
	|\Psi_{\rm vds}^{(1)}\rangle&=\mathscr{N}[\eta \ket{e,g}\otimes\ket{\rm vac}+\ket{g,g}\otimes|\psi_{\rm edge}^{(1)}\rangle]\equiv \mathscr{N}[\eta\, {a}^\dagger_{x_2+\varepsilon}\ket{\rm vac}+|\psi_{\rm edge}^{(1)}\rangle]=\ket{L},\label{PPNBS1}\\
	|\Psi_{\rm vds}^{(2)}\rangle&= \mathscr{N}[\eta \ket{g,e}\otimes\ket{\rm vac}+\ket{g,g}\otimes|\psi_{\rm edge}^{(2)}\rangle]\equiv \mathscr{N}[\eta \,{b}^\dagger_{x_1-\varepsilon}\ket{\rm vac}+|\psi_{\rm edge}^{(2)}\rangle]=\ket{R},\label{PPNBS2}
\end{align}
with the edge state
\begin{align}
	\delta<0:\ |\psi_{\rm edge}^{(1)}\rangle&=\frac{2\sqrt{|\delta|}}{1+|\delta|}\sum_{n=1}^{N}{\xi^{n-1}{a}_{x_2+n}^\dagger\ket{\rm vac}},\quad\ |\psi_{\rm edge}^{(2)}\rangle=\frac{2\sqrt{|\delta|}}{1+|\delta|}\sum_{n=1}^{N}{\xi^{n-1}{b}_{x_1+1-n}^\dagger\ket{\rm vac}},\label{edge1}\\
	\delta>0:\ |\psi_{\rm edge}^{(1)}\rangle&=\frac{2\sqrt{|\delta|}}{1+|\delta|}\sum_{n=1}^{N}{\xi^{1-n}{a}_{x_2+1-n}^\dagger\ket{\rm vac}},\ |\psi_{\rm edge}^{(2)}\rangle=\frac{2\sqrt{|\delta|}}{1+|\delta|}\sum_{n=1}^{N}{\xi^{1-n}{b}_{x_1+n}^\dagger\ket{\rm vac}},\label{edge2}
\end{align}
where $\mathscr{N}=-\text{g}/\sqrt{\text{g}^2+4J^2|\delta|}$. Due to $x_2<x_1$, i.e., the charger on the left and the QB on the right, for convenience, we use the notations $\ket{L}$ and $\ket{R}$ to represent the chiral edge-like states (i.e., VDSs) formed by the atom on the right and right coupled to the SSH chain, respectively, as depicted in Figs.\,\ref{fign6}(e) and \ref{fign6}(f) or Figs.\,\ref{fign10}. Although the two chiral edge-like states are orthogonal, i.e., $\braket{R}{L}=0$, Fig.\,\ref{fign10}(a) shows that their spatial envelopes overlap in the topologically nontrivial phase ($\delta<0$), whereas in the topologically trivial phase ($\delta>0$), the spatial envelopes do not overlap, as shown in Fig.\,\ref{fign10}(b). For consistency, we present the definition for the spatial envelope of the chiral edge-like state. First, since the photonic components of the VDS in the SSH model are confined solely to sublattice $A$ or $B$, when we treat the atomic component as the new photonic component on the edge, belonging to the same sublattice, the edge-like state is distributed solely in sublattice $A$ or $B$. After treating the atomic component as the new photonic component, we define the positions of the unit cells, including the new indices for the new photonic component, as spatial axes $z$, i.e.,
\begin{equation}
    \setlength\abovedisplayskip{5pt}
    \setlength\belowdisplayskip{5pt}
	z=\{1,2,\ldots, x_2, x_2+\varepsilon,\ldots,x_1-\varepsilon,x_1,\ldots,N-1,N\}.
\end{equation}
For the sake of intuitive analysis, we assume that the two atoms are connected at the center of the SSH chain, i.e., $1\ll x_2\sim x_1\ll N$. Based on these definitions, in the thermodynamic limit ($N\to\infty$), the spatial envelop of the edge-like state $\ket{L/R}$ is represented by the curve formed by the discrete points obtained from
\begin{equation}\label{envelope}
    \setlength\abovedisplayskip{6pt}
    \setlength\belowdisplayskip{6pt}
	\{(z_k, \mathcal{A}_{L/R}(z_k))\mid z_k\in z,\mathcal{A}_{L/R}(z_k)\neq0 \},\ \ \text{with}\ \  \mathcal{A}_{L/R}\equiv\mel{\rm vac}{a_{z_k}+b_{z_k}}{L/R},
\end{equation}
where $\mel{\rm vac}{b_{x_2+\varepsilon}}{L/R}=\mel{\rm vac}{a_{x_1-\varepsilon}}{L/R}=0$, which corresponds to the blue (red) dashed curve in Fig.\,\ref{fign10}.

\begin{figure}[t]
	\centering
	\includegraphics[width=17cm]{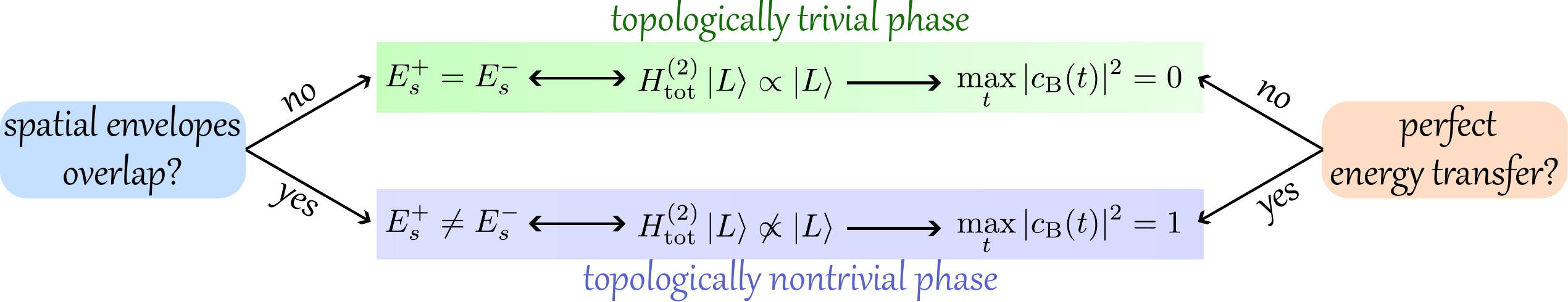}\\
	\caption{The relation between the spatial envelopes of the two edge-like states (i.e., $\ket{L}$ and $\ket{R}$) and the perfect energy transfer from the charger to the QB.
	}\label{fign12}
\end{figure}

Consequently, based on the discussion in Section \ref{SIVA}, we conclude that, when a two-level system couples to the periodic SSH chain with $N$ unit cells under resonance condition ($\Delta=0$), a chiral edge-like state (i.e., VDS) emerges in the the thermodynamic limit. Similarly, when two two-level systems couple to the SSH chain in the same conditions, the parity-positive and parity-negative bound states emerge, which can be approximately expressed as the positive and negative superposition of the two chiral edge-like states [see Eqs.~(\ref{Esvds}), (\ref{PPNBS1}), and (\ref{PPNBS2})]. Since $|\eta|\ll1$ under the Markovian limit, we have $\ket{L}\simeq\ket{e,g;\rm vac}$ and $\ket{R}\simeq\ket{g,e;\rm vac}$. Note that the parity-positive and parity-negative bound states satisfy $H_{\rm tot}^{(2)}\ket{\Psi_s^\pm}=(\omega_e+E_s^\pm)\ket{\Psi_s^\pm}$ and $\braket{\Psi_{s}^\mu}{\Psi_s^\nu}=\delta_{\mu,\nu}$ with $\mu,\nu\in\{+,-\}$, and we use the notations in the main text for brevity, i.e., $\ket{E_\pm}\equiv\ket{\Psi_s^\pm}$ and $E_\pm\equiv \omega_e+E_s^\pm$. Thus, the chiral edge-like states are also composed of the parity-positive and parity-negative bound states, i.e., $\ket{L}\propto\ket{E_+}+\ket{E_-}$ and $\ket{R}\propto \ket{E_+}-\ket{E_-}$. Specifically, in the topologically trivial phase, the spatial envelopes of the edge-like states do not overlap [see Fig.\,\ref{fign10}(b)], leading to a degeneracy of the parity-positive and parity-negative bound states (i.e., $E_s^+=E_s^-\Longrightarrow E_+=E_-$), which results in
\begin{align}\label{linear}
	H_{\rm tot}^{(2)}\ket{L/R}\propto E_+\ket{E_+}\pm E_-\ket{E_-}=E_+(\ket{E_+}\pm \ket{E_-})\propto\ket{L/R}.
\end{align}
This means that the edge-like states are also the bound states of the system. Thus, the dynamics of the charger and the QB are computed as 
\begin{align}
	\mel{\psi(t)}{{\sigma}_+^{\rm C}{\sigma}_-^{\rm C}}{\psi(t)}&=|\mel{e_1;\rm vac}{\text{exp}[-iH_{\rm tot}^{(2)}t]}{e_1;\rm vac}|^2\simeq |\mel{L}{\text{exp}[-iH_{\rm tot}^{(2)}t]}{L}|^2=|\exp(-iE_+t)\braket{L}{L}|^2=1,\\
	\mel{\psi(t)}{{\sigma}_+^{\rm B}{\sigma}_-^{\rm B}}{\psi(t)}&=|\mel{e_2;\rm vac}{\text{exp}[-iH_{\rm tot}^{(2)}t]}{e_1;\rm vac}|^2\simeq |\mel{R}{\text{exp}[-iH_{\rm tot}^{(2)}t]}{L}|^2=|\exp(-iE_+t)\braket{R}{L}|^2=0.
\end{align}
This demonstrate that the excitation in the charger is confined to the left edge (the component of the charger), completely suppressing energy transfer. However, in the topologically nontrivial, the spatial envelopes of the edge-like states overlap [see Fig.\,\ref{fign10}(a)], breaking the degeneracy of the parity-positive and parity-negative bound states (i.e., $E_s^+\neq E_s^-\Longrightarrow E_+\neq E_-$), which results in
\begin{align}\label{nonlinear}
	H_{\rm tot}^{(2)}\ket{L/R}\propto E_+\ket{E_+}\pm E_-\ket{E_-}=E_+[\ket{E_+}\pm (E_-/E_+)\ket{E_-}]\not\propto\ket{L/R}. 
\end{align}
This means that the edge-like states are not the bound states of the system. Thus, the dynamics are computed as
\begin{align}
	\mel{\psi(t)}{{\sigma}_+^{\rm C}{\sigma}_-^{\rm C}}{\psi(t)}&=|\mel{e_1;\rm vac}{\text{exp}[-iH_{\rm tot}^{(2)}t]}{e_1;\rm vac}|^2\simeq |\mel{L}{\text{exp}[-iH_{\rm tot}^{(2)}t]}{L}|^2\nonumber\\
	&=|\mel{L}{\exp(-iE_+t)}{E_+}+\mel{L}{\exp(-iE_-t)}{E_-}|^2/2\nonumber\\
	&=|\exp(-iE_+t)+\exp(-iE_-t)|^2/4=\cos^2[(E_--E_+)t/2],\\
	\mel{\psi(t)}{{\sigma}_+^{\rm B}{\sigma}_-^{\rm B}}{\psi(t)}&=|\mel{e_2;\rm vac}{\text{exp}[-iH_{\rm tot}^{(2)}t]}{e_1;\rm vac}|^2\simeq |\mel{R}{\text{exp}[-iH_{\rm tot}^{(2)}t]}{L}|^2\nonumber\\
	&=|\mel{R}{\exp(-iE_+t)}{E_+}+\mel{R}{\exp(-iE_-t)}{E_-}|^2/2\nonumber\\
	&=|\exp(-iE_+t)-\exp(-iE_-t)|^2/4=\sin^2[(E_--E_+)t/2].
\end{align}
This demonstrate that the excitation oscillates with  between the two edges (the components of the charger and the QB, respectively) with a period of $2\pi/|E_+-E_-|$, enabling perfect energy transfer. As shown in Fig.\,\ref{fign12}, we clearly show that the relation between the overlap of the spatial envelopes and the perfect energy transfer. Note that in Eqs.~(\ref{linear}) and (\ref{nonlinear}), we utilize the conclusion that when the spatial envelopes of the two edge-like states overlap, the parity-positive and parity-negative bound states are non-degenerate, i.e., $E_+\neq E_-$; conversely, when there is no overlap, the bound states are degenerate, i.e., $E_+=E_-$. Next, we will prove this pivotal conclusion. 

Specifically, to establish the connection between the overlap of the spatial envelopes of the edge-like states ($\ket{L}\equiv|\Psi_{\rm vds}^{(1)}\rangle$ and $\ket{R}\equiv|\Psi_{\rm vds}^{(2)}\rangle$) and the degeneracy of the parity-positive and parity-negative bound states ($\ket{E_+}$ and $\ket{E_-}$), we need to determine whether the edge-like states are the bound states of the system described by the Hamiltonian $H_{\rm tot}^{(2)}$. According to Eq.~(\ref{Esvds}), i.e., $|\Psi_{\rm vds}^{(1)}\rangle\propto \ket{E_+}+\ket{E_-}$ and $|\Psi_{\rm vds}^{(2)}\rangle\propto \ket{E_+}-\ket{E_-}$, we have
\begin{equation}
    \setlength\abovedisplayskip{5pt}
    \setlength\belowdisplayskip{5pt}
	H_{\rm tot}^{(2)}|\Psi_{\rm vds}^{(1)}\rangle\propto E_+\ket{E_+}+E_-\ket{E_-},\quad H_{\rm tot}^{(2)}|\Psi_{\rm vds}^{(2)}\rangle\propto E_+\ket{E_+}-E_-\ket{E_-}.
\end{equation}
It is clear that if the parity-positive and parity-negative bound states are degenerate (i.e., $E_+=E_-$), these two edge-like states are also bound states of the system; conversely, if the bound states are non-degenerate (i.e., $E_+\neq E_-$), the edge-like states are not the bound states of the system. Thus, the next step is to establish the relationship between the overlap of the spatial envelopes of the edge-like states and whether they correspond to bound states of the system. For the sake of convenience in calculation, we rewrite the Hamiltonian as
\begin{align}
	H_{\rm tot}^{(2)}&=\omega_{e}\sigma_+^{\rm B}\sigma_-^{\rm B}+\omega_{e}\sigma_+^{\rm C}\sigma_-^{\rm C}+H_{\rm bath}+\text{g}(\sigma_{-}^{\rm B}a_{x_1}^\dagger+\sigma_{-}^{\rm C}b_{x_2}^\dagger+\text{H.c.})\nonumber\\
	&=\omega_{e}\sigma_+^{\rm B}\sigma_-^{\rm B}+H_{\rm tot}^{\rm (C)}+\text{g}(\sigma_{-}^{\rm B}a_{x_1}^\dagger+\text{H.c.})\label{Htot2_1}\\
	&=\omega_{e}\sigma_+^{\rm C}\sigma_-^{\rm C}+H_{\rm tot}^{\rm (B)}+\text{g}(\sigma_{-}^{\rm C}b_{x_2}^\dagger+\text{H.c.}),\label{Htot2_2}
\end{align}
where ($\omega_e=\omega_c$)
\begin{align}
	H_{{\rm tot}}^{\rm (C)}&=\omega_e{\sigma}_+^{\rm C}{\sigma}_-^{\rm C}+H_{\rm bath}+\text{g}({\sigma}_-^{\rm C}{b}^\dagger_{x_2}+{b}^{}_{x_2}{\sigma}_+^{\rm C}),\ \  \text{with}\ \ H_{{\rm tot}}^{\rm (C)}|\Psi_{\rm vds}^{(1)}\rangle=\omega_e|\Psi_{\rm vds}^{(1)}\rangle,\\
	H_{{\rm tot}}^{\rm (B)}&=\omega_e{\sigma}_+^{\rm B}{\sigma}_-^{\rm B}+H_{\rm bath}+\text{g}({\sigma}_-^{\rm B}{a}^\dagger_{x_1}+{a}^{}_{x_1}{\sigma}_+^{\rm B}),\ \  \text{with}\ \ H_{{\rm tot}}^{\rm (B)}|\Psi_{\rm vds}^{(2)}\rangle=\omega_e|\Psi_{\rm vds}^{(2)}\rangle.
\end{align}
Here, $H_{{\rm tot}}^{\rm (B)}$ and $H_{{\rm tot}}^{\rm (C)}$ denote the Hamiltonians describing the QB and the charger, respectively, coupled to the periodic SSH chain with $N$ unit cells. Then, applying Eqs.~(\ref{Htot2_1}) and (\ref{Htot2_2}) to $\ket{L}$ and $\ket{R}$, respectively, we obtain
\begin{align}
	H_{\rm tot}^{(2)}\ket{L}&=H_{\rm tot}^{(2)}|\Psi_{\rm vds}^{(1)}\rangle=[\omega_e{\sigma}_+^{\rm B}{\sigma}_-^{\rm B}+H_{\rm tot}^{\rm (C)}+\text{g}({\sigma}_-^{\rm B}{a}^\dagger_{x_1}+{a}^{}_{x_1}{\sigma}_+^{\rm B})]|\Psi_{\rm vds}^{(1)}\rangle\nonumber\\
	&=\mathscr{N}[\omega_e{\sigma}_+^{\rm B}{\sigma}_-^{\rm B}+\text{g}({\sigma}_-^{\rm B}{a}^\dagger_{x_1}+{a}^{}_{x_1}{\sigma}_+^{\rm B})][\eta\ket{e,g}\otimes\ket{\rm vac}+\ket{g,g}\otimes|\psi_{\rm edge}^{(1)}\rangle]+\omega_e|\Psi_{\rm vds}^{(1)}\rangle\nonumber\\
	&=\mathscr{N}\text{g}\times{a}_{x_1}{\sigma}_+^{\rm B}\ket{L}+\omega_e\ket{L}\equiv\mathscr{N}\text{g}\times{b}_{x_{1}-\varepsilon}^\dagger{a}_{x_1}\ket{L}+\omega_e\ket{L},\label{HL}\\
	H_{\rm tot}^{(2)}\ket{R}&=H_{\rm tot}^{(2)}|\Psi_{\rm vds}^{(2)}\rangle=[\omega_e{\sigma}_+^{\rm C}{\sigma}_-^{\rm C}+H_{\rm tot}^{\rm (B)}+\text{g}({\sigma}_-^{\rm C}{b}^\dagger_{x_2}+{b}^{}_{x_2}{\sigma}_+^{\rm C})]|\Psi_{\rm vds}^{(2)}\rangle\nonumber\\
	&=\mathscr{N}[\omega_e{\sigma}_+^{\rm C}{\sigma}_-^{\rm C}+\text{g}({\sigma}_-^{\rm C}{b}^\dagger_{x_2}+{b}^{}_{x_2}{\sigma}_+^{\rm C})][\eta\ket{g,e}\otimes\ket{\rm vac}+\ket{g,g}\otimes|\psi_{\rm edge}^{(2)}\rangle]+\omega_e|\Psi_{\rm vds}^{(2)}\rangle\nonumber\\
	&=\mathscr{N}\text{g}\times{b}_{x_2}{\sigma}_+^{\rm C}\ket{R}+\omega_e\ket{R}\equiv\mathscr{N}\text{g}\times{a}_{x_{2}+\varepsilon}^\dagger{b}_{x_2}\ket{R}+\omega_e\ket{R}.\label{HR}
\end{align}
Once ${a}_{x_1}\ket{L}={b}_{x_2}\ket{R}=0$, we have $H_{\rm tot}^{(2)}\ket{L/R}\propto \ket{L/R}$; conversely, if ${a}_{x_1}\ket{L}\neq0$ and ${b}_{x_2}\ket{R}\neq0$, then $H_{\rm tot}^{(2)}\ket{L/R}\not\propto \ket{L/R}$. Notably, by observing whether the spatial envelopes of $\ket{L}$ and $\ket{R}$ overlap, we can clearly determine which condition is satisfied. To be specific, when the spatial envelopes do not overlap [see Fig.\,\ref{fign10}(b)], we have ${a}_{x_1}\ket{L}={b}_{x_2}\ket{R}=0$; however, when the spatial envelopes overlap [see Fig.\,\ref{fign10}(a)], we have ${a}_{x_1}\ket{L}\neq0$ and ${b}_{x_2}\ket{R}\neq0$. Or based on Eqs.~(\ref{PPNBS1}-\ref{edge2}), we can also analytically determine the condition in the thermodynamic limit ($N\to\infty$),
\begin{align}
	\lim\limits_{N\to\infty}{a}_{x_1}|\psi_{\rm edge}^{(1)}\rangle&=\frac{2\sqrt{|\delta|}}{1+|\delta|}\xi^{d}\ket{\rm vac}\times\lim\limits_{N\to\infty}\begin{cases}
		\xi^{-1} & \delta<0\\
		\xi^{1-N} & \delta>0
	\end{cases}=\frac{2\sqrt{|\delta|}}{1+|\delta|}\xi^{d-1}\Theta(-\delta)\ket{\rm vac},\\
	\lim\limits_{N\to\infty}{b}_{x_2}|\psi_{\rm edge}^{(2)}\rangle&=\frac{2\sqrt{|\delta|}}{1+|\delta|}\xi^{d}\ket{\rm vac}\times\lim\limits_{N\to\infty}\begin{cases}
		1 & \delta<0\\
		\xi^{1-N} & \delta>0
	\end{cases}=\frac{2\sqrt{|\delta|}}{1+|\delta|}\xi^{d}\Theta(-\delta)\ket{\rm vac}.
\end{align}
Consequently, Eqs.~(\ref{HL}) and (\ref{HR}) can be further simplified as
\begin{align}
	H_{\rm tot}^{(2)}\ket{L}=\mathscr{N}\text{g}\frac{2\sqrt{|\delta|}}{1+|\delta|}\xi^{d-1}\Theta(-\delta){b}_{x_1-\varepsilon}^\dagger\ket{\rm vac}+\omega_e\ket{L},\quad
	H_{\rm tot}^{(2)}\ket{R}=\mathscr{N}\text{g}\frac{2\sqrt{|\delta|}}{1+|\delta|}\xi^{d}\Theta(-\delta){a}_{x_2+\varepsilon}^\dagger\ket{\rm vac}+\omega_e\ket{R},
\end{align}
and this clearly indicates $H_{\rm tot}^{(2)}\ket{L/R}\propto\ket{L/R}$ and $H_{\rm tot}^{(2)}\ket{L/R}\not\propto\ket{L/R}$ in the topologically trivial and nontrivial phases, respectively. 

In summary, perfect energy transfer between the charger and the QB depends solely on the overlap of the spatial envelopes of these two chiral edge-like states. When the sign of the dimerization is reversed, for example, changing $\delta>0$ [see Fig.\,\ref{fign10}(b)] to $\delta<0$ [see Fig.\,\ref{fign10}(a)], the spatial envelope of the edge-like state undergoes a spatial inversion, thereby giving rise to the overlap between the chiral edge-like states. Note that for an ordinary non-topological waveguide (i.e., $\delta=0$), the conclusions derived above no longer apply. This is primarily because, under the resonant condition with $\delta=0$, the central bandgap of the bath closes, resulting in the two bound states (\ref{Esvds}) that originally existed in the system to disappear. Consequently, photons that were previously localized at the two atoms now disperse into the waveguide, leading to the complete loss of energy from both the quantum charger and the QB. Interestingly, despite the central bandgap is closed, the system can still support a dressed bound state in the continuum (BIC) under certain parametric conditions. Specifically, when $J_+=J_-=J$ (i.e., $\delta=0$), we can introduce a new notation, i.e., $a_j=c_{2j-1}$ and $b_j=c_{2j}$, to simplify the Hamiltonian $H_{\rm bath}$ (\ref{S17}), and this gives rise to a tight-binding model, which is given by
\begin{equation}
    \setlength\abovedisplayskip{5pt}
    \setlength\belowdisplayskip{5pt}
	H_{\rm bath}=\omega_c\sum_{j=1}^{2N}c_{j}^\dagger c_{j}^{}+J\sum_{j=1}^{2N}(c_{j}^\dagger c_{j+1}^{}+c_{j+1}^{\dagger} c_{j}^{})\equiv H_{\rm tb},
\end{equation}
where $c_{2N+1}=c_1$ corresponding to the periodic boundary condition. Then, based on $o_{x_{j,A}}=a_{x_j}=c_{2x_j-1}$ and $o_{x_{j,B}}=b_{x_j}=c_{2x_j}$, we have $o_{x_{j,\alpha}}=c_{x_{j,\alpha}}$ with $x_{j,\alpha}\equiv 2x_j-\delta_{\alpha,A}$. Under these definitions and parameter conditions, the total Hamiltonian (\ref{totHamiltonian}) can be simplified as \vspace{-1pt}
\begin{align}\label{totHamiltonian1}
	H_{\rm tot}^{(2)}=\omega_e(\sigma_+^{\rm B}\sigma_-^{\rm B}+\sigma_+^{\rm C}\sigma_-^{\rm C})+H_{\rm tb}+\text{g}(\sigma_-^{\rm B} c^\dagger_{x_{1,\alpha}}+c_{x_{1,\alpha}}\sigma_+^{\rm B}+\sigma_-^{\rm C} c^\dagger_{x_{2,\beta}}+c_{x_{2,\beta}}\sigma_+^{\rm C}).
\end{align}
Following Ref.\,\cite{vbs}, by solving the eigenvalue equation $H^{(2)}_{\rm tot}\ket{\Psi_{\rm BIC}}=\omega_e\ket{\Psi_{\rm BIC}}$, we find that the system can still support a BIC (also referred to as a VDS), provided that the following condition is satisfied:
\begin{align}\label{condition}
	\omega_{m}=\omega_c+2J\cos(\frac{m\pi}{L})=\omega_e,\quad \exists\ m\in\{1,2,\ldots,L-1\}\neq\emptyset,
\end{align}
which implies $L\ge2$, where $L=x_{1,\alpha}-x_{2,\beta}=2d+\delta_{\beta, A}-\delta_{\alpha, A}$. Provided that condition (\ref{condition}) is fulfilled, the BIC takes the form:
\begin{align}
	\ket{\Psi_{\rm BIC}}=\left(1+\frac{\Gamma\tau}{4}\right)^{-1/2}\left[\frac{{\sigma}_+^{\rm C}-(-1)^m{\sigma}_+^{\rm B}}{\sqrt{2}}+\sqrt{\frac{\Gamma}{v_{m}}}\sum_{n=1}^{L-1}\sin(\frac{mn\pi}{L}){c}_{x_{2,\beta}+n}^\dagger\right]\ket{g,g;\rm vac},
\end{align}
where $\Gamma=2\text{g}^2/v_{m}$, $\tau=2L/v_{m}$, and $v_{m}=-2J\sin(m\pi/L)$. Apart from solving the eigenvalue equation, the BIC can be also identify through the roots of the pole equation (\ref{poleequation}). As a result, when the condition (\ref{condition}) is satisfied, the residue at the bound-state energy is computed as
\begin{align}\label{residue}
	\text{Res}[\mathscr{C}^{(2)}(z), E_{\rm BIC}]=\frac{\text{g}^2G(x_{1,\alpha},x_{2,\beta};z)}{\mathrm{d}\mathscr{D}(z)/\mathrm{d}z}\bigg|_{z=E_{\rm BIC}}=\frac{2(-1)^{m+1}}{4+\Gamma\tau},
\end{align}
where $E_{\rm BIC}=\omega_e-\omega_c=2J\cos(m\pi/L)$. Since the energy of the BIC lies within the band, when using the self-energy expressions (\ref{S32}-\ref{S33}, \ref{S35}-\ref{S36}), the summation in the numerator should be treated as follows: $\sum_{p=\pm}f_p\to f_p$, where the index $p$ is consistently taken to be either positive or negative throughout. In this case, under the Markovian regime, no other bound states exist apart from the BIC. As a result, in the long-time limit, the dynamics of the QB is fully determined by the BIC contribution, leading to the result
\begin{align}\label{dynamics}
	\abs{c_{\rm B}(\infty)}^2=\abs{\Res[\mathscr{C}^{\rm ii}(z), E_{\rm BIC}]\exp(-iE_{\rm BIC} t)}^2=4/(4+\Gamma\tau)^2\approx1/4,
\end{align}
From the QB dynamics, we find that the presence of a single bound state is insufficient to establish effective Rabi oscillations between the charger and the QB. Consequently, for $|\Delta|<2J$, the interaction with the ordinary non-topological photonic chain does not give rise to an effective unitary dynamics governed by a Hamiltonian such as that in Eq.~(\ref{ddintF}). Although a single BIC cannot support the formation of an effective interaction, it can still facilitate partial energy---up to one fourth---transfer from the charger to the QB, which prevents the stored energy from vanishing entirely. According to Eq.~(\ref{dynamics}), the stored energy and ergotropy in the long-time limit are
\begin{align}
	\mathcal{E}(\infty)=\omega_e\abs{c_{\rm B}(\infty)}^2=\frac{4\omega_e}{(4+\Gamma\tau)^2}\approx\frac{\omega_e}{4}<\frac{\omega_e}{2}\Longrightarrow \mathcal{W}(\infty)=0.
\end{align}
Back to configuration I, we note that in Fig.2(a-d) of the main text, the system parameters are chosen as
\begin{align}
	x_{1,\alpha}=x_{1,A}=2x_1-1,\quad x_{2,\beta}=x_{2,B}=2x_2,\quad \Delta=\omega_e-\omega_c=0.
\end{align}
Thus, by solving condition (\ref{condition}), we have
\begin{align}
	2J\cos(\frac{m\pi}{L})=2J\cos\left[\frac{m\pi}{2(x_1-x_2)-1}\right]=0\Longrightarrow m=\frac{2(x_1-x_2)-1}{2}\notin\{1,2,\ldots,L-1\},
\end{align} 
Thus, for the case of $\delta=0$, the original two bound states---parity-positive and parity-negative bound states---vanish, and the system in configuration I does not support the BIC under the resonant condition, thereby resulting in an almost vanishing stored energy. For configuration II, since $L=x_{1,\alpha}-x_{2,\beta}=0$, it is evident that condition (\ref{condition}) is not satisfied. As a result, the BIC does not emerge in this configuration. Finally, since this section focuses primarily on the Markovian regime, the discussion about the non-Markovian regime is not a central focus here, but will be explored in the next section.

\vspace{-10pt}
\section{Charging Time in Markovian and Non-Markovian Regimes}
While common performance indicators such as store energy, ergotropy, and charging power are essential for characterizing the performance of quantum batteries, charging time is equally important. This often-overlooked metric plays a crucial role not only in quantum systems but also in classical systems. For example, in electric vehicles, charging time is a major bottleneck that limits their widespread adoption. Similarly, future fusion power plants will need to inject large amounts of energy rapidly in a short time to initiate reactions. In light of this, the focus of this section is on the charging time in our system, particularly in the Markovian and non-Markovian regimes. To ensure a comprehensive discussion, we will begin by defining charging time.

Specifically, a general Hamiltonian for describing the charging process can be written as
\begin{equation}\label{CT1}
    \setlength\abovedisplayskip{5pt}
    \setlength\belowdisplayskip{5pt}
	H(t)=H_{\rm B}+H_{\rm C}+\mathcal{V}(t),
\end{equation}
where the local Hamiltonian $H_{\rm B}$ and $H_{\rm C}$ characterize the QB part and the charger part, respectively. The charging operator $\mathcal{V}(t)$ incorporates all terms that control the switching of the energy injection, such as direct charger-battery interactions, indirect charger-bath-battery interactions, or a combination of both. Based on our model [see Eq.~(\ref{S1})], we adopt a combination of both interaction types. Accordingly, the general Hamiltonian (\ref{CT1}) can be rewritten as $H(t)=H_{\rm B}+H_{\rm C}+H_{\rm bath}+\mathcal{V}(t)$, with
\begin{equation}\label{CTV}
    \setlength\abovedisplayskip{5pt}
    \setlength\belowdisplayskip{5pt}
	H_{\rm B}=\omega_e\sigma_{+}^{\rm B}\sigma_{-}^{\rm B},\quad H_{\rm C}=\omega_e\sigma_{+}^{\rm C}\sigma_{-}^{\rm C}, \quad \mathcal{V}(t)=\theta(t)[\Omega_{12}^{\alpha\beta}(\sigma_+^{\text{B}}\sigma_-^{\text{C}}+\sigma_+^\text{C}\sigma_-^\text{B})+H_{\rm int}],
\end{equation}
where $\theta(t)$ is a classical parameter that represents the external control exerted on the entire system, and which is assumed to be given by a step function equal to 1 for $t\in[0, \tau]$ and zero elsewhere. At time $t=0$, the entire system is prepared in a product state $\ket{\psi(0)}=\ket{e_1;\rm vac}$ with the QB being ground state of $H_{\rm B}$, charger being the excited state of $H_{\rm C}$, and the bath being the vacuum state of $H_{\rm bath}$. We then suddenly turn on $\mathcal{V}$ and aim to inject as much energy as possible into the QB for a finite time interval $[0, \tau]$. Such a time interval $\tau$ is called the {\it charging time} of QB. Accordingly, denoting by $\rho(t)$ the evolved density matrix of the entire system at time $t$, the energy from the quantum charger injected into the QB can be thus expressed in terms of the stored energy at the end of the protocol: $\mathcal{E}(\tau)=\Tr[\rho(\tau)H_{\rm B}]=\Tr[\rho_{\rm B}(\tau)H_{\rm B}]$, where the reduced density matrix $\rho_{\rm B}(\tau)$ is given by Eq.~(\ref{S16}). In a QB, one is interested in minimizing the optimal charging time, where the {\it optimal charging time} $\tau_{\rm opt}$ is defined by 
\begin{align}\label{ctopt}
	\mathcal{E}(\tau_{\rm opt})=\max_{\tau>0}\mathcal{E}(\tau)=\omega_e\max_{\tau>0}|c_{\rm B}(\tau)|^2.
\end{align}

\subsection{Markovian regime: configurations I and II}
Here, we will focus primarily on minimizing the optimal charging time, which corresponds to the shortest time required for the complete transfer of all the energy from the charger to the QB in both configuration I and configuration II within the Markovian regime (i.e., the Born-Markov approximation). 

\textbf{Regarding the charging time in configuration \hyperref[fig5]{I}}, under the Born-Markov approximation, we have demonstrated in Section \ref{IVB} that, in the band-gap regime, the charger-bath-battery interaction results in an effective unitary dynamics governed by the Hamiltonian (\ref{ddint}). Especially under the resonance condition, where $\Delta=0$, the Hamiltonian can be explicitly expressed as given in Eq.~(\ref{ddintF}). Based on this, we can straightforwardly derive the reduced density matrix of the QB in the topologically nontrivial phase, i.e.,
\begin{align}
	\rho_{\rm B}(t)=\mqty[\cos[2](J_{12}^{AB}t)&0\\0&\sin[2](J_{12}^{AB}t)] \ \  \text{with}\ \  J_{12}^{AB}=\frac{\text{g}^2(-1)^d}{J(1+\delta)}\left(\frac{1+\delta}{1-\delta}\right)^{d},
\end{align} 
and the corresponding energy of the QB at time $t$ is given by
\begin{align}\label{SEI}
	\mathcal{E}_{\rm I}(t)=\Tr[\rho_{\rm B}(t)H_{\rm B}]=\omega_e \sin[2](J_{12}^{AB}t).
\end{align}
Then, according to Eqs.~(\ref{ctopt}) and (\ref{SEI}), we have
\begin{align}\label{moct}
	\mathcal{E}_{\rm I}(\tau_{\rm opt})=\max_{\tau>0}\mathcal{E}_{\rm I}(\tau)=\omega_e\Longrightarrow|J_{12}^{AB}|\tau_{\rm opt}=n\pi +\frac{\pi}{2}\Longrightarrow \min_{n\in\mathbb{Z}}[\tau_{\rm opt}^{\rm I}]=\frac{\pi}{2|J_{12}^{AB}|}=\frac{J(1+\delta)\pi}{2\text{g}^2}\left(\frac{1-\delta}{1+\delta}\right)^d.
\end{align}
Due to the Born-Markov approximation($\text{g}\ll\Delta_{\rm min}=2J\abs{\delta}$), the minimum optimal charging time is typically quite long, i.e., $\min[\tau_{\rm opt}^{\rm I}]\gg 1/J$. Nevertheless, this enables us to achieve a perfect long-range charging process ($d\gg1$) from the charger to the QB, albeit at the cost of much larger charging times. It is important to note that the optimal charging time increases exponentially with distance $d$. Therefore, to minimize the optimal charging time while achieving long-range charging, we must aim to make the base number $(1-\delta)/(1+\delta)$ approach 1, i.e., $\delta\to0^-$, all while maintaining the Born-Markov approximation. For simplicity, we introduce a dimensionless quantity $\epsilon=\text{g}/(2J\abs{\delta})$, which characterizes the Born-Markov approximation, i.e., $\epsilon\ll1$. Note that when $\epsilon$ is one or two order of magnitudes smaller than 1, such as $\epsilon=0.1$, $\epsilon=0.01$, or even small, the system can be approximated as being in the Markovian regime. Substituting the quantity $\epsilon$ into Eq.~(\ref{moct}), we obtain
\begin{align}
	\min[\tau_{\rm opt}^{\rm I}]=\frac{\pi}{J\epsilon^2}\frac{(1+\delta)}{8\delta^2}\left(\frac{1-\delta}{1+\delta}\right)^d,
\end{align}
which highlights the importance of selecting an appropriate value of $\delta$ for different distances $d$. As shown in Table~\ref{TABLE1}, a proper choice of $\delta$ can significantly shorten the charging time, particularly in long-range charging processes.
\begin{table}[htbp]
	\centering
	\renewcommand{\arraystretch}{1.5}
	\setlength{\tabcolsep}{10pt}
	\begin{tabular}{|c|c|c|c|c|c|}
		\hline
		\(\min[\tau_{\rm opt}^{\rm I}]\times J\epsilon^2/\pi\)& \(\delta=-0.5\) & \(\delta=-0.4\)&\(\delta=-0.3\)&\(\delta=-0.2\)&\(\delta=-0.1\) \\
		\hline
		\(d=1\)&\ \ \ \ 0.75\ \(\checkmark\) & 1.09375&1.80556 &3.75 &13.75 \\
		\hline
		\(d=5\)& 60.75 &32.4209 &21.4779 &\ \ \ \ 18.9844\ \(\checkmark\) &30.6834 \\
		\hline
		\(d=10\)& 14762.3 &2242.38 &474.481 &144.163 &\ \ \ \ 83.6863\ \(\checkmark\) \\
		\hline
	\end{tabular}
	\caption{The minimum optimal charging times $\min[\tau_{\rm opt}^{\rm I}]$ for different values of \(d\) and \(\delta\).}
	\label{TABLE1}
\end{table}

\textbf{Regarding the charging time in configuration \hyperref[fig5]{II}}, we have demonstrated in Section \ref{IIIB} that for $\Delta+\Omega=0$, both the dark state and the vacancy-like dressed state persist even in the dissipative bath, and these bound states are expressed in Eq.~(\ref{states}). Under the Born-Markov approximation, according to Eq.~(\ref{states}), we have
\begin{equation}\label{darkvds}
    \setlength\abovedisplayskip{5pt}
    \setlength\belowdisplayskip{5pt}
	\ket{\Psi_{\text{dark}}}=\frac{1}{\sqrt{2}}(\sigma_+^\mathrm{C}-\sigma_+^\mathrm{B})\ket{g,g;\text{vac}},\quad \ket{\Psi_{\rm vds}}=\frac{1}{\sqrt{2}}(\sigma_+^\mathrm{C}+\sigma_+^\mathrm{B})\ket{g,g;\text{vac}}+|\order{\text{g}^2/\Delta_{\rm min}^2}\rangle,
\end{equation}
which satisfies $H_{\rm eff}\ket{\Psi_{\text{dark}}}=(\omega_c-2\Omega)\ket{\Psi_{\text{dark}}}$ and $H_{\rm eff}\ket{\Psi_{\text{vds}}}=\omega_c\ket{\Psi_{\rm vds}}$. It is important to emphasize that here, we do not treat $\omega_c$ as the energy reference. Similarly, if the entire system is initially prepared in the state $\ket{\psi(0)}=\ket{e_1;\rm vac}$, according to Eq.~(\ref{darkvds}), $\ket{e_1;\rm vac}=[\ket{\Psi_{\text{dark}}}+\ket{\Psi_{\text{vds}}}]/\sqrt{2}+|\order{\text{g}^2/\Delta_{\rm min}^2}\rangle$ and $\ket{e_2;\rm vac}=[\ket{\Psi_{\text{vds}}}-\ket{\Psi_{\text{dark}}}]/\sqrt{2}+|\order{\text{g}^2/\Delta_{\rm min}^2}\rangle$ hold. Based on the density matrix (\ref{densityM}) at time $t$, we need to calculate $\exp(-iH_{\rm eff}t)\ket{\psi(0)}$, as follows:
\begin{align}
	\exp(-iH_{\rm eff}t)\ket{\psi(0)}&=\exp(-iH_{\rm eff}t)[\ket{\Psi_{\text{dark}}}+\ket{\Psi_{\text{vds}}}]/\sqrt{2}+\exp(-iH_{\rm eff}t)|\order{\text{g}^2/\Delta_{\rm min}^2}\rangle\nonumber\\
	&=\exp(-i\omega_ct)[\exp(2i\Omega t)\ket{\Psi_{\text{dark}}}+\ket{\Psi_{\text{vds}}}]/\sqrt{2}+\exp(-iH_{\rm eff}t)|\order{\text{g}^2/\Delta_{\rm min}^2}\rangle\nonumber\\
	&=\exp(-i\omega_et)[\cos(\Omega t)\ket{e_1}-i\sin(\Omega t)\ket{e_2}]\otimes\ket{\rm vac}+\exp(-iH_{\rm eff}t)|\order{\text{g}^2/\Delta_{\rm min}^2}\rangle.
\end{align}
This expression reveals that, within the Markovian regime, the evolution of the entire system, governed by the non-Hermitian Hamiltonian is equivalent to that governed by a Hermitian Hamiltonian. Therefore, under the Born-Markov approximation, it is easy to see that even in the dissipative bath, the combination of charger-battery and charger-bath-battery interactions leads to an effective unitary dynamics governed by the Hermitian Hamiltonian
\begin{align}
	H_{\rm sys}^{\rm eff}=\omega_e(\sigma_{+}^{\rm B}\sigma_{-}^{\rm B}+\sigma_{+}^{\rm C}\sigma_{-}^{\rm C})+\Omega(\sigma_{+}^{\rm B}\sigma_{-}^{\rm C}+\sigma_{+}^{\rm C}\sigma_{-}^{\rm B}).
\end{align}
In this context, we straightforwardly derive the reduced density matrix and the stored energy of the QB, i.e.,
\begin{align}\label{SEII}
	\rho_{\rm B}(t)=\mqty[\cos[2](\Omega t)&0\\0&\sin[2](\Omega t)],\quad \mathcal{E}_{\rm II}(t)=\Tr[\rho_{\rm B}(t)H_{\rm B}]=\omega_e\sin[2](\Omega t).
\end{align}
Then, according to Eqs.~(\ref{ctopt}) and (\ref{SEII}), we have
\begin{align}\label{IICT}
	\mathcal{E}_{\rm II}(\tau_{\rm opt})=\max_{\tau>0}\mathcal{E}_{\rm II}(\tau)=\omega_e\Longrightarrow|\Omega|\tau_{\rm opt}=n\pi+\frac{\pi}{2}\Longrightarrow \min_{n\in\mathbb{Z}}[\tau_{\rm opt}^{\rm II}]=\frac{\pi}{2|\Omega|}.
\end{align}
In configuration II, although long-range charging cannot be realized, the energy transfer from the charger to the QB remains unaffected by engineered photon loss in the photonic lattice, i.e., the single-sublattice dissipation---an advantage not present in configuration I. Moreover, the optimal charging time in configuration II is solely dependent on the direct coupling strength $\Omega$. This suggests that, even under the Born-Markov approximation, the charging time can be significantly reduced simply by adjusting the direct coupling strength.

\subsection{Non-Markovian regime: configurations I and II}
Now, we turn our attention to the non-Markovian regime, where the Born-Markov approximation breaks down, corresponding to the condition $\epsilon\mathrel{\slashed{\ll}}1$ (i.e., $\text{g}\mathrel{\slashed{\ll}}\Delta_{\rm min}$). In contrast to the straightforward expressions derived in the Markovian regime, such as those presented in Eqs.~(\ref{SEI}, \ref{SEII}), the results in the non-Markovian regime become considerably more complex, making it challenging to obtain precise analytical expressions. As a result, the following discussion will primarily focus on numerical results, with particular emphasis on the minimization of the optimal charging time in configurations I and II within the non-Markovian regime.
\begin{figure}
	\centering
	\includegraphics[width=17cm]{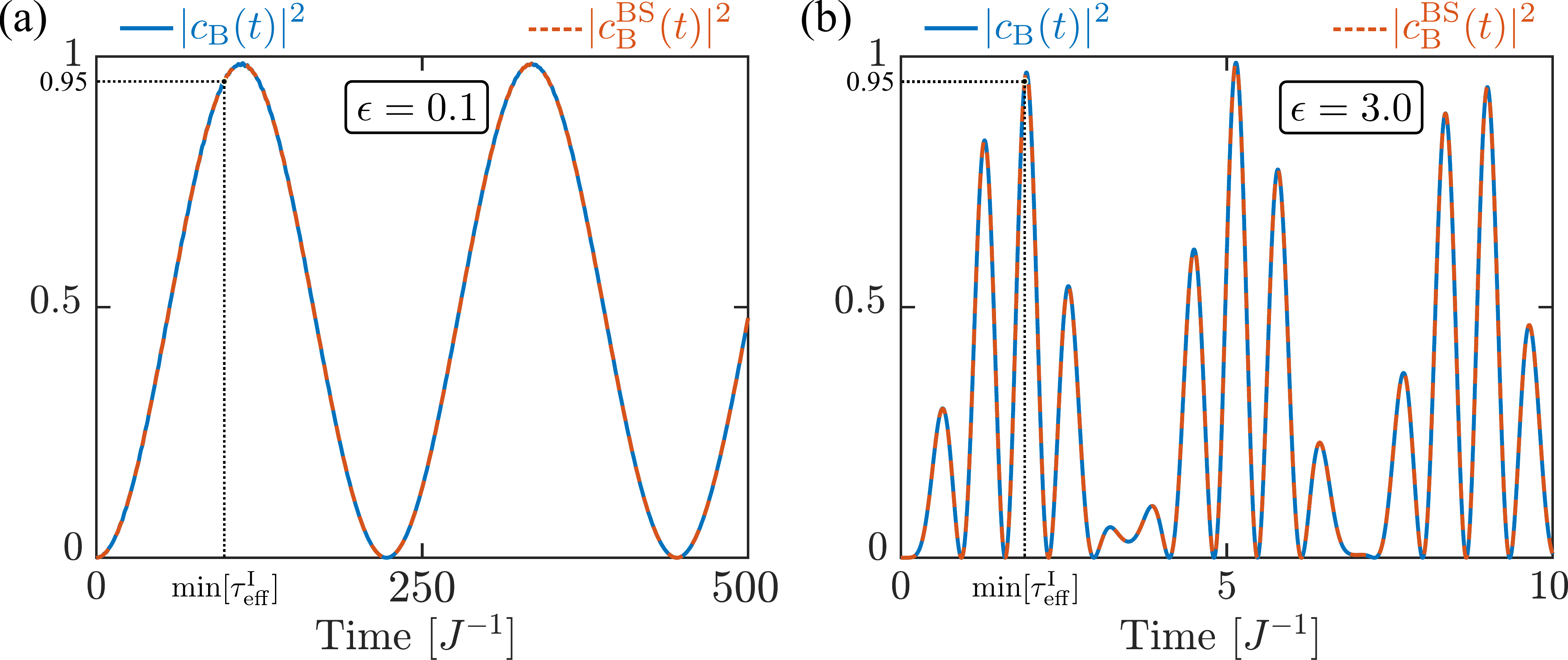}\\
	\caption{Atomic dynamics with full contributions (solid lines) and bound state contributions (dashed lines) in (a) the Markovian regime and (b) the non-Markovian regime. The system parameters are chosen as $\Delta=0$, $\delta=-0.8$, and $d=1$.
	}\label{fign7}
\end{figure}

\begin{table}[htbp]
	\centering
	\renewcommand{\arraystretch}{1.5}
	\setlength{\tabcolsep}{10pt}
	\begin{tabular}{|c|c|c|c|c|c|c|}
		\hline
		Characteristic& \multicolumn{4}{|c|}{\(E_i\neq E_s^\pm\)} &\(E_i= E_s^-\)&\(E_i= E_s^+\) \\
		\hline
		\(E_i\times J^{-1}\)& $-2.75037$ & $2.75037$ & $-2.66232$ & $2.66232$ & $-0.110345$ & $0.110345$\\
		\hline
		\(\Res[\mathscr{C}^{(2)}(z),E_i]\)& $-0.128931$& $0.128931$& $0.141889$ & $-0.141889$ & $-0.22421$  & $0.22421$\\
		\hline
	\end{tabular}
	\caption{Bound state energies $E_i$ and their residues $\text{Res}[\mathscr{C}^{(2)}(z),E_i]$ for configuration I in the non-Markovian regime. The system parameters are chosen as $\Delta=0$, $\delta=-0.8$, $d=2$, and $\epsilon=1.25$.}
	\label{TABLE2}
\end{table}

\textbf{Regarding the charging time in configuration \hyperref[fig5]{I}}, under the non-Markovian regime, some of the conclusions and discussions proved in Section \ref{IIIB}, such as those in Eqs.~(\ref{atomsR}, \ref{ddint}, \ref{ddintF}), will no longer hold. Specifically, regarding the conclusions derived in Eq.~(\ref{atomsR}), we can conclude that in the non-Markovian regime, the contribution of bound states with energies $E_i=E_s^\pm$ to the atomic dynamics is not the only significant one. In fact, the contributions from bound states with $E_i\neq E_s^\pm$ must also be considered can cannot be neglected, as shown by the residues corresponding to $E_i\neq E_s^\pm$ in Table~\ref{TABLE2}. Therefore, in the non-Markovian regime, the contributions from all bound states must be taken into account for the atomic dynamics, unlike in the Markovian regime, where only the two bound states with the most significant contributions are considered. For the sake of simplicity, we rewrite Eq.~(\ref{S41}) as
\begin{equation}
    \setlength\abovedisplayskip{6pt}
    \setlength\belowdisplayskip{6pt}
	c_{\mathrm{B}}(t)=c_{\rm B}^{\rm BS}(t)+c_{\rm B}^{\rm UP}(t)+c_{\rm B}^{\rm BCD}(t),
\end{equation}
where the first term $c_{\rm B}^{\rm BS}(t)=\sum_{E_k\in {\rm BSEs}}\text{Res}[\mathscr{C}^{(2)}(z),E_k]\exp(-iE_kt)$ denotes the contribution from the bound states, and the two other terms represent the contributions from the unstable poles and the branch cut detours. Generally, in the long-time limit or when $\Delta$ lies within the band gap and is far from the band edges [see Figs.\,\ref{fig3}(b, c) and \ref{fig3}(e, f)], the contributions from the branch cuts---specifically, the unstable poles and the branch cut detours---to the atomic dynamics are negligible compared to those from the bound states. Notably, when $\Delta=0$, as $\abs{\delta}$ approaches 1, the contributions from the branch cuts become even more negligible, provided that the system parameters $(g,\delta)$ are far from the vicinity of the phase boundaries $\ell_1$ and $\ell_2$ [see Eq.~(\ref{PB})]. As depicted in Figs.\,\ref{fign7}(a) and \ref{fign7}(b), the atomic dynamics with full contributions agree perfectly with those governed solely by the bound-state contributions, irrespective of whether the system resides in the Markovian or non-Markovian regime. Thus, in the following discussion, we will focus solely on the contributions from the bound states. For simplicity, we consider only the resonance condition ($\Delta=0$) discussed in the main text, and according to Eq.~(\ref{S64}), we have
\begin{equation}\label{cbbst}
    \setlength\abovedisplayskip{6pt}
    \setlength\belowdisplayskip{6pt}
	|c_{\rm B}^{\rm BS}(t)|^2=\bigg|\sum_{k=1}^M\text{Res}[\mathscr{C}^{(2)}(z),E_k]\exp(-iE_kt)\bigg|^2=4\bigg|\sum_{k=1}^{\lceil M/2\rceil}\text{Res}[\mathscr{C}^{(2)}(z),E_k]\sin(E_kt)\bigg|^2,
\end{equation}
where $M$ represents the number of bound states and $E_k$ are the bound state energies arranged in descending order. In the Markovian regime, only one term in Eq.~(\ref{cbbst}) contributes predominantly, with the others being neglected, which explains why the energy transfer exhibits Rabi oscillations. However, in the non-Markovian regime, we must consider all sine functions with frequencies $E_k$. Moreover, when $M\ge4$, the superposition of sine functions with different frequencies generally does not form a periodic function, and the squared modulus of the sum is bounded. This upper bound is given by
\begin{align}
	|c_{\rm B}^{\rm BS}(\tau_{\rm opt})|^2=\max_{\tau>0}|c_{\rm B}^{\rm BS}(\tau)|^2\simeq 4\bigg\{\sum_{k=1}^{\lceil M/2\rceil}\big|\text{Res}[\mathscr{C}^{(2)}(z),E_k]\big|\bigg\}^2\equiv\bar{c}_{\rm B}\approx\max_{\tau>0}|c_{\rm B}(\tau)|^2.
\end{align}
Note that while $c_{\rm B}^{\rm BS}(\tau_{\rm opt})$ can approach the upper bound $\bar{c}_{\rm B}$ arbitrarily closely, corresponding to the maximum stored energy [see Eq.~(\ref{ctopt})], this requires the optimal charging time $\tau_{\rm opt}$ to be extremely large and highly precise. Therefore, in the non-Markovian regime, to effectively balance both stored energy and charging time, we define the {\it efficient charging time} $\tau_{\rm eff}$ as the time required for the QB to reach 95\% of its full charge, i.e., $\mathcal{E}_{\rm I}(\tau_{\rm eff})/\omega_e\equiv95\%$, where this percentage can be adjusted based on specific requirements, typically set to a large value but below one. Here, we are still interested in minimizing the efficient charging time. Notably, in the Markovian regime, the efficient charging time and its corresponding stored energy are nearly identical to those of the optimal charging time, i.e., $\min[\tau_{\rm opt}^{\rm I}]\approx\min[\tau_{\rm eff}^{\rm I}]$ and $[\mathcal{E}_{\rm I}(\tau_{\rm opt})-\mathcal{E}_{\rm I}(\tau_{\rm eff})]/\omega_e\approx5\%$. In contrast, in the non-Markovian regime, the efficient charging time is significantly smaller than the optimal charging time, i.e., $\min[\tau_{\rm opt}^{\rm I}]\gg\min[\tau_{\rm eff}^{\rm I}]$, while the corresponding stored energies differ by no more than 5\%, i.e., $0<[\mathcal{E}_{\rm I}(\tau_{\rm opt})-\mathcal{E}_{\rm I}(\tau_{\rm eff})]/\omega_e<5\%$. This is why we define the efficient charging time specifically in the non-Markovian regime. Note that the above conclusion is clearly confirmed in Table~\ref{TABLE3}. Specifically, in the approximate Markovian regime (taking $\epsilon=0.1$ as an example), the minimum efficient charging time is only slightly different from the minimum optimal charging time, with the corresponding stored energy differing by no more than 4\%. In contrast, in the non-Markovian regime (excluding the interval where the upper bound falls below 95\%, as the definition of efficient charging time is invalid in this interval), the minimum efficient charging time can be reduced by an order of magnitude or more, even by tens to hundreds of times compared to the minimum optimal charging time, while the corresponding stored energy exhibits a derivation of no more than 5\%.

\begin{table}[htbp]
	\centering
	\renewcommand{\arraystretch}{1.5}
	\setlength{\tabcolsep}{4.5pt}
	\begin{tabular}{|c|c|c|c|c|c|c|c|c|c|c|}
		\hline
		\(\epsilon\)&0.1 &$0.221\sim0.493$&0.7&1.0&$1.302\sim1.911$ &2.2&2.5&2.8&3.1&3.4\\
		\hline
		\(\min[\tau_{\rm eff}^{\rm I}]\times J\)&98.54 &\ding{55}&8.53&1.60&\ding{55}&1.68&1.52&8.90&1.83&1.67\\
		\hline
		\(\min[\tau_{\rm opt}^{\rm I}]\times J\)&112.38 &\ding{55}&419.70&1563.4&\ding{55}&526.37&407.69&714.29&89.01&924.70\\
		\hline
		\(\bar{c}_{\rm B}\)&$98.5\%$&$<95\%$ &$96.9\%$&$97.8\%$&$<95\%$ &$98.3\%$&$99.0\%$&$99.4\%$&$99.5\%$&$99.6\%$\\
		\hline
	\end{tabular}
	\caption{The minimum efficient charging time $\min[\tau_{\rm eff}^{\rm I}]$, the minimum optimal charging time $\min[\tau_{\rm opt}^{\rm I}]$, and the upper bound $\bar{c}_{\rm B}$ for configuration I with various $\epsilon$'s. The system parameters are the same as in Fig.\,\ref{fign7}.}
	\label{TABLE3}
\end{table}

\textbf{Regarding the charging time in configuration \hyperref[fig5]{II}}, as discussed in Sections \ref{IIIB} and \ref{IIIC}, two dissipation-independent bound states---the dark state and the vacancy-like dressed state---exist in both the Markovian and non-Markovian regimes when $\Delta+\Omega=0$. The energies and the corresponding residues of these bound states are given by [see Eq.~(\ref{S68})] 
\begin{equation}\label{BS-R}
    \setlength\abovedisplayskip{1pt}
    \setlength\belowdisplayskip{5pt}
	\text{Res}[\mathscr{C}^{(2)}(z), E_{\rm dark}=-2\Omega]=-\frac{1}{2},\quad \text{Res}[\mathscr{C}^{(2)}(z), E_{\rm vds}=0]=\frac{J^2\abs{\delta}}{\text{g}^2+2J^2\abs{\delta}}.
\end{equation}
Additionally, two dissipation-dependent bound states, referred to as dissipative bound states, are also present, with their energies and corresponding residues specified by [see Eqs.~(\ref{DBSEs}) and (\ref{S76})]
\begin{align}\label{DBS-R}
	E_{\kappa,\pm}=-\frac{i}{4}\kappa\pm\sqrt{E_{0,\pm}^2-(\kappa/4)^2},\quad \text{Res}[\mathscr{C}^{(2)}(z),E_{\kappa,\pm}]=\text{g}^4\times\frac{1\pm[1-(4E_{0,\kappa}^2/\kappa)^2]^{-1/2}}{E_{0,\pm}^2[E_{0,\pm}^2-2J^2(1+\delta^2)]},
\end{align}
where $E_{0,\pm}=\pm\customsqrt{2J^2(1+\delta^2)+2\sqrt{\text{g}^4+J^4(1-\delta^2)^2}}$. Notably, in the Markovian regime ($\text{g}\ll 2J\abs{\delta}$ or equivalently, $\epsilon\ll1$), the atomic dynamics are dominated by the contribution from the two dissipation-independent bound states due to $\text{Res}[\mathscr{C}^{(2)}(z),E_{\rm dark}]=-1/2$ and $\text{Res}[\mathscr{C}^{(2)}(z),E_{\rm vds}]\to1/2$. As discussed earlier, the optimal charging time and the corresponding stored energy are characterized by Eq.~(\ref{IICT}). In contrast, in the non-Markovian regime, the contribution from the vacancy-like dressed state decreases progressively with increasing coupling strength $\text{g}$, while the contributions from the dissipative bound states and branch cuts become more significant on short timescales. The conditions under which the branch-cut contributions can be neglected have been discussed in Section \ref{IIIC}, with clear comparisons presented in Fig.\,\ref{fign9}. However, even when the contributions from the branch cuts are negligible, the inherent complexity [see Eq.~(\ref{DBS-R})] of the dissipative bound-state contributions render it infeasible to analytically determine the optimal charging time and the corresponding stored energy, especially when the system lies outside the strong dissipation or Markovian regimes---let alone in cases where the branch-cut contributions cannot be ignored. Therefore, in the following discussion, we primarily present numerical results to illustrate the impact of dissipation strength $\kappa$ and atom-bath coupling strength $\text{g}$ on transient atomic dynamics, optimal charging time, and the corresponding stored energy. 
\begin{figure}
	\centering
	\includegraphics[width=17cm]{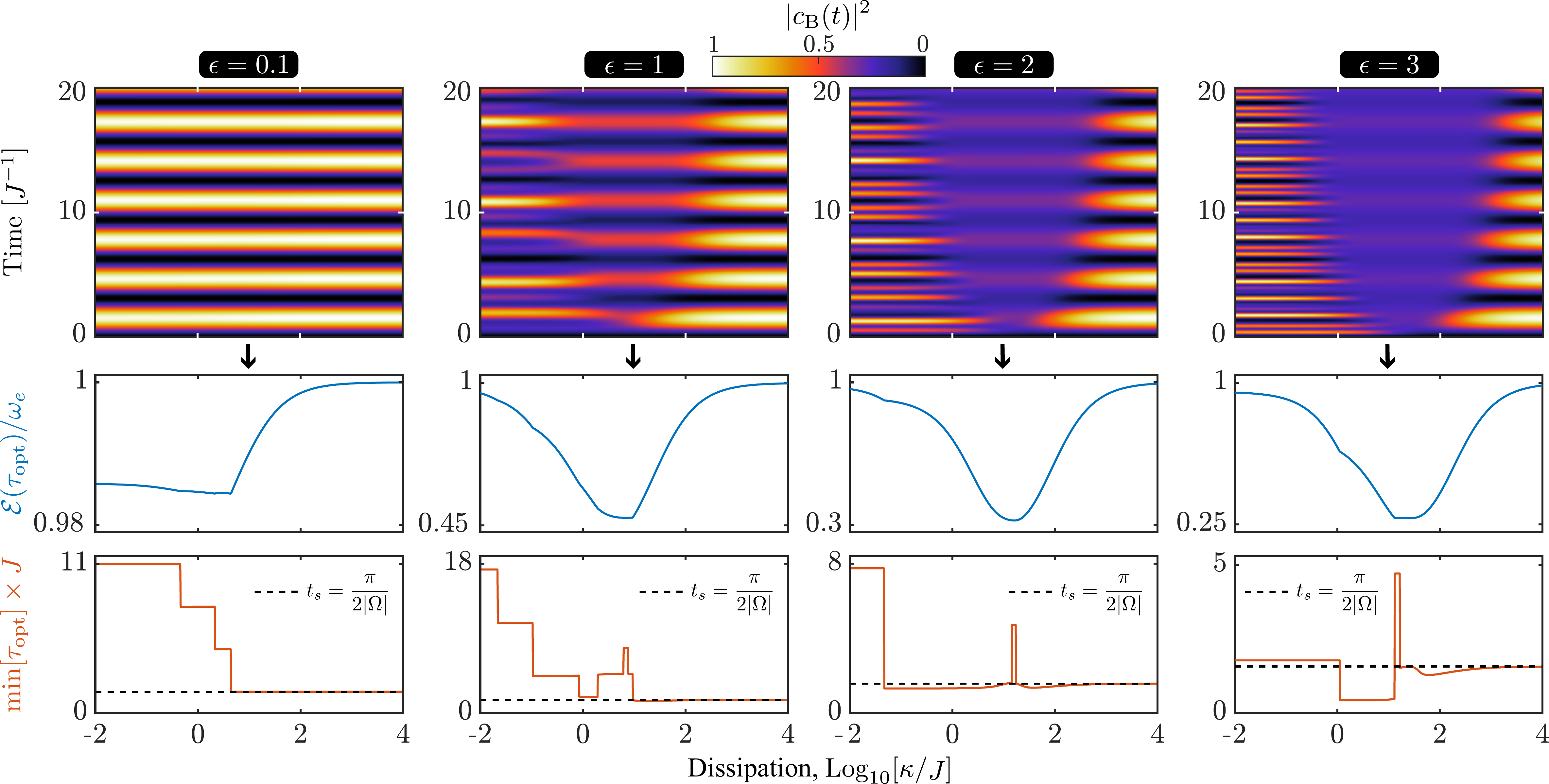}\\
	\caption{The atomic transient dynamics $c_{\rm B}(t)$, the maximum stored energy $\mathcal{E}_{\rm II}(\tau_{\rm opt})$, and the minimum optimal charging time $\min[\tau^{\rm II}_{\rm opt}]$ for configuration II as a function of the dissipation strength $\kappa$ with various $\epsilon$'s. The system parameters are chosen as $\Delta=-\Omega=J$, $\text{g}=2J\abs{\delta}\times\epsilon$, and $\delta=0.8$.
	}\label{fign11}
\end{figure}

Specifically, as shown in Fig.\,\ref{fign11}, in the approximately Markovian regime ($\epsilon=0.1\Longrightarrow \text{g}=1.6J$), we observe that the atomic dynamics remain largely unaffected by dissipation, primarily due to the fact that, in the Markovian regime---or even in the approximately Markovian regime---the atomic dynamics are predominantly governed by two dissipation-independent bound states. Nonetheless, as the system is approximately in the Markovian regime, the minimum optimal charging time and the corresponding stored energy exhibit moderate sensitivity to dissipation, with the effect being more pronounced on the minimum optimal charging time. As the dissipation strength increases further, the minimum optimal charging time decreases progressively, ultimately converging to  $\pi/(2|\Omega|)$, which corresponds to the minimum optimal charging time in the Markovian regime [see Eq.~(\ref{IICT})]. Simultaneously, the corresponding stored energy increase steadily, eventually reaching its maximum value, i.e., $\mathcal{E}(\tau_{\rm opt})=\omega_e$. Even in the non-Markovian regime, as depicted by $\epsilon=1\sim3$ in Fig.\,\ref{fign11}, this conclusion, i.e., $\min[\tau_{\rm opt}^{\rm II}]\to\pi/(2|\Omega|)$ and $\mathcal{E}_{\rm II}(\tau_{\rm opt})\to \omega_e$, remains valid under sufficiently strong dissipation (i.e., $\kappa/J\gtrsim 1000$), which has been analytically proven in Section \ref{IIIC} [see Eq.~(\ref{CBTF})]. Moreover, as $\epsilon$ increases (i.e., with stronger coupling strength $\text{g}$), the impact of dissipation on atomic dynamics becomes increasingly significant, as reflected in the variation trend of $\mathcal{E}_{\rm II}(\tau_{\rm opt})$ with dissipation. Notably, when $\epsilon_1>\epsilon_2$, e.g., $\epsilon=2>\epsilon=1$, it is observed that $\min_{\kappa}[\mathcal{E}_{\rm II}(\tau_{\rm opt}; \epsilon_1)]<\min_{\kappa}[\mathcal{E}_{\rm II}(\tau_{\rm opt}; \epsilon_2)]$, and the corresponding minimum optimal charging time also exhibits varying trends. Furthermore, larger values of $\epsilon$ require stronger dissipation for $\mathcal{E}_{\rm II}(\tau_{\rm opt})$ to approach one. This is primarily due to the increase in $E_{0,+}$ with $\text{g}$, which elevates the reference value for defining strong dissipation regime (i.e., $\kappa/E_{0,+}\gg1$).

Finally, in addition to the previously discussed influence of dissipation strength and coupling strength (or equivalently, $\epsilon$) on the atomic dynamics, the direct coupling strength $\Omega$ between the QB and the charger is also a critical factor. When $\Delta=-\Omega\neq0$, as demonstrated by Eqs.~(\ref{BS-R}) and (\ref{DBS-R}), the contributions from the vacancy-like dressed state and the dissipative bound states are unaffected by the direct coupling strength. Furthermore, to prove that the contribution from the branch cuts is also independent of the direct coupling strength, we begin by selecting some small closed contours, denoted as $\mathcal{C}_q$, that tightly enclose the branch cuts. According to Eq.~(\ref{INT2}), the contribution from the branch cuts is computed as ($\kappa\neq0$)
\begin{align}\label{BCTA}
	c_{\rm B}^{\rm BC}(t)=\int_{\mathcal{C}_q}\frac{\dd{z}}{2\pi i}\mathscr{C}^{(2)}(z)e^{-izt}=\int_{\mathcal{C}_q}\left[\frac{\exp(-izt)/2}{z-2\times\Sigma_{11}^{AA}(z)}-\frac{\exp(-izt)/2}{z+2\Omega}\right]\frac{\dd{z}}{2\pi i}=\int_{\mathcal{C}_q}\frac{\exp(-izt)}{z-2\times\Sigma_{11}^{AA}(z)}\frac{\dd{z}}{4\pi i},
\end{align}
where in the last step, as the branch cuts of the dissipative bath do not lie on the real axis, the integral of the second term is guaranteed to vanish according to residue theorem. It is important to note that for a non-dissipative bath (i.e., $\kappa=0$), when the bound-state energy of the dark state has been determined by solving the pole equation (\ref{S66}), the second integral term in Eq.~(\ref{BCTA}) must be explicitly excluded, as it corresponds to the contribution of the dark state. Consequently, the atomic dynamics with full contributions can be simplified as
\begin{align}
	c_{\rm B}(t)=c_{\rm B}^{\rm w}(t)+c_{\rm B}^{\rm wo}(t)=-\frac{1}{2}e^{2i\Omega t}+c_{\rm B}^{\rm wo}(t),
\end{align}
where $c_{\rm B}^{\rm w}(t)$ and $c_{\rm B}^{\rm wo}(t)$ are defined as the terms that are dependent and independent of the direct coupling strength $\Omega$, respectively. Note that $c_{\rm B}^{\rm wo}(0)=1/2$ due to $c_{\rm B}(0)=0$ and $c_{\rm B}^{\rm w}(0)=-1/2$. On the one hand, we find that as the direct coupling strength increases, the shortest time, i.e., $t_{\rm s}\equiv\pi/(2|\Omega|)$, required for $c_{\rm B}^{\rm w}(t)$ to transition from $-1/2$ to $1/2$ decreases proportionally. On the other hand, $c_{\rm B}^{\rm wo}(t)$ generally decreases over time due to the contributions from the dissipative bound states and branch cuts. Consequently, we have reason to believe that the value of $c_{\rm B}(t)$ at $t=t_{\rm s}$ increases as $\Omega$ grows, due to the reduction in $t_{\rm s}$, which in turn shortens the minimum optimal charging time and improves the corresponding stored energy. In fact, this point is further confirmed by the data presented in Table~\ref{TABLE4}, which shows that $\min[\tau_{\rm opt}^{\rm II}]$ decreases while $\mathcal{E}_{\rm II}(\tau_{\rm opt})$ increases steadily as $\Omega$ grows (or equivalently, as $t_{\rm s}$ decreases). We also find in Table~\ref{TABLE4} that for larger direct coupling strength, the minimum optimal charging time closely approximates the shortest time $t_m$, and the shortest transition time $t_{\rm s}$ is, in fact, equivalent to the minimum optimal charging time for configuration II in the Markovian regime, see Eq.~(\ref{IICT}). In summary, increasing the direct coupling strength not only reduces the optimal charging time but also enhances the corresponding stored energy, thereby further improving the charging power.

\begin{table}[htbp]
	\centering
	\renewcommand{\arraystretch}{1.5}
	\setlength{\tabcolsep}{4.5pt}
	\begin{tabular}{|c|c|c|c|c|c|c|c|c|c|c|}
		\hline
		\(\Omega/J\)&0.1 &1.0 &3.0&5.0&7.0&9.0&11.0&13.0&15.0&17.0\\
		\hline
		\(t_{\rm s}\times J\)&15.708\ldots &1.570\ldots&0.523\ldots&0.314\ldots&0.224\ldots&0.174\ldots&0.142\ldots&0.120\ldots&0.104\ldots&0.092\ldots\\
		\hline
		\(\min[\tau_{\rm opt}^{\rm II}]\times J\)&15.708 &1.539&0.36&0.256&0.198&0.161&0.134&0.115&0.101&0.09\\
		\hline
		\(\mathcal{E}_{\rm II}(\tau_{\rm opt})/\omega_e\)&$32.21\%$&$33.42\%$ &$38.68\%$&$58.65\%$&$71.9\%$ &$80.11\%$&$85.33\%$&$88.79\%$&$91.19\%$&$92.9\%$\\
		\hline
	\end{tabular}
	\caption{The minimum optimal charging time $\min[\tau_{\rm opt}^{\rm II}]$, the shortest transition time $t_{\rm s}$, and the corresponding stored energy $\mathcal{E}_{\rm II}(\tau_{\rm opt})$ for configuration II with various $\Omega$'s. The parameters are chosen as $\Delta=-\Omega$, $\kappa=10J$, $\delta=0.8$, and $\epsilon=2$.}
	\label{TABLE4}
\end{table}

\section{Full Cycle of Quantum Charging Protocol in the Scheme}
\begin{figure}
	\centering
	\includegraphics[width=17cm]{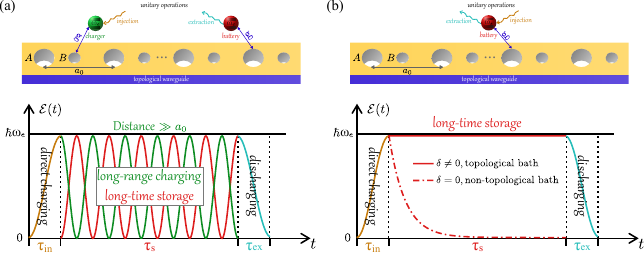}\\
	\caption{Upper panel of (a): QB protocol with a charger. The topological waveguide, as an emerging environment with a topological structure, could mediate a dipole-dipole interactions between the QB and the charger manifesting the formation of two exotic bound states such that the aging of the QB within the long-time storage process and the energy-delivery limitation over long distances are overcome. Lower panel of (a): Full cycle of the scheme. The initialization process is within $t\in[0,\tau_{\rm in})$. The combined charging and the storage processes are within $t\in[\tau_{\rm in},\tau_{\rm in}+\tau_{\rm s})$, where the charging occurs in a reversible way and the energy loss of the QB is fully overcome governed by the two exotic bound states. The energy extraction to the QB occurs within $t\in[\tau_{\rm in}+\tau_{\rm s}, \tau_{\rm in}+\tau_{\rm s}+\tau_{\rm ex})$. Upper panel of (b): QB protocol without a charger. Lower panel of (b): Full cycle of the scheme. The QB is charged with $t\in[0,\tau_{\rm in})$, and the storage process is within $t\in[\tau_{\rm in},\tau_{\rm in}+\tau_{\rm s})$. With the protection of the topological waveguide, which induces a topologically robust dressed state, the QB energy is immune to decoherence, enabling significant energy extraction during the discharging process within $t\in[\tau_{\rm in},\tau_{\rm s},\tau_{\rm in}+\tau_{\rm s}+\tau_{\rm ex})$. In this setup, the topological waveguide is realized as a one-dimensional coupled-cavity array, with adjacent cells separated by a distance $a_0$.
	}\label{fign1}
\end{figure}

In recent years, the design of various QB schemes has attracted significant attention, broadly categorized into charger-free \cite{PRXQuantum.5.030319, PhysRevLett.124.130601, PhysRevLett.125.040601}, and charger-assisted \cite{PhysRevLett.122.047702, PhysRevLett.120.117702, PhysRevLett.127.100601, PhysRevLett.132.210402} schemes. A key advantage of charger-assisted schemes lies in the charger's ability to facilitate the establishment of QB-charger entanglement and induce an effective long-range dipole-dipole interaction between the QB and the charger, thereby enabling the possibility of a perfect long-range charging process. Charger-free schemes, while simpler to implement, generally lack mechanisms to suppress decoherence effects induced by an external environment, resulting in energy loss of the QB during storage. However, our proposed scheme demonstrates that even for charger-free schemes, the presence of a topologically robust dressed state enables complete suppression of the decoherence effects, thereby avoiding the aging problem of the QB. Additionally, charger-assisted systems in our scheme not only achieve complete suppression of the aging effect but also enable non-contact remote charging through the topological waveguide. Notably, within a sufficiently long charging time, this approach can achieve perfect charging over ultra-long distances, a feat unattainable with traditional waveguide, such as photonic crystal waveguide \cite{lu2024, PhysRevLett.115.153901}, hollow metal waveguide \cite{PhysRevLett.132.090401}, and optical nanofibers \cite{PhysRevLett.110.243603}.

Here, we design a topological QB scheme where the charger formed by a two-level system and a topological waveguide together act as a power station, as depicted in Fig.\,\ref{fign1}(a). Our scheme leverages an emerging environment with a topological structure to mediate a topologically dependent dipole-dipole interaction between the charger and the QB. Specifically, this dipole-dipole interaction [see Eq.~(\ref{ddintF})], acting as a topological switch, is turned on in the topologically nontrivial phase and off in the trivial phase. When charging the QB is required, the topological switch is turned on by tuning the topological waveguide into its topologically nontrivial phase. This configuration enables the QB to be positioned at a distance, not necessarily close to the charger, i.e., $x_1-x_2\gg1$, with $\alpha=A$ and $\beta=B$, where the charging process can be automatically completed through nearly coherent energy transfer facilitated by the topological waveguide. Conversely, if charging is not needed, the topological switch can simply be turned off. In our scheme, although the topological environment induces decoherence in the two-level systems, it also plays a pivotal role in enabling efficient energy transfer from the charger to the QB. Specifically, in the Markovian regime (i.e., $\text{g}\ll 2J\abs{\delta}$), when the two-level systems resonate with the eigenfrequency of the environmental modes (i.e., $\Delta=0$), the topological waveguide in the topologically nontrivial phase mediates an effective dipole-dipole interaction between the QB and the charger, which enables perfect charging of the QB over ultra-long distances. This remarkable capability arises from the presence of parity-positive and parity-negative bound states with different energies in the system. These bound states are approximately formed by the positive and negative superpositions of the vacancy-like dressed states (also referred to as topologically robust dressed state) that emerge from the individual coupling of two atoms, i.e., the QB and the charger, to the topological waveguide. However, the topological waveguide in the topologically trivial phase cannot mediate the effective dipole-dipole interaction due to the degeneracy of the parity-positive and parity-negative bound states. This degeneracy renders the vacancy-like dressed states, formed by the individual coupling of each atom to the topological waveguide, also becoming the bound states of the system. Consequently, the energy stored in the charger remains confined, neither dissipating into the environment nor transferring to the QB. Interestingly, in the absence of the charger, see Fig.\,\ref{fign1}(b), the vacancy-like dressed state formed by the single QB coupled to the topological waveguide fully counteracts the destructive effect of the decoherence on the QB, thereby overcoming aging of the QB. Note that the underlying physical mechanism that prevents the energy of the two-level systems from dissipating into the topological environment is identical for the charger-free scheme and the charger-assisted scheme when the topological waveguide is in the topologically trivial phase. Notably, for a non-topological bath (i.e., $\delta=0$), corresponding to a tight-binding model, the aforementioned conclusions, including feasibility of perfect long-range charging and the suppression of the QB aging, will no longer remain valid.

Finally, as depicted in Fig.\,\ref{fign1}, the full cycle of the charging protocol in our scheme can be summarized as follows.
\begin{enumerate}
	\item Initialization Process. A $\pi$ laser pulse or coherent drive is applied to the charger in order to excite it from the ground state $\ket{g}$ to the excited state $\ket{e}$. Note that when the coherent drive is employed as a method for energy injection into the charger, it corresponds to the addition of a driving term to the interaction Hamiltonian $\mathcal{V}(t)$ in Eq.~(\ref{CTV}), i.e., $\mathcal{V}(t)\to \mathcal{V}(t)+[F(t)\sigma_{+}^{\rm C}+\text{H.c.}]$. Here, $F(t)=F\exp(-i\omega_et)\Theta(\tau_{\rm in}-t)$, where $F$ denotes the driving strength with $F=\order{J}$, and $\tau_{\rm in}=\pi/(2|F|)$ corresponds to the direct charging time of the charger.
	\item Charging Process. If charging of the QB is required, the QB is taken to the power station formed by the charger and topological waveguide. Specifically, the QB is positioned in sublattice $A$ of the $x_1$-th unit cell, while the charger is placed in sublattice $B$ of the $x_2$-th unit cell. When the topological switch is activated, meaning the topological waveguide is in the topologically nontrivial phase, a perfect long-range charging to the QB is automatically executed in a Rabi-like oscillation manner.
	\item Storage Process (Idle Process). In the charger-assisted scheme, energy can be stored in either the QB or the charger for an arbitrarily long time without energy loss, as the energy exchange between the charger and the QB is fully reversible. During this process, we propose here in two methods for energy storage. The first method involves allowing the energy to oscillate between the charger and the QB in Rabi-like oscillations, thereby sustaining energy storage. The second method entails turning off the topological switch at $t=\tau_{\rm in}+\min[\tau_{\rm opt}^{\rm I}]$, corresponding to the transition of the topological waveguide from the topologically nontrivial phase to the topologically trivial phase, thereby confining the energy entirely within the QB for long-time storage. Note that, the storage process depicted in Fig.\,\ref{fign1}(a) corresponds to the first method. In the charger-free scheme, once the QB is fully charged via the direct charging process, its energy can be stored in the QB for an arbitrarily long time, with no leakage into the topological environment. Consequently, the energy-loss problem in the storage process generally encountered by the traditional QB schemes is efficiently avoided.
	\item Discharging Process. If energy delivery from the QB is required, the QB is removed from the power station, and an interaction between the QB and the target system is activated, initiating the discharging of the QB. Note that the durations of the charging and discharging processes are substantially shorter than that of the storage process.
\end{enumerate}

Generally, throughout the lifespan of QB devices, a significant portion of time is spent in an idle state, where the QB is neither charged nor discharged. During these periods, ensuring that energy stored in the QB without loss becomes crucial for maintaining the overall efficiency and longevity of the device. In our QB protocol, whether in the charger-assisted scheme Fig.\,\ref{fign1}(a) or the charger-free scheme Fig.\,\ref{fign1}(b), the energy stored in the QB either undergoes flawless coherent exchange with the charger or remains perfectly preserved within the QB, with both processes incurring negligible energy loss until it is needed for energy extraction. By integrating a topological waveguide with quantum batteries, our scheme achieves efficient long-range charging while completely suppresses energy loss during idle periods, thereby effectively addressing the critical challenges of poor performance and short operational lifespan in QB devices.
\begin{figure}
	\centering
	\includegraphics[width=17cm]{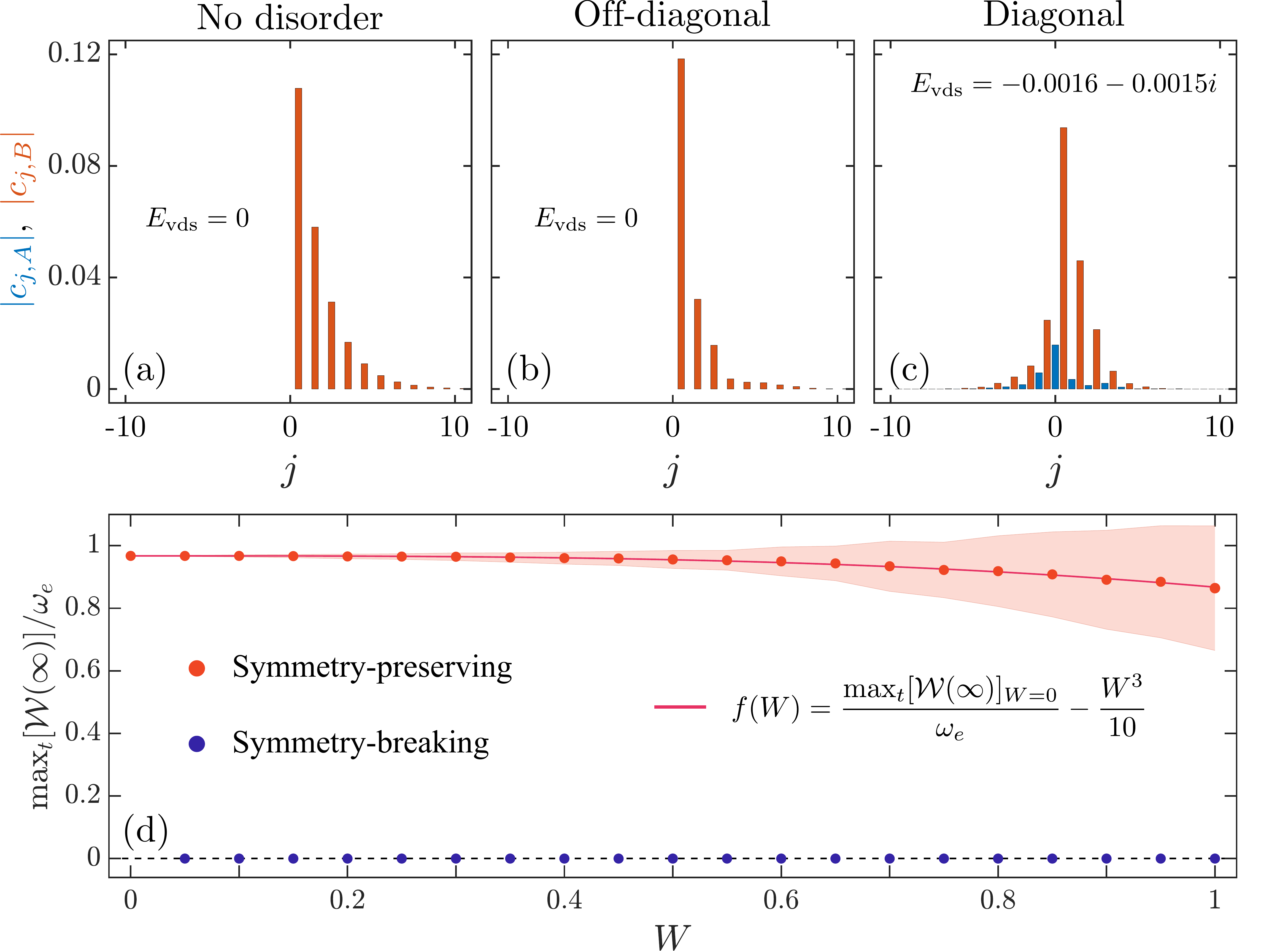}\\
	\caption{Panels (a-c) describe the properties of the vacancy-like dressed state with and without disorder. The absolute value of probability amplitudes $|c_{j,A}|$ are shown in blue, while the $|c_{j,B}|$ are shown in orange. Panel (a) corresponds to the model without disorder, panel (b) corresponds to the model with disorder in the couplings between cavities, and panel (c) corresponds to the model with disorder in the resonant frequencies of cavities. The disorder strength is set to $W=0.5$ in both cases with disorder. For each case, the value of the energy of the vacancy-like dressed state is shown at the inside of the plots, e.g., $E_{\text{vds}}=0$. Panel (d) shows the maximum ergotropy $\max_t[\mathcal{W}(\infty)]$ of QB for the two different models of disorder as a function of the disorder strength $W$. The red (blue) dots correspond to the average value computed with a total of $10^3$ instances of disorder for the symmetry-preserving (symmetry-breaking) case, and the shadow areas span their corresponding standard deviation. The pink line represents a fit. In all plots, the system parameters are chosen to be $\delta=0.3$, $\text{g}=0.1J$, $\Delta=-\Omega=0.2J$, $\kappa_a=10J$, $\kappa_b=0$, and $x_1=x_2=0$.
	}\label{fig8}
\end{figure}

\section{Effects of Symmetry-Preserving and Symmetry-Breaking Disorder}
In practical physical systems, disorder is inevitable and has profound effects on the performance of quantum batteries. In this section, we will discuss in detail the manifestation of disorder in the one-dimensional SSH model and its impact on vacancy-like dressed state and the performance of quantum batteries, particularly focusing on ergotropy.

Here, we primarily investigate the impact of two types of disorder: one that affects the cavities' free frequencies (diagonal), and the other that affects the tunneling amplitudes between them (off-diagonal). The former corresponds to the addition of random diagonal terms to the bath's Hamiltonian, modifying it as $H_{\text{bath}}\to H_{\text{bath}}+\sum_{j}(\epsilon_{a,j}a_j^\dagger a_j^{}+\epsilon_{b,j}b_j^\dagger b_j^{})$, thereby breaking the chiral symmetry of the original model. The latter corresponds to the addition of off-diagonal random terms, modifying the Hamiltonian as $H_{\text{bath}}\to H_{\text{bath}}+\sum_{j}(\epsilon_{1,j}a_j^\dagger b_j^{}+\epsilon_{2,j}b_j^\dagger a_{j+1}^{}+\text{H.c.})$, which preserves the chiral symmetry. We take the disorder parameters $\epsilon_{\nu,j}/J\ (\nu=a,b,1,2)$ from a uniform distribution within the range $[-W, W]$ for each $j$th unit cell, where $W$ represents the disorder strength. Additionally, we only focus on the configuration II [see Fig.\,\ref{fig5}(b)] and the maximum ergotropy. Therefore, as concluded in Sec.~\ref{IIIB}, we know that even in the presence of sublattice dissipation, the energy in the quantum charger can be almost completely transferred to QB through the dissipative topological bath, primarily due to the contributions from the dark state and vacancy-like dressed state, i.e., Eq.~(\ref{states}). We can be confident that the dark state is unaffected by above any type of disorder, due to the unique properties of the dark state, which decouple from the bath. Therefore, in the presence of disorder, changes in the vacancy-like dressed state are the sole factor affecting the performance of QB.

In the first row of Fig.\,\ref{fig8}, we plot the shape of the three vacancy-like dressed states for a situation without disorder and with off-diagonal (diagonal) disorder. Note that for the situation with diagonal disorder, although there is no vacancy-like dressed state, we still refer to it as such for convenience. For the symmetry-preserving case (i.e., no disorder or off-diagonal disorder), as shown in Figs.\,\ref{fig8}(a) and \ref{fig8}(b), we observe that the dressed bound state exhibits a unidirectional spatial profile and has components only on sublattice B. Additionally, compared to the clean system, in the case with off-diagonal disorder, the dressed bound-state energy remains zero, indicating that chiral symmetry ensures the presence of the dressed bound state, with slight changes in the absolute magnitude of the component of the bound state on sublattice B. In contrast, for the symmetry-breaking case (i.e., diagonal disorder), as shown in Fig.\,\ref{fig8}(c), we find that the state loses its unidirectional property and has a non-zero component on each sublattice. Furthermore, its energy also becomes a complex number with a non-zero imaginary part, i.e., $E_{\text{vds}}=-0.0016-0.0015i$, indicating that breaking chiral symmetry disrupts the presence of the dressed bound state. Consequently, for our system, we can conclude that as long as chiral symmetry is preserved, even in the presence of disorder, the maximum ergotropy can remain high due to the contributions of the dark state and the vacancy-like dressed state. Conversely, when chiral symmetry is broken, resulting in the disappearance of the vacancy-like dressed state, the maximum ergotropy drops to zero. In fact, the data of these orange and blue dots in Fig.\,\ref{fig8}(c) corroborates this conclusion. More importantly, as shown by the orange dots in Fig.\,\ref{fig8}(c), we find that the maximum ergotropy is strongly robust to off-diagonal disorder. As disorder strength $W$ increases, its average value obeys a power-law distribution, i.e., $f(W)=\max_t[\mathcal{W}(\infty)]_{W=0}/\omega_e-W^3/10$, where $\max_t[\mathcal{W}(\infty)]_{W=0}/\omega_e$ is given by Eq.~(\ref{analy_ergo}), as indicated by the pink line in Fig.\,\ref{fig8}(c). Even under strong off-diagonal disorder, such as $W=1$, the average value of maximum ergotropy can still exceed 4/5. However, as shown by the blue dots in Fig.\,\ref{fig8}(c), the maximum ergotropy is always zero for the symmetry-breaking case. 

\section{The Validity of the Rotating Wave Approximation}
In this section, we establish the validity of employing the rotating wave approximation (RWA) in the interaction Hamiltonian $H_{\rm int}$ between the system and topological waveguide through both analytical and numerical methods. As a well-known textbook-level statement, Jaynes-Cummings (JC) model is widely recognized as a specific limiting case of the Dicke model, which is expressed as
\begin{align}
	H_{\rm Dicke}=\omega_ca^\dagger a^{}+\sum_{n=1}^M\omega_e\sigma^\dagger_n\sigma_n^{}+\sum_{n=1}^M\text{g}(a^\dagger+a^{})(\sigma_n^\dagger + \sigma_n^{})\xlongrightarrow[\rm RWA]{M=1}H_{\rm JC}=\omega_ca^\dagger a^{}+\omega_e\sigma^\dagger\sigma^{}+\text{g}(a^\dagger\sigma^{}+\sigma^\dagger a^{}).
\end{align}
In our system, the interaction Hamiltonian $H_{\rm int}$ explicitly has applied the RWA; however, the conditions for this approximation have not been clearly stated. To verify its applicability, we will first derive these conditions analytically and then confirm their validity through numerical simulations.

In the following discussions, we take configuration I as an example, as illustrated in Fig.\,\ref{fig5}(a). Note that a similar analysis for configuration II yields the same conclusions. Here, the interaction Hamiltonian does not apply the RWA (denoted as $H_{\rm int}^{\rm wo}$), and thus, the total Hamiltonian can be written as
\begin{align}\label{S82}
	H_{\rm tot}^{\rm wo}=H_{\rm sys}+H_{\rm bath}+H_{\rm int}^{\rm wo},
\end{align}
where
\begin{align}
	H_{\rm sys}&=\omega_e\sigma_+^{\rm B}\sigma_-^{\rm B} + \omega_e\sigma_+^{\rm C}\sigma_-^{\rm C},\quad H_{\rm int}^{\rm wo} = \text{g}\left[\left(\sigma_-^{\rm B}+\sigma_+^{\rm B}\right) \left(a_{x_1}^{}+a_{x_1}^\dagger\right) + \left(\sigma_-^{\rm C}+\sigma_+^{\rm C}\right) \left(b_{x_2}^{}+b_{x_2}^\dagger\right)\right],\\
	H_{\text{bath}}&=\sum_{j=1}^{N}\omega_c(a_j^\dagger a_j^{}+b_j^\dagger b_j^{} )+J_+\sum_{j=1}^N (a_j^\dagger b_j^{} +  b_j^{\dagger} a_j^{} )+J_-\sum_{j=1}^{N-1}(b_{j}^\dagger a_{j+1}^{}+a_{j+1}^\dagger b_j^{}).\label{S84}
\end{align}
Note that for $N\gg1$, the results obtained under open boundary condition (OBC) agree perfectly with those using periodic boundary condition (PBC). Therefore, we have adopt OBC in Eq.~(\ref{S84}) for convenience in numerical simulations.

\subsection{Theoretical Derivation}
To employ the RWA in $H_{\rm int}^{\rm wo}$, we first transform into the interaction picture. The free Hamiltonian is defined as
\begin{align}
	H_0=\omega_e(\sigma_+^{\rm B}\sigma_-^{\rm B}+\sigma_+^{\rm C}\sigma_-^{\rm C})+\sum_{j=1}^{N}\omega_c(a_j^\dagger a_j^{}+b_{j}^{\dagger}b_j^{}),
\end{align}
and the evolution operator is denoted as $U_0(t)=\exp(-iH_0t)$. In the interaction picture, the corresponding interaction Hamiltonian $H_{\rm I}^{\rm wo}(t)$ is obtained through the following transformation:
\begin{align}\label{HInt}
	H_{\rm I}^{\rm wo}(t)&=U_0^\dagger(t)(H_{\rm tot}^{\rm wo}-H_0)U_0(t)=J_+\sum_{j=1}^N (a_j^\dagger b_j^{} +  b_j^{\dagger} a_j^{} )+J_-\sum_{j=1}^{N-1}(b_{j}^\dagger a_{j+1}^{}+a_{j+1}^\dagger b_j^{})\nonumber\\
	&+\text{g}\left[\sigma_-^{\rm B}a_{x_1}^{}e^{-i(\omega_e+\omega_c)t}+\sigma_-^{\rm B}a_{x_1}^{\dagger}e^{-i(\omega_e-\omega_c)t}+\sigma_-^{\rm C}b_{x_2}^{}e^{-i(\omega_e+\omega_c)t}+\sigma_-^{\rm C}b_{x_2}^{\dagger}e^{-i(\omega_e-\omega_c)t}+\text{H.c.} \right].
\end{align}
In weak-coupling condition, i.e., $\text{g}\ll \{\omega_c, \omega_e\}$, the fast-oscillating terms can be neglected as their time-averaged contributions are negligible. Consequently, we retain only the slower-varying terms with lower frequency $\omega_e-\omega_c$, while discarding those rapidly oscillating terms with higher frequency $\omega_e+\omega_c$. Specifically, the retained terms are $\sigma_-^{\rm B}a_{x_1}^{\dagger}e^{-i(\omega_e-\omega_c)t}$, $\sigma_+^{\rm B}a_{x_1}^{}e^{i(\omega_e-\omega_c)t}$, $\sigma_-^{\rm C}b_{x_2}^{\dagger}e^{-i(\omega_e-\omega_c)t}$, and $\sigma_+^{\rm C}b_{x_2}^{}e^{i(\omega_e-\omega_c)t}$. These terms vary slowly and are associated the near-resonant condition, i.e., $\omega_e\approx\omega_c$. After discarding the rapidly oscillating terms, i.e., $\sigma_-^{\rm B}a_{x_1}^{}$, $\sigma_+^{\rm B}a_{x_1}^{\dagger}$, $\sigma_-^{\rm C}b_{x_2}^{}$, and  $\sigma_+^{\rm C}b_{x_2}^\dagger$, the interaction Hamiltonian in Eq.~(\ref{HInt}) can be simplified as (denoted as $H_{\rm I}^{\rm w}(t)$)
\begin{align}
	H_{\rm I}^{\rm w}(t)=J_+\sum_{j=1}^N (a_j^\dagger b_j^{} +  b_j^{\dagger} a_j^{} )+J_-\sum_{j=1}^{N-1}(b_{j}^\dagger a_{j+1}^{}+a_{j+1}^\dagger b_j^{})+\text{g}\left[\sigma_-^{\rm B}a_{x_1}^{\dagger}e^{-i(\omega_e-\omega_c)t}+\sigma_-^{\rm C}b_{x_2}^{\dagger}e^{-i(\omega_e-\omega_c)t}+\text{H.c.} \right].
\end{align}
To eliminate the time-dependent phase factors $\exp[\pm i(\omega_e-\omega_c)t]$, an appropriate rotating reference frame can be chosen. The corresponding evolution operator for this frame is given by
\begin{align}
	U(t)=\exp[-i\Delta(\sigma_+^{\rm B}\sigma_-^{\rm B}+\sigma_+^{\rm C}\sigma_-^{\rm C})t],
\end{align}
where $\Delta=\omega_e-\omega_c$ is the detuning between the atomic transition frequency and the eigenfrequency of the cavities. Thus, in this rotating reference frame, the interaction Hamiltonian $H_{\rm I}^{\rm w}(t)$ is transformed into a time-independent Hamiltonian with the RWA, denoted as $H_{\rm tot}^{\rm w}$, which is expressed as
\begin{align}
	H_{\rm tot}^{\rm w}&=U(t)H_{\rm I}^{\rm w}(t)U^\dagger(t)+i[\partial_tU(t)]U^\dagger(t)\nonumber\\
	&=\Delta(\sigma_+^{\rm B}\sigma_-^{\rm B}+\sigma_+^{\rm C}\sigma_-^{\rm C})+J_+\sum_{j=1}^N (a_j^\dagger b_j^{} +  b_j^{\dagger} a_j^{} )+J_-\sum_{j=1}^{N-1}(b_{j}^\dagger a_{j+1}^{}+a_{j+1}^\dagger b_j^{})+\text{g}\left(\sigma_-^{\rm B}a_{x_1}^\dagger +\sigma_-^{\rm C}b_{x_2}^\dagger+\text{H.c.}  \right).\label{S89}
\end{align}
Consequently, when conditions $\text{g}\ll \{\omega_c, \omega_e\}$ and $\abs{\Delta}\ll \omega_e+\omega_c$ are satisfied, it is reasonable and valid to employ the rotating wave approximation in the interaction Hamiltonian $H_{\rm int}$ between the system and the topological waveguide.

\subsection{Numerical Simulations}
\begin{figure}
	\centering
	\includegraphics[width=17cm]{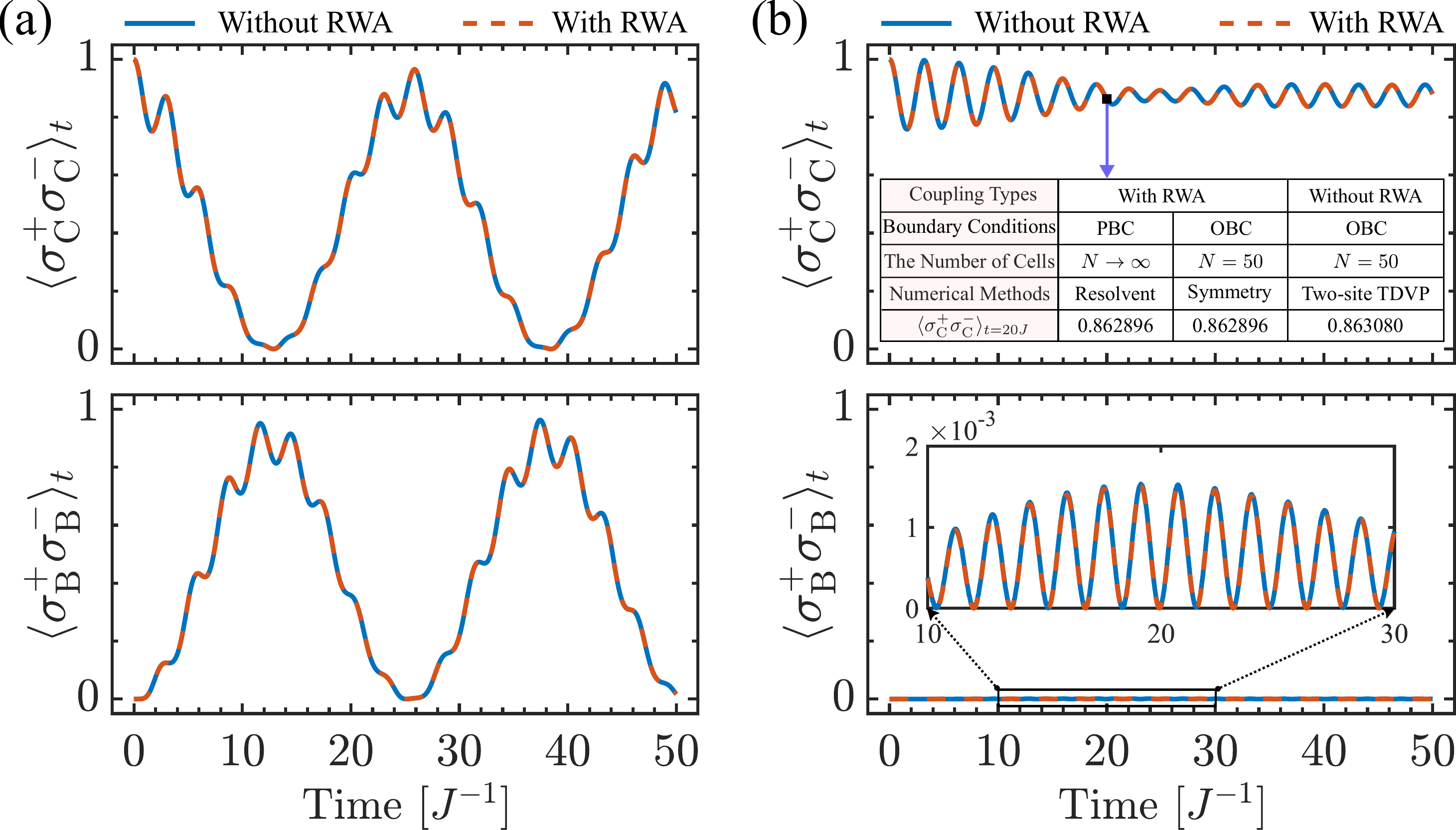}\\
	\caption{The evolution of the atomic excitation probability $\langle\sigma_{\rm C}^{+}\sigma_{\rm C}^{-}\rangle_{t}$ (Top panel) and $\langle\sigma_{\rm B}^{+}\sigma_{\rm B}^{-}\rangle_{t}$ (Bottom panel) without the RWA (blue solid line) and with the RWA (orange dashed line) in the topologically nontrivial [Left panel (a): $\delta=-0.9$] and trivial [Right panel (b): $\delta=0.9$] phases. The inset in the top panel (b) presents the numerical comparison for different coupling types at $t=20J$. The system parameters are set: $\omega_c=\omega_e=2\pi\times10\ {\rm GHz}, J=2\pi\times100\ {\rm MHz}$, ${\rm g}=0.5J$, $x_1=26$ and $x_2=25$.
	}\label{fig9}
\end{figure}

Under conditions where the RWA holds, we verify the accuracy of the theoretically derived Hamiltonian $H_{\rm tot}^{\rm w}$, by performing numerical simulations. Specifically, we employ a matrix-product state representation of the quantum state combined with the two-site time-dependent variational principle (TDVP) \cite{PhysRevLett.107.070601, PhysRevB.94.165116} to calculate the atomic dynamics governed by the full Hamiltonian $H_{\rm tot}^{\rm wo}$. Given that the initial state is restricted to the single-excitation subspace and $H_{\rm I}^{\rm w}$ conserves the total excitation number, implying that U(1) symmetry is satisfied, we project the entire system onto the single-excitation subspace based on this symmetry. This allows for efficient numerical simulations of atomic dynamics governed by the Hamiltonian $H_{\rm tot}^{\rm w}$. In contrast, the full Hamiltonian $H_{\rm tot}^{\rm wo}$, which includes counter-rotating terms, does not conserve the total excitation number. As a result, it requires the use of tensor networks to effectively simulate the dynamics of this Hamiltonian.

For realistic experimental parameters \cite{PhysRevX.11.011015}, we assume that the atomic and cavity frequencies are $\omega_c=\omega_e=2\pi\times 10\ \text{GHz}$, the average hopping amplitude in the topological bath is $J=2\pi\times 100\ \text{MHz}$, and the coupling strength between the atoms and the cavities is $\text{g}=0.5J$. Under these parameters, the rotating wave approximation is justified as $\text{g}=2\pi\times 50\ \text{MHz}\ll 2\pi\times 10\ \text{GHz}$. In Fig.\,\ref{fig9}, we present the evolution of the atomic excitation probability obtained from numerical simulations using both the Hamiltonian with the RWA (\ref{S89}) and without the RWA (\ref{S82}). In both the topologically nontrivial phase [see Fig.\,\ref{fig9}(a)] and the topologically trivial phase [see Fig.\,\ref{fig9}(b)], the dynamics exhibit the same trend for both Hamiltonians, confirming the validity of the Hamiltonian with the RWA. Additionally, the results at time $t=20J$, computed using the resolvent method under PBC and the thermodynamic limit ($N\to \infty$), show excellent agreement with those obtained via the $U(1)$ symmetric and the two-site TDVP methods under OBC and a finite number of unit cells ($N=50\gg 1$), as shown in the top inset of Fig.\,\ref{fig9}(b). These observations confirm the suitability of employing the JC coupling type in the interaction Hamiltonian between the system and the topological waveguide under parameters that satisfy the RWA.


\begin{thebibliography}{122}%
	\makeatletter
	\providecommand \@ifxundefined [1]{%
		\@ifx{#1\undefined}
	}%
	\providecommand \@ifnum [1]{%
		\ifnum #1\expandafter \@firstoftwo
		\else \expandafter \@secondoftwo
		\fi
	}%
	\providecommand \@ifx [1]{%
		\ifx #1\expandafter \@firstoftwo
		\else \expandafter \@secondoftwo
		\fi
	}%
	\providecommand \natexlab [1]{#1}%
	\providecommand \enquote  [1]{``#1''}%
	\providecommand \bibnamefont  [1]{#1}%
	\providecommand \bibfnamefont [1]{#1}%
	\providecommand \citenamefont [1]{#1}%
	\providecommand \href@noop [0]{\@secondoftwo}%
	\providecommand \href [0]{\begingroup \@sanitize@url \@href}%
	\providecommand \@href[1]{\@@startlink{#1}\@@href}%
	\providecommand \@@href[1]{\endgroup#1\@@endlink}%
	\providecommand \@sanitize@url [0]{\catcode `\\12\catcode `\$12\catcode
		`\&12\catcode `\#12\catcode `\^12\catcode `\_12\catcode `\%12\relax}%
	\providecommand \@@startlink[1]{}%
	\providecommand \@@endlink[0]{}%
	\providecommand \url  [0]{\begingroup\@sanitize@url \@url }%
	\providecommand \@url [1]{\endgroup\@href {#1}{\urlprefix }}%
	\providecommand \urlprefix  [0]{URL }%
	\providecommand \Eprint [0]{\href }%
	\providecommand \doibase [0]{https://doi.org/}%
	\providecommand \selectlanguage [0]{\@gobble}%
	\providecommand \bibinfo  [0]{\@secondoftwo}%
	\providecommand \bibfield  [0]{\@secondoftwo}%
	\providecommand \translation [1]{[#1]}%
	\providecommand \BibitemOpen [0]{}%
	\providecommand \bibitemStop [0]{}%
	\providecommand \bibitemNoStop [0]{.\EOS\space}%
	\providecommand \EOS [0]{\spacefactor3000\relax}%
	\providecommand \BibitemShut  [1]{\csname bibitem#1\endcsname}%
	\let\auto@bib@innerbib\@empty
	\bibitem [{\citenamefont {Alicki}\ and\ \citenamefont
		{Fannes}(2013)}]{PhysRevE.87.042123}%
	\BibitemOpen
	\bibfield  {author} {\bibinfo {author} {\bibfnamefont {R.}~\bibnamefont
			{Alicki}}\ and\ \bibinfo {author} {\bibfnamefont {M.}~\bibnamefont
			{Fannes}},\ }\href {https://doi.org/10.1103/PhysRevE.87.042123} {\bibfield
		{journal} {\bibinfo  {journal} {Phys. Rev. E}\ }\textbf {\bibinfo {volume}
			{87}},\ \bibinfo {pages} {042123} (\bibinfo {year} {2013})}\BibitemShut
	{NoStop}%
	\bibitem [{\citenamefont {Campaioli}\ \emph {et~al.}(2017)\citenamefont
		{Campaioli}, \citenamefont {Pollock}, \citenamefont {Binder}, \citenamefont
		{C\'eleri}, \citenamefont {Goold}, \citenamefont {Vinjanampathy},\ and\
		\citenamefont {Modi}}]{PhysRevLett.118.150601}%
	\BibitemOpen
	\bibfield  {author} {\bibinfo {author} {\bibfnamefont {F.}~\bibnamefont
			{Campaioli}}, \bibinfo {author} {\bibfnamefont {F.~A.}\ \bibnamefont
			{Pollock}}, \bibinfo {author} {\bibfnamefont {F.~C.}\ \bibnamefont {Binder}},
		\bibinfo {author} {\bibfnamefont {L.}~\bibnamefont {C\'eleri}}, \bibinfo
		{author} {\bibfnamefont {J.}~\bibnamefont {Goold}}, \bibinfo {author}
		{\bibfnamefont {S.}~\bibnamefont {Vinjanampathy}},\ and\ \bibinfo {author}
		{\bibfnamefont {K.}~\bibnamefont {Modi}},\ }\href
	{https://doi.org/10.1103/PhysRevLett.118.150601} {\bibfield  {journal}
		{\bibinfo  {journal} {Phys. Rev. Lett.}\ }\textbf {\bibinfo {volume} {118}},\
		\bibinfo {pages} {150601} (\bibinfo {year} {2017})}\BibitemShut {NoStop}%
	\bibitem [{\citenamefont {Seah}\ \emph {et~al.}(2021)\citenamefont {Seah},
		\citenamefont {Perarnau-Llobet}, \citenamefont {Haack}, \citenamefont
		{Brunner},\ and\ \citenamefont {Nimmrichter}}]{PhysRevLett.127.100601}%
	\BibitemOpen
	\bibfield  {author} {\bibinfo {author} {\bibfnamefont {S.}~\bibnamefont
			{Seah}}, \bibinfo {author} {\bibfnamefont {M.}~\bibnamefont
			{Perarnau-Llobet}}, \bibinfo {author} {\bibfnamefont {G.}~\bibnamefont
			{Haack}}, \bibinfo {author} {\bibfnamefont {N.}~\bibnamefont {Brunner}},\
		and\ \bibinfo {author} {\bibfnamefont {S.}~\bibnamefont {Nimmrichter}},\
	}\href {https://doi.org/10.1103/PhysRevLett.127.100601} {\bibfield  {journal}
		{\bibinfo  {journal} {Phys. Rev. Lett.}\ }\textbf {\bibinfo {volume} {127}},\
		\bibinfo {pages} {100601} (\bibinfo {year} {2021})}\BibitemShut {NoStop}%
	\bibitem [{\citenamefont {Zhu}\ \emph {et~al.}(2023)\citenamefont {Zhu},
		\citenamefont {Chen}, \citenamefont {Hasegawa},\ and\ \citenamefont
		{Xue}}]{PhysRevLett.131.240401}%
	\BibitemOpen
	\bibfield  {author} {\bibinfo {author} {\bibfnamefont {G.}~\bibnamefont
			{Zhu}}, \bibinfo {author} {\bibfnamefont {Y.}~\bibnamefont {Chen}}, \bibinfo
		{author} {\bibfnamefont {Y.}~\bibnamefont {Hasegawa}},\ and\ \bibinfo
		{author} {\bibfnamefont {P.}~\bibnamefont {Xue}},\ }\href
	{https://doi.org/10.1103/PhysRevLett.131.240401} {\bibfield  {journal}
		{\bibinfo  {journal} {Phys. Rev. Lett.}\ }\textbf {\bibinfo {volume} {131}},\
		\bibinfo {pages} {240401} (\bibinfo {year} {2023})}\BibitemShut {NoStop}%
	\bibitem [{\citenamefont {Centrone}\ \emph {et~al.}(2023)\citenamefont
		{Centrone}, \citenamefont {Mancino},\ and\ \citenamefont
		{Paternostro}}]{PhysRevA.108.052213}%
	\BibitemOpen
	\bibfield  {author} {\bibinfo {author} {\bibfnamefont {F.}~\bibnamefont
			{Centrone}}, \bibinfo {author} {\bibfnamefont {L.}~\bibnamefont {Mancino}},\
		and\ \bibinfo {author} {\bibfnamefont {M.}~\bibnamefont {Paternostro}},\
	}\href {https://doi.org/10.1103/PhysRevA.108.052213} {\bibfield  {journal}
		{\bibinfo  {journal} {Phys. Rev. A}\ }\textbf {\bibinfo {volume} {108}},\
		\bibinfo {pages} {052213} (\bibinfo {year} {2023})}\BibitemShut {NoStop}%
	\bibitem [{\citenamefont {Rossini}\ \emph {et~al.}(2020)\citenamefont
		{Rossini}, \citenamefont {Andolina}, \citenamefont {Rosa}, \citenamefont
		{Carrega},\ and\ \citenamefont {Polini}}]{PhysRevLett.125.236402}%
	\BibitemOpen
	\bibfield  {author} {\bibinfo {author} {\bibfnamefont {D.}~\bibnamefont
			{Rossini}}, \bibinfo {author} {\bibfnamefont {G.~M.}\ \bibnamefont
			{Andolina}}, \bibinfo {author} {\bibfnamefont {D.}~\bibnamefont {Rosa}},
		\bibinfo {author} {\bibfnamefont {M.}~\bibnamefont {Carrega}},\ and\ \bibinfo
		{author} {\bibfnamefont {M.}~\bibnamefont {Polini}},\ }\href
	{https://doi.org/10.1103/PhysRevLett.125.236402} {\bibfield  {journal}
		{\bibinfo  {journal} {Phys. Rev. Lett.}\ }\textbf {\bibinfo {volume} {125}},\
		\bibinfo {pages} {236402} (\bibinfo {year} {2020})}\BibitemShut {NoStop}%
	\bibitem [{\citenamefont {Gyhm}\ and\ \citenamefont
		{Fischer}(2023)}]{gyhm2023beneficial}%
	\BibitemOpen
	\bibfield  {author} {\bibinfo {author} {\bibfnamefont {J.-Y.}\ \bibnamefont
			{Gyhm}}\ and\ \bibinfo {author} {\bibfnamefont {U.~R.}\ \bibnamefont
			{Fischer}},\ }\href
	{https://pubs.aip.org/avs/aqs/article-abstract/6/1/012001/3061561/Beneficial-and-detrimental-entanglement-for}
	{\bibfield  {journal} {\bibinfo  {journal} {AVS Quantum Science}\ }\textbf
		{\bibinfo {volume} {6}} (\bibinfo {year} {2023})}\BibitemShut {NoStop}%
	\bibitem [{\citenamefont {Gyhm}\ \emph {et~al.}(2022)\citenamefont {Gyhm},
		\citenamefont {\ifmmode~\check{S}\else \v{S}\fi{}afr\'anek},\ and\
		\citenamefont {Rosa}}]{PhysRevLett.128.140501}%
	\BibitemOpen
	\bibfield  {author} {\bibinfo {author} {\bibfnamefont {J.-Y.}\ \bibnamefont
			{Gyhm}}, \bibinfo {author} {\bibfnamefont {D.}~\bibnamefont
			{\ifmmode~\check{S}\else \v{S}\fi{}afr\'anek}},\ and\ \bibinfo {author}
		{\bibfnamefont {D.}~\bibnamefont {Rosa}},\ }\href
	{https://doi.org/10.1103/PhysRevLett.128.140501} {\bibfield  {journal}
		{\bibinfo  {journal} {Phys. Rev. Lett.}\ }\textbf {\bibinfo {volume} {128}},\
		\bibinfo {pages} {140501} (\bibinfo {year} {2022})}\BibitemShut {NoStop}%
	\bibitem [{\citenamefont {Rosa}\ \emph {et~al.}(2020)\citenamefont {Rosa},
		\citenamefont {Rossini}, \citenamefont {Andolina}, \citenamefont {Polini},\
		and\ \citenamefont {Carrega}}]{rosa2020ultra}%
	\BibitemOpen
	\bibfield  {author} {\bibinfo {author} {\bibfnamefont {D.}~\bibnamefont
			{Rosa}}, \bibinfo {author} {\bibfnamefont {D.}~\bibnamefont {Rossini}},
		\bibinfo {author} {\bibfnamefont {G.~M.}\ \bibnamefont {Andolina}}, \bibinfo
		{author} {\bibfnamefont {M.}~\bibnamefont {Polini}},\ and\ \bibinfo {author}
		{\bibfnamefont {M.}~\bibnamefont {Carrega}},\ }\href
	{https://link.springer.com/article/10.1007/JHEP11(2020)067} {\bibfield
		{journal} {\bibinfo  {journal} {Journal of High Energy Physics}\ }\textbf
		{\bibinfo {volume} {2020}},\ \bibinfo {pages} {1} (\bibinfo {year}
		{2020})}\BibitemShut {NoStop}%
	\bibitem [{\citenamefont {Rodr\'{\i}guez}\ \emph {et~al.}(2023)\citenamefont
		{Rodr\'{\i}guez}, \citenamefont {Rosa},\ and\ \citenamefont
		{Olle}}]{PhysRevA.108.042618}%
	\BibitemOpen
	\bibfield  {author} {\bibinfo {author} {\bibfnamefont {C.}~\bibnamefont
			{Rodr\'{\i}guez}}, \bibinfo {author} {\bibfnamefont {D.}~\bibnamefont
			{Rosa}},\ and\ \bibinfo {author} {\bibfnamefont {J.}~\bibnamefont {Olle}},\
	}\href {https://doi.org/10.1103/PhysRevA.108.042618} {\bibfield  {journal}
		{\bibinfo  {journal} {Phys. Rev. A}\ }\textbf {\bibinfo {volume} {108}},\
		\bibinfo {pages} {042618} (\bibinfo {year} {2023})}\BibitemShut {NoStop}%
	\bibitem [{\citenamefont {Mazzoncini}\ \emph {et~al.}(2023)\citenamefont
		{Mazzoncini}, \citenamefont {Cavina}, \citenamefont {Andolina}, \citenamefont
		{Erdman},\ and\ \citenamefont {Giovannetti}}]{PhysRevA.107.032218}%
	\BibitemOpen
	\bibfield  {author} {\bibinfo {author} {\bibfnamefont {F.}~\bibnamefont
			{Mazzoncini}}, \bibinfo {author} {\bibfnamefont {V.}~\bibnamefont {Cavina}},
		\bibinfo {author} {\bibfnamefont {G.~M.}\ \bibnamefont {Andolina}}, \bibinfo
		{author} {\bibfnamefont {P.~A.}\ \bibnamefont {Erdman}},\ and\ \bibinfo
		{author} {\bibfnamefont {V.}~\bibnamefont {Giovannetti}},\ }\href
	{https://doi.org/10.1103/PhysRevA.107.032218} {\bibfield  {journal} {\bibinfo
			{journal} {Phys. Rev. A}\ }\textbf {\bibinfo {volume} {107}},\ \bibinfo
		{pages} {032218} (\bibinfo {year} {2023})}\BibitemShut {NoStop}%
	\bibitem [{\citenamefont {Konar}\ \emph {et~al.}(2024)\citenamefont {Konar},
		\citenamefont {Lakkaraju},\ and\ \citenamefont
		{Sen~(De)}}]{PhysRevA.109.042207}%
	\BibitemOpen
	\bibfield  {author} {\bibinfo {author} {\bibfnamefont {T.~K.}\ \bibnamefont
			{Konar}}, \bibinfo {author} {\bibfnamefont {L.~G.~C.}\ \bibnamefont
			{Lakkaraju}},\ and\ \bibinfo {author} {\bibfnamefont {A.}~\bibnamefont
			{Sen~(De)}},\ }\href {https://doi.org/10.1103/PhysRevA.109.042207} {\bibfield
		{journal} {\bibinfo  {journal} {Phys. Rev. A}\ }\textbf {\bibinfo {volume}
			{109}},\ \bibinfo {pages} {042207} (\bibinfo {year} {2024})}\BibitemShut
	{NoStop}%
	\bibitem [{\citenamefont {Zhang}\ \emph {et~al.}(2019)\citenamefont {Zhang},
		\citenamefont {Yang}, \citenamefont {Fu},\ and\ \citenamefont
		{Wang}}]{PhysRevE.99.052106}%
	\BibitemOpen
	\bibfield  {author} {\bibinfo {author} {\bibfnamefont {Y.-Y.}\ \bibnamefont
			{Zhang}}, \bibinfo {author} {\bibfnamefont {T.-R.}\ \bibnamefont {Yang}},
		\bibinfo {author} {\bibfnamefont {L.}~\bibnamefont {Fu}},\ and\ \bibinfo
		{author} {\bibfnamefont {X.}~\bibnamefont {Wang}},\ }\href
	{https://doi.org/10.1103/PhysRevE.99.052106} {\bibfield  {journal} {\bibinfo
			{journal} {Phys. Rev. E}\ }\textbf {\bibinfo {volume} {99}},\ \bibinfo
		{pages} {052106} (\bibinfo {year} {2019})}\BibitemShut {NoStop}%
	\bibitem [{\citenamefont {Yang}\ \emph {et~al.}(2023)\citenamefont {Yang},
		\citenamefont {Yang}, \citenamefont {Alimuddin}, \citenamefont {Salvia},
		\citenamefont {Fei}, \citenamefont {Zhao}, \citenamefont {Nimmrichter},\ and\
		\citenamefont {Luo}}]{PhysRevLett.131.030402}%
	\BibitemOpen
	\bibfield  {author} {\bibinfo {author} {\bibfnamefont {X.}~\bibnamefont
			{Yang}}, \bibinfo {author} {\bibfnamefont {Y.-H.}\ \bibnamefont {Yang}},
		\bibinfo {author} {\bibfnamefont {M.}~\bibnamefont {Alimuddin}}, \bibinfo
		{author} {\bibfnamefont {R.}~\bibnamefont {Salvia}}, \bibinfo {author}
		{\bibfnamefont {S.-M.}\ \bibnamefont {Fei}}, \bibinfo {author} {\bibfnamefont
			{L.-M.}\ \bibnamefont {Zhao}}, \bibinfo {author} {\bibfnamefont
			{S.}~\bibnamefont {Nimmrichter}},\ and\ \bibinfo {author} {\bibfnamefont
			{M.-X.}\ \bibnamefont {Luo}},\ }\href
	{https://doi.org/10.1103/PhysRevLett.131.030402} {\bibfield  {journal}
		{\bibinfo  {journal} {Phys. Rev. Lett.}\ }\textbf {\bibinfo {volume} {131}},\
		\bibinfo {pages} {030402} (\bibinfo {year} {2023})}\BibitemShut {NoStop}%
	\bibitem [{\citenamefont {Juli\`a-Farr\'e}\ \emph {et~al.}(2020)\citenamefont
		{Juli\`a-Farr\'e}, \citenamefont {Salamon}, \citenamefont {Riera},
		\citenamefont {Bera},\ and\ \citenamefont
		{Lewenstein}}]{PhysRevResearch.2.023113}%
	\BibitemOpen
	\bibfield  {author} {\bibinfo {author} {\bibfnamefont {S.}~\bibnamefont
			{Juli\`a-Farr\'e}}, \bibinfo {author} {\bibfnamefont {T.}~\bibnamefont
			{Salamon}}, \bibinfo {author} {\bibfnamefont {A.}~\bibnamefont {Riera}},
		\bibinfo {author} {\bibfnamefont {M.~N.}\ \bibnamefont {Bera}},\ and\
		\bibinfo {author} {\bibfnamefont {M.}~\bibnamefont {Lewenstein}},\ }\href
	{https://doi.org/10.1103/PhysRevResearch.2.023113} {\bibfield  {journal}
		{\bibinfo  {journal} {Phys. Rev. Res.}\ }\textbf {\bibinfo {volume} {2}},\
		\bibinfo {pages} {023113} (\bibinfo {year} {2020})}\BibitemShut {NoStop}%
	\bibitem [{\citenamefont {Gao}\ \emph {et~al.}(2022)\citenamefont {Gao},
		\citenamefont {Cheng}, \citenamefont {He}, \citenamefont {Mondaini},
		\citenamefont {Guan},\ and\ \citenamefont {Lin}}]{PhysRevResearch.4.043150}%
	\BibitemOpen
	\bibfield  {author} {\bibinfo {author} {\bibfnamefont {L.}~\bibnamefont
			{Gao}}, \bibinfo {author} {\bibfnamefont {C.}~\bibnamefont {Cheng}}, \bibinfo
		{author} {\bibfnamefont {W.-B.}\ \bibnamefont {He}}, \bibinfo {author}
		{\bibfnamefont {R.}~\bibnamefont {Mondaini}}, \bibinfo {author}
		{\bibfnamefont {X.-W.}\ \bibnamefont {Guan}},\ and\ \bibinfo {author}
		{\bibfnamefont {H.-Q.}\ \bibnamefont {Lin}},\ }\href
	{https://doi.org/10.1103/PhysRevResearch.4.043150} {\bibfield  {journal}
		{\bibinfo  {journal} {Phys. Rev. Res.}\ }\textbf {\bibinfo {volume} {4}},\
		\bibinfo {pages} {043150} (\bibinfo {year} {2022})}\BibitemShut {NoStop}%
	\bibitem [{\citenamefont {Zhang}\ \emph {et~al.}(2024)\citenamefont {Zhang},
		\citenamefont {Yang},\ and\ \citenamefont {Fei}}]{PhysRevA.109.042424}%
	\BibitemOpen
	\bibfield  {author} {\bibinfo {author} {\bibfnamefont {T.}~\bibnamefont
			{Zhang}}, \bibinfo {author} {\bibfnamefont {H.}~\bibnamefont {Yang}},\ and\
		\bibinfo {author} {\bibfnamefont {S.-M.}\ \bibnamefont {Fei}},\ }\href
	{https://doi.org/10.1103/PhysRevA.109.042424} {\bibfield  {journal} {\bibinfo
			{journal} {Phys. Rev. A}\ }\textbf {\bibinfo {volume} {109}},\ \bibinfo
		{pages} {042424} (\bibinfo {year} {2024})}\BibitemShut {NoStop}%
	\bibitem [{\citenamefont {Tirone}\ \emph {et~al.}(2024)\citenamefont {Tirone},
		\citenamefont {Salvia}, \citenamefont {Chessa},\ and\ \citenamefont
		{Giovannetti}}]{tirone2024quantum}%
	\BibitemOpen
	\bibfield  {author} {\bibinfo {author} {\bibfnamefont {S.}~\bibnamefont
			{Tirone}}, \bibinfo {author} {\bibfnamefont {R.}~\bibnamefont {Salvia}},
		\bibinfo {author} {\bibfnamefont {S.}~\bibnamefont {Chessa}},\ and\ \bibinfo
		{author} {\bibfnamefont {V.}~\bibnamefont {Giovannetti}},\ }\href
	{https://doi.org/10.21468/SciPostPhys.17.2.041} {\bibfield  {journal}
		{\bibinfo  {journal} {SciPost Phys.}\ }\textbf {\bibinfo {volume} {17}},\
		\bibinfo {pages} {041} (\bibinfo {year} {2024})}\BibitemShut {NoStop}%
	\bibitem [{\citenamefont {Shi}\ \emph {et~al.}(2022)\citenamefont {Shi},
		\citenamefont {Ding}, \citenamefont {Wan}, \citenamefont {Wang},\ and\
		\citenamefont {Yang}}]{PhysRevLett.129.130602}%
	\BibitemOpen
	\bibfield  {author} {\bibinfo {author} {\bibfnamefont {H.-L.}\ \bibnamefont
			{Shi}}, \bibinfo {author} {\bibfnamefont {S.}~\bibnamefont {Ding}}, \bibinfo
		{author} {\bibfnamefont {Q.-K.}\ \bibnamefont {Wan}}, \bibinfo {author}
		{\bibfnamefont {X.-H.}\ \bibnamefont {Wang}},\ and\ \bibinfo {author}
		{\bibfnamefont {W.-L.}\ \bibnamefont {Yang}},\ }\href
	{https://doi.org/10.1103/PhysRevLett.129.130602} {\bibfield  {journal}
		{\bibinfo  {journal} {Phys. Rev. Lett.}\ }\textbf {\bibinfo {volume} {129}},\
		\bibinfo {pages} {130602} (\bibinfo {year} {2022})}\BibitemShut {NoStop}%
	\bibitem [{\citenamefont {Andolina}\ \emph {et~al.}(2019)\citenamefont
		{Andolina}, \citenamefont {Keck}, \citenamefont {Mari}, \citenamefont
		{Campisi}, \citenamefont {Giovannetti},\ and\ \citenamefont
		{Polini}}]{PhysRevLett.122.047702}%
	\BibitemOpen
	\bibfield  {author} {\bibinfo {author} {\bibfnamefont {G.~M.}\ \bibnamefont
			{Andolina}}, \bibinfo {author} {\bibfnamefont {M.}~\bibnamefont {Keck}},
		\bibinfo {author} {\bibfnamefont {A.}~\bibnamefont {Mari}}, \bibinfo {author}
		{\bibfnamefont {M.}~\bibnamefont {Campisi}}, \bibinfo {author} {\bibfnamefont
			{V.}~\bibnamefont {Giovannetti}},\ and\ \bibinfo {author} {\bibfnamefont
			{M.}~\bibnamefont {Polini}},\ }\href
	{https://doi.org/10.1103/PhysRevLett.122.047702} {\bibfield  {journal}
		{\bibinfo  {journal} {Phys. Rev. Lett.}\ }\textbf {\bibinfo {volume} {122}},\
		\bibinfo {pages} {047702} (\bibinfo {year} {2019})}\BibitemShut {NoStop}%
	\bibitem [{\citenamefont {Francica}\ and\ \citenamefont
		{Dell'Anna}(2024)}]{PhysRevE.109.044119}%
	\BibitemOpen
	\bibfield  {author} {\bibinfo {author} {\bibfnamefont {G.}~\bibnamefont
			{Francica}}\ and\ \bibinfo {author} {\bibfnamefont {L.}~\bibnamefont
			{Dell'Anna}},\ }\href {https://doi.org/10.1103/PhysRevE.109.044119}
	{\bibfield  {journal} {\bibinfo  {journal} {Phys. Rev. E}\ }\textbf {\bibinfo
			{volume} {109}},\ \bibinfo {pages} {044119} (\bibinfo {year}
		{2024})}\BibitemShut {NoStop}%
	\bibitem [{\citenamefont {Monsel}\ \emph {et~al.}(2020)\citenamefont {Monsel},
		\citenamefont {Fellous-Asiani}, \citenamefont {Huard},\ and\ \citenamefont
		{Auff\`eves}}]{PhysRevLett.124.130601}%
	\BibitemOpen
	\bibfield  {author} {\bibinfo {author} {\bibfnamefont {J.}~\bibnamefont
			{Monsel}}, \bibinfo {author} {\bibfnamefont {M.}~\bibnamefont
			{Fellous-Asiani}}, \bibinfo {author} {\bibfnamefont {B.}~\bibnamefont
			{Huard}},\ and\ \bibinfo {author} {\bibfnamefont {A.}~\bibnamefont
			{Auff\`eves}},\ }\href {https://doi.org/10.1103/PhysRevLett.124.130601}
	{\bibfield  {journal} {\bibinfo  {journal} {Phys. Rev. Lett.}\ }\textbf
		{\bibinfo {volume} {124}},\ \bibinfo {pages} {130601} (\bibinfo {year}
		{2020})}\BibitemShut {NoStop}%
	\bibitem [{\citenamefont {Tirone}\ \emph {et~al.}(2023)\citenamefont {Tirone},
		\citenamefont {Salvia}, \citenamefont {Chessa},\ and\ \citenamefont
		{Giovannetti}}]{PhysRevLett.131.060402}%
	\BibitemOpen
	\bibfield  {author} {\bibinfo {author} {\bibfnamefont {S.}~\bibnamefont
			{Tirone}}, \bibinfo {author} {\bibfnamefont {R.}~\bibnamefont {Salvia}},
		\bibinfo {author} {\bibfnamefont {S.}~\bibnamefont {Chessa}},\ and\ \bibinfo
		{author} {\bibfnamefont {V.}~\bibnamefont {Giovannetti}},\ }\href
	{https://doi.org/10.1103/PhysRevLett.131.060402} {\bibfield  {journal}
		{\bibinfo  {journal} {Phys. Rev. Lett.}\ }\textbf {\bibinfo {volume} {131}},\
		\bibinfo {pages} {060402} (\bibinfo {year} {2023})}\BibitemShut {NoStop}%
	\bibitem [{\citenamefont {Tirone}\ \emph {et~al.}(2025)\citenamefont {Tirone},
		\citenamefont {Salvia}, \citenamefont {Chessa},\ and\ \citenamefont
		{Giovannetti}}]{tirone2023quantum}%
	\BibitemOpen
	\bibfield  {author} {\bibinfo {author} {\bibfnamefont {S.}~\bibnamefont
			{Tirone}}, \bibinfo {author} {\bibfnamefont {R.}~\bibnamefont {Salvia}},
		\bibinfo {author} {\bibfnamefont {S.}~\bibnamefont {Chessa}},\ and\ \bibinfo
		{author} {\bibfnamefont {V.}~\bibnamefont {Giovannetti}},\ }\href
	{https://doi.org/10.1103/PhysRevA.111.012204} {\bibfield  {journal} {\bibinfo
			{journal} {Phys. Rev. A}\ }\textbf {\bibinfo {volume} {111}},\ \bibinfo
		{pages} {012204} (\bibinfo {year} {2025})}\BibitemShut {NoStop}%
	\bibitem [{\citenamefont {Ferraro}\ \emph {et~al.}(2018)\citenamefont
		{Ferraro}, \citenamefont {Campisi}, \citenamefont {Andolina}, \citenamefont
		{Pellegrini},\ and\ \citenamefont {Polini}}]{PhysRevLett.120.117702}%
	\BibitemOpen
	\bibfield  {author} {\bibinfo {author} {\bibfnamefont {D.}~\bibnamefont
			{Ferraro}}, \bibinfo {author} {\bibfnamefont {M.}~\bibnamefont {Campisi}},
		\bibinfo {author} {\bibfnamefont {G.~M.}\ \bibnamefont {Andolina}}, \bibinfo
		{author} {\bibfnamefont {V.}~\bibnamefont {Pellegrini}},\ and\ \bibinfo
		{author} {\bibfnamefont {M.}~\bibnamefont {Polini}},\ }\href
	{https://doi.org/10.1103/PhysRevLett.120.117702} {\bibfield  {journal}
		{\bibinfo  {journal} {Phys. Rev. Lett.}\ }\textbf {\bibinfo {volume} {120}},\
		\bibinfo {pages} {117702} (\bibinfo {year} {2018})}\BibitemShut {NoStop}%
	\bibitem [{\citenamefont {Rossini}\ \emph {et~al.}(2019)\citenamefont
		{Rossini}, \citenamefont {Andolina},\ and\ \citenamefont
		{Polini}}]{PhysRevB.100.115142}%
	\BibitemOpen
	\bibfield  {author} {\bibinfo {author} {\bibfnamefont {D.}~\bibnamefont
			{Rossini}}, \bibinfo {author} {\bibfnamefont {G.~M.}\ \bibnamefont
			{Andolina}},\ and\ \bibinfo {author} {\bibfnamefont {M.}~\bibnamefont
			{Polini}},\ }\href {https://doi.org/10.1103/PhysRevB.100.115142} {\bibfield
		{journal} {\bibinfo  {journal} {Phys. Rev. B}\ }\textbf {\bibinfo {volume}
			{100}},\ \bibinfo {pages} {115142} (\bibinfo {year} {2019})}\BibitemShut
	{NoStop}%
	\bibitem [{\citenamefont {Peng}\ \emph {et~al.}(2021)\citenamefont {Peng},
		\citenamefont {He}, \citenamefont {Chesi}, \citenamefont {Lin},\ and\
		\citenamefont {Guan}}]{PhysRevA.103.052220}%
	\BibitemOpen
	\bibfield  {author} {\bibinfo {author} {\bibfnamefont {L.}~\bibnamefont
			{Peng}}, \bibinfo {author} {\bibfnamefont {W.-B.}\ \bibnamefont {He}},
		\bibinfo {author} {\bibfnamefont {S.}~\bibnamefont {Chesi}}, \bibinfo
		{author} {\bibfnamefont {H.-Q.}\ \bibnamefont {Lin}},\ and\ \bibinfo {author}
		{\bibfnamefont {X.-W.}\ \bibnamefont {Guan}},\ }\href
	{https://doi.org/10.1103/PhysRevA.103.052220} {\bibfield  {journal} {\bibinfo
			{journal} {Phys. Rev. A}\ }\textbf {\bibinfo {volume} {103}},\ \bibinfo
		{pages} {052220} (\bibinfo {year} {2021})}\BibitemShut {NoStop}%
	\bibitem [{\citenamefont {Fusco}\ \emph {et~al.}(2016)\citenamefont {Fusco},
		\citenamefont {Paternostro},\ and\ \citenamefont
		{De~Chiara}}]{PhysRevE.94.052122}%
	\BibitemOpen
	\bibfield  {author} {\bibinfo {author} {\bibfnamefont {L.}~\bibnamefont
			{Fusco}}, \bibinfo {author} {\bibfnamefont {M.}~\bibnamefont {Paternostro}},\
		and\ \bibinfo {author} {\bibfnamefont {G.}~\bibnamefont {De~Chiara}},\ }\href
	{https://doi.org/10.1103/PhysRevE.94.052122} {\bibfield  {journal} {\bibinfo
			{journal} {Phys. Rev. E}\ }\textbf {\bibinfo {volume} {94}},\ \bibinfo
		{pages} {052122} (\bibinfo {year} {2016})}\BibitemShut {NoStop}%
	\bibitem [{\citenamefont {Seidov}\ and\ \citenamefont
		{Mukhin}(2024)}]{PhysRevA.109.022210}%
	\BibitemOpen
	\bibfield  {author} {\bibinfo {author} {\bibfnamefont {S.~S.}\ \bibnamefont
			{Seidov}}\ and\ \bibinfo {author} {\bibfnamefont {S.~I.}\ \bibnamefont
			{Mukhin}},\ }\href {https://doi.org/10.1103/PhysRevA.109.022210} {\bibfield
		{journal} {\bibinfo  {journal} {Phys. Rev. A}\ }\textbf {\bibinfo {volume}
			{109}},\ \bibinfo {pages} {022210} (\bibinfo {year} {2024})}\BibitemShut
	{NoStop}%
	\bibitem [{\citenamefont {Crescente}\ \emph {et~al.}(2020)\citenamefont
		{Crescente}, \citenamefont {Carrega}, \citenamefont {Sassetti},\ and\
		\citenamefont {Ferraro}}]{PhysRevB.102.245407}%
	\BibitemOpen
	\bibfield  {author} {\bibinfo {author} {\bibfnamefont {A.}~\bibnamefont
			{Crescente}}, \bibinfo {author} {\bibfnamefont {M.}~\bibnamefont {Carrega}},
		\bibinfo {author} {\bibfnamefont {M.}~\bibnamefont {Sassetti}},\ and\
		\bibinfo {author} {\bibfnamefont {D.}~\bibnamefont {Ferraro}},\ }\href
	{https://doi.org/10.1103/PhysRevB.102.245407} {\bibfield  {journal} {\bibinfo
			{journal} {Phys. Rev. B}\ }\textbf {\bibinfo {volume} {102}},\ \bibinfo
		{pages} {245407} (\bibinfo {year} {2020})}\BibitemShut {NoStop}%
	\bibitem [{\citenamefont {Ahmadi}\ \emph {et~al.}(2024)\citenamefont {Ahmadi},
		\citenamefont {Mazurek}, \citenamefont {Horodecki},\ and\ \citenamefont
		{Barzanjeh}}]{PhysRevLett.132.210402}%
	\BibitemOpen
	\bibfield  {author} {\bibinfo {author} {\bibfnamefont {B.}~\bibnamefont
			{Ahmadi}}, \bibinfo {author} {\bibfnamefont {P.}~\bibnamefont {Mazurek}},
		\bibinfo {author} {\bibfnamefont {P.}~\bibnamefont {Horodecki}},\ and\
		\bibinfo {author} {\bibfnamefont {S.}~\bibnamefont {Barzanjeh}},\ }\href
	{https://doi.org/10.1103/PhysRevLett.132.210402} {\bibfield  {journal}
		{\bibinfo  {journal} {Phys. Rev. Lett.}\ }\textbf {\bibinfo {volume} {132}},\
		\bibinfo {pages} {210402} (\bibinfo {year} {2024})}\BibitemShut {NoStop}%
	\bibitem [{\citenamefont {Kim}\ \emph {et~al.}(2022)\citenamefont {Kim},
		\citenamefont {Murugan}, \citenamefont {Olle},\ and\ \citenamefont
		{Rosa}}]{PhysRevA.105.L010201}%
	\BibitemOpen
	\bibfield  {author} {\bibinfo {author} {\bibfnamefont {J.}~\bibnamefont
			{Kim}}, \bibinfo {author} {\bibfnamefont {J.}~\bibnamefont {Murugan}},
		\bibinfo {author} {\bibfnamefont {J.}~\bibnamefont {Olle}},\ and\ \bibinfo
		{author} {\bibfnamefont {D.}~\bibnamefont {Rosa}},\ }\href
	{https://doi.org/10.1103/PhysRevA.105.L010201} {\bibfield  {journal}
		{\bibinfo  {journal} {Phys. Rev. A}\ }\textbf {\bibinfo {volume} {105}},\
		\bibinfo {pages} {L010201} (\bibinfo {year} {2022})}\BibitemShut {NoStop}%
	\bibitem [{\citenamefont {Shaghaghi}\ \emph {et~al.}(2022)\citenamefont
		{Shaghaghi}, \citenamefont {Singh}, \citenamefont {Benenti},\ and\
		\citenamefont {Rosa}}]{Shaghaghi_2022}%
	\BibitemOpen
	\bibfield  {author} {\bibinfo {author} {\bibfnamefont {V.}~\bibnamefont
			{Shaghaghi}}, \bibinfo {author} {\bibfnamefont {V.}~\bibnamefont {Singh}},
		\bibinfo {author} {\bibfnamefont {G.}~\bibnamefont {Benenti}},\ and\ \bibinfo
		{author} {\bibfnamefont {D.}~\bibnamefont {Rosa}},\ }\href
	{https://doi.org/10.1088/2058-9565/ac8829} {\bibfield  {journal} {\bibinfo
			{journal} {Quantum Science and Technology}\ }\textbf {\bibinfo {volume}
			{7}},\ \bibinfo {pages} {04LT01} (\bibinfo {year} {2022})}\BibitemShut
	{NoStop}%
	\bibitem [{\citenamefont {Gemme}\ \emph {et~al.}(2023)\citenamefont {Gemme},
		\citenamefont {Andolina}, \citenamefont {Pellegrino}, \citenamefont
		{Sassetti},\ and\ \citenamefont {Ferraro}}]{batteries9040197}%
	\BibitemOpen
	\bibfield  {author} {\bibinfo {author} {\bibfnamefont {G.}~\bibnamefont
			{Gemme}}, \bibinfo {author} {\bibfnamefont {G.~M.}\ \bibnamefont {Andolina}},
		\bibinfo {author} {\bibfnamefont {F.~M.~D.}\ \bibnamefont {Pellegrino}},
		\bibinfo {author} {\bibfnamefont {M.}~\bibnamefont {Sassetti}},\ and\
		\bibinfo {author} {\bibfnamefont {D.}~\bibnamefont {Ferraro}},\ }\href
	{https://www.mdpi.com/2313-0105/9/4/197} {\bibfield  {journal} {\bibinfo
			{journal} {Batteries}\ }\textbf {\bibinfo {volume} {9}},\ \bibinfo {pages}
		{040197} (\bibinfo {year} {2023})}\BibitemShut {NoStop}%
	\bibitem [{\citenamefont {Andolina}\ \emph {et~al.}(2024)\citenamefont
		{Andolina}, \citenamefont {Erdman}, \citenamefont {No\'e}, \citenamefont
		{Pekola},\ and\ \citenamefont {Schir\`o}}]{andolina2024dicke}%
	\BibitemOpen
	\bibfield  {author} {\bibinfo {author} {\bibfnamefont {G.~M.}\ \bibnamefont
			{Andolina}}, \bibinfo {author} {\bibfnamefont {P.~A.}\ \bibnamefont
			{Erdman}}, \bibinfo {author} {\bibfnamefont {F.}~\bibnamefont {No\'e}},
		\bibinfo {author} {\bibfnamefont {J.}~\bibnamefont {Pekola}},\ and\ \bibinfo
		{author} {\bibfnamefont {M.}~\bibnamefont {Schir\`o}},\ }\href
	{https://doi.org/10.1103/PhysRevResearch.6.043128} {\bibfield  {journal}
		{\bibinfo  {journal} {Phys. Rev. Res.}\ }\textbf {\bibinfo {volume} {6}},\
		\bibinfo {pages} {043128} (\bibinfo {year} {2024})}\BibitemShut {NoStop}%
	\bibitem [{\citenamefont {Konar}\ \emph {et~al.}(2022)\citenamefont {Konar},
		\citenamefont {Lakkaraju}, \citenamefont {Ghosh},\ and\ \citenamefont
		{Sen(De)}}]{PhysRevA.106.022618}%
	\BibitemOpen
	\bibfield  {author} {\bibinfo {author} {\bibfnamefont {T.~K.}\ \bibnamefont
			{Konar}}, \bibinfo {author} {\bibfnamefont {L.~G.~C.}\ \bibnamefont
			{Lakkaraju}}, \bibinfo {author} {\bibfnamefont {S.}~\bibnamefont {Ghosh}},\
		and\ \bibinfo {author} {\bibfnamefont {A.}~\bibnamefont {Sen(De)}},\ }\href
	{https://doi.org/10.1103/PhysRevA.106.022618} {\bibfield  {journal} {\bibinfo
			{journal} {Phys. Rev. A}\ }\textbf {\bibinfo {volume} {106}},\ \bibinfo
		{pages} {022618} (\bibinfo {year} {2022})}\BibitemShut {NoStop}%
	\bibitem [{\citenamefont {Liu}\ \emph {et~al.}(2021)\citenamefont {Liu},
		\citenamefont {Shi}, \citenamefont {Shi}, \citenamefont {Wang},\ and\
		\citenamefont {Yang}}]{PhysRevB.104.245418}%
	\BibitemOpen
	\bibfield  {author} {\bibinfo {author} {\bibfnamefont {J.-X.}\ \bibnamefont
			{Liu}}, \bibinfo {author} {\bibfnamefont {H.-L.}\ \bibnamefont {Shi}},
		\bibinfo {author} {\bibfnamefont {Y.-H.}\ \bibnamefont {Shi}}, \bibinfo
		{author} {\bibfnamefont {X.-H.}\ \bibnamefont {Wang}},\ and\ \bibinfo
		{author} {\bibfnamefont {W.-L.}\ \bibnamefont {Yang}},\ }\href
	{https://doi.org/10.1103/PhysRevB.104.245418} {\bibfield  {journal} {\bibinfo
			{journal} {Phys. Rev. B}\ }\textbf {\bibinfo {volume} {104}},\ \bibinfo
		{pages} {245418} (\bibinfo {year} {2021})}\BibitemShut {NoStop}%
	\bibitem [{\citenamefont {Catalano}\ \emph {et~al.}(2024)\citenamefont
		{Catalano}, \citenamefont {Giampaolo}, \citenamefont {Morsch}, \citenamefont
		{Giovannetti},\ and\ \citenamefont {Franchini}}]{PRXQuantum.5.030319}%
	\BibitemOpen
	\bibfield  {author} {\bibinfo {author} {\bibfnamefont {A.}~\bibnamefont
			{Catalano}}, \bibinfo {author} {\bibfnamefont {S.}~\bibnamefont {Giampaolo}},
		\bibinfo {author} {\bibfnamefont {O.}~\bibnamefont {Morsch}}, \bibinfo
		{author} {\bibfnamefont {V.}~\bibnamefont {Giovannetti}},\ and\ \bibinfo
		{author} {\bibfnamefont {F.}~\bibnamefont {Franchini}},\ }\href
	{https://doi.org/10.1103/PRXQuantum.5.030319} {\bibfield  {journal} {\bibinfo
			{journal} {PRX Quantum}\ }\textbf {\bibinfo {volume} {5}},\ \bibinfo {pages}
		{030319} (\bibinfo {year} {2024})}\BibitemShut {NoStop}%
	\bibitem [{\citenamefont {Binder}\ \emph {et~al.}(2015)\citenamefont {Binder},
		\citenamefont {Vinjanampathy}, \citenamefont {Modi},\ and\ \citenamefont
		{Goold}}]{binder2015quantacell}%
	\BibitemOpen
	\bibfield  {author} {\bibinfo {author} {\bibfnamefont {F.~C.}\ \bibnamefont
			{Binder}}, \bibinfo {author} {\bibfnamefont {S.}~\bibnamefont
			{Vinjanampathy}}, \bibinfo {author} {\bibfnamefont {K.}~\bibnamefont
			{Modi}},\ and\ \bibinfo {author} {\bibfnamefont {J.}~\bibnamefont {Goold}},\
	}\href
	{https://iopscience.iop.org/article/10.1088/1367-2630/17/7/075015/meta}
	{\bibfield  {journal} {\bibinfo  {journal} {New Journal of Physics}\ }\textbf
		{\bibinfo {volume} {17}},\ \bibinfo {pages} {075015} (\bibinfo {year}
		{2015})}\BibitemShut {NoStop}%
	\bibitem [{\citenamefont {Garc\'{\i}a-Pintos}\ \emph
		{et~al.}(2020)\citenamefont {Garc\'{\i}a-Pintos}, \citenamefont {Hamma},\
		and\ \citenamefont {del Campo}}]{PhysRevLett.125.040601}%
	\BibitemOpen
	\bibfield  {author} {\bibinfo {author} {\bibfnamefont {L.~P.}\ \bibnamefont
			{Garc\'{\i}a-Pintos}}, \bibinfo {author} {\bibfnamefont {A.}~\bibnamefont
			{Hamma}},\ and\ \bibinfo {author} {\bibfnamefont {A.}~\bibnamefont {del
				Campo}},\ }\href {https://doi.org/10.1103/PhysRevLett.125.040601} {\bibfield
		{journal} {\bibinfo  {journal} {Phys. Rev. Lett.}\ }\textbf {\bibinfo
			{volume} {125}},\ \bibinfo {pages} {040601} (\bibinfo {year}
		{2020})}\BibitemShut {NoStop}%
	\bibitem [{\citenamefont {Kamin}\ \emph
		{et~al.}(2020{\natexlab{a}})\citenamefont {Kamin}, \citenamefont {Tabesh},
		\citenamefont {Salimi},\ and\ \citenamefont {Santos}}]{PhysRevE.102.052109}%
	\BibitemOpen
	\bibfield  {author} {\bibinfo {author} {\bibfnamefont {F.~H.}\ \bibnamefont
			{Kamin}}, \bibinfo {author} {\bibfnamefont {F.~T.}\ \bibnamefont {Tabesh}},
		\bibinfo {author} {\bibfnamefont {S.}~\bibnamefont {Salimi}},\ and\ \bibinfo
		{author} {\bibfnamefont {A.~C.}\ \bibnamefont {Santos}},\ }\href
	{https://doi.org/10.1103/PhysRevE.102.052109} {\bibfield  {journal} {\bibinfo
			{journal} {Phys. Rev. E}\ }\textbf {\bibinfo {volume} {102}},\ \bibinfo
		{pages} {052109} (\bibinfo {year} {2020}{\natexlab{a}})}\BibitemShut
	{NoStop}%
	\bibitem [{\citenamefont {Wang}\ \emph {et~al.}(2023)\citenamefont {Wang},
		\citenamefont {Liu}, \citenamefont {Wu}, \citenamefont {Fan},\ and\
		\citenamefont {Liu}}]{PhysRevA.108.062402}%
	\BibitemOpen
	\bibfield  {author} {\bibinfo {author} {\bibfnamefont {L.}~\bibnamefont
			{Wang}}, \bibinfo {author} {\bibfnamefont {S.-Q.}\ \bibnamefont {Liu}},
		\bibinfo {author} {\bibfnamefont {F.-l.}\ \bibnamefont {Wu}}, \bibinfo
		{author} {\bibfnamefont {H.}~\bibnamefont {Fan}},\ and\ \bibinfo {author}
		{\bibfnamefont {S.-Y.}\ \bibnamefont {Liu}},\ }\href
	{https://doi.org/10.1103/PhysRevA.108.062402} {\bibfield  {journal} {\bibinfo
			{journal} {Phys. Rev. A}\ }\textbf {\bibinfo {volume} {108}},\ \bibinfo
		{pages} {062402} (\bibinfo {year} {2023})}\BibitemShut {NoStop}%
	\bibitem [{\citenamefont {Kamin}\ \emph {et~al.}(2024)\citenamefont {Kamin},
		\citenamefont {Salimi},\ and\ \citenamefont
		{Arjmandi}}]{PhysRevA.109.022226}%
	\BibitemOpen
	\bibfield  {author} {\bibinfo {author} {\bibfnamefont {F.~H.}\ \bibnamefont
			{Kamin}}, \bibinfo {author} {\bibfnamefont {S.}~\bibnamefont {Salimi}},\ and\
		\bibinfo {author} {\bibfnamefont {M.~B.}\ \bibnamefont {Arjmandi}},\ }\href
	{https://doi.org/10.1103/PhysRevA.109.022226} {\bibfield  {journal} {\bibinfo
			{journal} {Phys. Rev. A}\ }\textbf {\bibinfo {volume} {109}},\ \bibinfo
		{pages} {022226} (\bibinfo {year} {2024})}\BibitemShut {NoStop}%
	\bibitem [{\citenamefont {Santos}(2021)}]{PhysRevE.103.042118}%
	\BibitemOpen
	\bibfield  {author} {\bibinfo {author} {\bibfnamefont {A.~C.}\ \bibnamefont
			{Santos}},\ }\href {https://doi.org/10.1103/PhysRevE.103.042118} {\bibfield
		{journal} {\bibinfo  {journal} {Phys. Rev. E}\ }\textbf {\bibinfo {volume}
			{103}},\ \bibinfo {pages} {042118} (\bibinfo {year} {2021})}\BibitemShut
	{NoStop}%
	\bibitem [{\citenamefont {Bruzewicz}\ \emph {et~al.}(2019)\citenamefont
		{Bruzewicz}, \citenamefont {Chiaverini}, \citenamefont {McConnell},\ and\
		\citenamefont {Sage}}]{bruzewicz2019trapped}%
	\BibitemOpen
	\bibfield  {author} {\bibinfo {author} {\bibfnamefont {C.~D.}\ \bibnamefont
			{Bruzewicz}}, \bibinfo {author} {\bibfnamefont {J.}~\bibnamefont
			{Chiaverini}}, \bibinfo {author} {\bibfnamefont {R.}~\bibnamefont
			{McConnell}},\ and\ \bibinfo {author} {\bibfnamefont {J.~M.}\ \bibnamefont
			{Sage}},\ }\href
	{https://pubs.aip.org/aip/apr/article/6/2/021314/570103/Trapped-ion-quantum-computing-Progress-and}
	{\bibfield  {journal} {\bibinfo  {journal} {Applied Physics Reviews}\
		}\textbf {\bibinfo {volume} {6}} (\bibinfo {year} {2019})}\BibitemShut
	{NoStop}%
	\bibitem [{\citenamefont {Forn-D{\'\i}az}\ \emph {et~al.}(2017)\citenamefont
		{Forn-D{\'\i}az}, \citenamefont {Garc{\'\i}a-Ripoll}, \citenamefont
		{Peropadre}, \citenamefont {Orgiazzi}, \citenamefont {Yurtalan},
		\citenamefont {Belyansky}, \citenamefont {Wilson},\ and\ \citenamefont
		{Lupascu}}]{forn2017ultrastrong}%
	\BibitemOpen
	\bibfield  {author} {\bibinfo {author} {\bibfnamefont {P.}~\bibnamefont
			{Forn-D{\'\i}az}}, \bibinfo {author} {\bibfnamefont {J.~J.}\ \bibnamefont
			{Garc{\'\i}a-Ripoll}}, \bibinfo {author} {\bibfnamefont {B.}~\bibnamefont
			{Peropadre}}, \bibinfo {author} {\bibfnamefont {J.-L.}\ \bibnamefont
			{Orgiazzi}}, \bibinfo {author} {\bibfnamefont {M.}~\bibnamefont {Yurtalan}},
		\bibinfo {author} {\bibfnamefont {R.}~\bibnamefont {Belyansky}}, \bibinfo
		{author} {\bibfnamefont {C.~M.}\ \bibnamefont {Wilson}},\ and\ \bibinfo
		{author} {\bibfnamefont {A.}~\bibnamefont {Lupascu}},\ }\href
	{https://www.nature.com/articles/nphys3905} {\bibfield  {journal} {\bibinfo
			{journal} {Nature Physics}\ }\textbf {\bibinfo {volume} {13}},\ \bibinfo
		{pages} {39} (\bibinfo {year} {2017})}\BibitemShut {NoStop}%
	\bibitem [{\citenamefont {Baumann}\ \emph {et~al.}(2010)\citenamefont
		{Baumann}, \citenamefont {Guerlin}, \citenamefont {Brennecke},\ and\
		\citenamefont {Esslinger}}]{baumann2010dicke}%
	\BibitemOpen
	\bibfield  {author} {\bibinfo {author} {\bibfnamefont {K.}~\bibnamefont
			{Baumann}}, \bibinfo {author} {\bibfnamefont {C.}~\bibnamefont {Guerlin}},
		\bibinfo {author} {\bibfnamefont {F.}~\bibnamefont {Brennecke}},\ and\
		\bibinfo {author} {\bibfnamefont {T.}~\bibnamefont {Esslinger}},\ }\href
	{https://www.nature.com/articles/nature09009} {\bibfield  {journal} {\bibinfo
			{journal} {Nature}\ }\textbf {\bibinfo {volume} {464}},\ \bibinfo {pages}
		{1301} (\bibinfo {year} {2010})}\BibitemShut {NoStop}%
	\bibitem [{\citenamefont {Devoret}\ and\ \citenamefont
		{Schoelkopf}(2013)}]{devoret2013superconducting}%
	\BibitemOpen
	\bibfield  {author} {\bibinfo {author} {\bibfnamefont {M.~H.}\ \bibnamefont
			{Devoret}}\ and\ \bibinfo {author} {\bibfnamefont {R.~J.}\ \bibnamefont
			{Schoelkopf}},\ }\href {https://www.science.org/doi/10.1126/science.1231930}
	{\bibfield  {journal} {\bibinfo  {journal} {Science}\ }\textbf {\bibinfo
			{volume} {339}},\ \bibinfo {pages} {1169} (\bibinfo {year}
		{2013})}\BibitemShut {NoStop}%
	\bibitem [{\citenamefont {Campaioli}\ \emph {et~al.}(2024)\citenamefont
		{Campaioli}, \citenamefont {Gherardini}, \citenamefont {Quach}, \citenamefont
		{Polini},\ and\ \citenamefont {Andolina}}]{RevModPhys.96.031001}%
	\BibitemOpen
	\bibfield  {author} {\bibinfo {author} {\bibfnamefont {F.}~\bibnamefont
			{Campaioli}}, \bibinfo {author} {\bibfnamefont {S.}~\bibnamefont
			{Gherardini}}, \bibinfo {author} {\bibfnamefont {J.~Q.}\ \bibnamefont
			{Quach}}, \bibinfo {author} {\bibfnamefont {M.}~\bibnamefont {Polini}},\ and\
		\bibinfo {author} {\bibfnamefont {G.~M.}\ \bibnamefont {Andolina}},\ }\href
	{https://doi.org/10.1103/RevModPhys.96.031001} {\bibfield  {journal}
		{\bibinfo  {journal} {Rev. Mod. Phys.}\ }\textbf {\bibinfo {volume} {96}},\
		\bibinfo {pages} {031001} (\bibinfo {year} {2024})}\BibitemShut {NoStop}%
	\bibitem [{\citenamefont {Binder}\ \emph {et~al.}(2018)\citenamefont {Binder},
		\citenamefont {Correa}, \citenamefont {Gogolin}, \citenamefont {Anders},\
		and\ \citenamefont {Adesso}}]{binderthermodynamics2018}%
	\BibitemOpen
	\bibfield  {author} {\bibinfo {author} {\bibfnamefont {F.}~\bibnamefont
			{Binder}}, \bibinfo {author} {\bibfnamefont {L.~A.}\ \bibnamefont {Correa}},
		\bibinfo {author} {\bibfnamefont {C.}~\bibnamefont {Gogolin}}, \bibinfo
		{author} {\bibfnamefont {J.}~\bibnamefont {Anders}},\ and\ \bibinfo {author}
		{\bibfnamefont {G.}~\bibnamefont {Adesso}},\ }\href@noop {} {\emph {\bibinfo
			{title} {Thermodynamics in the Quantum Regime: Fundamental Aspects and New
				Directions}}}\ (\bibinfo  {publisher} {Springer International Publishing},\
	\bibinfo {year} {2018})\BibitemShut {NoStop}%
	\bibitem [{\citenamefont {Southwell}(2008)}]{southwell2008quantum}%
	\BibitemOpen
	\bibfield  {author} {\bibinfo {author} {\bibfnamefont {K.}~\bibnamefont
			{Southwell}},\ }\href {https://www.nature.com/articles/4531003a} {\bibfield
		{journal} {\bibinfo  {journal} {Nature}\ }\textbf {\bibinfo {volume} {453}},\
		\bibinfo {pages} {1003} (\bibinfo {year} {2008})}\BibitemShut {NoStop}%
	\bibitem [{\citenamefont {Abah}\ \emph {et~al.}(2022)\citenamefont {Abah},
		\citenamefont {De~Chiara}, \citenamefont {Paternostro},\ and\ \citenamefont
		{Puebla}}]{PhysRevResearch.4.L022017}%
	\BibitemOpen
	\bibfield  {author} {\bibinfo {author} {\bibfnamefont {O.}~\bibnamefont
			{Abah}}, \bibinfo {author} {\bibfnamefont {G.}~\bibnamefont {De~Chiara}},
		\bibinfo {author} {\bibfnamefont {M.}~\bibnamefont {Paternostro}},\ and\
		\bibinfo {author} {\bibfnamefont {R.}~\bibnamefont {Puebla}},\ }\href
	{https://doi.org/10.1103/PhysRevResearch.4.L022017} {\bibfield  {journal}
		{\bibinfo  {journal} {Phys. Rev. Res.}\ }\textbf {\bibinfo {volume} {4}},\
		\bibinfo {pages} {L022017} (\bibinfo {year} {2022})}\BibitemShut {NoStop}%
	\bibitem [{\citenamefont {Saha}\ \emph {et~al.}(2023)\citenamefont {Saha},
		\citenamefont {Bhattacharyya}, \citenamefont {Sen},\ and\ \citenamefont
		{Sen}}]{saha2023harnessing}%
	\BibitemOpen
	\bibfield  {author} {\bibinfo {author} {\bibfnamefont {D.}~\bibnamefont
			{Saha}}, \bibinfo {author} {\bibfnamefont {A.}~\bibnamefont {Bhattacharyya}},
		\bibinfo {author} {\bibfnamefont {K.}~\bibnamefont {Sen}},\ and\ \bibinfo
		{author} {\bibfnamefont {U.}~\bibnamefont {Sen}},\ }\href
	{https://arxiv.org/abs/2309.15634} {\bibfield  {journal} {\bibinfo  {journal}
			{arXiv:2309.15634}\ } (\bibinfo {year} {2023})}\BibitemShut {NoStop}%
	\bibitem [{\citenamefont {Yang}\ \emph {et~al.}(2024)\citenamefont {Yang},
		\citenamefont {Shi}, \citenamefont {Wan}, \citenamefont {Zhang},
		\citenamefont {Wang},\ and\ \citenamefont {Yang}}]{PhysRevA.109.012204}%
	\BibitemOpen
	\bibfield  {author} {\bibinfo {author} {\bibfnamefont {H.-Y.}\ \bibnamefont
			{Yang}}, \bibinfo {author} {\bibfnamefont {H.-L.}\ \bibnamefont {Shi}},
		\bibinfo {author} {\bibfnamefont {Q.-K.}\ \bibnamefont {Wan}}, \bibinfo
		{author} {\bibfnamefont {K.}~\bibnamefont {Zhang}}, \bibinfo {author}
		{\bibfnamefont {X.-H.}\ \bibnamefont {Wang}},\ and\ \bibinfo {author}
		{\bibfnamefont {W.-L.}\ \bibnamefont {Yang}},\ }\href
	{https://doi.org/10.1103/PhysRevA.109.012204} {\bibfield  {journal} {\bibinfo
			{journal} {Phys. Rev. A}\ }\textbf {\bibinfo {volume} {109}},\ \bibinfo
		{pages} {012204} (\bibinfo {year} {2024})}\BibitemShut {NoStop}%
	\bibitem [{\citenamefont {Allahverdyan}\ \emph {et~al.}(2004)\citenamefont
		{Allahverdyan}, \citenamefont {Balian},\ and\ \citenamefont
		{Nieuwenhuizen}}]{allahverdyan2004maximal}%
	\BibitemOpen
	\bibfield  {author} {\bibinfo {author} {\bibfnamefont {A.~E.}\ \bibnamefont
			{Allahverdyan}}, \bibinfo {author} {\bibfnamefont {R.}~\bibnamefont
			{Balian}},\ and\ \bibinfo {author} {\bibfnamefont {T.~M.}\ \bibnamefont
			{Nieuwenhuizen}},\ }\href
	{https://iopscience.iop.org/article/10.1209/epl/i2004-10101-2/meta}
	{\bibfield  {journal} {\bibinfo  {journal} {Europhysics Letters}\ }\textbf
		{\bibinfo {volume} {67}},\ \bibinfo {pages} {565} (\bibinfo {year}
		{2004})}\BibitemShut {NoStop}%
	\bibitem [{\citenamefont {Song}\ \emph {et~al.}(2024)\citenamefont {Song},
		\citenamefont {Liu}, \citenamefont {Zhou}, \citenamefont {Yang},\ and\
		\citenamefont {An}}]{PhysRevLett.132.090401}%
	\BibitemOpen
	\bibfield  {author} {\bibinfo {author} {\bibfnamefont {W.-L.}\ \bibnamefont
			{Song}}, \bibinfo {author} {\bibfnamefont {H.-B.}\ \bibnamefont {Liu}},
		\bibinfo {author} {\bibfnamefont {B.}~\bibnamefont {Zhou}}, \bibinfo {author}
		{\bibfnamefont {W.-L.}\ \bibnamefont {Yang}},\ and\ \bibinfo {author}
		{\bibfnamefont {J.-H.}\ \bibnamefont {An}},\ }\href
	{https://doi.org/10.1103/PhysRevLett.132.090401} {\bibfield  {journal}
		{\bibinfo  {journal} {Phys. Rev. Lett.}\ }\textbf {\bibinfo {volume} {132}},\
		\bibinfo {pages} {090401} (\bibinfo {year} {2024})}\BibitemShut {NoStop}%
	\bibitem [{\citenamefont {Bai}\ and\ \citenamefont
		{An}(2021)}]{PhysRevLett.127.083602}%
	\BibitemOpen
	\bibfield  {author} {\bibinfo {author} {\bibfnamefont {S.-Y.}\ \bibnamefont
			{Bai}}\ and\ \bibinfo {author} {\bibfnamefont {J.-H.}\ \bibnamefont {An}},\
	}\href {https://doi.org/10.1103/PhysRevLett.127.083602} {\bibfield  {journal}
		{\bibinfo  {journal} {Phys. Rev. Lett.}\ }\textbf {\bibinfo {volume} {127}},\
		\bibinfo {pages} {083602} (\bibinfo {year} {2021})}\BibitemShut {NoStop}%
	\bibitem [{\citenamefont {Song}\ \emph {et~al.}(2017)\citenamefont {Song},
		\citenamefont {Yang}, \citenamefont {An},\ and\ \citenamefont
		{Feng}}]{song2017dissipation}%
	\BibitemOpen
	\bibfield  {author} {\bibinfo {author} {\bibfnamefont {W.}~\bibnamefont
			{Song}}, \bibinfo {author} {\bibfnamefont {W.}~\bibnamefont {Yang}}, \bibinfo
		{author} {\bibfnamefont {J.}~\bibnamefont {An}},\ and\ \bibinfo {author}
		{\bibfnamefont {M.}~\bibnamefont {Feng}},\ }\href
	{https://opg.optica.org/oe/fulltext.cfm?uri=oe-25-16-19226&id=370340}
	{\bibfield  {journal} {\bibinfo  {journal} {Optics Express}\ }\textbf
		{\bibinfo {volume} {25}},\ \bibinfo {pages} {19226} (\bibinfo {year}
		{2017})}\BibitemShut {NoStop}%
	\bibitem [{\citenamefont {Haldane}(1988)}]{PhysRevLett.61.2015}%
	\BibitemOpen
	\bibfield  {author} {\bibinfo {author} {\bibfnamefont {F.~D.~M.}\
			\bibnamefont {Haldane}},\ }\href
	{https://doi.org/10.1103/PhysRevLett.61.2015} {\bibfield  {journal} {\bibinfo
			{journal} {Phys. Rev. Lett.}\ }\textbf {\bibinfo {volume} {61}},\ \bibinfo
		{pages} {2015} (\bibinfo {year} {1988})}\BibitemShut {NoStop}%
	\bibitem [{\citenamefont {Armitage}\ \emph {et~al.}(2018)\citenamefont
		{Armitage}, \citenamefont {Mele},\ and\ \citenamefont
		{Vishwanath}}]{RevModPhys.90.015001}%
	\BibitemOpen
	\bibfield  {author} {\bibinfo {author} {\bibfnamefont {N.~P.}\ \bibnamefont
			{Armitage}}, \bibinfo {author} {\bibfnamefont {E.~J.}\ \bibnamefont {Mele}},\
		and\ \bibinfo {author} {\bibfnamefont {A.}~\bibnamefont {Vishwanath}},\
	}\href {https://doi.org/10.1103/RevModPhys.90.015001} {\bibfield  {journal}
		{\bibinfo  {journal} {Rev. Mod. Phys.}\ }\textbf {\bibinfo {volume} {90}},\
		\bibinfo {pages} {015001} (\bibinfo {year} {2018})}\BibitemShut {NoStop}%
	\bibitem [{\citenamefont {Nie}\ \emph {et~al.}(2021)\citenamefont {Nie},
		\citenamefont {Shi}, \citenamefont {Nori},\ and\ \citenamefont
		{Liu}}]{PhysRevApplied.15.044041}%
	\BibitemOpen
	\bibfield  {author} {\bibinfo {author} {\bibfnamefont {W.}~\bibnamefont
			{Nie}}, \bibinfo {author} {\bibfnamefont {T.}~\bibnamefont {Shi}}, \bibinfo
		{author} {\bibfnamefont {F.}~\bibnamefont {Nori}},\ and\ \bibinfo {author}
		{\bibfnamefont {Y.-x.}\ \bibnamefont {Liu}},\ }\href
	{https://doi.org/10.1103/PhysRevApplied.15.044041} {\bibfield  {journal}
		{\bibinfo  {journal} {Phys. Rev. Appl.}\ }\textbf {\bibinfo {volume} {15}},\
		\bibinfo {pages} {044041} (\bibinfo {year} {2021})}\BibitemShut {NoStop}%
	\bibitem [{\citenamefont {Bello}\ \emph {et~al.}(2019)\citenamefont {Bello},
		\citenamefont {Platero}, \citenamefont {Cirac},\ and\ \citenamefont
		{Gonz{\'a}lez-Tudela}}]{bello2019unconventional}%
	\BibitemOpen
	\bibfield  {author} {\bibinfo {author} {\bibfnamefont {M.}~\bibnamefont
			{Bello}}, \bibinfo {author} {\bibfnamefont {G.}~\bibnamefont {Platero}},
		\bibinfo {author} {\bibfnamefont {J.~I.}\ \bibnamefont {Cirac}},\ and\
		\bibinfo {author} {\bibfnamefont {A.}~\bibnamefont {Gonz{\'a}lez-Tudela}},\
	}\href {https://www.science.org/doi/10.1126/sciadv.aaw0297} {\bibfield
		{journal} {\bibinfo  {journal} {Science advances}\ }\textbf {\bibinfo
			{volume} {5}},\ \bibinfo {pages} {eaaw0297} (\bibinfo {year}
		{2019})}\BibitemShut {NoStop}%
	\bibitem [{\citenamefont {Tabares}\ \emph {et~al.}(2023)\citenamefont
		{Tabares}, \citenamefont {Mu\~noz de~las Heras}, \citenamefont {Tagliacozzo},
		\citenamefont {Porras},\ and\ \citenamefont
		{Gonz\'alez-Tudela}}]{PhysRevLett.131.073602}%
	\BibitemOpen
	\bibfield  {author} {\bibinfo {author} {\bibfnamefont {C.}~\bibnamefont
			{Tabares}}, \bibinfo {author} {\bibfnamefont {A.}~\bibnamefont {Mu\~noz
				de~las Heras}}, \bibinfo {author} {\bibfnamefont {L.}~\bibnamefont
			{Tagliacozzo}}, \bibinfo {author} {\bibfnamefont {D.}~\bibnamefont
			{Porras}},\ and\ \bibinfo {author} {\bibfnamefont {A.}~\bibnamefont
			{Gonz\'alez-Tudela}},\ }\href
	{https://doi.org/10.1103/PhysRevLett.131.073602} {\bibfield  {journal}
		{\bibinfo  {journal} {Phys. Rev. Lett.}\ }\textbf {\bibinfo {volume} {131}},\
		\bibinfo {pages} {073602} (\bibinfo {year} {2023})}\BibitemShut {NoStop}%
	\bibitem [{\citenamefont {Tian}\ \emph {et~al.}(2024)\citenamefont {Tian},
		\citenamefont {Wu},\ and\ \citenamefont
		{L\"u}}]{tian2024powerlawexponentialinteractioninducedquantum}%
	\BibitemOpen
	\bibfield  {author} {\bibinfo {author} {\bibfnamefont {G.}~\bibnamefont
			{Tian}}, \bibinfo {author} {\bibfnamefont {Y.}~\bibnamefont {Wu}},\ and\
		\bibinfo {author} {\bibfnamefont {X.-Y.}\ \bibnamefont {L\"u}},\ }\href
	{https://doi.org/10.1103/PhysRevResearch.6.033290} {\bibfield  {journal}
		{\bibinfo  {journal} {Phys. Rev. Res.}\ }\textbf {\bibinfo {volume} {6}},\
		\bibinfo {pages} {033290} (\bibinfo {year} {2024})}\BibitemShut {NoStop}%
	\bibitem [{\citenamefont {Lu}\ \emph {et~al.}(2014)\citenamefont {Lu},
		\citenamefont {Joannopoulos},\ and\ \citenamefont
		{Solja{\v{c}}i{\'c}}}]{lu2014topological}%
	\BibitemOpen
	\bibfield  {author} {\bibinfo {author} {\bibfnamefont {L.}~\bibnamefont
			{Lu}}, \bibinfo {author} {\bibfnamefont {J.~D.}\ \bibnamefont
			{Joannopoulos}},\ and\ \bibinfo {author} {\bibfnamefont {M.}~\bibnamefont
			{Solja{\v{c}}i{\'c}}},\ }\href {https://doi.org/10.1038/nphoton.2014.248}
	{\bibfield  {journal} {\bibinfo  {journal} {Nature photonics}\ }\textbf
		{\bibinfo {volume} {8}},\ \bibinfo {pages} {821} (\bibinfo {year}
		{2014})}\BibitemShut {NoStop}%
	\bibitem [{\citenamefont {Ozawa}\ \emph {et~al.}(2019)\citenamefont {Ozawa},
		\citenamefont {Price}, \citenamefont {Amo}, \citenamefont {Goldman},
		\citenamefont {Hafezi}, \citenamefont {Lu}, \citenamefont {Rechtsman},
		\citenamefont {Schuster}, \citenamefont {Simon}, \citenamefont {Zilberberg},\
		and\ \citenamefont {Carusotto}}]{RevModPhys.91.015006}%
	\BibitemOpen
	\bibfield  {author} {\bibinfo {author} {\bibfnamefont {T.}~\bibnamefont
			{Ozawa}}, \bibinfo {author} {\bibfnamefont {H.~M.}\ \bibnamefont {Price}},
		\bibinfo {author} {\bibfnamefont {A.}~\bibnamefont {Amo}}, \bibinfo {author}
		{\bibfnamefont {N.}~\bibnamefont {Goldman}}, \bibinfo {author} {\bibfnamefont
			{M.}~\bibnamefont {Hafezi}}, \bibinfo {author} {\bibfnamefont
			{L.}~\bibnamefont {Lu}}, \bibinfo {author} {\bibfnamefont {M.~C.}\
			\bibnamefont {Rechtsman}}, \bibinfo {author} {\bibfnamefont {D.}~\bibnamefont
			{Schuster}}, \bibinfo {author} {\bibfnamefont {J.}~\bibnamefont {Simon}},
		\bibinfo {author} {\bibfnamefont {O.}~\bibnamefont {Zilberberg}},\ and\
		\bibinfo {author} {\bibfnamefont {I.}~\bibnamefont {Carusotto}},\ }\href
	{https://doi.org/10.1103/RevModPhys.91.015006} {\bibfield  {journal}
		{\bibinfo  {journal} {Rev. Mod. Phys.}\ }\textbf {\bibinfo {volume} {91}},\
		\bibinfo {pages} {015006} (\bibinfo {year} {2019})}\BibitemShut {NoStop}%
	\bibitem [{\citenamefont {Anderson}\ \emph {et~al.}(2016)\citenamefont
		{Anderson}, \citenamefont {Ma}, \citenamefont {Owens}, \citenamefont
		{Schuster},\ and\ \citenamefont {Simon}}]{PhysRevX.6.041043}%
	\BibitemOpen
	\bibfield  {author} {\bibinfo {author} {\bibfnamefont {B.~M.}\ \bibnamefont
			{Anderson}}, \bibinfo {author} {\bibfnamefont {R.}~\bibnamefont {Ma}},
		\bibinfo {author} {\bibfnamefont {C.}~\bibnamefont {Owens}}, \bibinfo
		{author} {\bibfnamefont {D.~I.}\ \bibnamefont {Schuster}},\ and\ \bibinfo
		{author} {\bibfnamefont {J.}~\bibnamefont {Simon}},\ }\href
	{https://doi.org/10.1103/PhysRevX.6.041043} {\bibfield  {journal} {\bibinfo
			{journal} {Phys. Rev. X}\ }\textbf {\bibinfo {volume} {6}},\ \bibinfo {pages}
		{041043} (\bibinfo {year} {2016})}\BibitemShut {NoStop}%
	\bibitem [{\citenamefont {Haroche}\ and\ \citenamefont
		{Raimond}(2006)}]{10.1093/acprof:oso/9780198509141.001.0001}%
	\BibitemOpen
	\bibfield  {author} {\bibinfo {author} {\bibfnamefont {S.}~\bibnamefont
			{Haroche}}\ and\ \bibinfo {author} {\bibfnamefont {J.-M.}\ \bibnamefont
			{Raimond}},\ }\href@noop {} {\emph {\bibinfo {title} {Exploring the Quantum:
				Atoms, Cavities, and Photons}}}\ (\bibinfo  {publisher} {Oxford University
		Press},\ \bibinfo {year} {2006})\BibitemShut {NoStop}%
	\bibitem [{\citenamefont {Lodahl}\ \emph {et~al.}(2015)\citenamefont {Lodahl},
		\citenamefont {Mahmoodian},\ and\ \citenamefont
		{Stobbe}}]{RevModPhys.87.347}%
	\BibitemOpen
	\bibfield  {author} {\bibinfo {author} {\bibfnamefont {P.}~\bibnamefont
			{Lodahl}}, \bibinfo {author} {\bibfnamefont {S.}~\bibnamefont {Mahmoodian}},\
		and\ \bibinfo {author} {\bibfnamefont {S.}~\bibnamefont {Stobbe}},\ }\href
	{https://doi.org/10.1103/RevModPhys.87.347} {\bibfield  {journal} {\bibinfo
			{journal} {Rev. Mod. Phys.}\ }\textbf {\bibinfo {volume} {87}},\ \bibinfo
		{pages} {347} (\bibinfo {year} {2015})}\BibitemShut {NoStop}%
	\bibitem [{\citenamefont {Chang}\ \emph {et~al.}(2018)\citenamefont {Chang},
		\citenamefont {Douglas}, \citenamefont {Gonz\'alez-Tudela}, \citenamefont
		{Hung},\ and\ \citenamefont {Kimble}}]{RevModPhys.90.031002}%
	\BibitemOpen
	\bibfield  {author} {\bibinfo {author} {\bibfnamefont {D.~E.}\ \bibnamefont
			{Chang}}, \bibinfo {author} {\bibfnamefont {J.~S.}\ \bibnamefont {Douglas}},
		\bibinfo {author} {\bibfnamefont {A.}~\bibnamefont {Gonz\'alez-Tudela}},
		\bibinfo {author} {\bibfnamefont {C.-L.}\ \bibnamefont {Hung}},\ and\
		\bibinfo {author} {\bibfnamefont {H.~J.}\ \bibnamefont {Kimble}},\ }\href
	{https://doi.org/10.1103/RevModPhys.90.031002} {\bibfield  {journal}
		{\bibinfo  {journal} {Rev. Mod. Phys.}\ }\textbf {\bibinfo {volume} {90}},\
		\bibinfo {pages} {031002} (\bibinfo {year} {2018})}\BibitemShut {NoStop}%
	\bibitem [{\citenamefont {Barik}\ \emph {et~al.}(2018)\citenamefont {Barik},
		\citenamefont {Karasahin}, \citenamefont {Flower}, \citenamefont {Cai},
		\citenamefont {Miyake}, \citenamefont {DeGottardi}, \citenamefont {Hafezi},\
		and\ \citenamefont {Waks}}]{doi:10.1126/science.aaq0327}%
	\BibitemOpen
	\bibfield  {author} {\bibinfo {author} {\bibfnamefont {S.}~\bibnamefont
			{Barik}}, \bibinfo {author} {\bibfnamefont {A.}~\bibnamefont {Karasahin}},
		\bibinfo {author} {\bibfnamefont {C.}~\bibnamefont {Flower}}, \bibinfo
		{author} {\bibfnamefont {T.}~\bibnamefont {Cai}}, \bibinfo {author}
		{\bibfnamefont {H.}~\bibnamefont {Miyake}}, \bibinfo {author} {\bibfnamefont
			{W.}~\bibnamefont {DeGottardi}}, \bibinfo {author} {\bibfnamefont
			{M.}~\bibnamefont {Hafezi}},\ and\ \bibinfo {author} {\bibfnamefont
			{E.}~\bibnamefont {Waks}},\ }\href {https://doi.org/10.1126/science.aaq0327}
	{\bibfield  {journal} {\bibinfo  {journal} {Science}\ }\textbf {\bibinfo
			{volume} {359}},\ \bibinfo {pages} {666} (\bibinfo {year}
		{2018})}\BibitemShut {NoStop}%
	\bibitem [{\citenamefont {John}\ and\ \citenamefont
		{Wang}(1990)}]{PhysRevLett.64.2418}%
	\BibitemOpen
	\bibfield  {author} {\bibinfo {author} {\bibfnamefont {S.}~\bibnamefont
			{John}}\ and\ \bibinfo {author} {\bibfnamefont {J.}~\bibnamefont {Wang}},\
	}\href {https://doi.org/10.1103/PhysRevLett.64.2418} {\bibfield  {journal}
		{\bibinfo  {journal} {Phys. Rev. Lett.}\ }\textbf {\bibinfo {volume} {64}},\
		\bibinfo {pages} {2418} (\bibinfo {year} {1990})}\BibitemShut {NoStop}%
	\bibitem [{\citenamefont {Liu}\ and\ \citenamefont
		{Houck}(2017)}]{liu2017quantum}%
	\BibitemOpen
	\bibfield  {author} {\bibinfo {author} {\bibfnamefont {Y.}~\bibnamefont
			{Liu}}\ and\ \bibinfo {author} {\bibfnamefont {A.~A.}\ \bibnamefont
			{Houck}},\ }\href {https://doi.org/10.1038/nphys3834} {\bibfield  {journal}
		{\bibinfo  {journal} {Nature Physics}\ }\textbf {\bibinfo {volume} {13}},\
		\bibinfo {pages} {48} (\bibinfo {year} {2017})}\BibitemShut {NoStop}%
	\bibitem [{\citenamefont {Kim}\ \emph {et~al.}(2021)\citenamefont {Kim},
		\citenamefont {Zhang}, \citenamefont {Ferreira}, \citenamefont {Banker},
		\citenamefont {Iverson}, \citenamefont {Sipahigil}, \citenamefont {Bello},
		\citenamefont {Gonz\'alez-Tudela}, \citenamefont {Mirhosseini},\ and\
		\citenamefont {Painter}}]{PhysRevX.11.011015}%
	\BibitemOpen
	\bibfield  {author} {\bibinfo {author} {\bibfnamefont {E.}~\bibnamefont
			{Kim}}, \bibinfo {author} {\bibfnamefont {X.}~\bibnamefont {Zhang}}, \bibinfo
		{author} {\bibfnamefont {V.~S.}\ \bibnamefont {Ferreira}}, \bibinfo {author}
		{\bibfnamefont {J.}~\bibnamefont {Banker}}, \bibinfo {author} {\bibfnamefont
			{J.~K.}\ \bibnamefont {Iverson}}, \bibinfo {author} {\bibfnamefont
			{A.}~\bibnamefont {Sipahigil}}, \bibinfo {author} {\bibfnamefont
			{M.}~\bibnamefont {Bello}}, \bibinfo {author} {\bibfnamefont
			{A.}~\bibnamefont {Gonz\'alez-Tudela}}, \bibinfo {author} {\bibfnamefont
			{M.}~\bibnamefont {Mirhosseini}},\ and\ \bibinfo {author} {\bibfnamefont
			{O.}~\bibnamefont {Painter}},\ }\href
	{https://doi.org/10.1103/PhysRevX.11.011015} {\bibfield  {journal} {\bibinfo
			{journal} {Phys. Rev. X}\ }\textbf {\bibinfo {volume} {11}},\ \bibinfo
		{pages} {011015} (\bibinfo {year} {2021})}\BibitemShut {NoStop}%
	\bibitem [{\citenamefont {Sundaresan}\ \emph {et~al.}(2019)\citenamefont
		{Sundaresan}, \citenamefont {Lundgren}, \citenamefont {Zhu}, \citenamefont
		{Gorshkov},\ and\ \citenamefont {Houck}}]{PhysRevX.9.011021}%
	\BibitemOpen
	\bibfield  {author} {\bibinfo {author} {\bibfnamefont {N.~M.}\ \bibnamefont
			{Sundaresan}}, \bibinfo {author} {\bibfnamefont {R.}~\bibnamefont
			{Lundgren}}, \bibinfo {author} {\bibfnamefont {G.}~\bibnamefont {Zhu}},
		\bibinfo {author} {\bibfnamefont {A.~V.}\ \bibnamefont {Gorshkov}},\ and\
		\bibinfo {author} {\bibfnamefont {A.~A.}\ \bibnamefont {Houck}},\ }\href
	{https://doi.org/10.1103/PhysRevX.9.011021} {\bibfield  {journal} {\bibinfo
			{journal} {Phys. Rev. X}\ }\textbf {\bibinfo {volume} {9}},\ \bibinfo {pages}
		{011021} (\bibinfo {year} {2019})}\BibitemShut {NoStop}%
	\bibitem [{\citenamefont {Lodahl}\ \emph {et~al.}(2017)\citenamefont {Lodahl},
		\citenamefont {Mahmoodian}, \citenamefont {Stobbe}, \citenamefont
		{Rauschenbeutel}, \citenamefont {Schneeweiss}, \citenamefont {Volz},
		\citenamefont {Pichler},\ and\ \citenamefont {Zoller}}]{lodahl2017chiral}%
	\BibitemOpen
	\bibfield  {author} {\bibinfo {author} {\bibfnamefont {P.}~\bibnamefont
			{Lodahl}}, \bibinfo {author} {\bibfnamefont {S.}~\bibnamefont {Mahmoodian}},
		\bibinfo {author} {\bibfnamefont {S.}~\bibnamefont {Stobbe}}, \bibinfo
		{author} {\bibfnamefont {A.}~\bibnamefont {Rauschenbeutel}}, \bibinfo
		{author} {\bibfnamefont {P.}~\bibnamefont {Schneeweiss}}, \bibinfo {author}
		{\bibfnamefont {J.}~\bibnamefont {Volz}}, \bibinfo {author} {\bibfnamefont
			{H.}~\bibnamefont {Pichler}},\ and\ \bibinfo {author} {\bibfnamefont
			{P.}~\bibnamefont {Zoller}},\ }\href {https://doi.org/10.1038/nature21037}
	{\bibfield  {journal} {\bibinfo  {journal} {Nature}\ }\textbf {\bibinfo
			{volume} {541}},\ \bibinfo {pages} {473} (\bibinfo {year}
		{2017})}\BibitemShut {NoStop}%
	\bibitem [{\citenamefont {Mirhosseini}\ \emph {et~al.}(2018)\citenamefont
		{Mirhosseini}, \citenamefont {Kim}, \citenamefont {Ferreira}, \citenamefont
		{Kalaee}, \citenamefont {Sipahigil}, \citenamefont {Keller},\ and\
		\citenamefont {Painter}}]{mirhosseini2018superconducting}%
	\BibitemOpen
	\bibfield  {author} {\bibinfo {author} {\bibfnamefont {M.}~\bibnamefont
			{Mirhosseini}}, \bibinfo {author} {\bibfnamefont {E.}~\bibnamefont {Kim}},
		\bibinfo {author} {\bibfnamefont {V.~S.}\ \bibnamefont {Ferreira}}, \bibinfo
		{author} {\bibfnamefont {M.}~\bibnamefont {Kalaee}}, \bibinfo {author}
		{\bibfnamefont {A.}~\bibnamefont {Sipahigil}}, \bibinfo {author}
		{\bibfnamefont {A.~J.}\ \bibnamefont {Keller}},\ and\ \bibinfo {author}
		{\bibfnamefont {O.}~\bibnamefont {Painter}},\ }\href
	{https://doi.org/10.1038/s41467-018-06142-z} {\bibfield  {journal} {\bibinfo
			{journal} {Nature communications}\ }\textbf {\bibinfo {volume} {9}},\
		\bibinfo {pages} {3706} (\bibinfo {year} {2018})}\BibitemShut {NoStop}%
	\bibitem [{\citenamefont {Ferreira}\ \emph {et~al.}(2021)\citenamefont
		{Ferreira}, \citenamefont {Banker}, \citenamefont {Sipahigil}, \citenamefont
		{Matheny}, \citenamefont {Keller}, \citenamefont {Kim}, \citenamefont
		{Mirhosseini},\ and\ \citenamefont {Painter}}]{PhysRevX.11.041043}%
	\BibitemOpen
	\bibfield  {author} {\bibinfo {author} {\bibfnamefont {V.~S.}\ \bibnamefont
			{Ferreira}}, \bibinfo {author} {\bibfnamefont {J.}~\bibnamefont {Banker}},
		\bibinfo {author} {\bibfnamefont {A.}~\bibnamefont {Sipahigil}}, \bibinfo
		{author} {\bibfnamefont {M.~H.}\ \bibnamefont {Matheny}}, \bibinfo {author}
		{\bibfnamefont {A.~J.}\ \bibnamefont {Keller}}, \bibinfo {author}
		{\bibfnamefont {E.}~\bibnamefont {Kim}}, \bibinfo {author} {\bibfnamefont
			{M.}~\bibnamefont {Mirhosseini}},\ and\ \bibinfo {author} {\bibfnamefont
			{O.}~\bibnamefont {Painter}},\ }\href
	{https://doi.org/10.1103/PhysRevX.11.041043} {\bibfield  {journal} {\bibinfo
			{journal} {Phys. Rev. X}\ }\textbf {\bibinfo {volume} {11}},\ \bibinfo
		{pages} {041043} (\bibinfo {year} {2021})}\BibitemShut {NoStop}%
	\bibitem [{\citenamefont {Gu}\ \emph {et~al.}(2017)\citenamefont {Gu},
		\citenamefont {Kockum}, \citenamefont {Miranowicz}, \citenamefont {Liu},\
		and\ \citenamefont {Nori}}]{gu2017microwave}%
	\BibitemOpen
	\bibfield  {author} {\bibinfo {author} {\bibfnamefont {X.}~\bibnamefont
			{Gu}}, \bibinfo {author} {\bibfnamefont {A.~F.}\ \bibnamefont {Kockum}},
		\bibinfo {author} {\bibfnamefont {A.}~\bibnamefont {Miranowicz}}, \bibinfo
		{author} {\bibfnamefont {Y.-x.}\ \bibnamefont {Liu}},\ and\ \bibinfo {author}
		{\bibfnamefont {F.}~\bibnamefont {Nori}},\ }\href
	{https://doi.org/10.1016/j.physrep.2017.10.002} {\bibfield  {journal}
		{\bibinfo  {journal} {Physics Reports}\ }\textbf {\bibinfo {volume} {718}},\
		\bibinfo {pages} {1} (\bibinfo {year} {2017})}\BibitemShut {NoStop}%
	\bibitem [{\citenamefont {Morrone}\ \emph {et~al.}(2023)\citenamefont
		{Morrone}, \citenamefont {Rossi},\ and\ \citenamefont
		{Genoni}}]{PhysRevApplied.20.044073}%
	\BibitemOpen
	\bibfield  {author} {\bibinfo {author} {\bibfnamefont {D.}~\bibnamefont
			{Morrone}}, \bibinfo {author} {\bibfnamefont {M.~A.}\ \bibnamefont {Rossi}},\
		and\ \bibinfo {author} {\bibfnamefont {M.~G.}\ \bibnamefont {Genoni}},\
	}\href {https://doi.org/10.1103/PhysRevApplied.20.044073} {\bibfield
		{journal} {\bibinfo  {journal} {Phys. Rev. Appl.}\ }\textbf {\bibinfo
			{volume} {20}},\ \bibinfo {pages} {044073} (\bibinfo {year}
		{2023})}\BibitemShut {NoStop}%
	\bibitem [{\citenamefont {Pirmoradian}\ and\ \citenamefont
		{M\o{}lmer}(2019)}]{PhysRevA.100.043833}%
	\BibitemOpen
	\bibfield  {author} {\bibinfo {author} {\bibfnamefont {F.}~\bibnamefont
			{Pirmoradian}}\ and\ \bibinfo {author} {\bibfnamefont {K.}~\bibnamefont
			{M\o{}lmer}},\ }\href {https://doi.org/10.1103/PhysRevA.100.043833}
	{\bibfield  {journal} {\bibinfo  {journal} {Phys. Rev. A}\ }\textbf {\bibinfo
			{volume} {100}},\ \bibinfo {pages} {043833} (\bibinfo {year}
		{2019})}\BibitemShut {NoStop}%
	\bibitem [{\citenamefont {Zakavati}\ \emph {et~al.}(2021)\citenamefont
		{Zakavati}, \citenamefont {Tabesh},\ and\ \citenamefont
		{Salimi}}]{PhysRevE.104.054117}%
	\BibitemOpen
	\bibfield  {author} {\bibinfo {author} {\bibfnamefont {S.}~\bibnamefont
			{Zakavati}}, \bibinfo {author} {\bibfnamefont {F.~T.}\ \bibnamefont
			{Tabesh}},\ and\ \bibinfo {author} {\bibfnamefont {S.}~\bibnamefont
			{Salimi}},\ }\href {https://doi.org/10.1103/PhysRevE.104.054117} {\bibfield
		{journal} {\bibinfo  {journal} {Phys. Rev. E}\ }\textbf {\bibinfo {volume}
			{104}},\ \bibinfo {pages} {054117} (\bibinfo {year} {2021})}\BibitemShut
	{NoStop}%
	\bibitem [{\citenamefont {Xu}\ \emph {et~al.}(2021)\citenamefont {Xu},
		\citenamefont {Zhu}, \citenamefont {Zhang},\ and\ \citenamefont
		{Liu}}]{PhysRevE.104.064143}%
	\BibitemOpen
	\bibfield  {author} {\bibinfo {author} {\bibfnamefont {K.}~\bibnamefont
			{Xu}}, \bibinfo {author} {\bibfnamefont {H.-J.}\ \bibnamefont {Zhu}},
		\bibinfo {author} {\bibfnamefont {G.-F.}\ \bibnamefont {Zhang}},\ and\
		\bibinfo {author} {\bibfnamefont {W.-M.}\ \bibnamefont {Liu}},\ }\href
	{https://doi.org/10.1103/PhysRevE.104.064143} {\bibfield  {journal} {\bibinfo
			{journal} {Phys. Rev. E}\ }\textbf {\bibinfo {volume} {104}},\ \bibinfo
		{pages} {064143} (\bibinfo {year} {2021})}\BibitemShut {NoStop}%
	\bibitem [{\citenamefont {Shang}\ and\ \citenamefont
		{Li}(2024)}]{PhysRevApplied.21.044048}%
	\BibitemOpen
	\bibfield  {author} {\bibinfo {author} {\bibfnamefont {C.}~\bibnamefont
			{Shang}}\ and\ \bibinfo {author} {\bibfnamefont {H.}~\bibnamefont {Li}},\
	}\href {https://doi.org/10.1103/PhysRevApplied.21.044048} {\bibfield
		{journal} {\bibinfo  {journal} {Phys. Rev. Appl.}\ }\textbf {\bibinfo
			{volume} {21}},\ \bibinfo {pages} {044048} (\bibinfo {year}
		{2024})}\BibitemShut {NoStop}%
	\bibitem [{\citenamefont {Carrega}\ \emph {et~al.}(2020)\citenamefont
		{Carrega}, \citenamefont {Crescente}, \citenamefont {Ferraro},\ and\
		\citenamefont {Sassetti}}]{carrega2020dissipative}%
	\BibitemOpen
	\bibfield  {author} {\bibinfo {author} {\bibfnamefont {M.}~\bibnamefont
			{Carrega}}, \bibinfo {author} {\bibfnamefont {A.}~\bibnamefont {Crescente}},
		\bibinfo {author} {\bibfnamefont {D.}~\bibnamefont {Ferraro}},\ and\ \bibinfo
		{author} {\bibfnamefont {M.}~\bibnamefont {Sassetti}},\ }\href
	{https://iopscience.iop.org/article/10.1088/1367-2630/abaa01} {\bibfield
		{journal} {\bibinfo  {journal} {New Journal of Physics}\ }\textbf {\bibinfo
			{volume} {22}},\ \bibinfo {pages} {083085} (\bibinfo {year}
		{2020})}\BibitemShut {NoStop}%
	\bibitem [{\citenamefont {Yao}\ and\ \citenamefont
		{Shao}(2022)}]{PhysRevE.106.014138}%
	\BibitemOpen
	\bibfield  {author} {\bibinfo {author} {\bibfnamefont {Y.}~\bibnamefont
			{Yao}}\ and\ \bibinfo {author} {\bibfnamefont {X.~Q.}\ \bibnamefont {Shao}},\
	}\href {https://doi.org/10.1103/PhysRevE.106.014138} {\bibfield  {journal}
		{\bibinfo  {journal} {Phys. Rev. E}\ }\textbf {\bibinfo {volume} {106}},\
		\bibinfo {pages} {014138} (\bibinfo {year} {2022})}\BibitemShut {NoStop}%
	\bibitem [{\citenamefont {Barra}(2019)}]{PhysRevLett.122.210601}%
	\BibitemOpen
	\bibfield  {author} {\bibinfo {author} {\bibfnamefont {F.}~\bibnamefont
			{Barra}},\ }\href {https://doi.org/10.1103/PhysRevLett.122.210601} {\bibfield
		{journal} {\bibinfo  {journal} {Phys. Rev. Lett.}\ }\textbf {\bibinfo
			{volume} {122}},\ \bibinfo {pages} {210601} (\bibinfo {year}
		{2019})}\BibitemShut {NoStop}%
	\bibitem [{\citenamefont {Kamin}\ \emph
		{et~al.}(2020{\natexlab{b}})\citenamefont {Kamin}, \citenamefont {Tabesh},
		\citenamefont {Salimi}, \citenamefont {Kheirandish},\ and\ \citenamefont
		{Santos}}]{kamin2020non}%
	\BibitemOpen
	\bibfield  {author} {\bibinfo {author} {\bibfnamefont {F.}~\bibnamefont
			{Kamin}}, \bibinfo {author} {\bibfnamefont {F.}~\bibnamefont {Tabesh}},
		\bibinfo {author} {\bibfnamefont {S.}~\bibnamefont {Salimi}}, \bibinfo
		{author} {\bibfnamefont {F.}~\bibnamefont {Kheirandish}},\ and\ \bibinfo
		{author} {\bibfnamefont {A.~C.}\ \bibnamefont {Santos}},\ }\href
	{https://iopscience.iop.org/article/10.1088/1367-2630/ab9ee2} {\bibfield
		{journal} {\bibinfo  {journal} {New Journal of Physics}\ }\textbf {\bibinfo
			{volume} {22}},\ \bibinfo {pages} {083007} (\bibinfo {year}
		{2020}{\natexlab{b}})}\BibitemShut {NoStop}%
	\bibitem [{\citenamefont {Bai}\ and\ \citenamefont
		{An}(2020)}]{PhysRevA.102.060201}%
	\BibitemOpen
	\bibfield  {author} {\bibinfo {author} {\bibfnamefont {S.-Y.}\ \bibnamefont
			{Bai}}\ and\ \bibinfo {author} {\bibfnamefont {J.-H.}\ \bibnamefont {An}},\
	}\href {https://doi.org/10.1103/PhysRevA.102.060201} {\bibfield  {journal}
		{\bibinfo  {journal} {Phys. Rev. A}\ }\textbf {\bibinfo {volume} {102}},\
		\bibinfo {pages} {060201} (\bibinfo {year} {2020})}\BibitemShut {NoStop}%
	\bibitem [{\citenamefont {Hadipour}\ \emph {et~al.}(2025)\citenamefont
		{Hadipour}, \citenamefont {Yousefi}, \citenamefont {Mortezapour},
		\citenamefont {Miavaghi},\ and\ \citenamefont
		{Haseli}}]{hadipour2025amplifiedquantumbatterydynamical}%
	\BibitemOpen
	\bibfield  {author} {\bibinfo {author} {\bibfnamefont {M.}~\bibnamefont
			{Hadipour}}, \bibinfo {author} {\bibfnamefont {N.~N.}\ \bibnamefont
			{Yousefi}}, \bibinfo {author} {\bibfnamefont {A.}~\bibnamefont
			{Mortezapour}}, \bibinfo {author} {\bibfnamefont {A.~S.}\ \bibnamefont
			{Miavaghi}},\ and\ \bibinfo {author} {\bibfnamefont {S.}~\bibnamefont
			{Haseli}},\ }\href {https://arxiv.org/abs/2501.17526} {\bibfield  {journal}
		{\bibinfo  {journal} {arXiv:2501.17526}\ } (\bibinfo {year}
		{2025})}\BibitemShut {NoStop}%
	\bibitem [{\citenamefont {Su}\ \emph {et~al.}(1979)\citenamefont {Su},
		\citenamefont {Schrieffer},\ and\ \citenamefont {Heeger}}]{SSH}%
	\BibitemOpen
	\bibfield  {author} {\bibinfo {author} {\bibfnamefont {W.~P.}\ \bibnamefont
			{Su}}, \bibinfo {author} {\bibfnamefont {J.~R.}\ \bibnamefont {Schrieffer}},\
		and\ \bibinfo {author} {\bibfnamefont {A.~J.}\ \bibnamefont {Heeger}},\
	}\href {https://doi.org/10.1103/PhysRevLett.42.1698} {\bibfield  {journal}
		{\bibinfo  {journal} {Phys. Rev. Lett.}\ }\textbf {\bibinfo {volume} {42}},\
		\bibinfo {pages} {1698} (\bibinfo {year} {1979})}\BibitemShut {NoStop}%
	\bibitem [{\citenamefont {Su}\ \emph {et~al.}(1980)\citenamefont {Su},
		\citenamefont {Schrieffer},\ and\ \citenamefont {Heeger}}]{PhysRevB.22.2099}%
	\BibitemOpen
	\bibfield  {author} {\bibinfo {author} {\bibfnamefont {W.~P.}\ \bibnamefont
			{Su}}, \bibinfo {author} {\bibfnamefont {J.~R.}\ \bibnamefont {Schrieffer}},\
		and\ \bibinfo {author} {\bibfnamefont {A.~J.}\ \bibnamefont {Heeger}},\
	}\href {https://doi.org/10.1103/PhysRevB.22.2099} {\bibfield  {journal}
		{\bibinfo  {journal} {Phys. Rev. B}\ }\textbf {\bibinfo {volume} {22}},\
		\bibinfo {pages} {2099} (\bibinfo {year} {1980})}\BibitemShut {NoStop}%
	\bibitem [{\citenamefont {Gong}\ \emph
		{et~al.}(2022{\natexlab{a}})\citenamefont {Gong}, \citenamefont {Bello},
		\citenamefont {Malz},\ and\ \citenamefont {Kunst}}]{nhbath}%
	\BibitemOpen
	\bibfield  {author} {\bibinfo {author} {\bibfnamefont {Z.}~\bibnamefont
			{Gong}}, \bibinfo {author} {\bibfnamefont {M.}~\bibnamefont {Bello}},
		\bibinfo {author} {\bibfnamefont {D.}~\bibnamefont {Malz}},\ and\ \bibinfo
		{author} {\bibfnamefont {F.~K.}\ \bibnamefont {Kunst}},\ }\href
	{https://doi.org/10.1103/PhysRevLett.129.223601} {\bibfield  {journal}
		{\bibinfo  {journal} {Phys. Rev. Lett.}\ }\textbf {\bibinfo {volume} {129}},\
		\bibinfo {pages} {223601} (\bibinfo {year} {2022}{\natexlab{a}})}\BibitemShut
	{NoStop}%
	\bibitem [{\citenamefont {Gong}\ \emph
		{et~al.}(2022{\natexlab{b}})\citenamefont {Gong}, \citenamefont {Bello},
		\citenamefont {Malz},\ and\ \citenamefont {Kunst}}]{PhysRevA.106.053517}%
	\BibitemOpen
	\bibfield  {author} {\bibinfo {author} {\bibfnamefont {Z.}~\bibnamefont
			{Gong}}, \bibinfo {author} {\bibfnamefont {M.}~\bibnamefont {Bello}},
		\bibinfo {author} {\bibfnamefont {D.}~\bibnamefont {Malz}},\ and\ \bibinfo
		{author} {\bibfnamefont {F.~K.}\ \bibnamefont {Kunst}},\ }\href
	{https://doi.org/10.1103/PhysRevA.106.053517} {\bibfield  {journal} {\bibinfo
			{journal} {Phys. Rev. A}\ }\textbf {\bibinfo {volume} {106}},\ \bibinfo
		{pages} {053517} (\bibinfo {year} {2022}{\natexlab{b}})}\BibitemShut
	{NoStop}%
	\bibitem [{\citenamefont {Gonz\'alez-Tudela}\ and\ \citenamefont
		{Cirac}(2017{\natexlab{a}})}]{PhysRevA.96.043811}%
	\BibitemOpen
	\bibfield  {author} {\bibinfo {author} {\bibfnamefont {A.}~\bibnamefont
			{Gonz\'alez-Tudela}}\ and\ \bibinfo {author} {\bibfnamefont {J.~I.}\
			\bibnamefont {Cirac}},\ }\href {https://doi.org/10.1103/PhysRevA.96.043811}
	{\bibfield  {journal} {\bibinfo  {journal} {Phys. Rev. A}\ }\textbf {\bibinfo
			{volume} {96}},\ \bibinfo {pages} {043811} (\bibinfo {year}
		{2017}{\natexlab{a}})}\BibitemShut {NoStop}%
	\bibitem [{\citenamefont {Gong}\ \emph {et~al.}(2017)\citenamefont {Gong},
		\citenamefont {Higashikawa},\ and\ \citenamefont
		{Ueda}}]{PhysRevLett.118.200401}%
	\BibitemOpen
	\bibfield  {author} {\bibinfo {author} {\bibfnamefont {Z.}~\bibnamefont
			{Gong}}, \bibinfo {author} {\bibfnamefont {S.}~\bibnamefont {Higashikawa}},\
		and\ \bibinfo {author} {\bibfnamefont {M.}~\bibnamefont {Ueda}},\ }\href
	{https://doi.org/10.1103/PhysRevLett.118.200401} {\bibfield  {journal}
		{\bibinfo  {journal} {Phys. Rev. Lett.}\ }\textbf {\bibinfo {volume} {118}},\
		\bibinfo {pages} {200401} (\bibinfo {year} {2017})}\BibitemShut {NoStop}%
	\bibitem [{\citenamefont {Alicki}(1989)}]{PhysRevA.40.4077}%
	\BibitemOpen
	\bibfield  {author} {\bibinfo {author} {\bibfnamefont {R.}~\bibnamefont
			{Alicki}},\ }\href {https://doi.org/10.1103/PhysRevA.40.4077} {\bibfield
		{journal} {\bibinfo  {journal} {Phys. Rev. A}\ }\textbf {\bibinfo {volume}
			{40}},\ \bibinfo {pages} {4077} (\bibinfo {year} {1989})}\BibitemShut
	{NoStop}%
	\bibitem{supp}{See Supplemental Material at, which includes Refs. [99–108], for the derivation of exact dynamics in dissipative and non-dissipative topological waveguides, the calculation of phase boundaries, the solution of the Lindblad master equation in the single-excitation sector, an analytical proof of the physical mechanism behind perfect energy transfer, and detailed discussions on charging time, charging protocol, disorder effects, and the validity of the rotating wave approximation.}
	\bibitem [{\citenamefont {Daley}(2014)}]{qt}%
	\BibitemOpen
	\bibfield  {author} {\bibinfo {author} {\bibfnamefont {A.~J.}\ \bibnamefont
			{Daley}},\ }\href {https://doi.org/10.1080/00018732.2014.933502} {\bibfield
		{journal} {\bibinfo  {journal} {Advances in Physics}\ }\textbf {\bibinfo
			{volume} {63}},\ \bibinfo {pages} {77} (\bibinfo {year} {2014})}\BibitemShut
	{NoStop}%
	\bibitem [{\citenamefont {Guo}\ \emph {et~al.}(2009)\citenamefont {Guo},
		\citenamefont {Salamo}, \citenamefont {Duchesne}, \citenamefont {Morandotti},
		\citenamefont {Volatier-Ravat}, \citenamefont {Aimez}, \citenamefont
		{Siviloglou},\ and\ \citenamefont {Christodoulides}}]{PT}%
	\BibitemOpen
	\bibfield  {author} {\bibinfo {author} {\bibfnamefont {A.}~\bibnamefont
			{Guo}}, \bibinfo {author} {\bibfnamefont {G.~J.}\ \bibnamefont {Salamo}},
		\bibinfo {author} {\bibfnamefont {D.}~\bibnamefont {Duchesne}}, \bibinfo
		{author} {\bibfnamefont {R.}~\bibnamefont {Morandotti}}, \bibinfo {author}
		{\bibfnamefont {M.}~\bibnamefont {Volatier-Ravat}}, \bibinfo {author}
		{\bibfnamefont {V.}~\bibnamefont {Aimez}}, \bibinfo {author} {\bibfnamefont
			{G.~A.}\ \bibnamefont {Siviloglou}},\ and\ \bibinfo {author} {\bibfnamefont
			{D.~N.}\ \bibnamefont {Christodoulides}},\ }\href
	{https://doi.org/10.1103/PhysRevLett.103.093902} {\bibfield  {journal}
		{\bibinfo  {journal} {Phys. Rev. Lett.}\ }\textbf {\bibinfo {volume} {103}},\
		\bibinfo {pages} {093902} (\bibinfo {year} {2009})}\BibitemShut {NoStop}%
	\bibitem [{\citenamefont {Gonz{\'{a}}lez-Tudela}\ and\ \citenamefont
		{Cirac}(2018)}]{pa}%
	\BibitemOpen
	\bibfield  {author} {\bibinfo {author} {\bibfnamefont {A.}~\bibnamefont
			{Gonz{\'{a}}lez-Tudela}}\ and\ \bibinfo {author} {\bibfnamefont {J.~I.}\
			\bibnamefont {Cirac}},\ }\href {https://doi.org/10.22331/q-2018-10-01-97}
	{\bibfield  {journal} {\bibinfo  {journal} {{Quantum}}\ }\textbf {\bibinfo
			{volume} {2}},\ \bibinfo {pages} {97} (\bibinfo {year} {2018})}\BibitemShut
	{NoStop}%
	\bibitem [{\citenamefont {Asb{\'o}th}\ \emph {et~al.}(2016)\citenamefont
		{Asb{\'o}th}, \citenamefont {Oroszl{\'a}ny},\ and\ \citenamefont
		{P{\'a}lyi}}]{LRedge}%
	\BibitemOpen
	\bibfield  {author} {\bibinfo {author} {\bibfnamefont {J.~K.}\ \bibnamefont
			{Asb{\'o}th}}, \bibinfo {author} {\bibfnamefont {L.}~\bibnamefont
			{Oroszl{\'a}ny}},\ and\ \bibinfo {author} {\bibfnamefont {A.}~\bibnamefont
			{P{\'a}lyi}},\ }\href
	{https://link.springer.com/content/pdf/10.1007/978-3-319-25607-8.pdf}
	{\bibfield  {journal} {\bibinfo  {journal} {Lecture notes in physics}\
		}\textbf {\bibinfo {volume} {919}},\ \bibinfo {pages} {166} (\bibinfo {year}
		{2016})}\BibitemShut {NoStop}%
	\bibitem [{\citenamefont {Ciccarello}(2011)}]{edge}%
	\BibitemOpen
	\bibfield  {author} {\bibinfo {author} {\bibfnamefont {F.}~\bibnamefont
			{Ciccarello}},\ }\href {https://doi.org/10.1103/PhysRevA.83.043802}
	{\bibfield  {journal} {\bibinfo  {journal} {Phys. Rev. A}\ }\textbf {\bibinfo
			{volume} {83}},\ \bibinfo {pages} {043802} (\bibinfo {year}
		{2011})}\BibitemShut {NoStop}%
	\bibitem [{\citenamefont {Lu}\ \emph {et~al.}(2025)\citenamefont {Lu},
		\citenamefont {Wu},\ and\ \citenamefont {L\"u}}]{lu2024}%
	\BibitemOpen
	\bibfield  {author} {\bibinfo {author} {\bibfnamefont {Z.-G.}\ \bibnamefont
			{Lu}}, \bibinfo {author} {\bibfnamefont {Y.}~\bibnamefont {Wu}},\ and\
		\bibinfo {author} {\bibfnamefont {X.-Y.}\ \bibnamefont {L\"u}},\ }\href
	{https://doi.org/10.1103/PhysRevLett.134.013602} {\bibfield  {journal}
		{\bibinfo  {journal} {Phys. Rev. Lett.}\ }\textbf {\bibinfo {volume} {134}},\
		\bibinfo {pages} {013602} (\bibinfo {year} {2025})}\BibitemShut {NoStop}%
	\bibitem [{\citenamefont {Young}\ \emph {et~al.}(2015)\citenamefont {Young},
		\citenamefont {Thijssen}, \citenamefont {Beggs}, \citenamefont
		{Androvitsaneas}, \citenamefont {Kuipers}, \citenamefont {Rarity},
		\citenamefont {Hughes},\ and\ \citenamefont
		{Oulton}}]{PhysRevLett.115.153901}%
	\BibitemOpen
	\bibfield  {author} {\bibinfo {author} {\bibfnamefont {A.~B.}\ \bibnamefont
			{Young}}, \bibinfo {author} {\bibfnamefont {A.~C.~T.}\ \bibnamefont
			{Thijssen}}, \bibinfo {author} {\bibfnamefont {D.~M.}\ \bibnamefont {Beggs}},
		\bibinfo {author} {\bibfnamefont {P.}~\bibnamefont {Androvitsaneas}},
		\bibinfo {author} {\bibfnamefont {L.}~\bibnamefont {Kuipers}}, \bibinfo
		{author} {\bibfnamefont {J.~G.}\ \bibnamefont {Rarity}}, \bibinfo {author}
		{\bibfnamefont {S.}~\bibnamefont {Hughes}},\ and\ \bibinfo {author}
		{\bibfnamefont {R.}~\bibnamefont {Oulton}},\ }\href
	{https://doi.org/10.1103/PhysRevLett.115.153901} {\bibfield  {journal}
		{\bibinfo  {journal} {Phys. Rev. Lett.}\ }\textbf {\bibinfo {volume} {115}},\
		\bibinfo {pages} {153901} (\bibinfo {year} {2015})}\BibitemShut {NoStop}%
	\bibitem [{\citenamefont {Reitz}\ \emph {et~al.}(2013)\citenamefont {Reitz},
		\citenamefont {Sayrin}, \citenamefont {Mitsch}, \citenamefont {Schneeweiss},\
		and\ \citenamefont {Rauschenbeutel}}]{PhysRevLett.110.243603}%
	\BibitemOpen
	\bibfield  {author} {\bibinfo {author} {\bibfnamefont {D.}~\bibnamefont
			{Reitz}}, \bibinfo {author} {\bibfnamefont {C.}~\bibnamefont {Sayrin}},
		\bibinfo {author} {\bibfnamefont {R.}~\bibnamefont {Mitsch}}, \bibinfo
		{author} {\bibfnamefont {P.}~\bibnamefont {Schneeweiss}},\ and\ \bibinfo
		{author} {\bibfnamefont {A.}~\bibnamefont {Rauschenbeutel}},\ }\href
	{https://doi.org/10.1103/PhysRevLett.110.243603} {\bibfield  {journal}
		{\bibinfo  {journal} {Phys. Rev. Lett.}\ }\textbf {\bibinfo {volume} {110}},\
		\bibinfo {pages} {243603} (\bibinfo {year} {2013})}\BibitemShut {NoStop}%
	\bibitem [{\citenamefont {Haegeman}\ \emph {et~al.}(2011)\citenamefont
		{Haegeman}, \citenamefont {Cirac}, \citenamefont {Osborne}, \citenamefont
		{Pi\ifmmode~\check{z}\else \v{z}\fi{}orn}, \citenamefont {Verschelde},\ and\
		\citenamefont {Verstraete}}]{PhysRevLett.107.070601}%
	\BibitemOpen
	\bibfield  {author} {\bibinfo {author} {\bibfnamefont {J.}~\bibnamefont
			{Haegeman}}, \bibinfo {author} {\bibfnamefont {J.~I.}\ \bibnamefont {Cirac}},
		\bibinfo {author} {\bibfnamefont {T.~J.}\ \bibnamefont {Osborne}}, \bibinfo
		{author} {\bibfnamefont {I.}~\bibnamefont {Pi\ifmmode~\check{z}\else
				\v{z}\fi{}orn}}, \bibinfo {author} {\bibfnamefont {H.}~\bibnamefont
			{Verschelde}},\ and\ \bibinfo {author} {\bibfnamefont {F.}~\bibnamefont
			{Verstraete}},\ }\href {https://doi.org/10.1103/PhysRevLett.107.070601}
	{\bibfield  {journal} {\bibinfo  {journal} {Phys. Rev. Lett.}\ }\textbf
		{\bibinfo {volume} {107}},\ \bibinfo {pages} {070601} (\bibinfo {year}
		{2011})}\BibitemShut {NoStop}%
	\bibitem [{\citenamefont {Haegeman}\ \emph {et~al.}(2016)\citenamefont
		{Haegeman}, \citenamefont {Lubich}, \citenamefont {Oseledets}, \citenamefont
		{Vandereycken},\ and\ \citenamefont {Verstraete}}]{PhysRevB.94.165116}%
	\BibitemOpen
	\bibfield  {author} {\bibinfo {author} {\bibfnamefont {J.}~\bibnamefont
			{Haegeman}}, \bibinfo {author} {\bibfnamefont {C.}~\bibnamefont {Lubich}},
		\bibinfo {author} {\bibfnamefont {I.}~\bibnamefont {Oseledets}}, \bibinfo
		{author} {\bibfnamefont {B.}~\bibnamefont {Vandereycken}},\ and\ \bibinfo
		{author} {\bibfnamefont {F.}~\bibnamefont {Verstraete}},\ }\href
	{https://doi.org/10.1103/PhysRevB.94.165116} {\bibfield  {journal} {\bibinfo
			{journal} {Phys. Rev. B}\ }\textbf {\bibinfo {volume} {94}},\ \bibinfo
		{pages} {165116} (\bibinfo {year} {2016})}\BibitemShut {NoStop}%
	\bibitem [{\citenamefont {Gonz\'alez-Tudela}\ and\ \citenamefont
		{Cirac}(2017{\natexlab{b}})}]{PhysRevLett.119.143602}%
	\BibitemOpen
	\bibfield  {author} {\bibinfo {author} {\bibfnamefont {A.}~\bibnamefont
			{Gonz\'alez-Tudela}}\ and\ \bibinfo {author} {\bibfnamefont {J.~I.}\
			\bibnamefont {Cirac}},\ }\href
	{https://doi.org/10.1103/PhysRevLett.119.143602} {\bibfield  {journal}
		{\bibinfo  {journal} {Phys. Rev. Lett.}\ }\textbf {\bibinfo {volume} {119}},\
		\bibinfo {pages} {143602} (\bibinfo {year} {2017}{\natexlab{b}})}\BibitemShut
	{NoStop}%
	\bibitem [{\citenamefont {Cohen-Tannoudji}\ \emph {et~al.}(1998)\citenamefont
		{Cohen-Tannoudji}, \citenamefont {Dupont-Roc},\ and\ \citenamefont
		{Grynberg}}]{cohen1998atom}%
	\BibitemOpen
	\bibfield  {author} {\bibinfo {author} {\bibfnamefont {C.}~\bibnamefont
			{Cohen-Tannoudji}}, \bibinfo {author} {\bibfnamefont {J.}~\bibnamefont
			{Dupont-Roc}},\ and\ \bibinfo {author} {\bibfnamefont {G.}~\bibnamefont
			{Grynberg}},\ }\href@noop {} {\emph {\bibinfo {title} {Atom-photon
				interactions:~basic processes and applications}}}\ (\bibinfo  {publisher}
	{John Wiley \& Sons},\ \bibinfo {year} {1998})\BibitemShut {NoStop}%
	\bibitem [{\citenamefont {M{\'e}nard}\ \emph {et~al.}(2015)\citenamefont
		{M{\'e}nard}, \citenamefont {Guissart}, \citenamefont {Brun}, \citenamefont
		{Pons}, \citenamefont {Stolyarov}, \citenamefont {Debontridder},
		\citenamefont {Leclerc}, \citenamefont {Janod}, \citenamefont {Cario},
		\citenamefont {Roditchev} \emph {et~al.}}]{menard2015coherent}%
	\BibitemOpen
	\bibfield  {author} {\bibinfo {author} {\bibfnamefont {G.~C.}\ \bibnamefont
			{M{\'e}nard}}, \bibinfo {author} {\bibfnamefont {S.}~\bibnamefont
			{Guissart}}, \bibinfo {author} {\bibfnamefont {C.}~\bibnamefont {Brun}},
		\bibinfo {author} {\bibfnamefont {S.}~\bibnamefont {Pons}}, \bibinfo {author}
		{\bibfnamefont {V.~S.}\ \bibnamefont {Stolyarov}}, \bibinfo {author}
		{\bibfnamefont {F.}~\bibnamefont {Debontridder}}, \bibinfo {author}
		{\bibfnamefont {M.~V.}\ \bibnamefont {Leclerc}}, \bibinfo {author}
		{\bibfnamefont {E.}~\bibnamefont {Janod}}, \bibinfo {author} {\bibfnamefont
			{L.}~\bibnamefont {Cario}}, \bibinfo {author} {\bibfnamefont
			{D.}~\bibnamefont {Roditchev}}, \emph {et~al.},\ }\href
	{https://doi.org/10.1038/nphys3508} {\bibfield  {journal} {\bibinfo
			{journal} {Nature Physics}\ }\textbf {\bibinfo {volume} {11}},\ \bibinfo
		{pages} {1013} (\bibinfo {year} {2015})}\BibitemShut {NoStop}%
	\bibitem{supp1}{When treating the atomic component as a new photonic component at the edge, the vacancy-like dressed state---formed by a single two-level system coupled to a periodic SSH chain---can be regarded as an edge-like state due to its similarity to edge state. In our study, introducing the edge-like state allows for a well-defined spatial envelope, thereby enabling a precise mathematical characterization of the overlap between these envelopes.}
	\bibitem [{\citenamefont {Leonforte}\ \emph {et~al.}(2021)\citenamefont
		{Leonforte}, \citenamefont {Carollo},\ and\ \citenamefont
		{Ciccarello}}]{vbs}%
	\BibitemOpen
	\bibfield  {author} {\bibinfo {author} {\bibfnamefont {L.}~\bibnamefont
			{Leonforte}}, \bibinfo {author} {\bibfnamefont {A.}~\bibnamefont {Carollo}},\
		and\ \bibinfo {author} {\bibfnamefont {F.}~\bibnamefont {Ciccarello}},\
	}\href {https://doi.org/10.1103/PhysRevLett.126.063601} {\bibfield  {journal}
		{\bibinfo  {journal} {Phys. Rev. Lett.}\ }\textbf {\bibinfo {volume} {126}},\
		\bibinfo {pages} {063601} (\bibinfo {year} {2021})}\BibitemShut {NoStop}%
	\bibitem [{\citenamefont {Quach}\ and\ \citenamefont
		{Munro}(2020)}]{PhysRevApplied.14.024092}%
	\BibitemOpen
	\bibfield  {author} {\bibinfo {author} {\bibfnamefont {J.~Q.}\ \bibnamefont
			{Quach}}\ and\ \bibinfo {author} {\bibfnamefont {W.~J.}\ \bibnamefont
			{Munro}},\ }\href {https://doi.org/10.1103/PhysRevApplied.14.024092}
	{\bibfield  {journal} {\bibinfo  {journal} {Phys. Rev. Appl.}\ }\textbf
		{\bibinfo {volume} {14}},\ \bibinfo {pages} {024092} (\bibinfo {year}
		{2020})}\BibitemShut {NoStop}%
	\bibitem [{\citenamefont {Popkov}\ and\ \citenamefont
		{Presilla}(2021)}]{PhysRevLett.126.190402}%
	\BibitemOpen
	\bibfield  {author} {\bibinfo {author} {\bibfnamefont {V.}~\bibnamefont
			{Popkov}}\ and\ \bibinfo {author} {\bibfnamefont {C.}~\bibnamefont
			{Presilla}},\ }\href {https://doi.org/10.1103/PhysRevLett.126.190402}
	{\bibfield  {journal} {\bibinfo  {journal} {Phys. Rev. Lett.}\ }\textbf
		{\bibinfo {volume} {126}},\ \bibinfo {pages} {190402} (\bibinfo {year}
		{2021})}\BibitemShut {NoStop}%
	\bibitem [{\citenamefont {Hatano}\ and\ \citenamefont
		{Nelson}(1996)}]{PhysRevLett.77.570}%
	\BibitemOpen
	\bibfield  {author} {\bibinfo {author} {\bibfnamefont {N.}~\bibnamefont
			{Hatano}}\ and\ \bibinfo {author} {\bibfnamefont {D.~R.}\ \bibnamefont
			{Nelson}},\ }\href {https://doi.org/10.1103/PhysRevLett.77.570} {\bibfield
		{journal} {\bibinfo  {journal} {Phys. Rev. Lett.}\ }\textbf {\bibinfo
			{volume} {77}},\ \bibinfo {pages} {570} (\bibinfo {year} {1996})}\BibitemShut
	{NoStop}%
	\bibitem [{\citenamefont {Shi}\ \emph {et~al.}(2016)\citenamefont {Shi},
		\citenamefont {Wu}, \citenamefont {Gonz\'alez-Tudela},\ and\ \citenamefont
		{Cirac}}]{PhysRevX.6.021027}%
	\BibitemOpen
	\bibfield  {author} {\bibinfo {author} {\bibfnamefont {T.}~\bibnamefont
			{Shi}}, \bibinfo {author} {\bibfnamefont {Y.-H.}\ \bibnamefont {Wu}},
		\bibinfo {author} {\bibfnamefont {A.}~\bibnamefont {Gonz\'alez-Tudela}},\
		and\ \bibinfo {author} {\bibfnamefont {J.~I.}\ \bibnamefont {Cirac}},\ }\href
	{https://doi.org/10.1103/PhysRevX.6.021027} {\bibfield  {journal} {\bibinfo
			{journal} {Phys. Rev. X}\ }\textbf {\bibinfo {volume} {6}},\ \bibinfo {pages}
		{021027} (\bibinfo {year} {2016})}\BibitemShut {NoStop}%
	\bibitem [{\citenamefont {Shi}\ \emph {et~al.}(2018)\citenamefont {Shi},
		\citenamefont {Wu}, \citenamefont {Gonz{\'a}lez-Tudela},\ and\ \citenamefont
		{Cirac}}]{shi2018effective}%
	\BibitemOpen
	\bibfield  {author} {\bibinfo {author} {\bibfnamefont {T.}~\bibnamefont
			{Shi}}, \bibinfo {author} {\bibfnamefont {Y.~H.}\ \bibnamefont {Wu}},
		\bibinfo {author} {\bibfnamefont {A.}~\bibnamefont {Gonz{\'a}lez-Tudela}},\
		and\ \bibinfo {author} {\bibfnamefont {J.~I.}\ \bibnamefont {Cirac}},\ }\href
	{https://iopscience.iop.org/article/10.1088/1367-2630/aae4a9/meta} {\bibfield
		{journal} {\bibinfo  {journal} {New Journal of Physics}\ }\textbf {\bibinfo
			{volume} {20}},\ \bibinfo {pages} {105005} (\bibinfo {year}
		{2018})}\BibitemShut {NoStop}%
	\bibitem [{\citenamefont {Shi}\ \emph {et~al.}(2011)\citenamefont {Shi},
		\citenamefont {Fan},\ and\ \citenamefont {Sun}}]{PhysRevA.84.063803}%
	\BibitemOpen
	\bibfield  {author} {\bibinfo {author} {\bibfnamefont {T.}~\bibnamefont
			{Shi}}, \bibinfo {author} {\bibfnamefont {S.}~\bibnamefont {Fan}},\ and\
		\bibinfo {author} {\bibfnamefont {C.~P.}\ \bibnamefont {Sun}},\ }\href
	{https://doi.org/10.1103/PhysRevA.84.063803} {\bibfield  {journal} {\bibinfo
			{journal} {Phys. Rev. A}\ }\textbf {\bibinfo {volume} {84}},\ \bibinfo
		{pages} {063803} (\bibinfo {year} {2011})}\BibitemShut {NoStop}%
	\bibitem [{\citenamefont {Mayo}\ and\ \citenamefont
		{Roncaglia}(2022)}]{PhysRevA.105.062203}%
	\BibitemOpen
	\bibfield  {author} {\bibinfo {author} {\bibfnamefont {F.}~\bibnamefont
			{Mayo}}\ and\ \bibinfo {author} {\bibfnamefont {A.~J.}\ \bibnamefont
			{Roncaglia}},\ }\href {https://doi.org/10.1103/PhysRevA.105.062203}
	{\bibfield  {journal} {\bibinfo  {journal} {Phys. Rev. A}\ }\textbf {\bibinfo
			{volume} {105}},\ \bibinfo {pages} {062203} (\bibinfo {year}
		{2022})}\BibitemShut {NoStop}%
	\bibitem [{\citenamefont {Schollw{\"o}ck}(2011)}]{schollwock2011density}%
	\BibitemOpen
	\bibfield  {author} {\bibinfo {author} {\bibfnamefont {U.}~\bibnamefont
			{Schollw{\"o}ck}},\ }\href
	{https://www.sciencedirect.com/science/article/pii/S0003491610001752}
	{\bibfield  {journal} {\bibinfo  {journal} {Annals of physics}\ }\textbf
		{\bibinfo {volume} {326}},\ \bibinfo {pages} {96} (\bibinfo {year}
		{2011})}\BibitemShut {NoStop}%
	\bibitem [{\citenamefont {Tanimura}(2020)}]{tanimura2020numerically}%
	\BibitemOpen
	\bibfield  {author} {\bibinfo {author} {\bibfnamefont {Y.}~\bibnamefont
			{Tanimura}},\ }\href
	{https://pubs.aip.org/aip/jcp/article/153/2/020901/76291} {\bibfield
		{journal} {\bibinfo  {journal} {The Journal of chemical physics}\ }\textbf
		{\bibinfo {volume} {153}} (\bibinfo {year} {2020})}\BibitemShut {NoStop}%
\end{thebibliography}
\end{document}